\documentclass[12pt,a4paper]{report}
\usepackage[utf8]{inputenc}
\usepackage[T1]{fontenc}
\usepackage{geometry}
\usepackage{titlesec}
\usepackage{array}
\usepackage{lscape}
\usepackage{caption}
\usepackage{multicol}
\usepackage{enumitem}
\usepackage{longtable}
\usepackage{graphicx}
\usepackage{adjustbox}
\usepackage{amsfonts}
\usepackage{amssymb}
\usepackage{tcolorbox}
\usepackage{fontenc}
\usepackage{amsmath}
\usepackage{algorithm}
\usepackage{algorithmic}
\usepackage{pifont}

\usepackage{fancyhdr}
\usepackage{lmodern}
\usepackage{pstricks,pst-node} 
\usepackage{pgf,tikz}
\usetikzlibrary{arrows}
\usepackage[T1]{fontenc}
\makeindex
\usepackage{fullpage}
\usepackage{xcolor}
\usepackage{listings}
\definecolor{mygreen}{rgb}{0,0.6,0}
\definecolor{mydarkpink}{rgb}{0.91,0.39,0.61}

\definecolor{mygreen}{rgb}{0,0.6,0}
\definecolor{mygray}{rgb}{0.5,0.5,0.5}
\definecolor{mymauve}{rgb}{0.58,0,0.82}
\definecolor{myRed}{RGB}{200, 0, 0}

\titleformat{\chapter}[display]
  {\normalfont\fontsize{22}{22}\bfseries}
  {Chapter~\thechapter}
  {0pt}
  {\vspace{5mm}}

\titleformat{\section}
  {\normalfont\fontsize{15}{15}\bfseries}
  {\thesection}
  {1em}
  {}

\titleformat{\subsection}
  {\normalfont\fontsize{13}{13}\bfseries}
  {\thesubsection}
  {1em}
  {}

\usepackage{amsmath}
\usepackage{fontawesome5}
\usepackage{pifont}
\usepackage{newunicodechar}
\usepackage{array}
\newunicodechar{✓}{\ding{51}}
\newunicodechar{✗}{\ding{55}}
\usepackage[nottoc,numbib]{tocbibind}
\usepackage{acro}
\usepackage{hyperref}
\usepackage{appendix}
\usepackage{xurl}
\usepackage{nomencl}
\usepackage[dash]{dashundergaps}
\hypersetup{colorlinks=true,linkcolor=black,filecolor=magenta,urlcolor=black,pdfpagemode=FullScreen}
\makenomenclature
\begin{document}
\begin{titlepage}
\centering
\vspace*{0.7cm}
\rule{\linewidth}{0.3mm}
{\huge \bfseries
SkyEye: When Your Vision Reaches Beyond IAM Boundary Scope in AWS Cloud \\[0.2cm] 
}
\rule{\linewidth}{0.3mm}
\\[0.5cm]
\href{https://github.com/0x7a6b4c/SkyEye}{\faGithubSquare\ https://github.com/0x7a6b4c/SkyEye}\\[0.5cm]
\vspace{7mm}
\noindent
\begin{minipage}{0.7\textwidth}
    \vspace{-3mm}
  \begin{flushleft} \large
    \centering
    \vspace{1.2cm}
    \begin{center}
        \textsc{Minh Hoang} \textbf{Nguyen}\textsuperscript{1}\\
        {\small Université Libre de Bruxelles (ULB)} \\
        {\small \href{mailto:mhoangnguyen.work@gmail.com}{\underline{mhoangnguyen.work@gmail.com}}}\\
        \vspace{10mm}
        \textsc{Anh Minh} \textbf{Ho}\textsuperscript{2}\\
        {\small Technische Universität Darmstadt (TU Darmstadt)} \\
        {\small \href{mailto:anhminhho2409@gmail.com }{\underline{anhminhho2409@gmail.com}}}\\
        \vspace{10mm}
        \textsc{Bao Son} \textbf{To}\textsuperscript{3}\\
        {\small KrisShop, Singapore Airlines} \\
        {\small \href{mailto:tbson2000@gmail.com}{\underline{tbson2000@gmail.com}}}
    \end{center}
    \vspace{6mm}
    \small \noindent\textsuperscript{1}First Author\hspace{5mm}\textsuperscript{2}Second Author\hspace{5mm}\textsuperscript{3}Third Author\\
    \vspace{3mm}
    Minh Hoang Nguyen is the Corresponding Author.
  \end{flushleft}
\end{minipage}
\begin{minipage}{0.4\textwidth}
  \begin{flushright} \large
  \end{flushright}
\end{minipage}\\[1cm]
\vspace*{1.8cm}
{\large \textbf{2025}}\\[0.2cm]
\centering
\begin{tabular}{ll}
\large \textsc{26th June 2025}
\end{tabular}

\end{titlepage}
\pagenumbering{roman}
\tableofcontents
\listoftables
\listoffigures
\makenomenclature
\newpage
\renewcommand{\nomname}{Symbol List}
\nomenclature{\textbf{API}}{Application Programming Interface}
\nomenclature{\textbf{SDK}}{Software Development Kit}
\nomenclature{\textbf{IAM}}{Identity and Access Management}
\nomenclature{\textbf{AWS}}{Amazon Web Services}
\nomenclature{\textbf{IP}}{Internet Protocol}
\nomenclature{\textbf{EC2}}{Elastic Compute Cloud}
\nomenclature{\textbf{STS}}{Security Token Service}
\nomenclature{\textbf{ARN}}{Amazon Resource Name}
\nomenclature{\textbf{SQS}}{Simple Queue Service}
\nomenclature{\textbf{CSPM}}{Cloud Security Posture Management}
\nomenclature{\textbf{CIEM}}{Cloud Infrastructure Entitlement Management}
\nomenclature{\textbf{MFA}}{Multi-Factor Authentication}
\nomenclature{\textbf{SCP}}{Service Control Policy}
\nomenclature{\textbf{SSO}}{Single-Sign On}
\nomenclature{\textbf{S3}}{Simple Storage Service}
\nomenclature{\textbf{SAML}}{Security Assertion Markup Language}
\nomenclature{\textbf{ABAC}}{Attribute-Based Access Control}
\nomenclature{\textbf{JSON}}{JavaScript Object Notation}
\nomenclature{\textbf{CLI}}{Command Line Interface}
\nomenclature{\textbf{OU}}{Organizational Unit}
\nomenclature{\textbf{U2F}}{Universal 2nd Factor}
\nomenclature{\textbf{PCI-DSS}}{Payment Card Industry Data Security Standard}
\nomenclature{\textbf{FedRAMP}}{Federal Risk and Authorization Management Program}

\printnomenclature
\newpage
\pagenumbering{arabic}
\addcontentsline{toc}{chapter}{Abstract}
\newpage
\chapter*{Abstract}
In recent years, cloud security has emerged as a primary concern for enterprises due to the increasing trend of migrating internal infrastructure and applications to cloud environments. This shift is driven by the desire to reduce the high costs and maintenance fees associated with traditional on-premise infrastructure. By leveraging cloud capacities such as high availability and scalability, companies can achieve greater operational efficiency and flexibility. However, this migration also introduces new security challenges. Ensuring the protection of sensitive data, maintaining compliance with regulatory requirements, and mitigating the risks of cyber threats are critical issues that must be addressed. Identity and Access Management (IAM) constitutes the critical security backbone of most cloud deployments, particularly within AWS environments. As organizations adopt AWS to scale applications and store data, the need for a thorough, methodical, and precise enumeration of IAM configurations grows exponentially. Enumeration refers to the systematic mapping and interrogation of identities, permissions, and resource authorizations with the objective of gaining situational awareness. By understanding the interplay between users, groups, and their myriads of policies, whether inline or attached managed policies, security professionals need to enumerate and identify misconfigurations, reduce the risk of unauthorized privilege escalation, and maintain robust compliance postures. This paper will present SkyEye, a cooperative multi-principal IAM enumeration framework, which comprises cutting-edge enumeration models in supporting complete situational awareness regarding the IAMs of provided AWS credentials, crossing the boundary of principal-specific IAM entitlement vision to reveal the complete visionary while insufficient authorization is the main challenge.\\\\
\textbf{Keywords}: Amazon Web Services (AWS), Identity and Access Management (IAM), Privilege Escalation, Enumeration, Reconnaissance, Offensive Security, Cloud Security, Penetration Testing.
\newpage
\chapter{Introduction}
\section{Background}
The rapid migration of critical corporate infrastructure to cloud computing platforms, most notably Amazon Web Services (AWS) has redefined how organizations control and secure access to resources. Unlike traditional on‑premises setups, where firewalls and network segmentation provided primary security perimeters, cloud environments adopt an identity-centric model: Identity and Access Management (IAM) is the central gatekeeper. Every cloud operation spinning up an EC2 instance, accessing an S3 bucket, or managing RDS databases depends on IAM to determine who is authorized to do what, when, and how.

At its core, IAM facilitates secure delegation of permissions across individual users, service roles, and automated scripts. However, this flexibility introduces complexity. Policies can easily become overly permissive, especially in environments that use multi-account architectures, federated identity providers, or automated resource provisioning pipelines. These intricate relationships between policies, trust, and temporary credentials, forming challenging attack surfaces for adversaries.

\subsection{Real-World Cloud Security Incidents}
Cloud IAM misconfigurations have directly contributed to some of the most significant security breaches in recent history \cite{spark_2023}:

\begin{itemize}
    \item \textbf{Capital One Breach (2019)}: A misconfigured web application firewall allowed unauthorized access to AWS credentials tied to an EC2 metadata role. The attacker used those credentials to retrieve sensitive data from approximately 106 million customer records. This breach prompted a regulatory fine of around 80 million dollars \cite{capital_one_2022} \cite{wired_2019}.
    \item \textbf{LAPSUS\$ Attacks (2022)}: The LAPSUS\$ acker group exploited stolen or weakly protected cloud credentials, which enabled them to infiltrate major tech firms including Okta, Microsoft, and Nvidia. Their campaigns highlighted the risks introduced by token theft, phishing, and multi-factor authentication bypasses \cite{lapsus_2022} \cite{bbc_2022}.
\end{itemize}

These incidents highlight that while cloud providers maintain physical security, the responsibility for reinforced IAM configuration lies with customers, which is a gap that sophisticated attackers can readily exploit.

\subsection{The Complexity of Modern IAM Environments}
Multiple factors contribute to the inherent complexity of IAM in contemporary cloud environments. Enterprises frequently partition workloads across several AWS accounts, utilizing AWS Organizations and cross-account roles to manage access at scale. This segmentation, while beneficial for administrative separation and security, introduces intricate trust relationships that are challenging to audit and maintain. The integration of federated identities which are often via external identity providers, and the widespread use of AWS Security Token Service (STS) for temporary credentials further expand the potential attack surface.\\

A prevalent issue in such environments is the over-permissioning of roles. This may arise from reliance on AWS managed policies, which are intentionally broad to accommodate diverse use cases, or from misapplication of permission boundaries and conditions. As a consequence, principals may inadvertently receive privileges exceeding their operational requirements. Dynamic trust relationships, such as those formed through chained role assumptions (for example, an IAM role used by AWS Lambda invoking services that subsequently assume additional roles), complicate the task of privilege analysis. These scenarios often evade detection by static policy analysis tools, leaving organizations exposed to subtle privilege escalation pathways.\\

Recent studies have demonstrated that IAM misconfiguration is not an isolated phenomenon, but rather a systemic issue affecting a significant proportion of cloud deployments. Systematically exploitable vulnerabilities, often arising from the interplay of multiple policies and trust relationships, present persistent challenges to effective cloud security management.

\subsection{Detection and Remediation Gaps}
Existing governance solutions, including Cloud Security Posture Management (CSPM) platforms and Cloud Infrastructure Entitlement Management (CIEM) tools, primarily focus on the identification of static misconfigurations. These tools are effective at flagging explicit issues such as overly permissive IAM policies, lack of multi-factor authentication (MFA), or the presence of unused credentials. However, they frequently lack the capability to model and emulate contextual multi-step privilege escalation chains based on realistic adversary behaviors, which may only become apparent when detection mechanisms for indicators of compromise are evaluated in a live operational context.\\

The evaluation of IAM policies in AWS is deliberately multifaceted, incorporating the following layers:
\begin{itemize}
    \item Identity-based policies
    \item Resource-based policies
    \item Service Control Policies (SCPs) in AWS Organizations \cite{aws_security_blog_2023}
    \item Permission boundaries
    \item Session-based credentials \cite{cornel_model_checking_2023}
\end{itemize}
Without a holistic approach that considers these layers in aggregate, it is infeasible to accurately determine the effective permissions granted to any principal. Tools developed for static IAM misconfiguration detection, such as those by \cite{detecting_anomalous_misconfig_2022}, have proven valuable for identifying latent policy errors, yet they remain limited in their ability to model dynamic attack paths incorporated by multiple principals that arise during runtime.

\subsection{Why IAM Enumeration Matters?}
Comprehensive risk assessment of IAM configurations requires more than static policy review. From a red teaming perspective, it necessitates the active enumeration of permissions to ascertain the practical capabilities of identities under realistic conditions. This approach supports:

\begin{itemize}
\item \textbf{Proactive defense:} Identifying and remediating privilege escalation vectors prior to exploitation.
\item \textbf{Realistic threat simulation:} Enhancing red team exercises and penetration testing with actual cloud session contexts.
\item \textbf{Regulatory compliance:} Ensuring that granted permissions align with least-privilege requirements as enforced in production.
\end{itemize}

Recent research, such as “Effective IAM Exploitation Cascade Detection” \cite{usenix_2023}, corroborates the feasibility and value of identifying multi-step escalation paths in complex IAM configurations.

\subsection{Academic and Industrial Perspectives}
Recent academic studies support the viability and importance of live IAM context analysis:
\begin{itemize}
    \item “Detecting Anomalous Misconfigurations in AWS IAM” \cite{detecting_anomalous_misconfig_2022} demonstrates how specialized detectors can highlight risky IAM patterns live .
    \item “Efficient IAM Greybox Penetration Testing” \cite{cornel_efficient_iam_2023} proposes optimized IAM model querying techniques for identifying privilege escalation with minimal footprint
\end{itemize}
These contributions underscore both the demand for, and the viability of, advanced enumeration frameworks capable of detecting escalation risk in real time.\\

IAM misconfiguration is not merely an abstract or academic concern; it represents a pervasive and actionable threat, as demonstrated by high-profile breaches such as those involving Capital One and LAPSUS\$. Although static analysis and compliance validation tools are essential for maintaining a robust security posture, they fall short in addressing the dynamic, context-dependent nature of privileges in cloud environments. These tools often lack the deep contextual awareness and attack modeling necessary to detect complex privilege escalation paths. \\

Academic studies and experimental frameworks demonstrate that live, multi-step IAM enumeration are not only achievable but essential for securing cloud environments. This identified gap in existing tooling underscores the need for a new generation of offensive-capable IAM enumeration framework. Such frameworks are designed to proactively uncover and remediate risky IAM configurations while simultaneously enhancing detection and response mechanisms, thus strengthening cloud security posture before a breach can occur.\\

\section{Problem Statement - SkyEye's Motivation}
As enterprises increasingly migrate critical workloads to public cloud platforms like AWS, securing identity and access management (IAM) and consolidating cloud breach detection and response mechanisms have become a defining challenge. Despite AWS providing powerful IAM capabilities, the shared responsibility model places the burden of reinforced configuration on customers. This has resulted in persistent and systemic gaps between policy definitions and actual behavior, exposing organizations to substantial risks.

\subsection{IAM Misconfigurations}
IAM misconfigurations are widely recognized as one of the leading causes of cloud security breaches. According to IBM’s Cost of a Data Breach Report 2023, cloud-based breaches incurred the highest costs among incident types, with a substantial portion attributed to misconfigured cloud identity permissions \cite{ibm_cost_data_breach_2023}.
Misconfigurations often stem from:
\begin{itemize}
    \item Overly permissive IAM roles left in production for convenience or legacy reasons.
    \item Complex cloud environment structures with interdependent trust relationships.
    \item Hardcoded credentials accidentally exposed in source repositories.
    \item Federated identity bridges (e.g., SSO with corporate identity systems) introducing unforeseen trust boundaries.
\end{itemize}

\subsection{The Visibility Gap}
Most cloud security tooling and frameworks \cite{Pacu2021, CloudPEASS, CloudFox, ScoutSuite, Cloudsplaining, enumerate-iam} focus on exposing the privileges tied to their harvested AWS credentials by relying solely on the independent vision of those credentials, which is capable of enumerating its authorization only if all the required privileges has been assigned to an independent AWS user principal. While useful, these tools often:
\begin{itemize}
    \item Fail to discover a complete IAM vision context that produces significant false negatives that hiding the critical attack vectors or privilege escalation
    \item Ignore temporal aspects, such as temporary credentials generated by assumable roles with powerful privileges
    \item Ignore the transitive role assumption chain that providing the door to generate temporary credentials for indirectly assumable roles
    \item Lack understanding of context, particularly across multiple principals in a cloud environment
\end{itemize}
A particularly persistent aspect of the visibility gap stems from the limitations of single-principal IAM enumeration approaches. In the initial reconnaissance phase, penetration testers or defenders may harvest multiple AWS credentials. However, existing enumeration tools generally operate on a per-principal basis, performing either single-principal or independent multi-principal enumerations without coordination between principals. This traditional strategy inevitably leads to significant false negatives: specific permissions or escalation paths that are only discoverable when multiple principals cooperate remain hidden. As a result, neither adversaries nor defenders can attain a holistic understanding of the true IAM exposure, since no single principal possesses the complete entitlement vision.\\

In the recent ENISA Threat Landscape 2024, it noted that IAM-related misconfigurations and privilege escalation paths remain underexplored by traditional cloud security solutions, creating persistent visibility gaps for both red-teamers and security defenders \cite{enisa_threat_landscape_2024}.

\subsection{Complexity in Practice}
The difficulty lies not just in writing secure IAM policies, but in understanding how they interact at runtime:
\begin{itemize}
    \item Modern architectures like serverless computing introduce ephemeral roles.
    \item Resource-based policies (e.g., S3 bucket policies) introduce additional policy evaluation layers beyond the principal’s direct IAM role.
    \item AssumeRole chains allow privilege escalation if misconfigured or misunderstood.
\end{itemize}
The problem is compounded by single-principal enumeration models, which cannot fully reconstruct the complex web of trust relationships and transitive permissions that emerge in dynamic environments. For example, privilege escalation vectors may exist where no single principal possesses all required permissions, but a combination of principals does, yet single-principal enumeration is blind to such scenarios. Similarly, transitive role assumptions and cross-account trust chains often remain undetected, underestimating the effective attack surface.\\

With the current static IAM validation tools and principal-isolated enumeration, security teams could not achieve a complete picture of IAM configuration. A context-aware, adversary-perspective, cooperative multi-principal enumeration framework is necessary to address the following:
\begin{itemize}
    \item Cooperate multi principals in enumerating the complete picture of IAM configuration for targeted AWS credentials, eliminating the disadvantages of principal-specific IAM enumeration which is restricted by standalone IAM entitlement vision.
    \item Map out privilege escalation pathways across services and accounts, including those requiring cooperation between principals.
    \item Enable proactive threat hunting for identity abuse scenarios that exploit combinations of entitlements.
    \item Support red teaming with realistic attack path modeling, encompassing cross-principal and transitive privilege chains.
    \item Validate principle-of-least-privilege assertions under adversarial conditions.
\end{itemize}
Without such an approach, cloud security defenders or red-teamers operate with incomplete visibility in highly interconnected, dynamic environments. Furthermore, the prevalence and impact of IAM misconfigurations in recent high-profile incidents underscore the inadequacy of static analysis and single-principal enumeration alone. Dynamic, context-sensitive, and cooperative IAM enumeration tools are indispensable for accurately assessing the real-world implications of granted permissions.

\subsection{Research objectives \& Contributions}
To resolve these limitations, this thesis introduces SkyEye, a novel cooperative multi-principal IAM enumeration framework. The original idea of SkyEye came from the difficulty that occurs with the single-principal IAM enumeration approach: when multiple AWS credentials are obtained, traditional enumeration cannot reveal the aggregate or cooperative entitlements. SkyEye’s Cross-Principal IAM Enumeration Model (CPIEM) enables advanced enumeration by simultaneously involving and coordinating multiple principals, thereby exposing the complete IAM visibility of each principal in context. This cooperation allows:
\begin{itemize}
    \item Systematic discovery of cross-principal privilege escalation chains.
    \item Exhaustive mapping of resource access paths that would be otherwise missed.
    \item Realistic modeling of adversarial actions leveraging all available credentials.
\end{itemize}

Additionally, SkyEye introduces the Transitive Cross-Role Enumeration Model (TCREM), which systematically tracks roles that a principal can assume directly or indirectly, recursively following AssumeRole chains. This model enables dynamic expansion of the enumeration scope to include all in-scope IAM roles, thus complementing and enriching the IAM vision context of multiple principals.
\newpage
\chapter{Background Knowledge}
\section{What is AWS Identity and Access Management?}
AWS Identity and Access Management (IAM) is a comprehensive service provided by Amazon Web Services (AWS) that allows us to manage access to AWS services and resources securely \cite{aws_iam_user_guide}. IAM allows you to control who is authenticated (signed in) and authorized (has permissions) to use resources. With IAM, we can create and manage AWS users and groups, and use permissions to allow and deny their access to AWS resources.

\subsection{History and Evolution of AWS IAM}
AWS Identity and Access Management (IAM) emerged in the early 2010s as a response to the growing need for robust security controls in multi-tenant cloud environments, where the ability to delegate and audit privileges became paramount \cite{aws_iam_user_guide}. Although its official release date is commonly referenced as 2011, the underlying principle of adopting least-privilege, separation of duties, and tighter governance had been taking shape even prior to public announcement. Initially, AWS users were restricted to a single root account with unrestricted privileges, posing potential security and operational risks. With IAM, AWS introduced the concept of distinct, fine-grained identities (users, groups, and roles) to address these concerns. This shift allowed organizations to create multiple user identities, each bound by well-defined permissions and governed by policies, effectively implementing the principle of least privilege in AWS environments \cite{aws_iam_user_guide} \cite{aws_iam_federation}.\\

Over subsequent years, AWS has iteratively enhanced IAM to account for expanding customer use cases and complex regulatory requirements \cite{aws_organizations_iam}. Notable milestones include the introduction of federated identities (enabling single sign-on with identity providers such as SAML 2.0 and OpenID Connect), managed policies (providing reusable sets of permissions), and grammatical improvements to policy language for clearer, more auditable rule sets \cite{aws_iam_federation} \cite{aws_iam_policies}. Enhancements such as service-linked roles enabled AWS services to interact on behalf of a user with meticulously scoped permissions, while permission boundaries made it practical to enforce upper limits on user privileges \cite{aws_iam_policies}. Attribute-based access control (ABAC) extended IAM’s capability to evaluate complex attributes (e.g., tags, organizational units), supporting more dynamic, context-aware authorization logic in large-scale deployments. Collectively, these innovations highlight AWS’s trajectory planning toward increasingly granular, automated, and compliance-driven access control frameworks \cite{aws_policy_evaluation_logic}.

\subsection{Core Concepts and Terminology}
AWS IAM distinguishes between authentication (verifying the unique identity of users or services) and authorization (determining the set of actions permitted to those entities) \cite{aws_iam_user_guide}. At its core, IAM leverages several foundational elements:

\begin{itemize}
\item \textbf{Users} represent tangible (human) or programmatic identities that require authenticated access to AWS services. Each user is assigned unique credentials: passwords or access keys, and is bound by the principle of least privilege to reduce the exposed threat surface \cite{aws_iam_users}.
\item \textbf{Groups} serve as logical collections of users, enabling streamlined administration by centralizing permission sets which is often in managed policies, that can be applied or removed in a single action \cite{aws_iam_groups}.
\item \textbf{Roles} constitute assumable identities, allowing AWS services, external users, or other AWS accounts to inherit permissions temporarily. Roles are integral for cross-account access, federated login, and machine-driven tasks, as they mitigate permanent credential storage and emphasize short-lived credentials \cite{aws_iam_roles}.
\end{itemize}

These identity constructs function through policies with JSON-based documents specifying allow or deny rules for individual actions and resources \cite{aws_iam_policies}. Policies may be customer-managed, AWS-managed, or inline, and they subscribe to a default-deny strategy, adhering to explicit permission grants or denials under AWS’s policy evaluation process \cite{aws_policy_evaluation_logic}. The interplay of policies, principals, and resources underscores IAM’s overarching security design, which aligns to recognized best practices:

\begin{itemize}
\item \textbf{Principals:} Authenticated entities (e.g., IAM user, IAM role, federated user) permitted to make requests.
\item \textbf{Resources:} AWS services or objects (e.g., Amazon S3, Amazon EC2, AWS Lambda) targeted by access requests.
Actions: Specific operations authorized or disallowed (e.g., s3:ListBucket, ec2:DescribeInstances).
\item \textbf{Conditions:} Additional contextual checks (often based on IP, time of day, or user attributes) that refine authorization rules.
\item \textbf{Temporary Security Credentials:} Short-lived credentials tied to roles or federated access, mitigating the risk of long-term credential compromise \cite{aws_iam_federation}.
\end{itemize}

IAM’s continued evolution also integrates advanced governance layers, such as AWS Organizations for multi-account management, enabling consistent enforcement of policies across a fleet of accounts \cite{aws_organizations_iam}. By coupling IAM’s role-based constructs with condition-based and attribute-based controls, organizations can better adhere to regulatory obligations, maintain separation of duties, and instantiate defense-in-depth models \cite{aws_organizations_iam} \cite{aws_policy_evaluation_logic}. This holistic design philosophy positions IAM as a versatile, extensible foundation for cloud security, exemplifying AWS’s strategic emphasis on continuous refinement of key identity and access mechanisms in alignment with modern enterprise demands.

\section{Core IAM Entities}
AWS Identity and Access Management (IAM) comprises several principal entities essential for controlling authentication and authorization flows within Amazon Web Services (AWS). These entities include users, groups, roles, and the various policy configurations that govern operational privileges. When orchestrated appropriately, these constructs uphold security requirements such as least privilege, defense in depth, and regulatory compliance obligations. The subsequent discussion focuses on the multitude of policy mechanisms provided by IAM, illustrating how each policy type imposes specific constraints and thereby contributes to holistic access governance.
\subsection{IAM Users}
An IAM user is an entity that we create in the AWS environment to represent the person or application that interacts with AWS resources. Users can log into the AWS Management Console, interact with AWS services through the AWS CLI, and use AWS APIs. Each IAM user is associated with a unique set of credentials and permissions that illustrate what actions the user can perform. Users are often created for individual employees or applications that require direct access to AWS resources \cite{aws_iam_users}.
\subsection{IAM Groups}
IAM groups are the collections of IAM users, which are often used to centrally manage the privileges of a group of users. We can use groups to simplify the management of permissions for multiple users \cite{aws_iam_groups}. Instead of assigning permissions to each user individually, we can assign permissions to a group, and all users in that group will inherit those permissions. This makes it easier to manage permissions for users with similar access needs, such as teams or departments within an organization.
\subsection{IAM Roles}
An IAM role is an IAM identity that we can create in our AWS account that has specific permissions. The roles are intended to be assumed by trusted entities, such as IAM users, applications, or AWS services \cite{aws_iam_roles}. Unlike IAM users, roles do not have long-term credentials (passwords or access keys) associated with them. Instead, when we request to assume a role, we are provided with temporary security credentials. Roles are particularly useful for granting access to resources across different AWS accounts or for allowing AWS services to interact with each other on your behalf.
\subsection{IAM Policies}
IAM policies are formal statements, expressed in JSON, that define granular permissions associated with particular IAM entities \cite{iam_json_policy_reference}. By specifying the conditions under which certain operations are allowed or denied, policies enable secure management of multifaceted AWS environments. This policy-driven model is anchored by a default-deny approach, wherein all requests are implicitly denied unless explicitly allowed. The structured schema of IAM policies further simplifies audits and compliance reporting by providing a declarative representation of permissible actions.
\begin{itemize}
\item \textbf{Embedded Inline Policies}\\
Inline policies in AWS Identity and Access Management (IAM) are policies that are directly embedded within a specific IAM user, group, or role. These policies maintain a strict one-to-one relationship with the entity they are attached to, meaning they are specifically tailored to the needs of that single user, group, or role. Inline policies are particularly useful when we need to define unique permissions for a specific entity and not intended to be shared with others \cite{aws_managed_policies_and_inline_policies}. For instance, if a particular IAM user requires special permissions that no other user needs, an inline policy is a suitable choice. When the user, group, or role to which an inline policy is attached is deleted, the inline policy is also deleted. This tight coupling ensures that the unique permissions granted by the inline policy are removed when the entity is no longer needed, thus enhancing security by minimizing the risk of orphaned policies that could be misused.
\item \textbf{Attached Managed Policies}\\
Attached Managed policies in AWS IAM are reusable policy documents that can be attached to multiple IAM users, groups, or roles \cite{aws_managed_policies_and_inline_policies}. This policy provides several advantages, including ease of management, as any updates to a managed policy automatically propagate to all entities that the policy is attached to. This ensures consistency and simplifies the process of updating permissions across multiple users, groups, or roles. There are two types of managed policies: AWS-managed policies and customer-managed policies.
\begin{itemize}
\item \textbf{AWS-managed policies:} These policies are pre-defined by AWS and designed to provide permissions for common use cases, making it easier for administrators to grant necessary permissions without writing policies from scratch \cite{aws_managed_policies_and_inline_policies}.
\item \textbf{Customer-managed policies:} on the other hand, these policies are created and maintained by the AWS account administrators. These policies offer greater flexibility and customization, allowing organizations to define specific permissions tailored to their unique requirements \cite{aws_managed_policies_and_inline_policies}.
\item \textbf{Customer-managed policy versioning:} Customer-managed policy versioning is a feature that allows administrators to manage and maintain different versions of their custom IAM policies \cite{versioning_iam_policy}. When we create a customer managed policy, AWS allows us to update and refine it over time without losing the previous versions. Each time we make a change to a policy, a new version is created and stored, with AWS supporting up to five versions per managed policy, including the current version. This capability is particularly useful for auditing, compliance, and troubleshooting, as it provides a historical record of policy changes and ensures that administrators can track how permissions have evolved over time.
\end{itemize}
\item \textbf{Permissions Boundaries}\\
Permissions boundaries serve as a secondary layer of containment on top of standard policies, restricting the maximum privileges an IAM entity can attain, regardless of other attached policies \cite{permissions_boundaries_iam_entities}. This mechanism operates in conjunction with the principle of least privilege, ensuring that no single user or role can escalate its privileges beyond what the boundary permits. Permissions boundaries thereby reinforce secure delegation models, enabling delegated administrators to define or manage policies without the risk of granting excessive permissions to themselves or others.
\item \textbf{Service Control Policies (SCPs)}\\
Service Control Policies (SCPs) are enforced at the organization or organizational unit level through AWS Organizations \cite{aws_service_control_policies}. Unlike standard IAM policies, SCPs do not grant permissions. Instead, they act as overarching filters that define allowable operations, effectively constraining the maximum effective permissions within an organizational hierarchy. By applying SCPs, enterprises can institute restrictive baselines that align with top-level compliance mandates, ensuring that individual accounts cannot override organizational security boundaries.
\item \textbf{Resource-Based Policies}\\
Resource-based policies are embedded within specific AWS resources (for example, Amazon S3 buckets or Amazon SNS topics), permitting cross-account access or fine-tuned sharing of those resources \cite{iam_json_policy_reference}. These policies define who can perform which actions on the resource and under what conditions. Resource-based policies differ from identity-based policies (inline or managed) by stationing the permission structure alongside the resource in question rather than tying it to a principal. This alignment is advantageous in cross-account scenarios, simplifying the secure granting of resource access to external entities.
\item \textbf{Session Policies}\\
Session policies are temporary policies passed when principal entities assume roles via the AWS Security Token Service (STS) \cite{aws_iam_policies}. These transient, context-specific policies layer atop the existing identity-based permissions, further limiting the maximum permissions that a session can acquire. Session policies enable use cases such as short-lived privilege escalations for break-glass scenarios or environment-specific restrictions during continuous integration and deployment processes. By employing session policies, organizations can improve real-time governance, restrict credential lifetimes, and implement dynamic access control constructs.
\end{itemize}

\section{AWS IAM Organizational Structure and Scoping}
AWS provides a hierarchical organizational model encompassing accounts, organizational units, and groups of resources to facilitate structured access control and multi-account administration. This overarching model is primarily managed through AWS Organizations, enabling consolidated billing, centralized governance, and robust identity and access management controls. Such layered structures are particularly relevant for large enterprises or government agencies requiring stringent isolation of workloads, cost visibility, and compliance enforcement across multiple AWS accounts.

\subsection{AWS Accounts and Organizational Units (OUs)}
AWS accounts serve as fundamental security and billing boundaries, delineating resource ownership and responsibility while enabling precise cost tracking \cite{aws_accounts}. Each account contains its own collection of services (for instance, Amazon EC2, Amazon S3) and is subject to the identity, access, and networking configurations defined therein. Placing workloads in separate accounts bolsters the defense-in-depth model by preventing unauthorized lateral movement across environments. Furthermore, the use of separate accounts aids in isolating development, staging, and production environments, simplifying regulatory compliance and incident containment.\\

Organizational Units (OUs) are logical containers that group AWS accounts under a hierarchical structure within AWS Organizations \cite{aws_organizations_ous}. By segmenting accounts into OUs, administrators can apply Service Control Policies (SCPs) to enforce baseline security measures at the organizational or unit level. This approach streamlines policy management, ensuring consistent governance mandates (such as encryption requirements or restricted AWS regions) across multiple accounts. OUs also facilitate simpler access auditing, as SCPs logically cascade to all member accounts, leaving minimal room for unauthorized deviation from organizational policy.

\subsection{AWS Organizations and Cross-Account Access}

AWS Organizations provides a centralized console to manage multiple AWS accounts under a single master (also referred to as “management”) account, thus unifying billing and security oversight \cite{aws_organizations_intro}. Through AWS Organizations, administrators can create new accounts programmatically, migrate existing ones, and apply organization-wide policy constraints via SCPs. These capabilities reduce operational overhead by promoting consistency in identity configuration, logging, governance, and cost management across accounts.\\

Cross-account access capabilities within AWS Organizations are established through roles, trust policies, and resource-based permissions \cite{aws_cross_account_roles}. By configuring an IAM role in one account with a trust policy that references a principal entity in another account, AWS administrators can enable secure resource sharing without duplicating user credentials. This trust-based mechanism underscores the principle of least privilege, as cross-account roles typically grant only the minimum necessary level of authority. Enterprises frequently utilize such cross-account constructs for shared services (e.g., logging, monitoring) or for delegated administration of centralized resources, reinforcing the advantages of both segregation of duties and cost accountability.

\subsection{Delegated Administration and Trust Relationships}

Delegated administration refers to the distribution of administrative privileges to specific subgroups or accounts within an organization, so that these subgroups can manage certain AWS services or resources without requiring access to the root account's credentials \cite{aws_delegated_admin}. This model bolsters resiliency and security by compartmentalizing privileges among trusted administrators, thereby reducing the blast radius of a potential breach or misconfiguration. In practice, delegated administration is implemented through IAM roles and corresponding trust relationships that define which principals can assume an administrative role.\\

Trust relationships are integral to secure cross-account interactions, as they define the principal entities permitted to assume an IAM role in a target account \cite{aws_cross_account_roles}. Administrators construct trust policies, typically in JSON, specifying the conditions under which access is granted. This includes referencing the source account or user, the allowed role to be assumed, and optional condition-based controls such as Multi-Factor Authentication (MFA). By carefully crafting trust relationships, organizations uphold security best practices, preserving the integrity of cross-account workflows while aligning with compliance imperatives stemming from frameworks such as ISO 27001, PCI DSS, and FedRAMP.

\section{IAM Policy Language and Evaluation}

AWS IAM policies regulate access decisions through JSON-based documents that adhere to a specific schema, defining the permissions granted or denied to authenticated and authorized entities. Unlike traditional access-control models, IAM policies embrace fine-grained permissions, incorporate explicit deny mechanics, and enable context-aware constraints through conditions. This multifaceted approach reflects AWS’s commitment to least privilege, ensuring that cloud environments remain both secure and adaptable to evolving organizational needs.

\subsection{Policy Document Structure (JSON)}

An IAM policy document comprises a series of JSON statements, each of which includes relevant attributes for governing access rules \cite{iam_json_policy_reference}. Typical attributes in the JSON schema include “Version,” which declares the policy language syntax, and “Statement,” an array of objects that encapsulate specific permission directives. In many cases, policy documents also include “Id” elements to facilitate policy auditing or reference mapping in large-scale deployments. For ease of maintenance, policy authors often rely on AWS-managed policy templates or reuse common statements in customer-managed policies. This uniform structure ensures that security teams and automated tools can parse, validate, and enforce policies consistently across diverse AWS services and accounts.

\subsection{Policy Elements (Effect, Action, Resource, Condition, Principal)}

At the core of each statement within an IAM policy document are five key elements: Effect, Action, Resource, Condition, and Principal, each of which specifies a distinct facet of permission logic \cite{aws_iam_policy_elements}. The “Effect” element declares whether the statement grants (“Allow”) or denies (“Deny”) the specified permissions. “Action” enumerates the API calls or operations governed by the rule (for example, “s3:GetObject” or “ec2:StartInstances”). The “Resource” field identifies the AWS assets or services the effect applies to, frequently using Amazon Resource Names (ARNs) to pinpoint targets. In more advanced policies, the “Condition” element refines permissions based on contextual keys, including IP address ranges, dates, or user attributes. Finally, the “Principal” element indicates the user, role, or entity that is subject to the policy, which can include cross-account or external Federated Identities where trust relationships are in effect.

\subsection{Policy Evaluation Logic (Explicit Deny, Allow, Implicit Deny)}

IAM policy evaluation follows a sequential mechanism where AWS first applies an implicit deny to all unreferenced actions and resources, effectively defaulting to “no access” \cite{aws_policy_evaluation_logic}. If a policy statement explicitly denies an action, that instruction supersedes any opposing “allow” statement. This explicit deny principle is especially critical for implementing overarching security guardrails, ensuring that certain operations remain inaccessible even if inadvertently allowed in a subordinate policy. The final step in policy evaluation confirms whether the request is allowed by at least one relevant policy statement; if not, the implicit deny persists. Thus, a combination of explicit deny, allow statements, and a pervasive default-deny posture fortifies the system against misconfigurations, ensuring that access privileges remain carefully controlled.

\subsection{Condition Keys and Advanced Policy Constructs}

Condition keys augment the granularity of IAM by enabling context-sensitive permission decisions, often mandated by strict compliance or multi-tenant architectures \cite{aws_iam_condition_logic}. The “Condition” element can reference built-in AWS keys such as aws:SourceIp, aws:CurrentTime, or aws:SecureTransport, or custom keys defined through AWS services. An illustrative example materializes when restricting AWS Management Console logins to a specific IP range or requiring Multi-Factor Authentication (MFA) for critical API operations. Beyond individual conditions, policy authors may combine multiple keys using logical operators, increasing the precision of authorization models.\\

Advanced constructs also include attribute-based access control (ABAC), which leverages resource tags or user attributes to assign permissions dynamically \cite{aws_abac}. This pattern significantly reduces administrative complexity, as a well-defined tagging schema can replace numerous static policies. Coupled with other AWS innovations, such as prefixes and policy templates, these advanced condition keys and policy features ensure that organizations maintain both the flexibility and rigor required in modern, large-scale cloud security architectures.

\section{Authentication and Authorization in AWS}

Authentication and authorization in AWS constitute the primary pillars of Identity and Access Management (IAM). By requiring users and services to establish their identities and limiting their permissions through IAM policies, AWS ensures robust protection of cloud assets. IAM implements various authentication methods, including the AWS Management Console, Command Line Interface (CLI), Software Development Kits (SDKs), and direct Application Programming Interface (API) calls, all governed by the principle of least privilege. These authentication mechanisms affirm user identity, after which IAM policies determine the permissible scope of operations on AWS services and resources, thus creating a foundational layer for securing cloud infrastructures conforming to industry best practices.

\subsection{Authentication Mechanisms (Console, CLI, SDK, API)}

AWS supports multiple authentication avenues to accommodate diverse operational scenarios and security postures. The AWS Management Console, a web-based graphical user interface, is often favored for interactive tasks such as configuring services, previewing logs, or performing administrative functions. For developers and DevOps teams, the AWS CLI and SDKs offer programmatic access to AWS resources, enabling script-based or application-driven administration of services like Amazon S3, Amazon EC2, and Amazon RDS \cite{aws_cli}. Interaction at the API level provides direct access to AWS infrastructure via HTTP requests, affording fine-grained control in automated and custom integration workflows \cite{aws_apis}. Each of these authentication methods relies on credentials, either long-lived or temporary, to verify the user identity and link ensuing API calls to the corresponding IAM policies.

\subsection{Temporary Security Credentials (STS, AssumeRole, Federation)}

Temporary security credentials mitigate some of the risks posed by long-lived credentials by defining a constrained lifespan for access tokens, thus reducing their exposure window \cite{aws_temp_security_credentials}. AWS Security Token Service (STS) underpins this concept by issuing time-limited credentials upon request, enabling entities to perform only the actions allowed under the associated IAM policies. Common STS scenarios include AssumeRole, which facilitates cross-account access and delegating privileges to AWS services without exposing sensitive credentials \cite{aws_assume_role_methods}. Additionally, IAM supports identity federation, granting temporary AWS access to users authenticating through external identity providers, such as Active Directory Federation Services or SAML 2.0-based solutions \cite{aws_identity_providers_federation}. This federated model allows organizations to preserve existing user directories and authentication workflows while enforcing AWS policies and restricting session duration in alignment with security best practices.

\subsection{Multi-Factor Authentication (MFA)}

Multi-Factor Authentication (MFA) offers an elevated layer of security by requiring users to present an additional factor beyond their password or access key \cite{aws_mfa_iam}. IAM supports multiple forms of MFA, including virtual MFA applications (e.g., Google Authenticator), hardware MFA tokens, and Universal 2nd Factor (U2F) devices \cite{aws_mfa_types_identity_center}. Administrators can mandate MFA at sign-in or at the execution of sensitive API operations, tightening defenses against compromised passwords. Notably, combining MFA with conditional IAM policies and short-lived credentials helps enforce strict access requirements while maintaining operational flexibility. Numerous regulatory frameworks, such as PCI DSS and FedRAMP, endorse MFA as a baseline criterion for safeguarding cloud environments, making MFA adoption a central component of robust cloud security strategies.
\newpage
\chapter{Related Works - Prior-Art Models and Frameworks}
\section{Introduction}
This chapter will begin with an introduction to recent research, models and frameworks, tools on cloud security, with a particular focus on cloud IAM enumeration and cloud privilege escalation. We will first examine framework and tools to gain a deep understanding about its capabilities, methodologies and core logic, as well as its current drawbacks and gap. Next, we will further analyze its limitation to improve by proposing and developing a new framework that integrates our cutting-edge models.

\section{Tools/Frameworks Analysis}

\begin{landscape}
\begingroup
\scriptsize
\begin{longtable}[htbp]{|>{\raggedright\arraybackslash}p{4.5cm}|>{\raggedright\arraybackslash}p{5cm}|>{\raggedright\arraybackslash}p{5.5cm}|>{\raggedright\arraybackslash}p{6.5cm}|}
\hline
\textbf{Description} & \textbf{Tool/Framework Capabilities} & \textbf{Methodologies and Core Logic} & \textbf{Drawbacks/Gaps} \\
\hline
PACU is an open-source AWS exploitation framework designed to assist penetration testers and red teamers in auditing the security of Amazon Web Services environments. Developed by Rhino Security Labs, PACU is structured as a modular post-exploitation framework, primarily focusing on exploiting misconfigurations or weak implementations of IAM policies, roles, and AWS services. Unlike general-purpose AWS recon tools, PACU is purpose-built for offensive security, offering modules that can perform privilege escalation, data exfiltration, lateral movement, and other offensive tasks against AWS environments. The framework is written in Python and is interactive, allowing users to load and execute individual modules as needed. PACU has become a staple within the offensive cloud security community for AWS-focused operations, particularly when an adversary has already obtained initial access via compromised API credentials or misconfigurations. 
& PACU’s main capability revolves around interacting with the AWS API using supplied credentials to enumerate, exploit, and escalate within the cloud environment. The tool is designed to operate using a single set of AWS credentials (single-principal context), meaning that each execution session focuses on exploiting what that specific principal (user, role, or service) has access to. The tool supports numerous modules categorized by function, such as: Recon modules (e.g., enumerating IAM roles, permissions, policies, S3 buckets); Privilege escalation modules (e.g., identifying actions that can escalate privileges through misconfigured IAM policies or trust relationships); Persistence modules (e.g., creating new access keys); Exploitation modules (e.g., exfiltrating data or manipulating services). A core capability unique to PACU is its ability to simulate IAM permission policies using the SimulatePrincipalPolicy API call, allowing it to test what actions might succeed for the given credentials without actually executing them (helpful in stealthy testing). However, while PACU has modules for enumerating AssumeRole relationships, it doesn’t correlate or automatically exploit multi-principal chains unless specifically scripted by the user. Its primary focus is what this credential can do rather than building an escalation map across principals and services.
& PACU is built as a modular, Python-based framework with a core loop where the user loads modules interactively to perform specific tasks. Each module is structured as a Python file with standardized entry points (main() function) and arguments passed from the interactive shell. The modular design allows contributors and red teamers to add their custom exploit or enumeration techniques seamlessly. The core methodology revolves around: Enumerating resources and permissions available to the provided AWS credentials; Testing permissions using the SimulatePrincipalPolicy API when applicable; Executing real API calls to validate findings or perform attacks; Caching results in PACU’s internal database for quick reference between modules. Each exploitation step depends on the validity of cached data or dynamic enumeration, but there’s no automated recursive traversal of transitive permissions (e.g., what happens after assuming a role). The tool leaves these decisions to the operator’s discretion, giving flexibility but requiring expert knowledge.
& Despite having a strong collection of exploitation modules, PACU is lacking in a number of crucial areas for offensive operations involving modern cloud security. The single-principal constraint is its most noticeable drawback; it does not automatically identify or list chains of escalation across multiple principals, and each execution context is limited to the permissions of a single user or role. Comparing this to actual cloud attacks, where lateral movement frequently entails cross-principal privilege escalation through resource policies or trust relationships, reveals a substantial disparity. For example, the Capital One breach in 2019 involved a chain of misconfigurations where the attacker exploited an SSRF vulnerability to obtain temporary AWS credentials and then exploited excessive IAM permissions to escalate privileges, access sensitive S3 buckets, and ultimately exfiltrate data. Tools like PACU, if operated in isolation with single-principal methodology, would not have automatically discovered this entire privilege chain without extensive manual analysis and scripting by the operator. Additionally, PACU does not implement any form of fuzzing or permutation testing to discover unknown or non-explicit attack paths. It relies almost entirely on known actions and standard AWS APIs. As AWS evolves rapidly, static or scripted approaches lag behind the real world privilege escalation paths discovered by sophisticated adversaries. Another critical drawback is that PACU does not model trust relationships across multiple AWS accounts or organizations, severely limiting its utility in scenarios involving cross-account privilege escalation, which is increasingly relevant in large enterprise environments. Without automation to build privilege graphs or enumerate transitive paths, PACU’s utility is confined primarily to post compromise exploitation of known configurations, not active discovery of hidden privilege paths.\\
\hline
\caption{PACU - The AWS Exploitation Framework \cite{Pacu2021}}
\label{tab:PACU}
\end{longtable}
\endgroup
\end{landscape}

\begin{landscape}
\begingroup
\scriptsize
\setlength{\LTpre}{2mm} 
\setlength{\LTpost}{0pt}
\begin{longtable}[htbp]{|>{\raggedright\arraybackslash}p{4.5cm}|>{\raggedright\arraybackslash}p{5cm}|>{\raggedright\arraybackslash}p{5.5cm}|>{\raggedright\arraybackslash}p{6.5cm}|}
\hline
\textbf{Description} & \textbf{Tool/Framework Capabilities} & \textbf{Methodologies and Core Logic} & \textbf{Drawbacks/Gaps} \\
\hline
CloudPEASS is part of the broader PEASS (Privilege Escalation Awesome Scripts SUITE) project, originally created to help with privilege escalation enumeration on Windows and Linux systems. CloudPEASS specifically targets cloud environments, focusing on both AWS and Azure privilege escalation paths. The tool is open-source and written in Go, making it portable and efficient for execution across different platforms. CloudPEASS aims to automate the enumeration of potential privilege escalation opportunities by analyzing the permissions assigned to the authenticated principal (user or role) and matching them against a known set of escalation techniques. Unlike general-purpose AWS security tools, CloudPEASS emphasizes misconfigured privilege relationships and service-specific escalation paths, such as those involving S3 buckets, Lambda functions, or IAM role trusts. It provides ready-made detection for escalation opportunities based on community curated knowledge of known privilege abuse techniques. While still evolving, CloudPEASS has become popular for red teaming and post-exploitation assessments of cloud infrastructure. 
& CloudPEASS focuses exclusively on enumerating potential privilege escalation vectors within the cloud environment it is operating in. For AWS, this means identifying IAM policies, roles, groups, and services that could potentially be leveraged for unauthorized access or escalation. The tool works in a single-principal context, meaning it analyzes only the currently authenticated credentials, and it does not perform active exploitation but rather an enumeration and reporting tool only. Capabilities include: Enumerating IAM permissions, roles, and policies; Identifying known privilege escalation paths by matching discovered permissions with a list of known exploitation techniques (similar to how LinPeass identifies Linux privilege escalation vectors); Providing clear recommendations for potential attacks based on known patterns (e.g., “You have the permission to create/update Lambda functions. This could potentially lead to privilege escalation if leveraged properly.”); Cross-referencing actions like PassRole, UpdateFunctionCode, or AssumeRole with trust relationships to highlight paths to escalation.; Some support for discovering cross-service privilege abuse opportunities (e.g., S3 and Lambda combinations). However, like PACU, CloudPEASS operates on a static list of known techniques and doesn’t discover unknown attack paths or novel misconfigurations that deviate from the established privilege escalation library it relies on.
& The methodology behind CloudPEASS can be broken down into three primary steps: 1. Permission Enumeration: Using the authenticated AWS API credentials, CloudPEASS enumerates all assigned IAM policies, including inline and managed policies. It inspects service permissions, attached roles, and group memberships; 2. Pattern Matching with Known Exploits: The gathered permissions are then checked against a curated list of known privilege escalation scenarios maintained in the CloudPEASS codebase. This static list maps permissions to known abuses, such as the ability to: Create or update Lambda functions; Attach roles to EC2 instances; Update CodeBuild projects; Use iam:PassRole with various AWS services; Modify resource policies (e.g., granting public access to S3 buckets); Output Reporting: The tool generates a human-readable output, often recommending next steps for exploiting the discovered paths. The core logic is deterministic: it does not experiment or test with real API calls, nor does it leverage fuzzing or enumeration of relationships beyond what the APIs explicitly return. It relies on predefined relationships rather than dynamically discovering new privilege chains.
& The static, signature-driven method of privilege escalation detection used by CloudPEASS is one of its main drawbacks. This makes the tool extremely effective at identifying well-known vectors of privilege abuse, but it also makes it vulnerable to new or untested escalation techniques. For example, CloudPEASS would not detect a new privilege escalation path involving a mysterious combination of AWS services or configurations unless it was explicitly updated with the new information. Like PACU, CloudPEASS operates entirely in the context of a single-principal view of the AWS account. It doesn’t build escalation graphs across multiple principals, meaning it won’t automatically discover chained privilege escalation paths, such as “User A can assume Role B, which can then modify Lambda C, which then executes with Privileged Role D.” These multi-hop escalation chains are exactly the kind of privilege abuse that attackers increasingly exploit in complex cloud environments, and CloudPEASS does not currently handle such scenarios. Another major gap is that CloudPEASS does not perform any enumeration of transitive trust relationships or resource-based policy attacks (e.g., S3 bucket policies that reference other accounts or principals). Similarly, it does not attempt to brute-force or fuzz unknown attack surfaces, focusing strictly on what’s already defined in its knowledge base. If CloudPEASS had been used in the Capital One breach scenario, it may have identified individual misconfigurations, such as over-permissioned IAM policies. But it would not have been able to map the complete attack path automatically, nor would it have discovered cross-service trust relationships contributing to the breach.\\
\hline
\caption{CloudPEASS - Cloud Privilege Escalation Awesome Script Suite \cite{CloudPEASS}}
\label{tab:CloudPEASS}
\end{longtable}
\endgroup
\end{landscape}

\begin{landscape}
\begingroup
\scriptsize
\begin{longtable}[htbp]{|>{\raggedright\arraybackslash}p{4.5cm}|>{\raggedright\arraybackslash}p{5cm}|>{\raggedright\arraybackslash}p{5.5cm}|>{\raggedright\arraybackslash}p{6.5cm}|}
\hline
\textbf{Description} & \textbf{Tool/Framework Capabilities} & \textbf{Methodologies and Core Logic} & \textbf{Drawbacks/Gaps} \\
\hline
enumerate-iam is a specialized open-source tool developed by nccgroup, designed specifically to identify privilege escalation paths within AWS IAM configurations. Unlike broader cloud security tools, enumerate-iam narrows its focus to one key objective: finding privilege escalation possibilities within IAM policies and roles. Written in Python, it’s relatively lightweight and straightforward to deploy on any environment where Python is supported. The tool gained traction in the security community for its automation of privilege escalation path detection, offering penetration testers and cloud security teams a more structured way to discover risky permission assignments. Enumerate-iam is often recommended alongside larger AWS assessment frameworks like PACU for privilege analysis. It is especially useful for quickly identifying potential abuse scenarios in environments where roles, users, or groups have seemingly innocuous permissions that could lead to privilege escalation via privilege chaining though only within the scope of single-principal analysis.
& enumerate-iam performs static analysis on the IAM policies accessible to a given set of AWS credentials. It then matches these discovered permissions against a curated database of known privilege escalation methods. The results are categorized into various exploit techniques such as: Creating new policies that escalate permissions; Attaching elevated policies to existing roles or groups; Assuming roles with higher privileges; Updating or creating new Lambda functions tied to privileged roles. One notable capability of enumerate-iam is its built-in understanding of the relationships between AWS service permissions and privilege escalation scenarios. It automatically evaluates service-level permissions like: iam:CreatePolicyVersion; iam:AttachUserPolicy; lambda:UpdateFunctionCode; ec2:RunInstances combined with iam:PassRole. Enumerate-iam, however, functions on a per-identity basis, analyzing only the currently authenticated AWS principal (user, group, or assumed role). Although it can detect hazardous permissions, it does not investigate multi-principal or transitive escalation paths. Moreover, it does not carry out or test exploit paths actively, it simply lists and documents possible escalation vectors. Due to this static analysis method, it is easy to predict but somewhat restricted when confronted with inventive opponents using unorthodox methods of attack.
& The core methodology of enumerate-iam revolves around permission analysis combined with privilege escalation knowledge mapping. The tool does not use graph-based or fuzzing methodologies; rather, it builds a direct mapping of AWS permissions to possible privilege escalation paths. Here’s a simplified breakdown of its logic: \textbf{Permission Gathering:} Uses AWS API calls like: ListAttachedUserPolicies, ListInlinePolicies and SimulatePrincipalPolicy to gather effective permissions for the authenticated principal; \textbf{Matching to Exploits:} The tool compares discovered permissions against a predefined list of privilege escalation techniques. This list is derived from known AWS privilege escalation documentation, such as those curated by Rhino Security Labs and NCC Group; \textbf{Result Reporting:} Matches are reported to the user, categorized by technique, with references to the type of escalation possible. The strength of enumerate-iam is its accurate matching for known escalation vectors, but its analysis stops at identifying singular paths for the authenticated principal only. It lacks analysis of how those permissions may interoperate with others in the AWS environment.
& enumerate-iam’s primary weakness lies in its inability to analyze privilege escalation chains across multiple principals or to interpret complex, transitive permission relationships that arise from resource-based policies. For instance, if User A can assume Role B, and Role B can modify a Lambda tied to Role C (with admin privileges), enumerate-iam would fail to map that entire sequence unless each hop explicitly grants recognizable direct escalation privileges. Additionally, enumerate-iam relies on a fixed set of known escalation techniques. If a novel privilege escalation pathway is discovered, the tool won’t detect it until the escalation database is manually updated. This makes it reactive rather than proactive. It’s also limited in visibility to what the current principal can enumerate. If iam:Get* permissions aren’t granted, the tool cannot fully map potential privilege abuse. Adversaries could still exploit those pathways if they know about them through external discovery or prior compromise of other principals. Regarding fuzzing or complex attack path enumeration, enumerate-iam provides no dynamic enumeration of unforeseen privilege combinations. In incidents like the Capital One breach, which involved complex misconfigurations across several AWS services, enumerate-iam alone would likely have missed critical privilege paths unless those permissions were explicitly on the tool’s radar. It performs admirably for basic privilege escalation hygiene, but does not scale to real-world, multi-hop, or privilege-chain scenarios.\\
\hline
\caption{enumerate-iam - Enumerate AWS IAM Permissions \cite{enumerate-iam}}
\label{tab:enumerateiam}
\end{longtable}
\endgroup
\end{landscape}

\begin{landscape}
\begingroup
\scriptsize
\begin{longtable}[htbp]{|>{\raggedright\arraybackslash}p{4.5cm}|>{\raggedright\arraybackslash}p{5cm}|>{\raggedright\arraybackslash}p{5.5cm}|>{\raggedright\arraybackslash}p{6.5cm}|}
\hline
\textbf{Description} & \textbf{Tool/Framework Capabilities} & \textbf{Methodologies and Core Logic} & \textbf{Drawbacks/Gaps} \\
\hline
CloudFox is a command-line tool developed by Bishop Fox, primarily targeted at helping penetration testers and security professionals enumerate AWS environments during assessments. Unlike tools that focus narrowly on privilege escalation (like enumerate-iam), CloudFox provides a broader spectrum of AWS enumeration, enabling red teamers to discover interesting targets within AWS environments such as EC2 instances, S3 buckets, secrets, and IAM roles. CloudFox, which is implemented in Go, prioritizes speed, parallel execution, and concise output. CloudFox’s primary objective is to identify the attack surface, which involves locating potentially exploitable resources and misconfigurations that could assist adversaries in pivoting or escalating privileges during cloud security assessments. It neither directly carries out exploitation nor simulates privilege escalation like PACU does; rather, it aids in the systematic and modular enumeration of misconfigurations and relationships. Due to its extensible nature, numerous practitioners utilize CloudFox as a preliminary reconnaissance tool before applying exploitation or privilege escalation frameworks.
& CloudFox’s strength lies in its comprehensive AWS reconnaissance abilities. It can enumerate: EC2 Instances with interesting metadata; Lambda Functions, including identifying over-permissive roles attached; Secrets stored in SSM Parameter Store or Secrets Manager; IAM Role Trust Policies, including cross-account trust relationships; IAM User policies and Group memberships; Resource-based policies that may allow access escalation, particularly on S3, Lambda, or API Gateway. One of its most powerful features is its analysis of trust relationships, which can provide leads for privilege escalation via role assumption. This is especially useful for identifying scenarios where a low-privileged principal might pivot by exploiting poorly configured AssumeRole permissions or misconfigured trust policies. However, CloudFox primarily operates by enumerating resources accessible by a single principal. Like enumerate-iam, it does not stitch multiple principals’ permissions together dynamically, meaning complex privilege escalation chains remain hidden unless an operator manually traces them using CloudFox’s output. CloudFox also doesn’t actively execute attacks or privilege escalation attempts but it provides the visibility necessary for analysts to pursue exploitation manually. Additionally, it shines in highlighting cross-account trust policies, something several older AWS enumeration tools neglected.
& CloudFox follows a modular execution pattern, with each command or “module” tailored to a specific AWS service or discovery focus. The core logic typically follows this pattern for each resource type: 1. Service API Interrogation, which invokes various AWS APIs to enumerate resources the authenticated principal has access to (e.g ec2:DescribeInstances, s3:ListBuckets, iam:GetRolePolicy, lambda:ListFunctions,  sts:GetCallerIdentity; 2. Trust Relationship Extraction: For IAM roles or resource-based policies, CloudFox will parse trust policies to identify cross-account trusts and trust relationships with EC2, Lambda, or other AWS services; 3. Output Structuring for Attackers via: Interesting instances or resources, Potential paths to privilege escalation or lateral movement, Weak permissions (e.g., S3 buckets open to authenticated users); 4. Parallel Execution and Performance: Built in Go, CloudFox can query APIs efficiently, handling large environments faster than older Python-based enumeration tools. While the tool exposes paths to possible privilege escalation (e.g., AssumeRole opportunities), it relies on the user’s interpretation and manual effort to build escalation chains or exploits from the data provided.
& While CloudFox is robust in reconnaissance, it exhibits several limitations in advanced privilege escalation detection: \textbf{1. Single-Principal Enumeration Limitation:} CloudFox only enumerates resources visible to the currently authenticated principal. It doesn’t attempt to enumerate policies from the perspective of other principals unless manually authenticated as those identities; \textbf{2. No Automatic Privilege Chaining:} CloudFox does not automatically discover multi-hop escalation paths. (e.g If User A can assume Role B, which allows creating a Lambda tied to Role C (with higher privileges), CloudFox will enumerate trust relationships and policy permissions, but it requires manual analysis to map the entire escalation path.); \textbf{3. No Fuzzing or Enumeration of Unknown Attack Paths:} CloudFox relies on AWS permissions structures and visibility but does not perform dynamic or speculative privilege enumeration. This prevents discovery of zero-day misconfigurations or unexpected privilege escalations unless explicitly defined in IAM policies; \textbf{4. No Active Attack Simulation:} Unlike tools such as Stratus Red Team, CloudFox does not attempt to simulate or test attack paths. This places a cognitive load on analysts to translate discovered data into actionable exploitation strategies
\\
\hline
\caption{\centering CloudFox - Automate Situational Awareness for Cloud Penetration Tests \cite{CloudFox}}
\label{tab:CloudFox}
\end{longtable}
\endgroup
\end{landscape}

\begin{landscape}
\begingroup
\scriptsize
\begin{longtable}[htbp]{|>{\raggedright\arraybackslash}p{4.5cm}|>{\raggedright\arraybackslash}p{5cm}|>{\raggedright\arraybackslash}p{5.5cm}|>{\raggedright\arraybackslash}p{6.5cm}|}
\hline
\textbf{Description} & \textbf{Tool/Framework Capabilities} & \textbf{Methodologies and Core Logic} & \textbf{Drawbacks/Gaps} \\
\hline
ScoutSuite is a popular multi-cloud security auditing tool developed by NCC Group. Originally designed for AWS, it has since expanded to support Azure, GCP, and Alibaba Cloud. Written in Python, ScoutSuite offers security practitioners and auditors a comprehensive view of cloud account configurations, with a strong emphasis on identifying misconfigurations that could expose cloud resources to risks. Unlike tools such as CloudFox or PACU, which are more offensive-security focused, ScoutSuite sits at the intersection of offensive and defensive cloud security auditing. Its core mission is to enable a holistic, configuration-based assessment of a cloud environment by collecting resource metadata and presenting it through a web-based or JSON-based report. It excels at identifying weaknesses such as publicly exposed S3 buckets, over-permissive IAM roles, and unsecured databases. Due to its vendor-agnostic support, ScoutSuite has become a go-to tool for broad compliance checks and security reviews across diverse cloud architectures.
& ScoutSuite’s primary capability is static cloud security posture analysis (CSPM). It retrieves configuration metadata from the target cloud account and generates a detailed security report. These reports highlight: 1. Publicly accessible resources (e.g., S3 buckets, EC2 security groups with 0.0.0.0/0 rules); 2. IAM risks, such as overly permissive policies or wildcard resource grants; 3. Misconfigurations in storage, compute, networking, and identity services; 4. Cross-account access risks. ScoutSuite supports multiple cloud platforms, including AWS, Microsoft Azure, Google Cloud Platform (GCP) and Alibaba Cloud. While on AWS specifically, ScoutSuite focuses its abilities on S3 bucket policies, Security Groups, IAM policies and attached roles, EC2 instance meta data exposure, Route53 configurations and Lambda configurations. However, ScoutSuite works from a configuration audit perspective, rather than a principal-based enumeration approach like enumerate-iam or PACU. It does not focus on what an individual principal can do, but instead what risks exist at the account or service level. Further, ScoutSuite is read-only by design. It does not attempt exploitation or active testing but remains strictly auditorial, making it ideal for compliance checks, blue teaming, and risk assessments.
& ScoutSuite follows a three-phase methodology to perform its security auditing: 1. API Collection Phase: The tool uses cloud-native SDKs (e.g., boto3 for AWS) to pull account metadata through API calls authenticated with the provided cloud credentials (e.g s3:listBuckets, ec2:DescribeSecurityGroups, iam:ListPolicies, iam:GetAccountSummary); 2. Data Aggregation Phase: Collected data is aggregated into JSON files representing cloud services’ configurations. ScoutSuite then correlates services to highlight policy misconfigurations, resources with public exposure and dangerous permissions patterns; 3. Analysis and Reporting Phase: ScoutSuite cross-references the gathered data against predefined risk patterns and security best practices. These heuristics are rule-based, covering known misconfigurations such as public S3 buckets with s3:GetObject\} allowed for *\} (all), security groups allowing SSH or RDP from anywhere, and IAM policies granting *:* privileges. The output is rendered in HTML or JSON. The HTML format offers clickable drill-downs into individual services, IAM roles, or specific configuration concerns. Unlike offensive tools, ScoutSuite’s engine does not predict potential privilege escalation paths unless they are clearly defined as misconfigurations.
& Despite its strengths in static auditing, ScoutSuite has notable limitations for advanced privilege enumeration or offensive security assessments: \textbf{1. Configuration-Only Focus:} ScoutSuite evaluates resource configurations, but does not perform active privilege testing. If a privilege escalation path exists that relies on runtime conditions (e.g., chaining multiple IAM roles across services), ScoutSuite won’t detect it unless it’s reflected directly in static configurations; \textbf{2. Single Snapshot, Not Dynamic Enumeration:} ScoutSuite does not adapt dynamically to discover emerging attack paths. If configurations change during a live assessment, or if attacker-controlled elements inject risky configurations (e.g., via SSRF or malicious resource creation), ScoutSuite will miss them unless re-run manually; \textbf{3. No Multi-Principal Analysis:} Like PACU and CloudFox, ScoutSuite is bound by the permissions of the credential used for analysis. It cannot simulate or enumerate privileges of other principals unless authenticated separately for each; \textbf{4. No Enumeration or Fuzzing of Unknown Paths:} Critically, ScoutSuite does not enumerate complex privilege escalation chains, nor does it support speculative testing or fuzzing of IAM permissions to uncover obscure or undocumented pathways for privilege escalation. It primarily detects misconfigurations against known heuristics or compliance rules; \textbf{5. Capital One Case Context:} In relation to breaches like Capital One, where the attack path involved SSRF exploitation to access EC2 instance metadata and leverage IAM roles, ScoutSuite would likely have flagged the public exposure risks (e.g., insecure web application security groups) but would not have detected the privilege escalation chain itself. Its static nature makes it better suited for preventative controls than reactive analysis of novel privilege abuse scenarios.
\\
\hline
\caption{ScoutSuite - Multi-Cloud Security Auditing Tool \cite{ScoutSuite}}
\label{tab:ScoutSuite}
\end{longtable}
\endgroup
\end{landscape}

\begin{landscape}
\begingroup
\scriptsize
\begin{longtable}[htbp]{|>{\raggedright\arraybackslash}p{4.5cm}|>{\raggedright\arraybackslash}p{5cm}|>{\raggedright\arraybackslash}p{5.5cm}|>{\raggedright\arraybackslash}p{6.5cm}|}
\hline
\textbf{Description} & \textbf{Tool/Framework Capabilities} & \textbf{Methodologies and Core Logic} & \textbf{Drawbacks/Gaps} \\
\hline
Stratus Red Team is an offensive security tool tailored specifically for cloud-native attack simulation in AWS, Azure, and GCP environments. Developed by Datadog security engineers, Stratus Red Team draws inspiration from tools like MITRE ATT\&CK but specializes in cloud-native adversary emulation. Its core purpose is to simulate real-world attack techniques against cloud infrastructure to help defenders validate and improve detection capabilities. Unlike ScoutSuite or PACU, which focus on identifying misconfigurations or enumerating permissions, Stratus Red Team simulates adversarial behavior directly within cloud environments. This makes it more comparable to tools like Atomic Red Team, but specialized for cloud scenarios. The tool includes predefined attack scenarios ("attack techniques"), which users can execute to mimic behaviors like privilege escalation, data exfiltration, or credential harvesting. The tool is declarative and modular, with support for executing, describing, and cleaning up attack simulations, thus making it suitable for red teaming exercises and defensive security validation alike.
& Stratus Red Team’s capabilities revolve around simulation of known attack techniques, broken down by cloud provider and mapped to MITRE ATT\&CK tactics and techniques. It enables security teams to proactively test detection pipelines and incident response processes by executing controlled offensive behaviors in a repeatable, documented way. Capabilities include: 1. Cloud-specific Attack Simulations - AWS: IAM privilege escalation, data exfiltration via EC2 instance roles, access key exposure - GCP: Service account token abuse- Azure: Role assignment attacks, identity privilege escalation; 2. Technique Taxonomy: Techniques are categorized by ATT\&CK tactics (e.g., Credential Access, Privilege Escalation, Defense Evasion); 3. Automation-friendly Interface with simple CLI solution; 4. Cleanup Features: after each test, users can invoke cleanup commands to revert cloud environments to their original state. Unlike tools like enumerate-iam or CloudFox, Stratus Red Team does not aim to enumerate permissions or map attack surfaces. Its sole focus is simulating specific attack chains that are known or highly probable in cloud environments. Additionally, Stratus Red Team includes telemetry generation for SOC teams to validate whether security tools like GuardDuty, CloudTrail, or custom SIEM rules are triggering properly upon simulation
& Stratus Red Team employs a scenario-based, declarative methodology for attack simulation. The tool is designed around modular "attack techniques" definitions written in YAML. These definitions specify: 1. Preconditions: What cloud infrastructure must exist before the attack can be executed; 2. Attack Execution Logic: Python or shell code that performs the attack, often via cloud SDKs or CLI tools; 3. Cleanup Logic: Commands or scripts to remove traces of the attack after execution. Methodologies breakdown: Technique Declaration (YAML-based): use yaml file to declare command with ease of use and format; Execution Framework: Internally, the tool parses the YAML technique file and runs the corresponding Python or Bash scripts that execute the cloud-native API calls to perform the attack.; Observability Hooks: Many techniques include hooks for validating detection controls. Users can integrate this into SIEM pipelines or custom security telemetry collection; State Cleanup: Once the attack is tested, Stratus Red Team cleans up temporary resources, IAM roles, or permissions created during the simulation to restore the cloud account to its prior state; Modularity: Users can add their own custom techniques or modify existing ones for specific threat modeling exercises. This focus on execution realism makes Stratus Red Team ideal for validating whether real-world cloud attacks would succeed undetected in a target environment.
& Despite its strengths in simulating known cloud attack behaviors, Stratus Red Team has limitations when evaluated from the perspective of complex privilege enumeration or proactive discovery of unknown attack paths: 1. Single-Technique Granularity: The simulation framework is bounded by predefined attack scenarios. It cannot enumerate or fuzz privileges dynamically to discover novel escalation chains. Its utility is bounded by what is already known and encoded as attack techniques in its repository; 2. No Principal Enumeration or Fuzzing: Unlike enumerate-iam, which can brute-force IAM actions to discover what a principal may or may not do, Stratus Red Team does not explore what actions an unknown principal can perform. Instead, it requires explicit permissions to execute the simulations successfully; 3. Assumes Predefined Knowledge of Attack Techniques: By focusing solely on known techniques, Stratus Red Team cannot handle “unknown" privilege chains or attack paths that haven’t been formally cataloged in frameworks like ATT\&CK. This is a significant gap when conducting research-oriented or exploratory adversarial analysis; 4. No Automated Discovery or Attack Path Mapping: Where tools like CloudFox assist in manually mapping attack surfaces, Stratus Red Team does not provide automated discovery of accessible resources, policy inheritance patterns, or chained privilege escalation vectors. In summary, Stratus Red Team excels as a detection validation framework, but lacks exploratory capability, privilege fuzzing, and dynamic attack path generation, leaving critical gaps in cloud adversary research and unknown attack path discovery.
\\
\hline
\caption{\centering Stratus Red Team (DataDog) - Granular, Actionable Adversary Emulation for the Cloud \cite{StratusRedTeam}}
\label{tab:StratusRedTeam}
\end{longtable}
\endgroup
\end{landscape}

\begin{landscape}
\begingroup
\scriptsize
\begin{longtable}[htbp]{|>{\raggedright\arraybackslash}p{4cm}|>{\raggedright\arraybackslash}p{5.5cm}|>{\raggedright\arraybackslash}p{4.7cm}|>{\raggedright\arraybackslash}p{7.5cm}|}
\hline
\textbf{Description} & \textbf{Tool/Framework Capabilities} & \textbf{Methodologies and Core Logic} & \textbf{Drawbacks/Gaps} \\
\hline
Cloudsplaining is an open-source security tool developed by Salesforce that focuses on analyzing AWS IAM policies to identify permissions risks and privilege escalations. It is designed to help security teams understand and remediate over-permissive IAM policies, which are one of the most common root causes of cloud security breaches. Whereas tools like ScoutSuite provide broad cloud misconfiguration analysis, Cloudsplaining exclusively focuses on the fine-grained analysis of IAM policies. Its primary purpose is to detect dangerous permissions that could be exploited by attackers or inadvertently misused by internal users. The tool generates human-readable reports highlighting risky actions like iam:PassRole, ec2:CreateTags, or policies granting wildcard \texttt{*:*} permissions. It is often used in security audits, CI/CD pipelines, or pre-deployment policy reviews to enforce least-privilege practices. It bridges a critical need in cloud security by transforming the often opaque structure of AWS IAM into actionable risk insights.
& Cloudsplaining’s capabilities revolve around analyzing AWS IAM policies for potential security risks. It inspects both inline and managed policies and maps discovered permissions against dangerous actions or escalation vectors. Core Capabilities:1. Identification of Privilege Escalation Risks: Detects risky combinations of permissions that could allow for privilege escalation (e.g., combining iam:PassRole with EC2 operations to gain administrator access); 2. Overly Broad Permissions Detection: Flags policies with wildcard actions; 3. Exposure of Sensitive Data: Identifies permissions like s3:GetObject or kms:Decrypt that can be used to access sensitive data; 4. IAM Policy Review Automation: Can be integrated into CI/CD pipelines to enforce IAM policy review before deployment; 5. Comprehensive Reporting: Generates HTML reports or JSON outputs, enabling review by both developers and security teams; 6. Limitation of Scope: Cloudsplaining is not a real-time monitoring tool. It only works against static IAM policies. It also does not execute or test whether the permissions are active or exploitable by a given identity in a live cloud account. Unlike PACU or enumerate-iam, Cloudsplaining doesn’t operate interactively or dynamically. It complements those tools by providing pre-deployment or audit-time security posture analysis, rather than active exploitation or detection validation.
& The core methodology of Cloudsplaining is static analysis of AWS IAM policies using policy document parsing and rule-based pattern matching to identify dangerous configurations. Its internal logic consists of the following stages: 1. IAM Policy Parsing: Cloudsplaining reads JSON-formatted IAM policies from AWS accounts or local files. It then decomposes these policies into individual statements, breaking them down into: Effect (Allow or Deny), Action(s), Resource(s) and Condition(s) (if present); 2. Dangerous Actions Database: Maintains a curated list of high-risk IAM actions (e.g iam:PassRole, ec2:RunInstances, lambda:CreateFunction, eks:CreateCluster), Actions are grouped by risk category: Privilege Escalation, Resource Exposure and Infrastructure Modification; 3. Risk Rule Application: For each parsed policy, Cloudsplaining applies predefined detection rules to match against risky combinations or patterns. It's noted that some compound rules evaluate sequences of permissions rather than just single actions
& While Cloudsplaining is highly effective for static IAM analysis, it also has notable gaps and drawbacks in addressing the full scope of privilege escalation and IAM complexity in cloud environments: 1. Static Analysis Only: Cloudsplaining cannot validate whether the risky permissions are actually accessible or exploitable by a particular principal. It does not cross-reference policies with real-time effective permissions of users, roles, or groups. This is a crucial gap since Cloudsplaining's analysis does not interpret permission boundaries, resource-based policies, or SCPs (Service Control Policies), which can effectively deny permissions; 2. No Multi-Principal Enumeration: It evaluates individual policies in isolation but does not enumerate across multiple principals to build a comprehensive picture of cross-entity privilege escalation paths. While tools like enumerate-iam can actively test whether a principal can perform an action, which Cloudsplaining cannot do; 3. No Enumeration of Role Chaining: Privilege escalation vectors often involve chaining roles or using intermediate privilege levels to gain elevated access. However, Cloudsplaining does not simulate or analyze these privilege chains dynamically; 4. Cannot Detect Complex Attack Paths: Contextual exploitation paths, such as abuse via unintended network exposure (e.g., SSRF), are outside of Cloudsplaining’s scope; 5. No Fuzzing or Enumeration for Discovery: Cloudsplaining only works with known, defined risks. It cannot fuzz unknown API actions or brute-force IAM capabilities to uncover hidden permission paths. As a result, it is best suited for governance, compliance, and policy audits but not exploratory adversary simulation or unknown privilege discovery. In summary, Cloudsplaining is extremely effective at detecting obvious, static IAM risks, but its lack of dynamic analysis, fuzzing, principal enumeration, and attack path simulation leaves significant coverage gaps for offensive security use cases or proactive discovery of unknown privilege escalation risks.
\\
\hline
\caption{\centering Cloudsplaining (Salesforce) - AWS IAM Security Assessment tool that identifies violations of least privilege \cite{Cloudsplaining}}
\label{tab:Cloudsplaining}
\end{longtable}
\endgroup
\end{landscape}

\section{Comparative Analysis: SkyEye Approach}
To assess state-of-the-art: PACU, CloudFox, Scoutsuite, Cloudsplaining, CloudPEASS, enumerate-iam, we adopt six capability dimensions that materially affect an assessor’s ability to discover and validate identity-driven risk in AWS cloud:
\begin{enumerate}
    \item \textbf{Principal Scope:} Single-principal vs. multi-principal analysis (ability to reason across many identities simultaneously)
    \item \textbf{Transitive Role Assumption Chain:} Detection of multi-hop paths\\e.g., A Principal $\rightarrow$ Directly-Assumable ole(B) $\rightarrow$ Indirectly-Assumable Role(C)
    \item \textbf{Discovery Mechanism:} Static rules, active enumeration, fuzzing for unknown combinations, and/or execution simulation.
    \item \textbf{Temporal / Version Awareness:} Sensitivity to policy versions, session policies, and ephemeral/conditional access.
    \item \textbf{MITRE ATT\&CK Contextual Risk Mapping:} Mapping to MITRE ATT\&CK for Cloud or similar frameworks for prioritization.
    \item \textbf{Cross-Policy / Resource Aggregation:} Unifying identity-based policies, resource-based policies, permission boundaries, SCPs, and service configuration into a single effective-permission view.
\end{enumerate}

These dimensions correspond directly to where false negatives occur in practice: when identities are considered in isolation; when transitive chains are not explored; when the method is limited to known patterns; when temporal/session context is ignored; when results lack actionable prioritization; or when policy/resource layers are not aggregated.

\subsection{Comparative Matrix Table}
\begin{table}[htbp]
\hspace{-1cm}
\scriptsize
\begin{tabular}{|p{2.7cm}|p{1.2cm}|p{1.5cm}|p{1.5cm}|p{1.5cm}|p{1.5cm}|p{1.5cm}|p{2.2cm}|}
\hline
\textbf{Dimension} & \textbf{Pacu} & \textbf{enumerate-iam} & \textbf{CloudFox} & \textbf{ScoutSuite} & \textbf{Cloud splaining} & \textbf{Cloud PEASS} & \textbf{SkyEye\textsuperscript{Proposed}} \\ \hline
Principal Scope & Single & Single & Single & Account config & Policy-centric & Single
(Batchable) & Multi-principal \\ \hline
Transitive Role Assumption Chain & Partial & No & No & No & No & No & Yes \\ \hline
Discovery Mechanism & Active & Active + Fuzzing & Active Recon & Static rules & Static rules & Active + Simulation & Active + Fuzzing + Simulation \\ \hline
Temporal / Version Awareness & Low & Low & Low & Low & Low & Low & High \\ \hline
MITRE ATT\&CK Contextual Mapping & Low & Low & Low & Low & Medium & Low - Medium* & High \\ \hline
Cross-Policy / Resource Aggregation & Low & Low & Low & Medium & Low & Low - Medium* & High \\ \hline
\end{tabular}
\caption{Comparison of Frameworks on Selected Dimensions}
\label{tab:cloud_tools_dimensions}
\end{table}

Notes: 
\begin{itemize}
    \item \textbf{Active:} Means live AWS API calls; “fuzzing” means hypothesis-generation for unknown action/resource/condition permutations; “Scenario-bound” means coverage is limited to predefined simulations.
    \item \textbf{Low-Medium:} CloudPEASS sometimes surfaces escalation tactics using HackTricks-style heuristics, but it does not natively align findings to ATT\&CK with scoring or path semantics. Additionally, CloudPEASS reasons primarily from the principal’s effective permissions; it may hint at resource-policy angles for known techniques, but it doesn’t unify identity, resource policies, permission boundaries, and SCPs into a single evaluator.
\end{itemize}

\subsection{The State-of-Art Work is Strong, yet Insufficient}

\paragraph{Principal Scope (Single vs. Multi-Principal):}
Prevailing tools for AWS permission analysis (e.g., Pacu, enumerate-iam, CloudFox, ScoutSuite, Stratus Red Team, Cloudsplaining) typically evaluate permissions on a per-principal basis, or from independent account snapshots without correlating entitlements across multiple identities. In contemporary multi-account estates, this siloed view obscures privilege that only emerges when permissions are composed across principals. Consequently, escalation chains remain latent even when each identity is fully enumerated in isolation, driving false negatives in complex environments. SkyEye should therefore include a multi-principal graph engine that ingests many principals concurrently and computes cross-identity paths to reveal emergent privilege.

\paragraph{Transitive Role Assumption Chain:}
Existing tools do not fully model recursive role assumption and transitive permissions; Pacu offers only partial, largely manual support. Because many real-world attack paths are multi-hop (A $\rightarrow$ B $\rightarrow$ C), single-hop or static analyses surface merely local opportunities while missing global privilege that becomes reachable after one or more assumptions. This mismatch with adversary tradecraft, which iteratively tests assumptions, systematically underestimates effective, reachable privilege. SkyEye should implement bounded DFS/BFS over the role-trust graph with stateful capability accumulation (i.e., “after assuming B, the action set changes; now test C”) to capture transitivity.

\paragraph{Discovery Mechanism (Known Patterns vs. Fuzzing):}
Tools such as enumerate-iam and Pacu issue known actions, while Cloudsplaining and ScoutSuite rely on static rules and Stratus Red Team on predefined scenarios; none systematically fuzz permutations of actions, resources, and conditions to discover unknown-unknowns. Given the rapid evolution of AWS features, condition keys, and policy interactions, rule-only approaches incur structural false negatives until rules are updated. To overcome this lag, SkyEye should incorporate a constrained search-and-prune fuzzing engine that proposes and tests plausible action/resource/condition tuple prioritized by service semantics, using safe, non-destructive probes to uncover novel permission effects.

\paragraph{Temporal and Version Awareness:}
Prior tooling largely treats policies as static artifacts and seldom accounts for session policies, permission boundaries, service control policies (SCPs), policy versioning, or time-bounded credentials. Yet effective permissions are frequently contextual and time-dependent (e.g., temporary elevation during deployment windows), so static snapshots can materially misrepresent operational reality and risk. SkyEye should therefore support versioned policy ingestion, session-context simulation (e.g., AssumeRole with PolicyArns/Policy), and temporal differencing to compute effective-permissions across relevant time windows.

\paragraph{Contextual Risk Mapping and Prioritization:}
With the partial exception of Stratus Red Team’s ATT\&CK alignment, most tools provide limited or ad-hoc prioritization, producing voluminous outputs that impede triage and stakeholder communication. Without consistent semantics that relate findings to adversary behaviors and business impact (e.g., credential access vs. read-only), teams disproportionately expend effort on low-value issues. SkyEye should provide first-class MITRE ATT\&CK mapping and impact scoring that incorporate path length, blast radius, and data sensitivity to enable scalable, risk-informed prioritization.

\paragraph{Cross-Policy and Resource Aggregation:}
Current approaches struggle to jointly evaluate identity-based and resource-based policies (e.g., S3, KMS, Lambda, SQS/SNS) alongside permission boundaries and SCPs within a single effective-permissions model. Omitting any layer yields incorrect allow/deny inferences and conceals escalation edges, such as cross-account access granted by a resource policy that circumvents a restrictive identity policy. SkyEye should implement a unified evaluator that composes all relevant policy layers for each step in a chain (identity $\rightarrow$ resource $\rightarrow$ boundary $\rightarrow$ SCP) and re-evaluates the post-assumption state after every hop.

\subsection{Partial Coverage Creates Structural False Negatives}
Let the effective capability of a principal $\rho$ be $C(\rho)$, which depends on identity policies I, resources policies R, boundaries B, and org controls S:
\begin{equation}
C(\rho) = \text{Eval}(I_\rho, R, B_\rho, S)
\end{equation}

Let a chain be a sequence $(\rho_0 \to \rho_1 \to \dots \to \rho_k)$ where each transition is feasible under $C(\rho_i)$. Single-principal methods estimate only $C(\rho_0)$. Single-hop methods esitmate at most $C(\rho_1)$. True reachable capability is:

\begin{equation}
C^*(p_0) = \bigcup_{i=0}^k C(p_i)
\end{equation}

Failure to enumerate chains (multi-principal, multi-hop) produces a strict subset $\hat{C}(\rho_0) \subset C^*(\rho_0)$, i.e., structural false negatives. This is not an implementation bug; it is a methodological limitation. Prior tools, by design, compute at $\hat{C}$, not $C^*$

\subsection{Why Each State-of-the-Art Works Fall Short (Per Dimension)}

\paragraph{Pacu:} Strong interactive exploitation but user-driven and single-principal. No systematic transitive enumeration, limited aggregation, no fuzzing. Good for proof-of-concept exploitation; weak for comprehensive coverage.

\paragraph{enumerate-iam:} Accurate action testing for one identity; no chaining, no fuzzing, limited context. Excellent for answering “what can this do?”; insufficient for “what can be reached through this?”

\paragraph{CloudFox:} Fast recon and trust-policy visibility; no chaining engine, no effective-permission evaluator. Great at “what exists?”; weak at “how do these compose into a path?”

\paragraph{ScoutSuite:} Broad misconfiguration snapshot; static only, no privilege path construction, limited policy aggregation depth. Ideal for compliance; not for adversarial path discovery.

\paragraph{Cloudsplaining:} Excellent static risk heuristics; isolated policies, no effective-permission evaluation, no transitivity. Ideal gate in CI; not sufficient for reachability analysis.

\paragraph{CloudPEASS:} Practical for single-principal privilege-escalation enumeration using SimulatePrincipalPolicy (when allowed) and read-only brute enumeration (List/Get/Describe) as fallback. It maps discovered permissions to a catalog of known escalation techniques (e.g., iam:PassRole + service updates, Lambda code injection, CodeBuild abuse). However, it does not construct multi-principal or multi-hop chains (no recursive AssumeRole traversal with stateful capability accumulation), offers no fuzzing to discover novel, undocumented combinations, and lacks temporal awareness (session policies, ephemeral elevation windows, policy version diffs). Its hints at resource-policy angles remain heuristic rather than a unified evaluator of identity policies, resource policies, permission boundaries, and SCPs per step. Net effect: CloudPEASS is strong at recognizing known single-hop patterns for the current identity but systematically under-approximates true reachable privilege in complex estates where risk emerges from cross-principal composition, transitive role chaining, and contextual/temporal factors it does not model.

The gaps are methodological (not merely feature gaps) by approaching with the single-principal focus, lack of transitive reasoning, missing fuzzing and simulation, inadequate temporal / context modeling, and weak aggregation. As the result, these produce predictable, repeatable blind spots in real cloud environments.

\subsection{Design Targets Derived from the Gaps}
To directly answer the deficiencies above, SkyEye should be framed and evaluated against the following explicit objectives:
\begin{enumerate}
    \item \textbf{Multi-Principal Graph Construction:} Build a directed graph $G = (V, E)$ where V are principals/resources and E encode assume-role edges, resources-policy grants, and service-linked transitions. Support coss-account edges.
    \item \textbf{Transitive Path Enumeration with State:} Perform bounded search over G where each hop updates a stateful capability set (recompute $C(p_{i+1})$ ). Detect cycles and enforce depth/timeouts.
    \item \textbf{Fuzzing-Augmented Discovery:} Augment known checks with hypothesis generators that propose action/resource/condition permutations, prioritized by service semantics and prior observations; probe with safe, non-destructive calls
    \item \textbf{Temporal \& Version Awareness:} Ingest policy version; model session policies and short-lived credentials; compute diffs $\Delta_C$ across time windows to surface ephemeral elevation.
    \item \textbf{ATT\&CK-Aligned Prioritization:} Map discovered edges and complete paths to ATT\&CK techniques (e.g., credential access, persistence, ex-filtration), derive a path score combining impact, path length, blast radius, sensitivity, and detectability.
    \item \textbf{Full-Stack Aggregation:} Compose identity policies, resource policies, permission boundaries, SCPs, condition keys, and region/service constraints per hop, not just at origin.
\end{enumerate}

\subsection{Pratical Implications}
\begin{itemize}
    \item \textbf{For Red Team:} Prior stacks are excellent for tactics (recon, static auditing, scenario simulation) but lack a strategy engine for identity path-finding. SkyEye provides that engine.
    \item \textbf{For Blue Team:} Outputs mapped to ATT\&CK with quantified blast radius enable defensible prioritization and regression testing (e.g., “did the risky path disappear after remediation?”).
    \item \textbf{For Governance:} Static gates (Cloudsplaining) remain valuable, but SkyEye’s reachability analysis closes the loop: not just “is the policy risky?” but “can the attacker reach the risky capability from plausible starting points?”
\end{itemize}

\section{Conclusion}
The comprehensive examination of prior-art models and frameworks for AWS IAM enumeration and privilege escalation reveals a rapidly evolving yet still fundamentally constrained landscape. Most existing frameworks, including PACU, CloudPEASS, CloudFox, and enumerate-iam concentrate on single-principal IAM enumeration, confining their perspective to the permissions and visibility of the authenticated identity. This approach restricts the automated discovery of multi-hop or chained escalation paths, which are increasingly exploited in complex cloud environments through transitive trust and interplay between resource-based and identity-based policies.\\

The reliance on static, signature-driven detection further limits the adaptability of these tools. While matching known permissions to curated escalation techniques is effective for documented risks, it fails to identify novel or emerging attack vectors, especially as AWS services evolve rapidly and adversarial tactics become more innovative. Additionally, the lack of automated privilege graphing and cross-principal enumeration severely hampers the discovery of complex escalation scenarios involving multiple principals to reveal the complete vision. Manual investigation can theoretically address these gaps, but it is impractical at the scale of modern cloud deployments.\\

Audit-focused and compliance-focused framework, such as ScoutSuite and Cloudsplaining, are valuable for identifying policy misconfigurations and enforcing least-privilege practices. However, their static, configuration-centric designs are not intended to simulate dynamic, real-world attack chains or assess the practical exploitability of discovered risks. Cloud attack simulation framework such as Stratus Red Team excels at validating detection and response processes for known attacks, but their utility is bounded by predefined scenarios and lacks proactive discovery of undocumented escalation paths.\\

In summary, the current generation of IAM enumeration and privilege escalation framework is indispensable for baseline security and compliance, but falls short in enabling comprehensive adversary simulation and proactive defense. The absence of dynamic, context-aware analysis, automated privilege graphing, and exploratory methodologies represents a significant gap that is increasingly exploited in real-world breaches, as evidenced by incidents like Capital One. Addressing these challenges requires a new re-invention of frameworks that integrate context-aware privilege analysis, automated cooperative enumeration by cross-principal approach, and hybrid methodologies combining static and dynamic techniques. Such advancements will be essential for defenders and red teamers alike, providing deeper, actionable insights and enabling more resilient cloud security postures. The following chapters will introduce a new framework with our proposed models designed to fill these critical gaps in the current landscape.
\newpage
\chapter{SkyEye Framework and Proposed Models}
In this chapter, we will present the SkyEye - a cutting-edge cooperative multi-principal IAM enumeration framework, along with its proposed and developed models, which significantly demonstrate how SkyEye differs from prior-art models and frameworks in enumerating IAMs of cloud user principals from the black-box approach.

\section{Design Overview of SkyEye Framework}
\subsection{The Core Idea and Motivation}
The rapid proliferation of cloud computing, in particular, the adoption of AWS Identity and Access Management (IAM), has fundamentally transformed the landscape of access control, privilege management, and attack surface exposure in enterprise environments. As organizations increasingly leverage multi-principal, multi-role architectures across vast AWS estates, the complexity and opacity of IAM configurations have grown disproportionately. This complexity introduces significant risk: incomplete visibility into permissions, latent privilege escalation paths, and policy misconfigurations are now among the most common root causes of cloud security incidents.\\

Traditional IAM enumeration techniques, rooted in single-principal enumeration or isolated session fuzzing, are not suitable to this reality. These approaches are inherently limited by the visibility granted to any one set of credentials. As a result, they routinely underestimate the true extent of accessible privileges, overlook transitive trust relationships, and fail to synthesize the aggregate risk posed by multiple, distributed permissions across users and roles. The consequence is a persistent gap between what security teams believe is possible within their cloud environment, and what adversaries or insider threats can actually achieve.

\subsection{The Main Objectives}
SkyEye is introduced in this thesis as a direct response to these emerging challenges. Its overarching objective is to enable comprehensive, accurate, and actionable IAM enumeration within AWS cloud environments, empowering both offensive and defensive security teams to discover, analyze, and mitigate privilege-based risks at scale.\\

Specifically, SkyEye Framework aims to:

\begin{itemize}
    \item Minimize false negatives in privilege and policy discovery through cooperative, multi-principal enumeration.
    \item Reveal hidden privilege escalation paths by modeling both direct and indirect trust relationships between users and roles.
    \item Provide operationally actionable intelligence by mapping discovered permissions to real-world adversarial tactics (via MITRE ATT\&CK) and assigning risk levels.
    \item Support both research and enterprise use cases through an extensible, modular, and automation-friendly architecture.
\end{itemize}

\subsection{Architectural Design Principles}
To fulfill these objectives, SkyEye’s design is guided by the following architectural principles:

\paragraph{a. Cooperative Multi-Principal Enumeration and Data Synthesis\\}

At the core of SkyEye is the principle of cooperative multi-principal enumeration, which is a paradigm shift from traditional, siloed principal-specific IAM enumeration. By orchestrating parallel enumeration sessions across all available AWS credentials and roles, SkyEye aggregates and synthesizes IAM data from every perspective, constructing a unified, high-fidelity map of the effective permissions landscape.

\paragraph{b. Modularity and Extensibility\\}

SkyEye is built with a modular architecture, where each major model (e.g., Cross-Principal IAM Enumeration Model - CPIEM, Transitive Cross-Role Enumeration Model - TCREM, Deep Comparison Model, etc.) is implemented as a discrete component. This modularity enables rapid extension such as the addition of new enumeration modes, support for other cloud providers, or integration with external analytics platforms, while maintaining codebase clarity and maintainability.

\paragraph{c. Concurrency and Performance Optimization\\}

Given the potentially vast number of principals and roles in an enterprise AWS environment, SkyEye leverages multi-threading and asynchronous operations to maximize enumeration speed and minimize operational overhead. Intelligent optimization, such as early termination when highly privileged permissions (e.g., iam:GetAccountAuthorizationDetails) are detected, reduces redundant API calls and limits detectable traces in audit logs.

\paragraph{d. Risk-Driven and Adversarial-Aware Intelligence\\}

SkyEye integrates a comprehensive dataset mapping all enumerated AWS actions to MITRE ATT\&CK tactics, techniques, and sub-techniques, and classifies each action by severity and abuse methodology. This mapping bridges the gap between raw permission enumeration and real-world attack simulation or defense, enabling users to prioritize remediation, simulate adversarial behavior, and inform incident response.

\paragraph{e. Transparency and Automation\\}

All enumeration processes, logic chains, and results are fully transparent and output in machine-readable formats [JSON], supporting downstream integration, auditability, and automated analysis or remediation workflows.

\subsection{How SkyEye addresses key challenges?}
SkyEye’s design directly addresses the core challenges endemic to cloud IAM enumeration and security:

\begin{enumerate}
    \item \textbf{Fragmented Permission Visibility}: By coordinating simultaneous sessions across all credentials and roles, SkyEye reconstructs the full privilege topology, even when key permissions are distributed across multiple principals. This approach mitigates the “blind spots” inherent in single-principal models.
    \item \textbf{Transitive Privilege Escalation Detection:} Through the Transitive Cross-Role Enumeration Model (TCREM), SkyEye recursively maps and simulates all possible trust relationships, surfacing both direct and indirect privilege escalation paths that would otherwise remain hidden.
    \item \textbf{Operational Efficiency and Stealth:} By terminating enumeration early when full-account visibility is detected and minimizing redundant operations, SkyEye improves both performance and stealth, reducing the risk of detection by defenders or cloud service providers.
    \item \textbf{Actionable, Context-Rich Intelligence:} SkyEye’s integration of MITRE ATT\&CK mapping, severity classification, and abuse methodologies ensures that every discovered permission is not just catalogued, but contextualized for practical attack simulation or risk mitigation.
    \item \textbf{Extensibility for Research and Practice:} The modular, open design of SkyEye supports rapid adaptation to new research advances (e.g., support for new clouds, machine learning-based risk ranking) and integration into enterprise security pipelines.
\end{enumerate}

\subsection{Impact and Scope Expectation}
By operationalizing these design principles, SkyEye delivers significant value to both the research and practitioner communities:
\begin{itemize}
    \item \textbf{For Security Researchers:} SkyEye provides a rigorous, extensible framework for studying privilege relationships, policy misconfigurations, and attack vectors in complex AWS environments, enabling new research on privilege escalation, defense-in-depth, and cloud security analytics.
    \item \textbf{For Practitioners (Red/Blue Teams, Auditors):} SkyEye enables more accurate, efficient, and actionable IAM risk assessments, supporting both proactive defense (policy hardening, least privilege enforcement) and offensive validation (attack simulation, red teaming).
    \item \textbf{For the Broader Cloud Security Ecosystem:} SkyEye establishes a new standard for IAM enumeration, risk mapping, and adversarial modeling in cloud environments, bridging the gap between detection and defense, and laying a robust foundation for future innovation.
\end{itemize}

In summary, SkyEye’s architectural vision is to transcend the inherent limitations of conventional IAM enumeration, establishing a blueprint for resilient, comprehensive, and actionable cloud access governance in the era of multi-principal, multi-cloud operations.

\section{Cross-Principal IAM Enumeration Model (CPIEM)}
From this section, we will delve deeper into proposed models and capabilities of SkyEye framework. The original idea of the Cross-Principal IAM Enumeration Model (CPIEM) [Fig. \ref{fig:42}, \ref{fig:43}] came from the difficulty that occurs with the single-principal IAM enumeration approach. In the enumeration phase of the penetration testing process, penetration testers often gather multiple AWS credentials in the format: AccessKey, SecretKey, Session Token. However, it could only perform separate-principal or single-principal IAM enumeration from each user session, leading to false negatives due to limitation of principal-specific IAM entitlement vision. To resolve this limitation, the Cross-Principal IAM Enumeration Model (CPIEM) [Fig. \ref{fig:42}, \ref{fig:43}]  was proposed and developed to efficiently perform advanced IAM enumeration across multiple user principals within the AWS Account Id, to complement each user's IAM vision context. By coordinating available sessions of each valid credential simultaneously, it can:

\begin{itemize}
\item Discover hidden permissions
\item Reveal a more accurate and complete IAM policy landscape for each IAM entity
\item Minimize false negatives that typically occur with single-principal IAM enumeration
\end{itemize}

But before delving deeper into the enumeration model, we will firstly start with the discussions regarding single-principal or separate-principal IAM enumeration capabilities.

\subsection{What is Single-Principal IAM Enumeration Model (SiPIEM)?}
\sloppy Single-Principal IAM Enumeration is the capability to scan single AWS credentials separately, utilizing principal-specific IAMs of one single user principal at a time. In recently published framework and tool, it was often integrated the capability of utilizing relevant IAMs permissions such as: \texttt{iam:ListGroupsForUser}, \texttt{iam:ListGroups}, \texttt{iam:GetGroup}, \texttt{iam:ListUserPolicies}, \texttt{iam:GetUserPolicy}, \texttt{iam:ListAttachedUserPolicy}, \texttt{iam:ListAttachedGroupPolicy}, \texttt{iam:ListEntitiesForPolicy}, \texttt{iam:GetGroupPolicy}, \texttt{iam:ListGroupPolicies}, \texttt{iam:ListRoles}, \texttt{iam:ListRolePolicies}, \texttt{iam:GetRolePolicy}, \texttt{iam:ListAttachedRolePolicy}, \texttt{iam:ListPolicyVersions}, \texttt{iam:GetPolicy}, \texttt{iam:GetPolicyVersion}, etc. to expose the complete IAM vision context of the current user from black-box perspective. Moreover, it also coordinates the fuzzing capability and the permissions simulation capability by leveraging iam:SimulatePrincipalPolicy permission to reveal the IAM vision context. In SkyEye framework, we build upon the single-principal IAM enumeration model discussed in Chapter 3. Our proposed framework integrates these models as supplement components while introducing novel mechanisms for the cross-principal IAM enumeration model. The fuzzing capability and permissions simulation capability are also integrated into this single-principal IAM enumeration model only; the reason behind is due to a huge amount of time required to finalize these tasks for multiple AWS credentials.\\

\begin{figure}[htbp]
    \centering
    \includegraphics[width=1.0\linewidth]{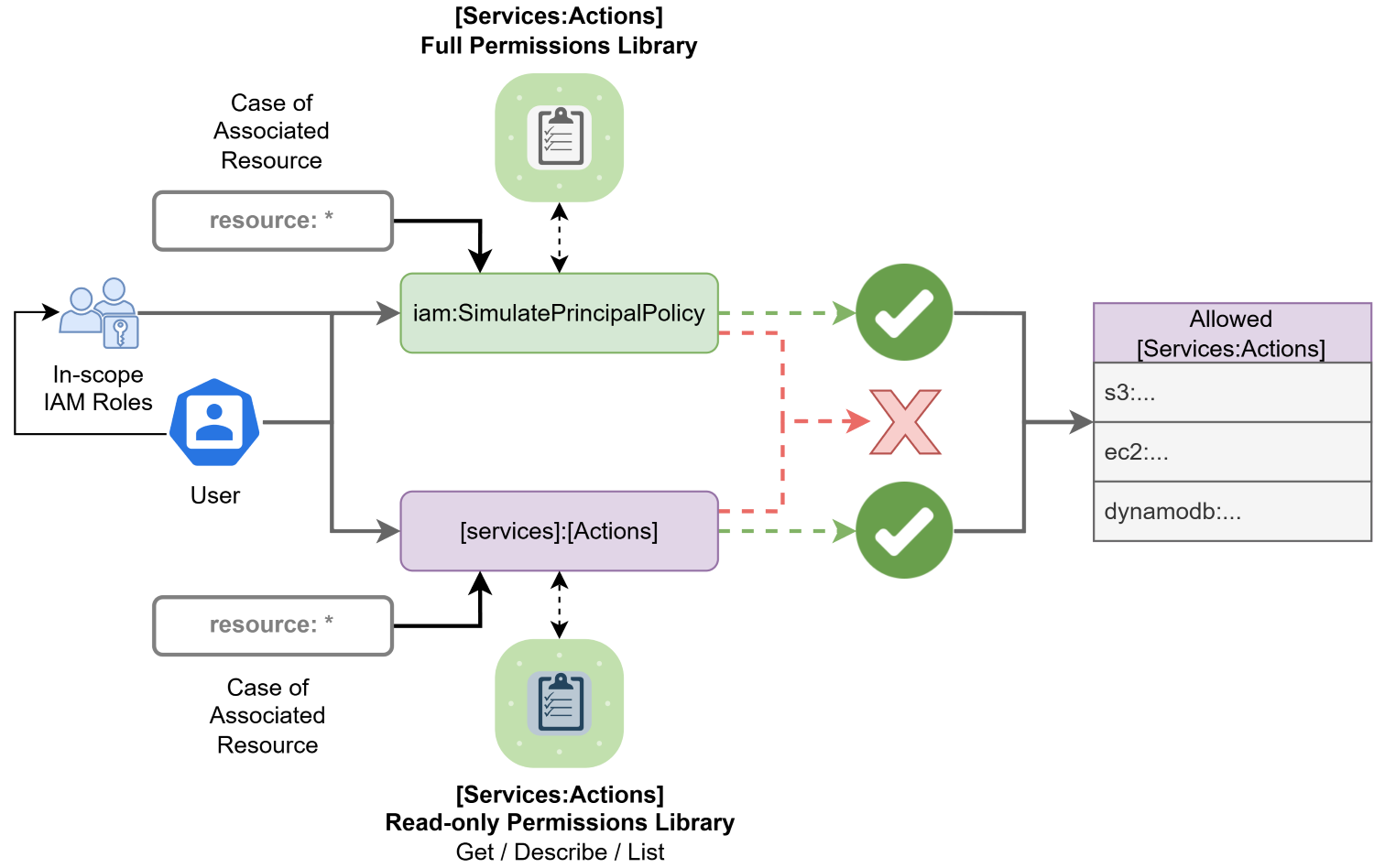}
    \caption{Permissions Simulation and Fuzzing models}
    \label{fig:41}
\end{figure}

\sloppy Permissions simulation capability in [Fig. \ref{fig:41}] will be performed by leveraging the \texttt{iam:SimulatePrincipalPolicy} permission. This permission will support the enumeration process by simulating how a set of IAM policies attached to an IAM entity works with a list of API operations and AWS resources to determine the policies' effective permissions. The entity can be an IAM user, group, or role; and if a user is specified, then the simulation also includes all of the policies that are attached to groups that the user belongs to. This model will check if the user principal’s session is allowed to perform iam:SimulatePrincipalPolicy, if yes, the model will leverage this permission to simulate all AWS actions which are nearly 20,000 actions, to understand which actions the user principal can perform. Moreover, since iam:SimulatePrincipalPolicy can only simulate the user principal and inherited permissions from in-scope IAM groups that the user belongs to, it lacks the capability of simulating the inherited permissions from the in-scope IAM roles that the user could perform assumption directly or indirectly. The model will actively incorporate with the Transitive Cross-Role Enumeration Model (TCREM) - which is one of the core models of SkyEye and will be discussed in next section, to gain the understanding of in-scope IAM roles, and leverage iam:SimulatePrincipalPolicy to target those in-scope IAM roles to return a most complete result.\\

If the user principal does not have sufficient permission to perform iam:SimulatePrincipalPolicy, the model will switch directly to initialize the fuzzing capability [Fig. \ref{fig:41}]. The fuzzing capability will be performed by actively invoking the AWS API of nearly 8000 AWS read-only actions to understand which actions the user principal can perform. Only AWS read-only actions will be undertaken in the fuzzing capability, due to the fact that almost all the read-only actions will not require the essential parameters and values to be provided before the execution. 

\subsection{What is Separate-Principal IAM Enumeration Model (SePIEM)?}
Similarly to Single-Principal IAM Enumeration, Separate-Principal IAM Enumeration can support scanning by principal-specific IAMs, but is extended further by the capability of scanning multiple AWS credentials separately. In SkyEye framework, we also build upon the idea of single-principal IAM enumeration model discussed in Chapter 3 as a supplement component, to diversify the capabilities of the framework.

\subsection{What is the limitation of the single-principal or separate-principal IAM enumeration model?}
In this section, we will point out several scenarios where single-principal or separate-principal IAM enumeration models are highly restricted in enumerating a complete IAM vision context.
\paragraph{Scenario A: Retrieval of inline policies for user principal\\}
- \textbf{IAM Action Chain:} iam:ListUserPolicies $\rightarrow$ iam:GetUserPolicy
\begin{itemize}
\item \textbf{User A} is permitted to perform iam:ListUserPolicies for a specified user, allowing this uer to list the names of all inline policies attached to that user. However, User A does not have iam:GetUserPolicy, so cannot retrieve the actual policy documents.
\item \textbf{User B} is permitted to perform iam:GetUserPolicy, allowing this user to retrieve the policy document if given the username and policy name. However, User B does not have iam:ListUserPolicies, so does not know which policy names exist for a user.
\item Therefore, User A can list the policy names but not see their content, and User B can retrieve the policy document but does not know which policies to request.
\end{itemize}

\textbf{Conclusion:} Due to the separation of IAM permissions across users, neither user operating independently can retrieve the complete set of inline policies (including detailed documents) for a user; both actions are required in sequence, but split across users. The lack of a single user's ability to execute the full chain - discovering policy name of user's inline policies, and enumerating and retrieving full policy documents - prevents complete visibility.

\paragraph{Scenario B: Retrieval of attached managed policies for user principal\\}
- \textbf{IAM Action Chain:} iam:ListAttachedUserPolicies $\rightarrow$ iam:ListPolicyVersions or iam:GetPolicy $\rightarrow$ iam:GetPolicyVersion
\begin{itemize}
\item \textbf{User B} has permissions to perform iam:ListAttachedUserPolicies, allowing User B to list all managed policies attached to a specific IAM user. However, User B cannot perform iam:GetPolicy or iam:ListPolicyVersions and iam:GetPolicyVersion, so cannot retrieve the detailed documents or at least current version of those policies.
\item \textbf{User A} has permissions to perform iam:ListPolicyVerion, allowing User A to retrieve a managed policy's metadata and current version if provided with the policy ARN. However, User A cannot perform iam:ListAttachedUserPolicies, so cannot enumerate which policies are attached to any user, nor can User A retrieve complete documents if provided policy ARN along with its version due to the lack of iam:GetPolicyVersion.
\item \textbf{User C} has permission to perform iam:GetPolicyVersion, enabling User C to retrieve the complete documents of a specific policy version if provided with the ARN of policy and version identifier. However, User C cannot enumerate attached policies or retrieve policy metadata to understand its current version.
\item As a result, User B can enumerate which managed policies are attached to a user, but cannot retrieve their contents or versions. User A can retrieve policy metadata and its current version, but cannot enumerate which policies are attached to users, nor retrieve complete documents of given policy ARN and its current version. User C can retrieve the contents of a specific managed policy version, but only if provided with the policy ARN and the current version ID. Therefore, no single user can enumerate and retrieve the full content of all managed policies attached to the groups to which a user belongs to.
\end{itemize}

\textbf{Conclusion:} Due to the separation of IAM permissions across users, neither user operating independently can enumerate and retrieve the complete set of all attached managed policies (including detailed documents) for a user. The lack of a single user's ability to execute the full chain - discovering ARN of managed policies attached to a given user, and retrieving full policy documents of those attached managed policies - prevents complete visibility.

\paragraph{Scenario C: Retrieval of inline policies for in-scope IAM groups\\}
- \textbf{IAM Action Chain:} iam:ListGroupsForUser or [iam:ListGroups and iam:GetGroup] $\rightarrow$ iam:ListGroupPolicies $\rightarrow$ iam:GetGroupPolicy
\begin{itemize}
\item \textbf{User B} is permitted to perform iam:ListGroupPolicies and iam:GetGroupPolicy, enabling User B to enumerate the inline policies attached to a specified IAM group and retrieve the document of a specific inline policy embedded within that group. However, User B does not have permission to perform iam:ListGroupsForUser, and therefore cannot determine the IAM groups to which they themselves (or any other user) belong.
\item \textbf{User A} is permitted to perform iam:ListGroupsForUser, enabling User A to enumerate all IAM groups to which a specified IAM user belongs. However, User A does not have permission to perform iam:ListGroupPolicies or iam:GetGroupPolicy, so it cannot list or retrieve the inline policies attached to any IAM group.
\item As a result, User B can enumerate and retrieve inline policies for any group, but does not know which groups they belong to (i.e., IAM groups a particular user is a member of). User A, on the other hand, can determine group membership for users, but cannot enumerate or retrieve inline policies (including complete documents) for those groups. Therefore, User B cannot retrieve any information about inline policies for the IAM groups to which they themselves belong, because they lack the ability to determine their own group membership. Similarly, User A can only determine to which groups a user belongs to, but cannot retrieve the details of inline policies for those groups.
\end{itemize}

\textbf{Conclusion:} Due to the separation of IAM permissions across users, neither user operating independently can enumerate and retrieve the complete set of inline policies (including detailed documents) for all in-scope IAM groups (i.e., the groups to which users they have access belong). The lack of a single user's ability to execute the full chain - discovering group membership and then enumerating policy names of in-scope groups' inline policies, and retrieving full policy documents - prevents complete visibility.

\paragraph{Scenario D: Retrieval of attached managed policies for in-scope IAM groups\\}
- \textbf{IAM Action Chain:} iam:ListGroupsForUser or [iam:ListGroups and iam:GetGroup] $\rightarrow$ iam:ListAttachedGroupPolicies $\rightarrow$ iam:ListPolicyVersion or iam:GetPolicy \\
$\rightarrow$ iam:GetPolicyVersion
\begin{itemize}
\item \sloppy \textbf{User E} can perform iam:ListGroupsForUser, enabling them to list all IAM groups to which a specified user belongs. However, User E cannot perform \texttt{iam:ListAttachedGroupPolicies} and so cannot determine which managed policies are attached to those groups.
\item \textbf{User B} can perform iam:ListAttachedGroupPolicies, allowing them to list all managed policies attached to a specified group, but cannot enumerate which groups a user belongs to (iam:ListGroupsForUser).
\item \textbf{User A} can perform iam:GetPolicy and iam:GetPolicyVersion, enabling them to retrieve policy documents and its versions, but cannot enumerate which groups the user principal belongs to or which policies belong to which groups.
\item As a result, User E can enumerate the IAM groups to which a specified user belongs, but cannot determine which managed policies are attached to those groups. User B can enumerate all managed policies attached to a specified group, but cannot identify which groups are associated with a specific user. User A can retrieve the complete documents and its current version of managed policies if provided with the ARN of policy, but cannot enumerate groups or determine group membership. Therefore, no single user can enumerate and retrieve the complete set of managed policy documents attached to the groups to which a user belongs.
\end{itemize}

\textbf{Conclusion:} Due to the separation of IAM permissions across users, neither user operating independently can enumerate and retrieve the complete set of attached managed policies (including detailed documents) for all in-scope IAM groups (i.e., the groups to which users they have access to belong). The lack of a single user's ability to execute the full chain - discovering group membership and then discovering ARN of managed policies attached to in-scope IAM groups, and retrieving full policy documents of those attached managed policies - prevents complete visibility.

\paragraph{Scenario E: Retrieval of inline policies for in-scope IAM roles\\}
- \textbf{IAM Action Chain:} iam:ListRoles $\rightarrow$ iam:ListRolePolicies $\rightarrow$ iam:GetRolePolicy
\begin{itemize}
\item \textbf{User A} can perform iam:ListRoles, allowing them to enumerate all IAM roles in the account, but cannot list or retrieve inline policies for those roles.
\item \textbf{User C} can perform iam:ListRolePolicies, allowing them to list the names of inline policies for a specific role, but cannot enumerate all roles (iam:ListRoles) or retrieve policy documents.
\item \textbf{User D} can perform iam:GetRolePolicy, enabling them to retrieve the document of a specific inline policy for a given role and policy name, but cannot list roles or policies.
\item As a result, User A can enumerate all in-scope IAM roles in the account, but cannot list or retrieve inline policies (including complete documents) for those roles. User C can list the names of inline policies attached to a specific role, but cannot enumerate all in-scope IAM roles or retrieve policy documents. User D can retrieve the complete document of inline policies for a given role and policy name, but cannot list roles or enumerate policy names. Therefore, no single user can enumerate and retrieve the complete content of all inline policies attached to all in-scope IAM roles.
\end{itemize}

\textbf{Conclusion:} Due to the separation of IAM permissions across users, neither user operating independently, can enumerate and retrieve the complete set of inline policies (including detailed documents) for all in-scope IAM roles (i.e., the roles that users they have access to can assume). The lack of a single user's ability to execute the full chain - discovering in-scope IAM roles that the user can assume, and then enumerating policy names of in-scope roles' inline policies, and retrieving full policy documents - prevents complete visibility.

\paragraph{Scenario F: Retrieval of attached managed policies for in-scope IAM roles\\}
- \textbf{IAM Action Chain:} iam:ListRoles $\rightarrow$ iam:ListAttachedRolePolicies $\rightarrow$ iam:ListPol-
icyVersion or iam:GetPolicy $\rightarrow$ iam:GetPolicyVersion
\begin{itemize}
\item \textbf{User B} can perform iam:ListRoles, allowing enumeration of all IAM roles. However, User B cannot list the attached policies for these roles.
\item \textbf{User D} can perform iam:ListAttachedRolePolicies, enabling them to list all managed policies attached to a specified role, but cannot enumerate roles or retrieve policy documents.
\item \textbf{User A} can perform iam:GetPolicy and iam:GetPolicyVersion, enabling them to retrieve the document and versions of a managed policy if given the ARN, but cannot list roles or attached policies.
\item As a result, User B can enumerate all in-scope IAM roles in the account, but cannot list the managed policies attached to those roles, nor retrieve the complete documents of those managed policies. User D can list all managed policies attached to a specified role, but cannot enumerate all in-scope IAM roles or retrieve the complete documents of those managed policies. User A can retrieve the complete documents and its current version of managed policies if provided with the ARN of policy, but cannot enumerate in-scope IAM roles or determine which policies are attached to in-scope IAM roles. Therefore, no single user can enumerate and retrieve the full content of all managed policies attached to all in-scope IAM roles.
\end{itemize}

\textbf{Conclusion:} Due to the separation of IAM permissions across users, neither user operating independently, can enumerate the complete set of attached managed policies (including detailed documents) for all in-scope IAM roles (i.e., the roles that users to which they have access can assume). The lack of a single user's ability to execute the full chain - discovering in-scope IAM roles that the user can assume, and then discovering ARN of managed policies attached to in-scope IAM roles, and retrieving full policy documents of those attached managed policies, prevents complete visibility.

\paragraph{Scenario G: iam:GetAccountAuthorizationDetails to complement other users' IAM entitlement visibility\\}
- \textbf{IAM Action Chain:} iam:GetAccountAuthorizationDetails
\begin{itemize}
\item \textbf{User B} cannot perform any IAM actions to reveal its IAM entitlement visibility
\item \textbf{User C} cannot perform any IAM actions to reveal its IAM entitlement visibility
\item \textbf{User A} can perform iam:GetAccountAuthorizationDetails , empowering them to retrieve all information of IAMs in their AWS Account Id, including their IAM context.
\item As a result, User A can reveal its full IAM context, but User B and User C cannot reveal any information regarding permissions that they are allowed to perform, and resources that they are allowed to interact with. 
\end{itemize}

\textbf{Conclusion:} Due to the separation of IAM permissions across users, if users are operating independently, User A cannot support User B and User C to enumerate the complete set of their IAM entitlement visibility.\\

These 7 scenarios illustrate the limitations of the traditional single-principal IAM enumeration approach in a cloud environment. By not having sufficient permissions in each phase of the IAM chain, it could lead to the failure of fully revealing any of the inline policies and attached managed policies for IAM users, in-scope IAM groups, in-scope IAM roles, in-scope IAM policies during the enumeration process.

\subsection{How SkyEye Framework and CPIEM mitigate these limitations?}

\begin{figure}[htbp]
    \centering
    \includegraphics[width=1.0\linewidth]{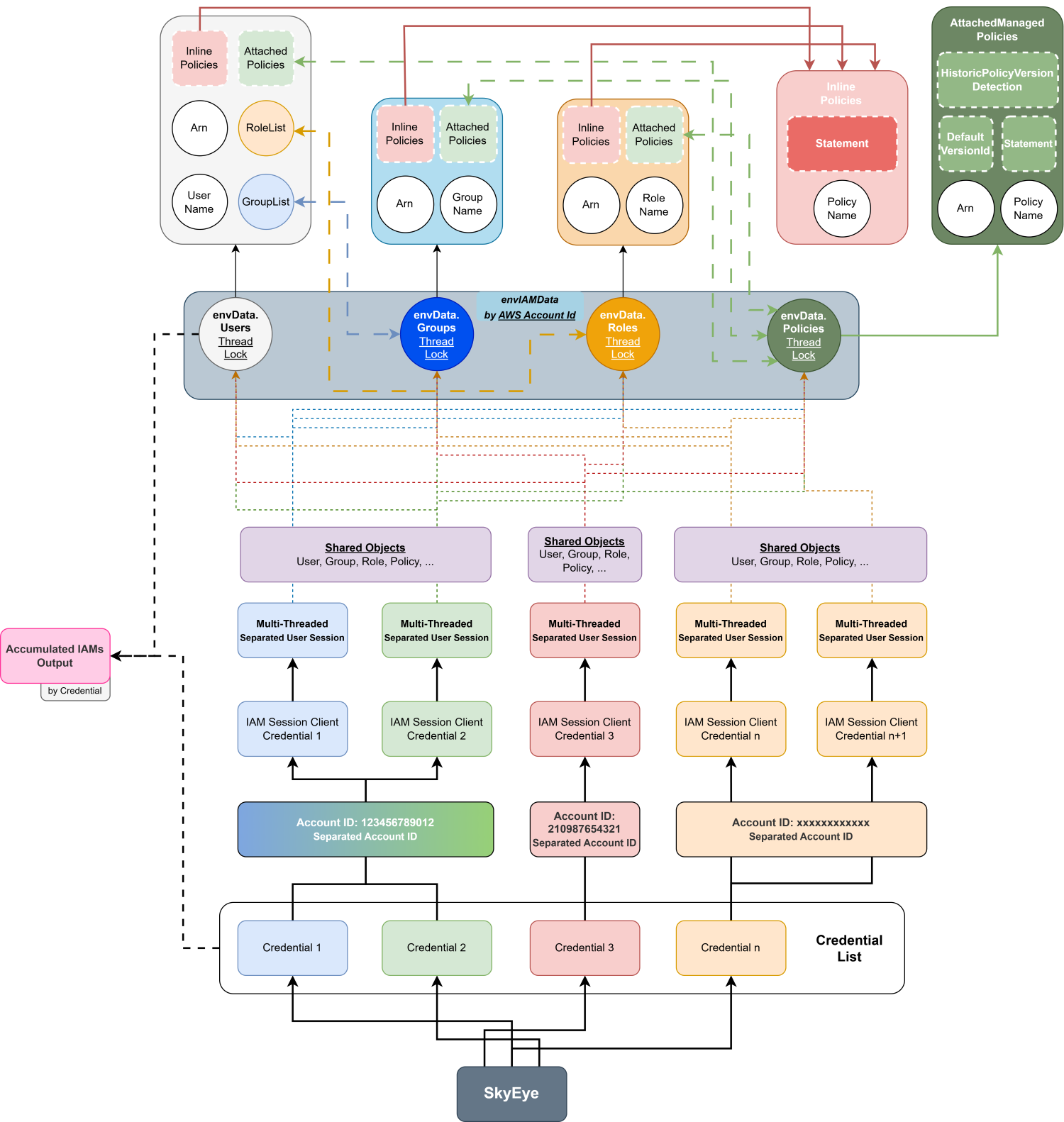}
    \caption{Core of SkyEye - Cross-Principal IAM Enumeration Model (CPIEM)}
    \label{fig:42}
\end{figure}

\begin{figure}[htbp]
    \centering
    \includegraphics[width=1.0\linewidth]{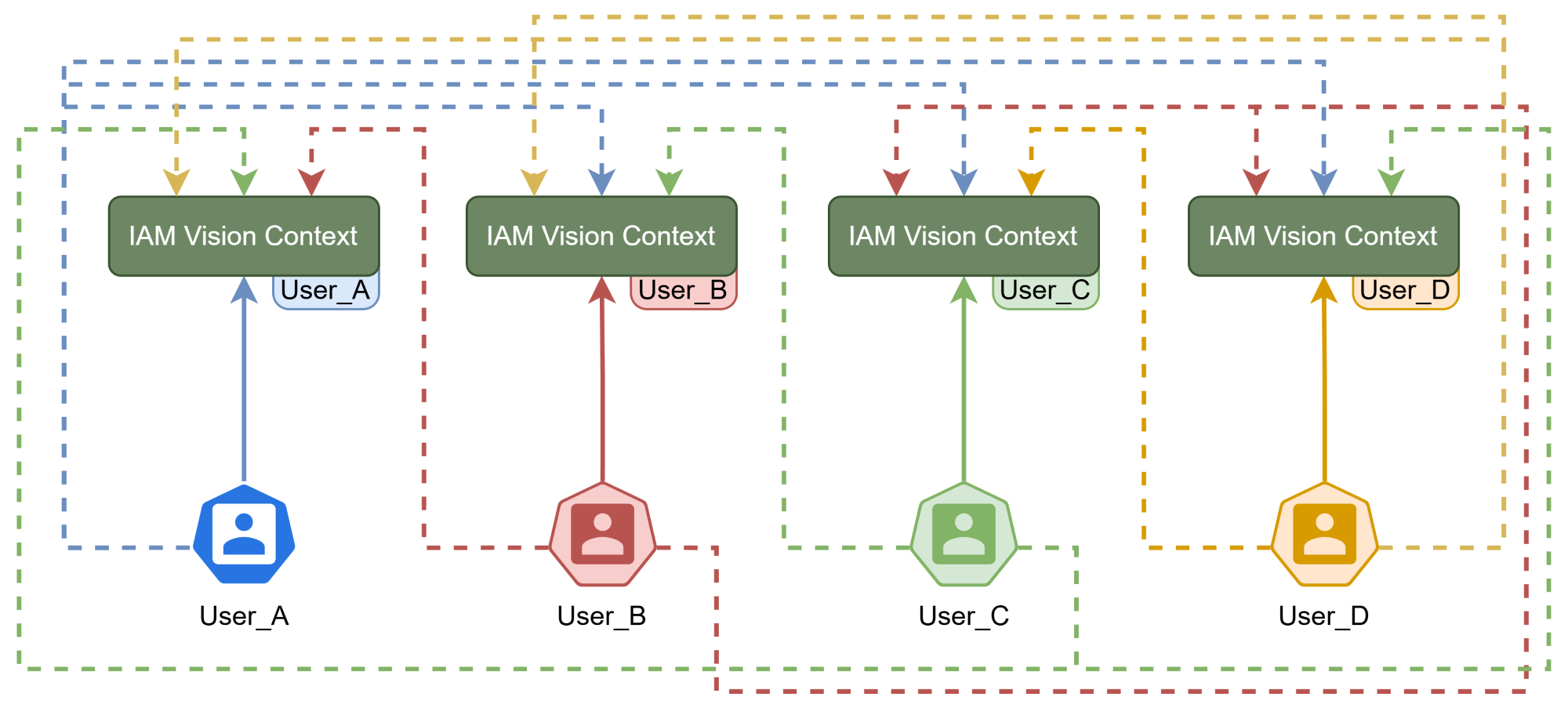}
    \caption{The Interconnection in Cross-Principal IAM Enumeration Model (CPIEM)}
    \label{fig:43}
\end{figure}

\begin{figure}[ht]
    \centering
    \includegraphics[width=1.0\linewidth]{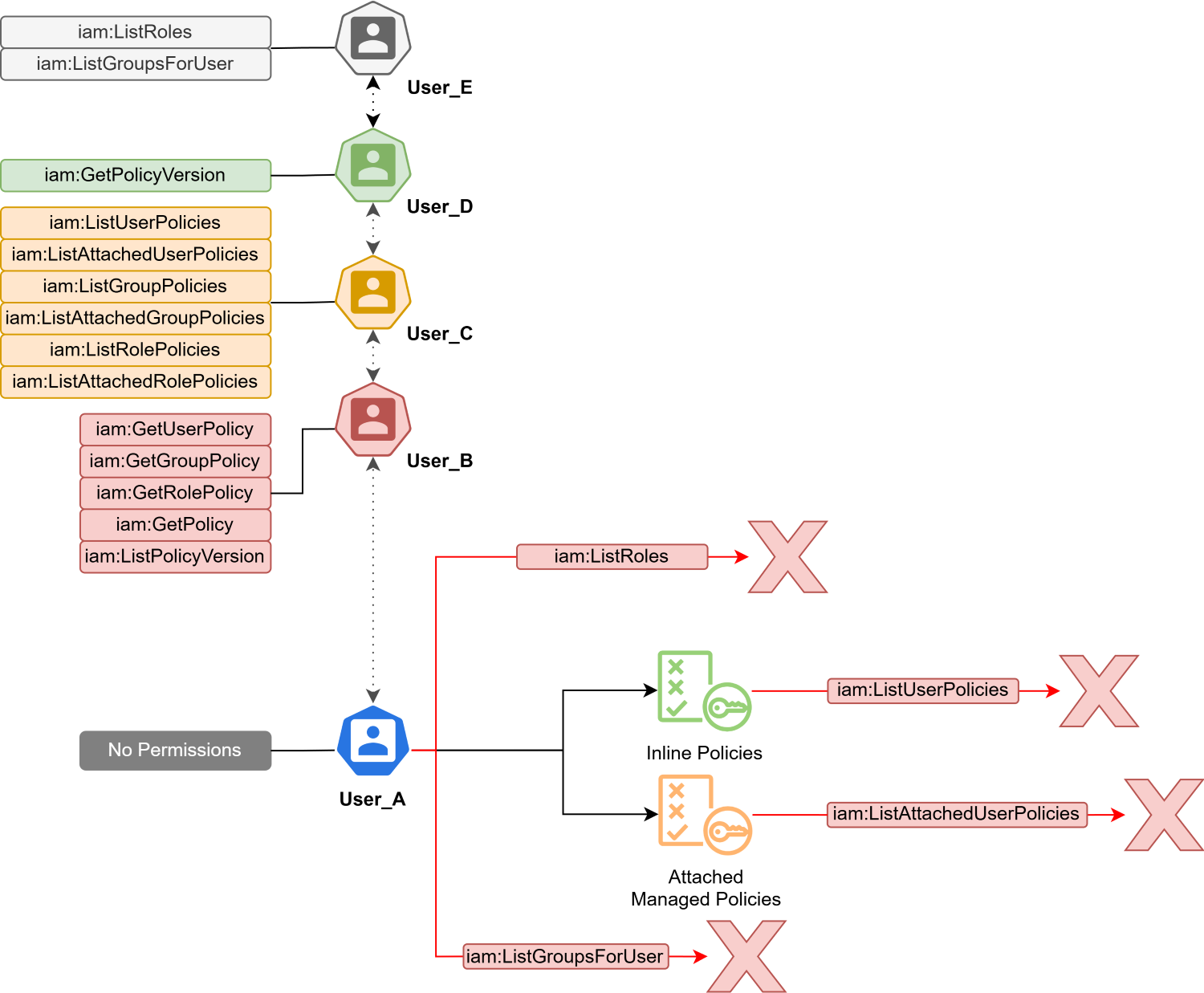}
    \caption{Cross-Principal IAM Enumeration Model - Example Scenario - Stage 1}
    \label{fig:44}
\end{figure}

\begin{figure}[htbp]
    \centering
    \includegraphics[width=1.0\linewidth]{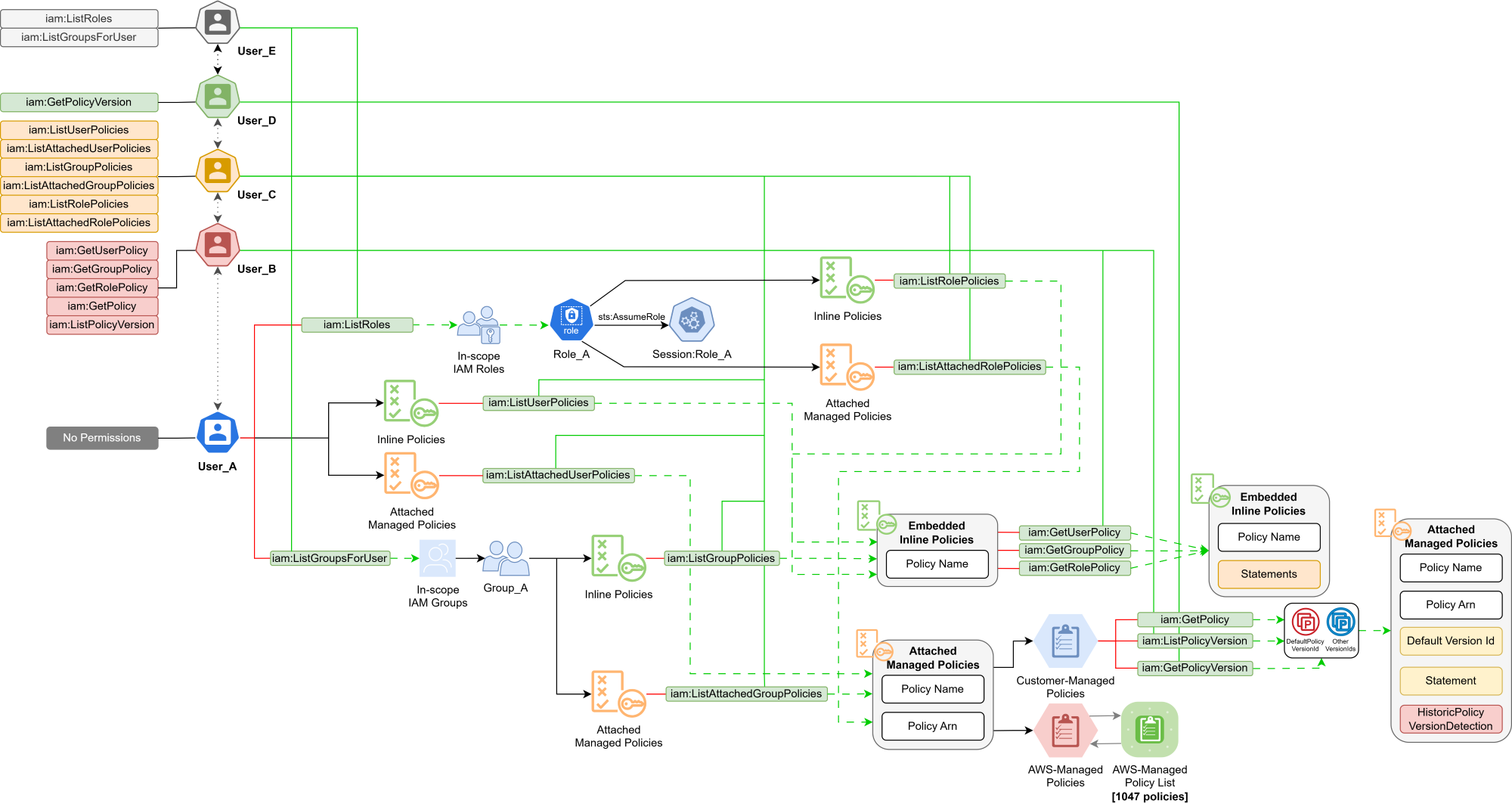}
    \caption{Cross-Principal IAM Enumeration Model - Example Scenario - Stage 2}
    \label{fig:45}
\end{figure}

Instead of depending on the self-access IAM entitlement visibility of single user to reveal its IAM context, and to fully understand what permissions and what resources that the user is allowed to perform and interact with, sometimes leading to false negatives when user could perform some specific permissions to specific resources but could not have the situational awareness on that, the cross-principal IAM enumeration model in [Fig. \ref{fig:42}, \ref{fig:43}], which is the core capability of SkyEye framework, is designed to tackle this limitation by involving and correlating simultaneously multiple valid credentials to continually expose the complete IAM visibility of each user principal.\\\\
Initially, SkyEye will validate the provided AWS credentials, and split them into separate "AWS Account Id" clusters, ensuring that only the users from similar AWS Account Id, will be involved in complementing other users' IAM visibility from similar account id.\\\\
SkyEye will then construct multi-threaded mechanism to run the enumeration from each user’s session simultaneously and perform cross-principal enumerations across ARNs of each user, and interact with the shared envIAMData objects - separated by AWS Account Id, to store the IAM Users, IAM Groups, IAM Roles, IAM Policies gathered from each user's session in run-time.\\\\
Shared envIAMData objects will act as independently shared storage for multiple users from each different AWS Account Id, to support each user’s session in adding new IAM objects or complementing existing ones if missing components are identified in those objects before initiating the complementation process.\\\\
In addition to the IAM chains to fully reveal inline policies and attached managed policies for IAM users, in-scope IAM groups, in-scope IAM roles, in-scope IAM policies, during the enumeration process, if the iam:GetAccountAuthorizationDetails permission is detected in run-time to be executable by at least one user’s session from similar AWS Account Id, the model will immediately terminate all user’s session come from that AWS Account Id, and utilize the iam:GetAccountAuthorizationDetails permission to retrieve full IAM context of that AWS Account Id, and distribute the correspondent result to the user principal that involved in the IAM enumeration for that AWS Account Id. This approach will reduce significantly 95\% of the entire scanning process, and result in a most sufficient IAM output for each user principal involved while not producing redundant API invocation, potentially leading to detectable traces in the log.\\\\
The efficacy of the SkyEye CPIEM model is best demonstrated through a representative operational scenario [Fig. \ref{fig:44}, \ref{fig:45}], which underscores its utility in advanced IAM enumeration. Consider a context in which the enumeration framework, such as SkyEye, processes active sessions for five distinct user principals: User\_A, User\_B, User\_C, User\_D, and User\_E, each instantiated with unique AWS credential pairs. Each principal is provisioned with a discrete, non-overlapping subset of IAM permissions, as follows:\\

\begin{itemize}
\item \textbf{User\_A:} Lacks explicit permissions to perform any IAM actions.
\item \textbf{User\_B:} iam:GetUserPolicy, iam:GetGroupPolicy, iam:GetRolePolicy, iam:GetPolicy, and iam:ListPolicyVersion
\item \textbf{User\_C:} iam:ListAttachedGroupPolicies, iam:ListAttachedUserPolicies, iam:ListAttachedRolePolicies, iam:ListGroupPolicies, iam:ListUserPolicies, iam:ListRolePolicies
\item \textbf{User\_D:} iam:GetPolicyVersion
\item \textbf{User\_E:} iam:ListRoles, iam:ListGroupsForUser
\end{itemize}

In a traditional enumeration paradigm where each set of credentials is leveraged in isolation, and the enumeration process is constrained either to a single principal or to disjoint, parallel sessions per principal, the discoverability of the broader IAM topology becomes inherently limited. The inability to synthesize and correlate permissions across user boundaries leads to a fragmented security posture analysis, potentially obscuring critical relationships and privilege escalation vectors within the AWS environment.\\

The CPIEM model overcomes these limitations by enabling concurrent, synchronized enumeration sessions across all available user principals. This cooperative multi-principal model facilitates the aggregation and cross-correlation of disparate permission sets, thereby augmenting the granularity and completeness of the overall IAM visibility. For instance, while User\_A cannot directly access or enumerate any IAM resources, the intersection and union of the permissions held by User\_B, User\_C, User\_D, and User\_E when orchestrated in concert, can collectively reconstruct a comprehensive access control map, not only for User\_A but for each principal under scrutiny.\\

Furthermore, the model’s capability to dynamically integrate the results of each enumeration session ensures a holistic perspective, wherein the knowledge gleaned from one principal’s permissions enriches the contextual understanding of others. In defensive security, this approach significantly enhances the identification of implicit trust relationships, hidden privilege chains, and potential security misconfigurations, which are core concerns in advanced IAM threat modeling and defense. By operationalizing cross-principal enumeration and correlation, the CPIEM model advances the state of the art in IAM reconnaissance, supporting both offensive and defensive cybersecurity postures within complex cloud ecosystems.

\section{Transitive Cross-Role Enumeration Model (TCREM)}
Each "user" principal might have the permission to assume some specific roles and retrieve the temporary session tokens to act on behalf of those roles. Each "role" principal might also have the permission to assume the other roles, and to act on behalf of those roles through temporary session tokens.\\\\
The term \textbf{In-scope IAM Roles} in \textit{Transitive Cross-Role Enumeration Model (TCREM)} is defined by:
\begin{itemize}
    \item The roles that can be assumed \underline{directly} by provided AWS credentials:
    \begin{itemize}
        \item \textbf{User A} $\rightarrow$ \textbf{Role A}
        \item \textbf{User A} $\rightarrow$ \textbf{Role B}
    \end{itemize}

    \item The roles that can be assumed \underline{indirectly} by the roles that can be assumed by provided AWS credentials:
    \begin{itemize}
        \item \textbf{User A} $\rightarrow$ \textbf{Role A}
        \begin{itemize}
            \item Role A $\rightarrow$ \textbf{Role E}
            \begin{itemize}
                \item Role E $\rightarrow$ \textbf{Role F}
                \item[] \hspace{1.5em} Role F $\rightarrow$ \textbf{Role I}
            \end{itemize}
            \item Role A $\rightarrow$ \textbf{Role G}
            \begin{itemize}
                \item Role G $\rightarrow$ \textbf{Role H}
            \end{itemize}
        \end{itemize}
        \item \textbf{User A} $\rightarrow$ \textbf{Role B}
    \end{itemize}
\end{itemize}

\noindent$\rightarrow$ \textbf{In-Scope IAM Roles:} Role A, Role B, Role E, Role F, Role I, Role G, Role H

\begin{figure}[htbp]
    \centering
    \includegraphics[width=1.0\linewidth]{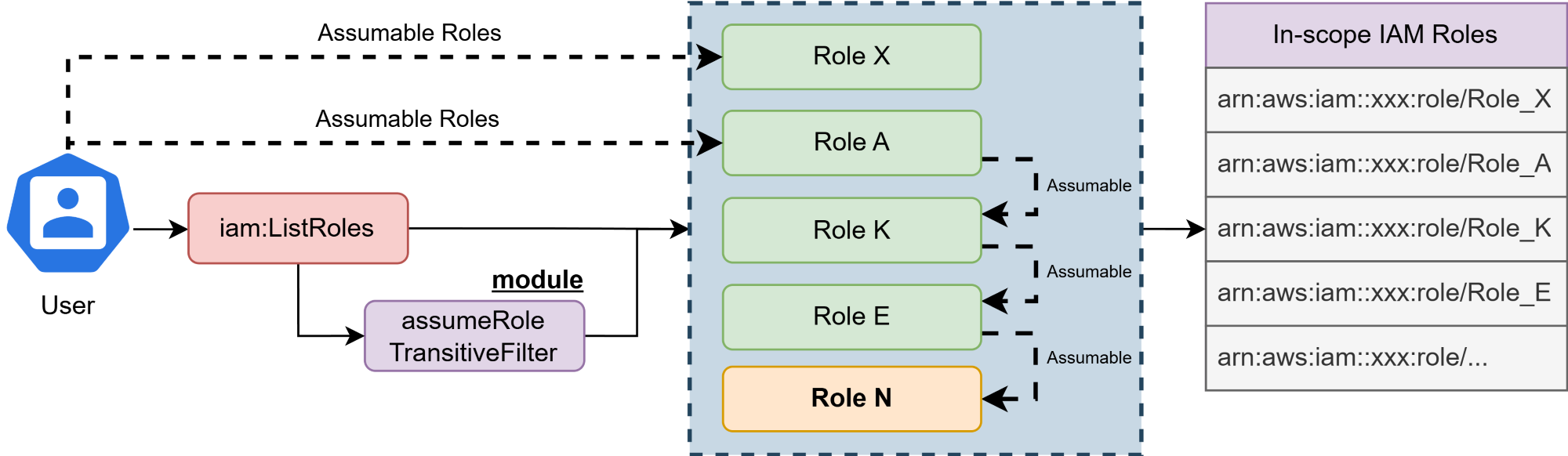}
    \caption{Core of SkyEye - Transitive Cross-Role Enumeration Model (TCREM)}
    \label{fig:46}
\end{figure}

Transitive Cross-Role Enumeration Model (TCREM) [Fig. \ref{fig:46}] is proposed and developed with the capability of gathering in-scope IAM roles, performing the direct assumption from user principal, or indirect assumption from the roles that can be assumed by the user principal, to act on behalf of in-scope IAM roles, and simultaneously complementing to the entire scanning output, subsequently contributing to the reduction of false negatives, and improving the overall accuracy of the IAMs output. Each role is an independent principal with associated permissions assigned to, which can be leveraged in complementing to the overall enumeration of IAM users, groups, roles, policies that have a strong bond to the targeting AWS credentials.\\

Transitive Cross-Role Enumeration Model will be integrated into:

\begin{itemize}
\item Single-Principal IAM Enumeration Model (SiPIEM): In-scope IAM roles will only involve complementing single user principal’s IAM vision context
\item Separate-Principal IAM Enumeration Model (SePIEM): In-scope IAM roles will only involve complementing each user principal’s IAM vision context separately
\item Cross-Principal IAM Enumeration Model (CPIEM): In-scope IAM roles come from each user principal, will involve in complementing not only original user principal's IAM vision context, but also other user principals' IAM vision context 
\end{itemize}

During run-time, if iam:GetAccountAuthorizationDetails permission is detected to be executable by at least one role’s session from similar AWS Account Id, the model will immediately terminate all the session come from that AWS Account Id, and utilize the iam:GetAccountAuthorizationDetails permission to retrieve full IAM vision context of that AWS Account Id, and distribute the correspondent result to the user principal that involved in the IAM enumeration for that AWS Account Id. This approach will reduce significantly 95\% of the entire scanning process, and result in a most sufficient IAM output for each involved user principals while not producing redundant API invocation, potentially leading to detectable traces in logging.\\

\begin{figure}[htbp]
    \centering
    \includegraphics[width=1.0\linewidth]{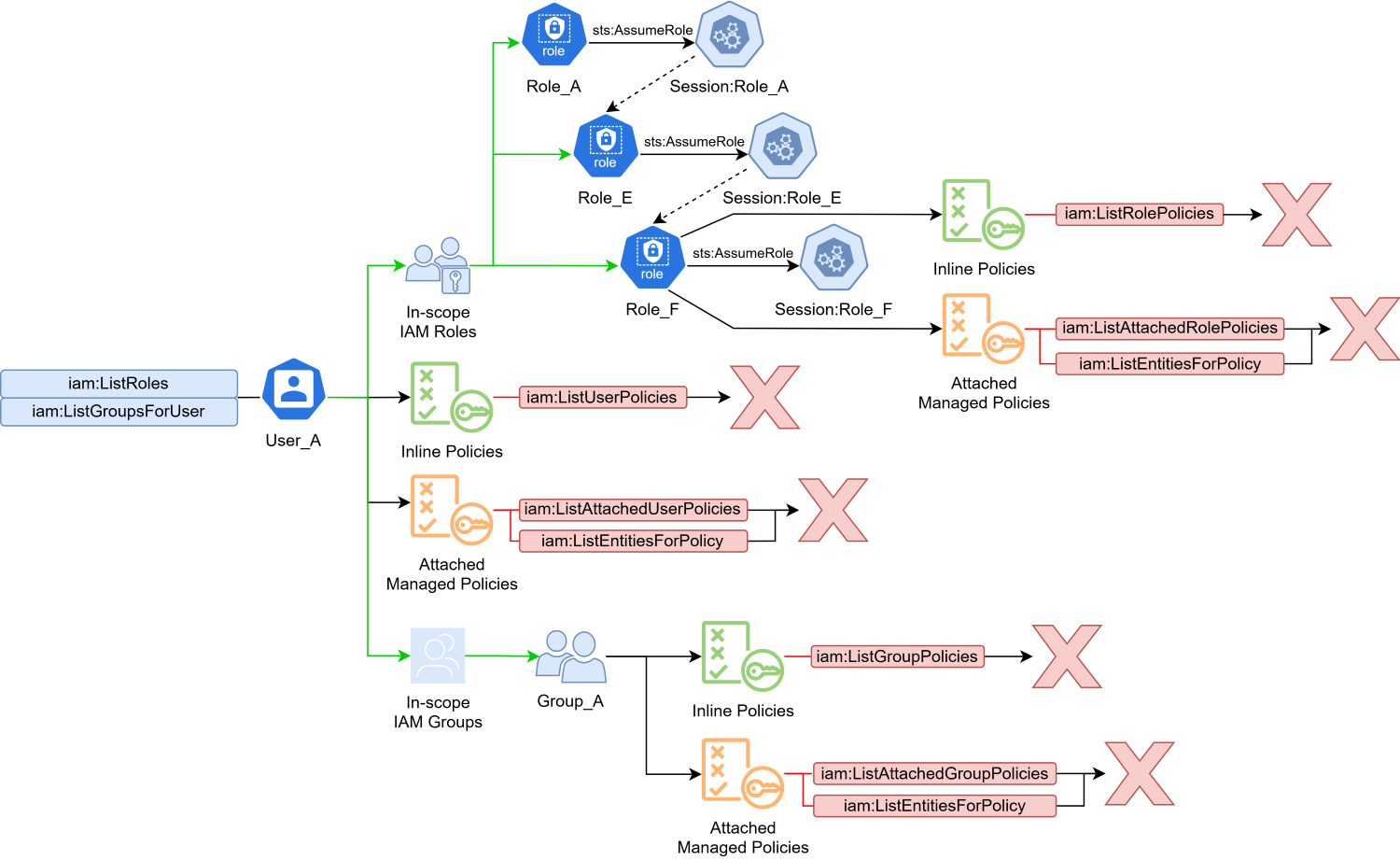}
    \caption{Transitive Cross-Role Enumeration Model - Example Scenario - Stage 1}
    \label{fig:47}
\end{figure}

\begin{figure}[htbp]
    \centering
    \includegraphics[width=1.0\linewidth]{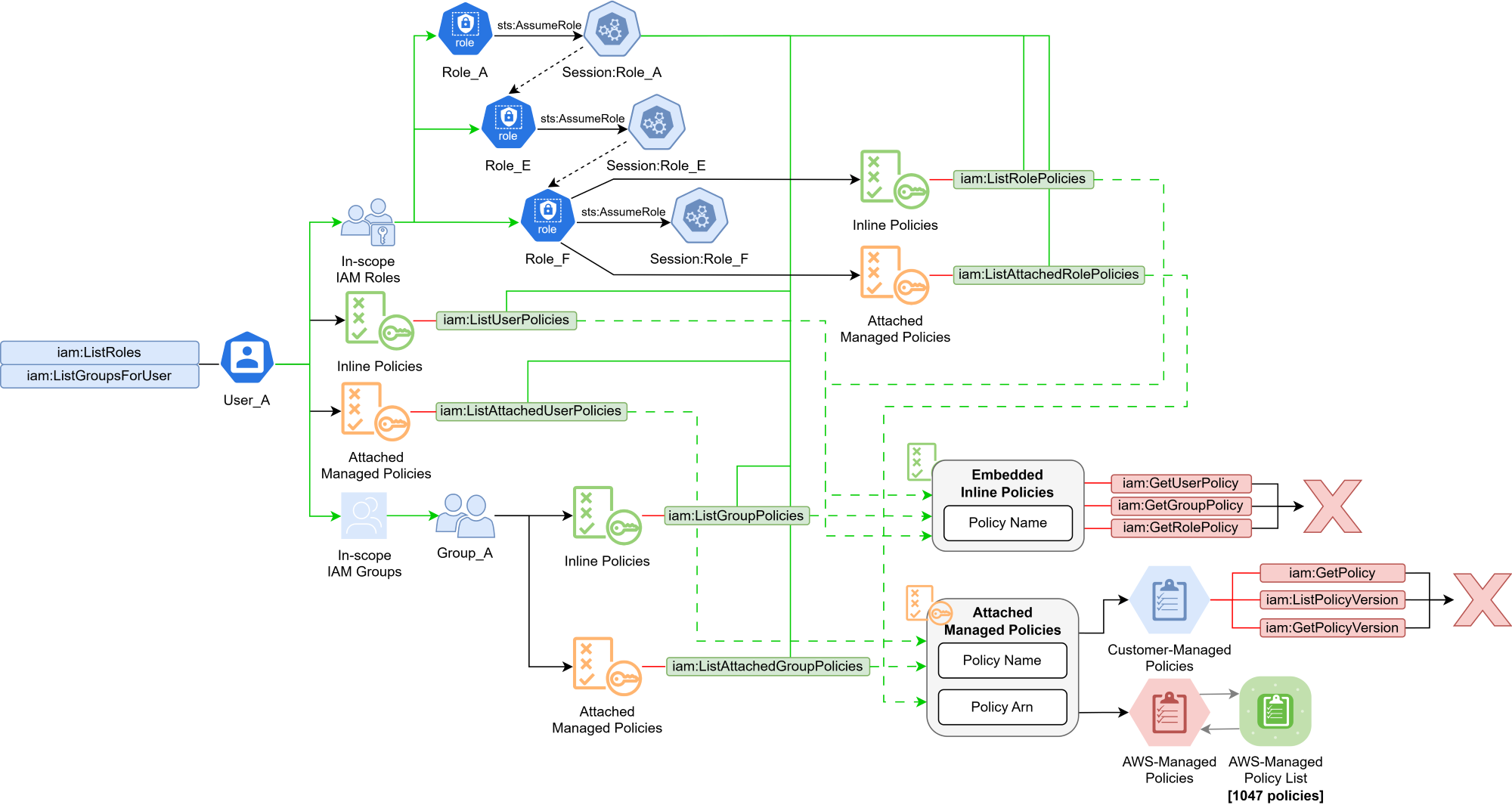}
    \caption{Transitive Cross-Role Enumeration Model - Example Scenario - Stage 2}
    \label{fig:48}
\end{figure}

\begin{figure}[htbp]
    \centering
    \includegraphics[width=1.0\linewidth]{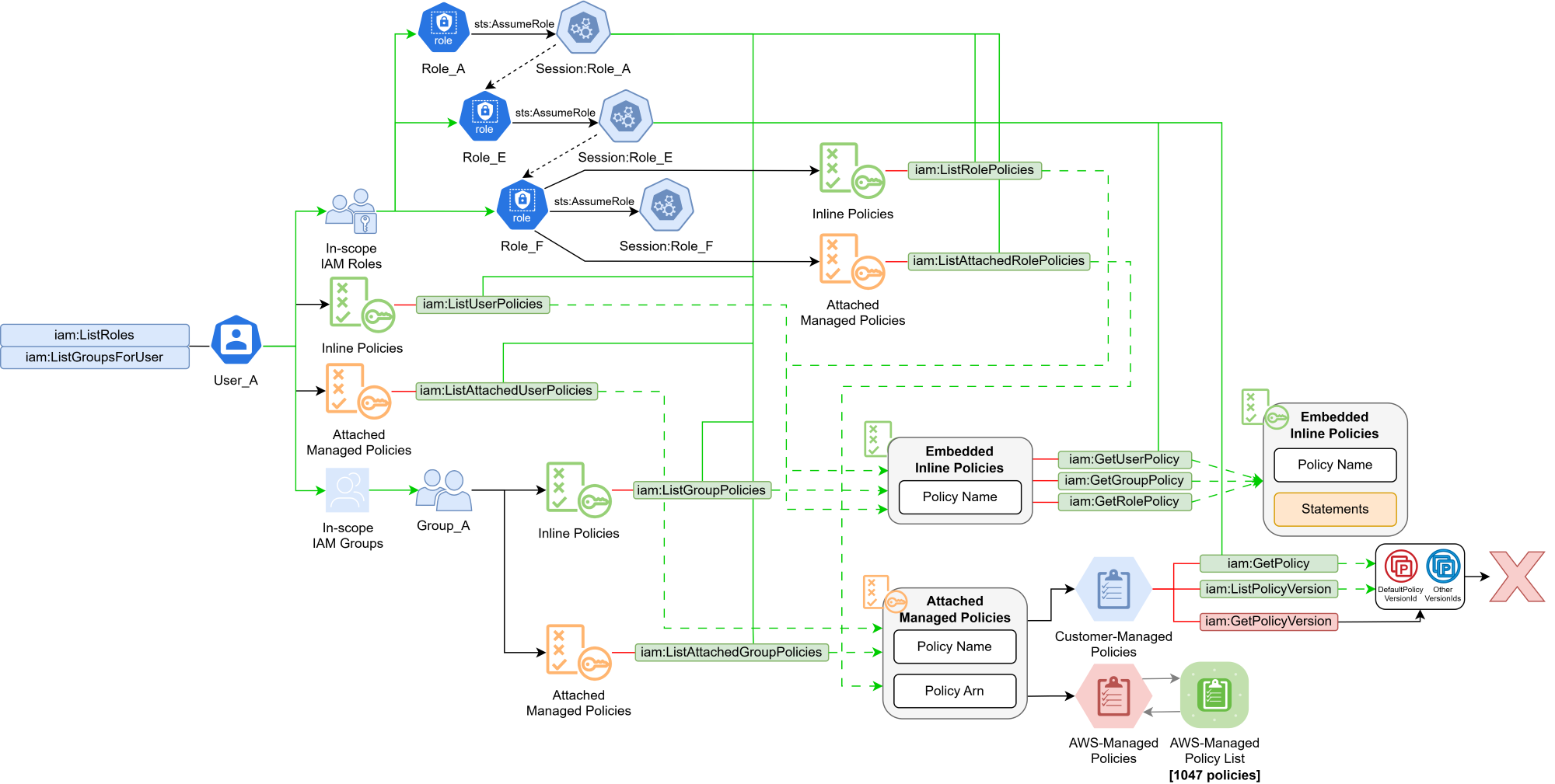}
    \caption{Transitive Cross-Role Enumeration Model - Example Scenario - Stage 3}
    \label{fig:49}
\end{figure}

\begin{figure}[htbp]
    \centering
    \includegraphics[width=1.0\linewidth]{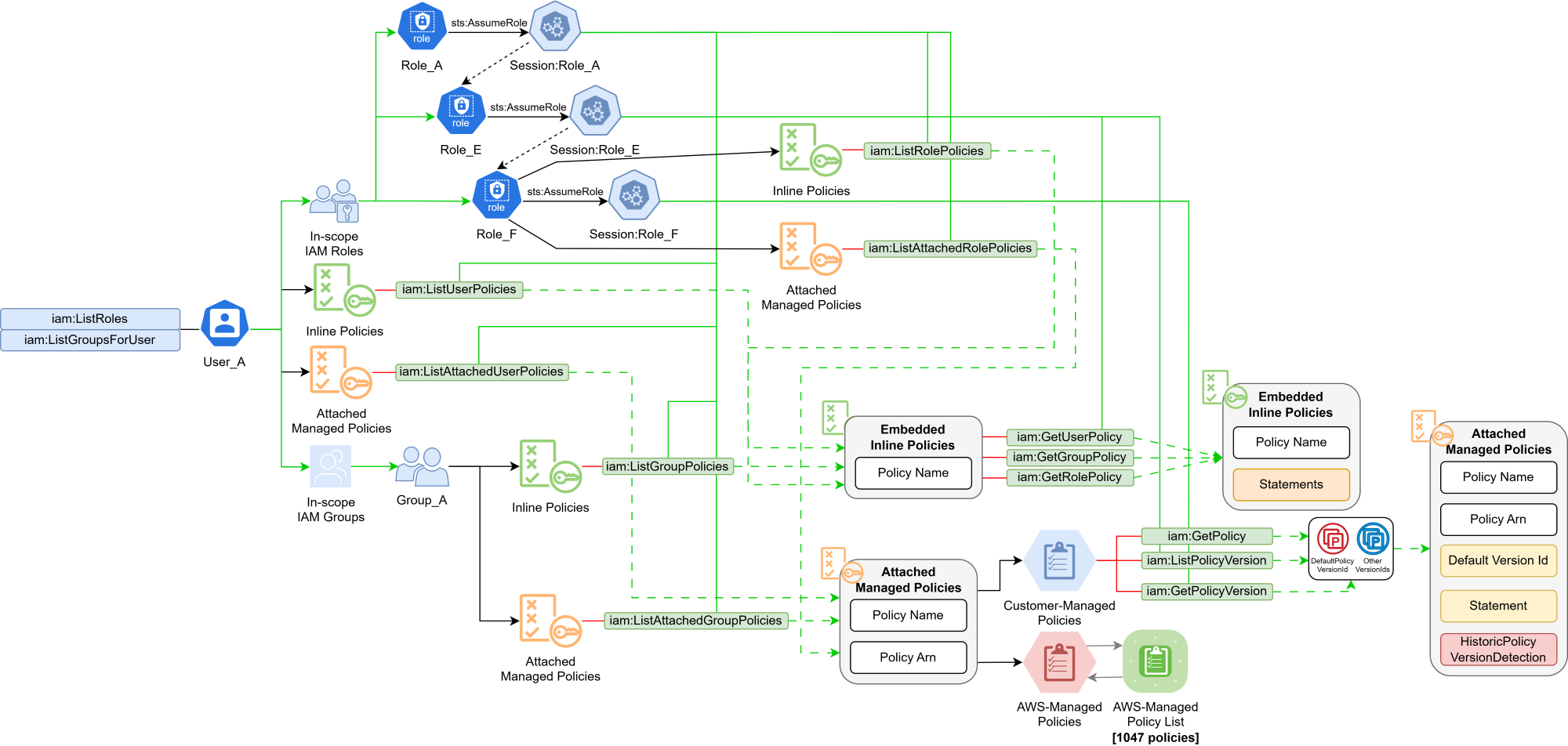}
    \caption{Transitive Cross-Role Enumeration Model - Example Scenario - Stage 4}
    \label{fig:410}
\end{figure}

The operational advantages of the TCREM model are most effectively illuminated through the analysis of a representative scenario [Fig. \ref{fig:47}, \ref{fig:48}, \ref{fig:49}, \ref{fig:410}] that encapsulates advanced IAM enumeration techniques within a contemporary AWS environment. In this context, let us consider User\_A, an identity equipped with the capability to assume a set of in-scope IAM roles: Role\_A, Role\_E, and Role\_F, either through direct or via transitive trust relationships. Each principal, whether user or role, is endowed with a distinct and non-overlapping subset of IAM permissions, delineated as follows:

\begin{itemize}
    \item \textbf{User\_A:} Possesses capabilities such as iam:ListRoles and iam:ListGroupsForUser, enabling enumeration of associations with in-scope IAM roles and in-scope IAM groups
    \item \textbf{Role\_A:} Authorized to enumerate policy associations across roles, users, and groups through permissions including iam:ListRolePolicies, iam:ListAttachedRolePolicies, iam:ListUserPolicies, iam:ListAttachedUserPolicies, iam:ListGroupPolicies, and iam:ListAttachedGroupPolicies
    \item \textbf{Role\_E:} Holds advanced policy retrieval permissions including iam:GetUserPolicy, iam:GetGroupPolicy, iam:GetRolePolicy, iam:GetPolicy, and iam:ListPolicyVersions
    \item \textbf{Role\_F:} Entitled to retrieve specific policy document by the versions via iam:GetPolicyVersion
\end{itemize}

Traditionally, in penetration testing or red team operations, enumeration efforts are often constrained to the context of a single identity session. While User\_A is able to enumerate certain role and group metadata, their view remains incomplete, as the permissions and policy insights granted to Role\_A, Role\_E, and Role\_F are inaccessible unless those roles are actively assumed and enumerated. This approach prevents the correlation and aggregation of permissions across assumed roles, thereby hindering the construction of a comprehensive, system-wide IAM topology. Such limitations can lead to an incomplete understanding of privilege boundaries, trust relationships, and potential privilege escalation vectors within the AWS environment.\\

\begin{figure}[ht]
    \centering
    \includegraphics[width=1.0\linewidth]{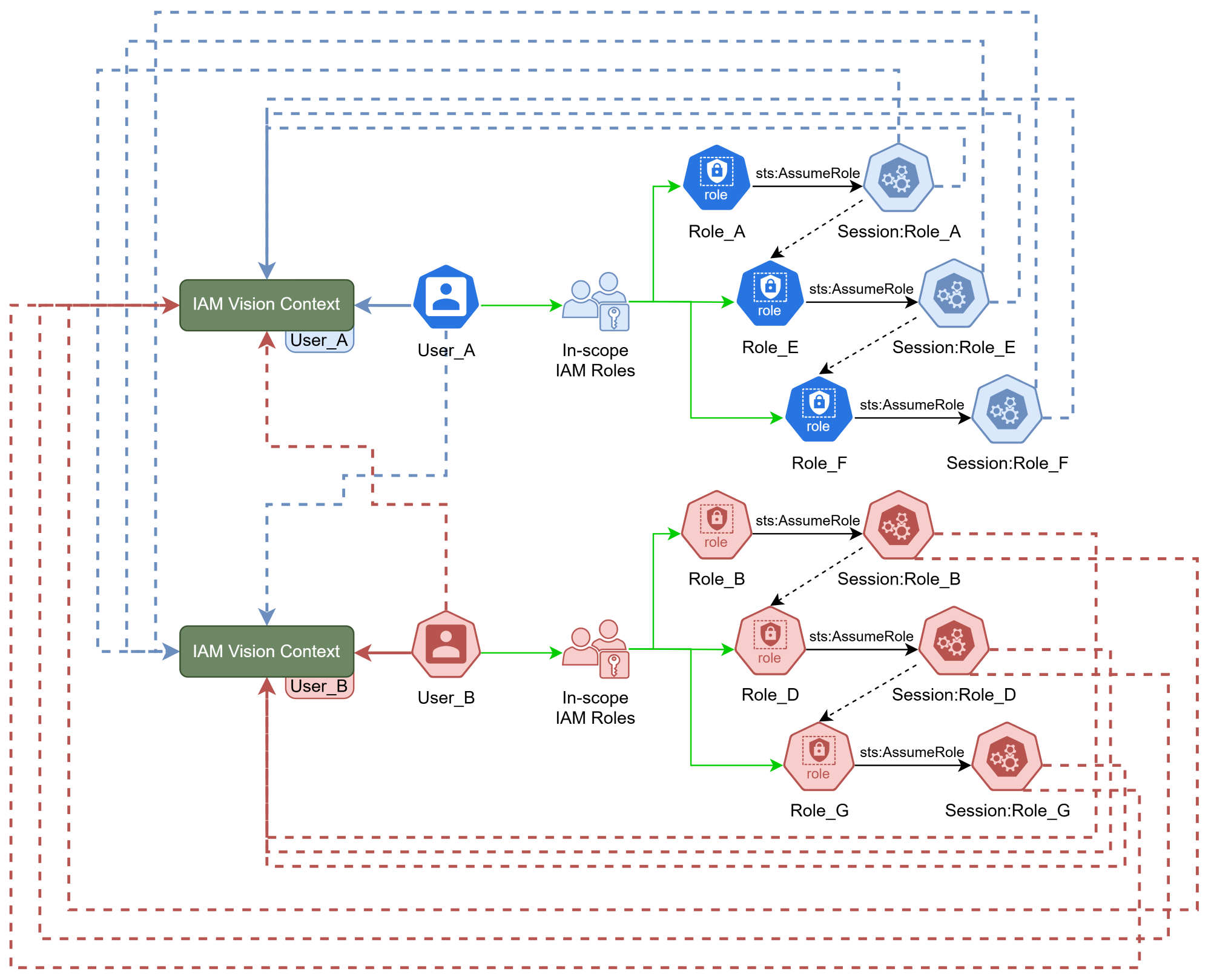}
    \caption{The Interconnection between Users in CPIEM and Roles in TCREM}
    \label{fig:411}
\end{figure}

The TCREM model fundamentally advances this paradigm by operationalizing simultaneously transitive cross-role enumeration. When integrated, the model leverages User\_A’s privileges to discover in-scope IAM roles, and autonomously assume each in-scope IAM role and instantiates temporary sessions for Role\_A, Role\_E, and Role\_F concurrently alongside the primary User\_A session. These parallel enumeration processes enable the synthesis of permissions and policy data across both the originating user principal and all assumable roles, therefore, constructing an integrated and multidimensional view of the IAM environment.\\

Crucially, this approach not only augments the IAM visibility for User\_A by aggregating permissions and policy insights from the assumed roles, but also enables the integration with cross-principal IAM enumeration model (CPIEM), as demonstrated in [Fig. \ref{fig:411}]. In scenarios involving multiple user principals, each with discrete trust relationships and role assumption capabilities, the TCREM model orchestrates enumeration sessions for all user identities and their respective assumable roles. This cooperative enumeration methodology empowers a holistic assessment of the IAM landscape, facilitating the discovery of complex privilege chains, indirect privilege escalation pathways, and latent policy misconfigurations that would otherwise remain undetected under a single-principal enumeration model.\\

In summary, the TCREM model represents a significant advancement in IAM enumeration methodology, enabling security practitioners and penetration testers to transcend the inherent limitations of isolated principal analysis. Through its support for concurrent and transitive enumeration, the model fosters a more precise and exhaustive understanding of access control dynamics, privilege escalation relationships, and the overall security posture of AWS IAM deployments. This makes SkyEye as an indispensable framework for both offensive security assessments and defensive IAM governance within complex cloud environments.\\

\section{IAM Deep Enumeration Capabilities}
This section presents an advanced structure for enumerating IAM user principals, in-scope IAM groups, and in-scope IAM roles, illustrating the logical chains of actions necessary to reveal the complete IAM vision context.
\subsection{Retrieval of In-Scope IAM Groups and In-Scope IAM Roles for User Principals}

\begin{figure}[htbp]
    \centering
    \includegraphics[width=1.0\linewidth]{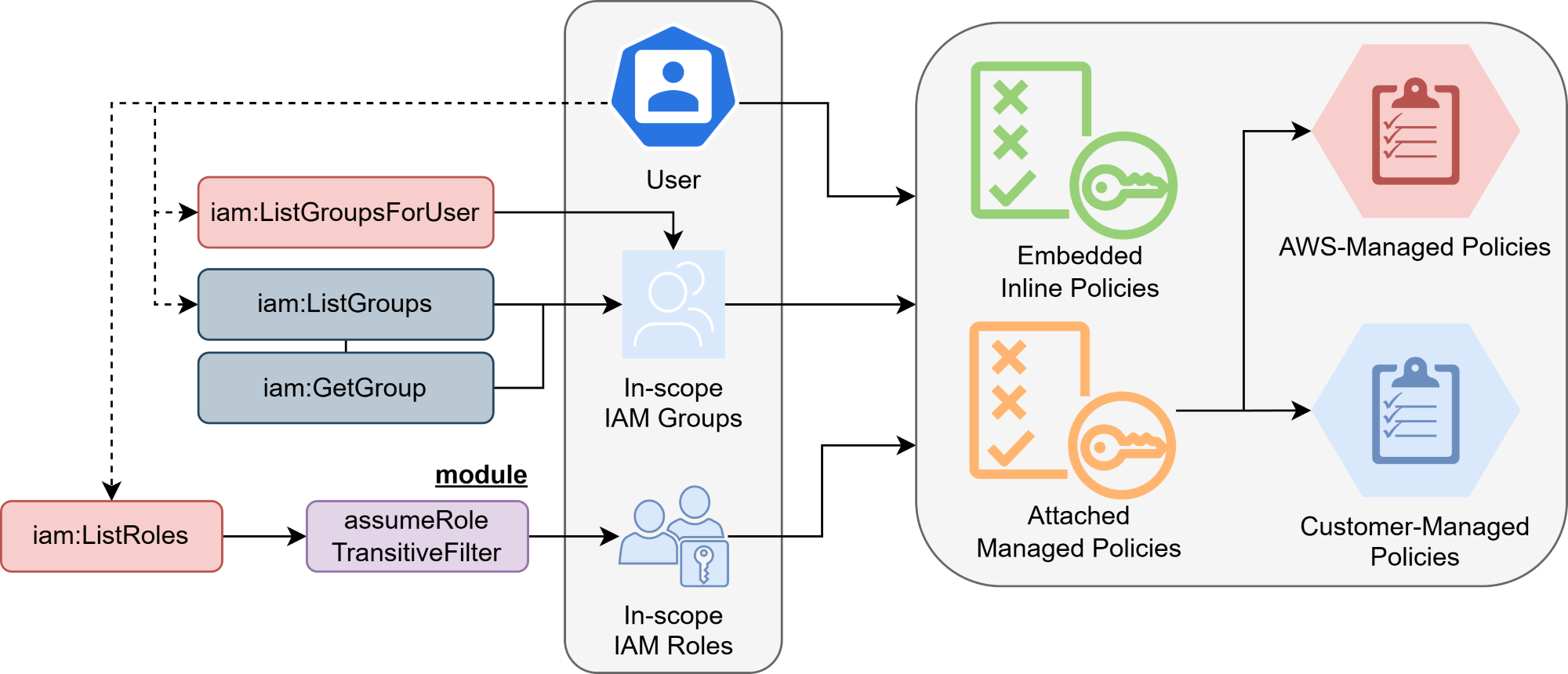}
    \caption{How to define in-scope IAM groups and in-scope IAM roles?}
    \label{fig:412}
\end{figure}

Before delving deeper into the enumeration of inline policies and attached managed policies for related IAM entities, it is necessary to define the related IAM entities in the scope of the targeting user principals. In the previous section of Transitive Cross-Role Enumeration Model (TCREM), we discussed how SkyEye will define in-scope IAM roles. In-scope IAM roles is a set of IAM roles that can be assumed directly or indirectly by the user principals, which contributes to a broader picture of what is the complete set of permissions and resources that the user principals can interact with. On the other hand, in-scope IAM groups are the IAM groups that the user principal directly belongs to, and from which it inherits permissions into its own IAM policies.\\

To retrieve the complete picture of in-scope IAM groups for user principal, it is necessary to have at least one of these two sets of permissions: iam:ListGroupsForUser; or iam:ListGroups and iam:GetGroup. The iam:ListGroupsForUser permission will directly return the IAM groups that the provided user principal ARN belongs to, while iam:ListGroups and iam:GetGroup will indirectly list all IAM groups and their membership, requiring to further filter to retrieve the complete set of in-scope IAM groups.\\

In term of in-scope IAM roles, it is necessary to have iam:ListRoles permission to retrieve the complete list of IAM roles, and perform the filtering on AssumeRolePolicyDocument of each role to disclose which role can be assumed directly by user principal or indirectly by the roles that can be assumed by user principal, to accumulate into the complete set of in-scope IAM roles.\\

\subsection{Retrieval of Inline Policies for User Principals}

\begin{figure}[htbp]
    \centering
    \includegraphics[width=1.0\linewidth]{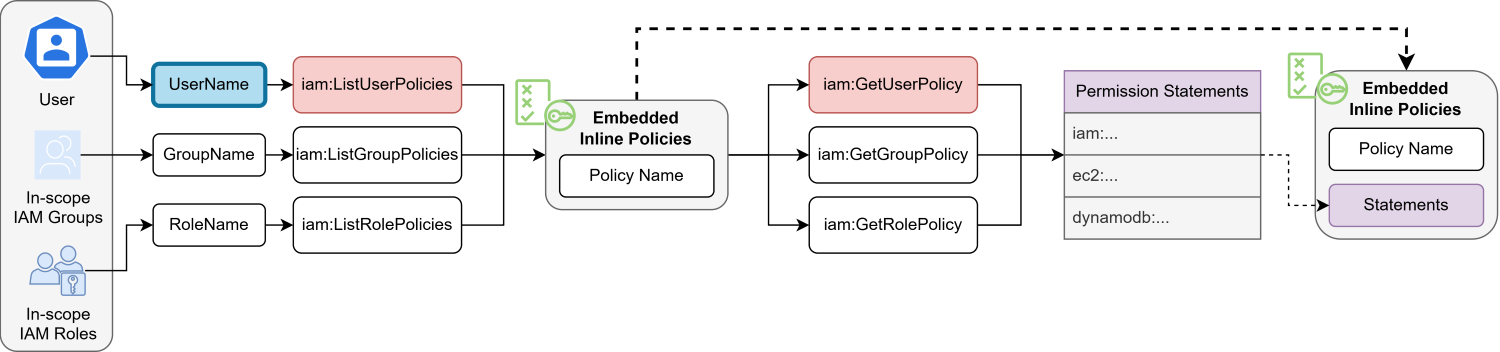}
    \caption{The Retrieval of User Principal's Inline Policies}
    \label{fig:413}
\end{figure}

The enumeration of user principal inline policies begins with the fundamental action iam:ListUserPolicies. This call enumerates the set of policy names explicitly attached to a given user principal. Because inline policies are scoped to that specific user identity, they are frequently overlooked in large environments where common permissions are often handled through managed policies. However, inline policies can grant powerful privileges and might be used in exceptional cases that deviate from standard best practices.\\

Once the set of inline policy names is acquired through iam:ListUserPolicies, the next step involves invoking iam:GetUserPolicy. This latter operation retrieves the actual policy document associated with each policy name enumerated. Through this two-action chain, SkyEye gain direct visibility into the textual policy statements. By structuring the retrieval process in discrete steps, SkyEye can automate the enumeration and analysis of user-specific policies that may impose excessive or contradictory permissions, and ensure the complete IAM vision across the permissions from user-specific inline policy scope.\\

\subsection{Retrieval of Attached Managed Policies for User Principals}

\begin{figure}[htbp]
    \centering
    \includegraphics[width=1.0\linewidth]{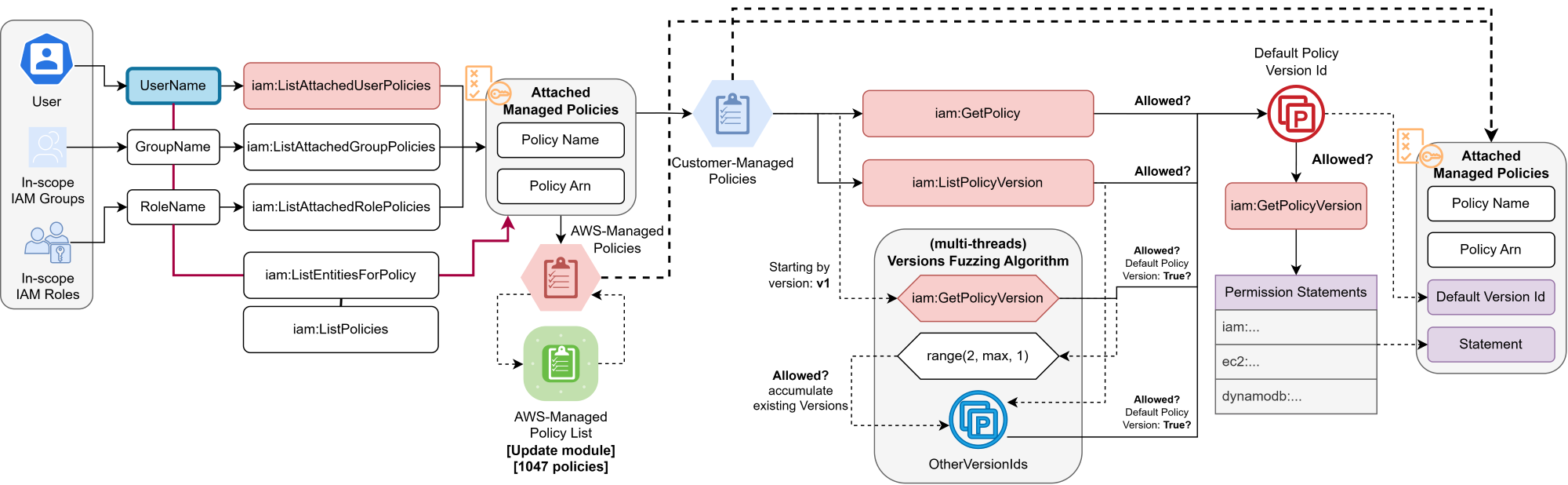}
    \caption{The Retrieval of User Principal's Attached Managed Policies}
    \label{fig:414}
\end{figure}

While inline policies are user-specific, attached managed policies represent a more scalable approach to permission administration within AWS. Enumerating these policies for a particular user involves initiating iam:ListAttachedUserPolicies, an action that returns an array of managed policy ARNs attached directly to the user. In typical organizations, these managed policies might be official AWS-managed policies (e.g., AdministratorAccess or AmazonS3ReadOnlyAccess) or custom organizational policies intended for role-based access control paradigms.\\

Most managed policies maintain one or more versions. Hence, after capturing the policy ARNs via iam:ListAttachedUserPolicies, it is crucial to delve deeper using either iam:ListPolicyVersions or iam:GetPolicy in conjunction with iam:GetPolicyVersion. Each Policy ARN retrieved by iam:ListAttachedUserPolicies, is then processed via either iam:ListPolicyVersions or iam:GetPolicy to determine the policy’s versioning state and to identify the default active version of the customer-managed policy. Subsequently, iam:GetPolicyVersion provides the structured JSON policy document of a default active version. This final step completes the chain by disclosing the complete IAM vision context across the permissions and resources that the user could perform.\\

As an alternative solution to resolve the specific case that both the permissions iam:ListPolicyVersions and iam:GetPolicy are not permitted, resulting in the lack of understanding about the current active version of customer-managed policies. SkyEye introduces the “Version Fuzzing Algorithms” which will support the fuzzing capabilities throughout the policy versions of targeting customer-managed policy ARNs, if only iam:GetPolicyVersion is permitted.\\

\subsection{Retrieval of Inline Policies for In-Scope IAM Groups}

\begin{figure}[htbp]
    \centering
    \includegraphics[width=1.0\linewidth]{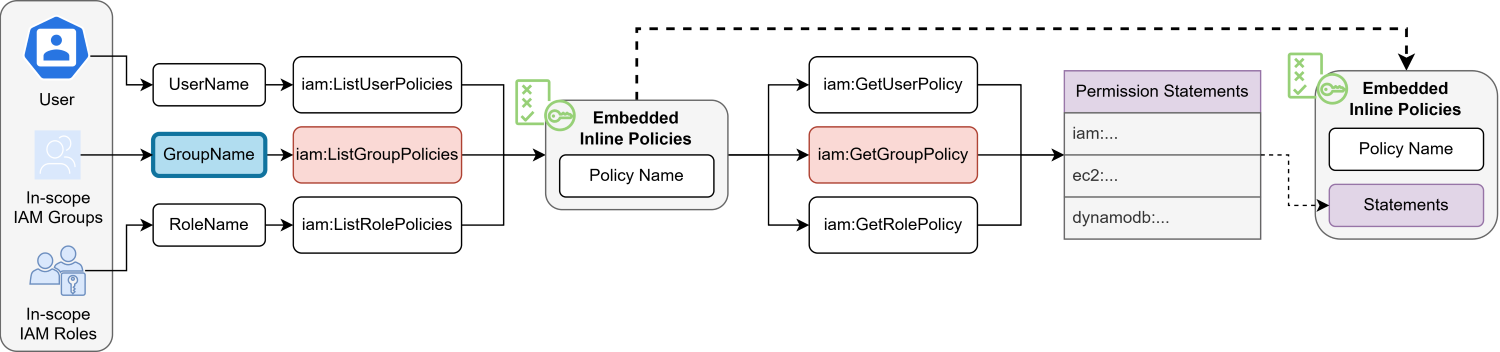}
    \caption{The Retrieval of In-scope IAM Groups' Inline Policies}
    \label{fig:415}
\end{figure}

Beyond user-centric investigations, robust IAM enumeration necessarily extends to group-level analysis. Within AWS, group memberships can significantly change an individual’s effective permission set. Consequently, the first step in enumerating in-scope IAM groups which is defined as the groups that a particular user principal belongs to, often begins with either iam:ListGroupsForUser or a combination of iam:ListGroups followed by iam:GetGroup to retrieve the situational awareness about the in-scope IAM groups. The direct approach iam:ListGroupsForUser yields the list of groups to which the user belongs. Alternatively, if iam:ListGroupsForUser is not permitted to perform, SkyEye will switch to invoke iam:ListGroups and iam:GetGroup systematically to enumerate all existing groups and confirms membership based on the targeting user principals.\\

\sloppy Once the in-scope groups have been identified, the chain proceeds to iam:ListGroupPolicies for each group. This action enumerates the names of inline policies residing at the group level. Next, for each policy name discovered, the call iam:GetGroupPolicy retrieves the underlying policy document. This chain ensures that every inline policy statement nested within group membership is processed, providing the full scope of relevant permissions. It is critical in multi-account or multi-group scenarios where ephemeral group memberships might be leveraged, intentionally or inadvertently, to circumvent standard user-level constraints. By detailing the chain from group identification to policy retrieval, SkyEye ensures the complete IAM vision context across the permissions inherited from in-scope IAM groups, supporting the overall situational awareness of current IAM context assigned to the targeting user principals.\\

\subsection{Retrieval of Attached Managed Policies for In-Scope IAM Groups}

\begin{figure}[htbp]
    \centering
    \includegraphics[width=1.0\linewidth]{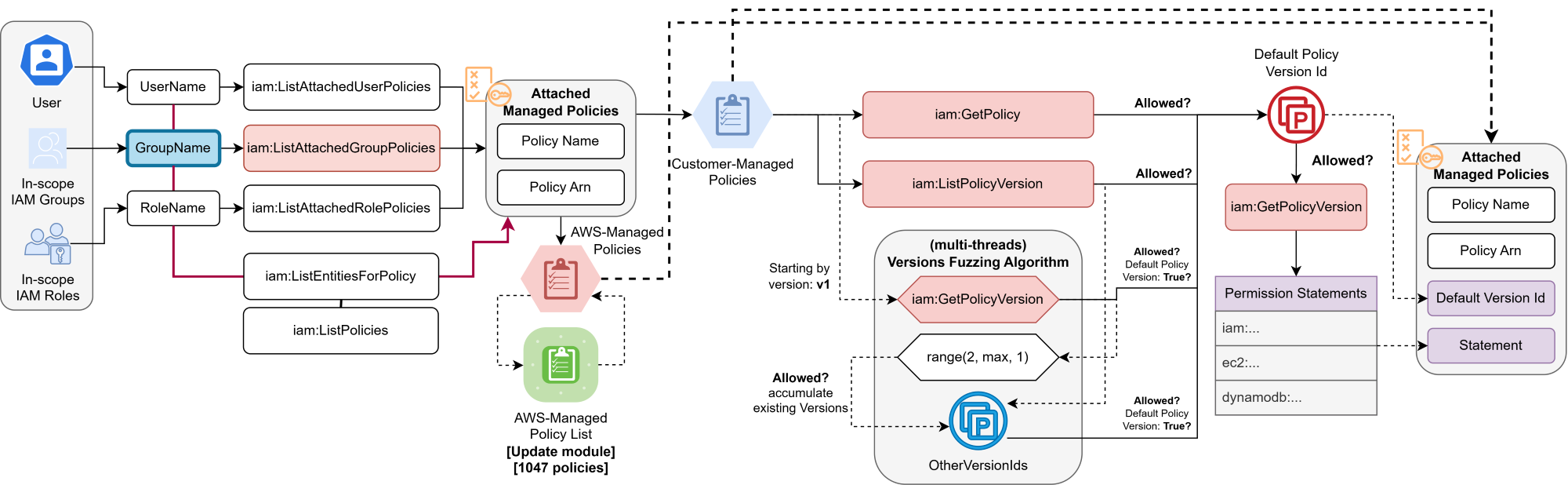}
    \caption{The Retrieval of In-scope IAM Groups' Attached Managed Policies}
    \label{fig:416}
\end{figure}

Equivalent to user principals, IAM groups may also have attached managed policies. These can range from AWS-supplied offerings, typically used to facilitate administrative tasks (e.g., service-level read/write access), to organization-managed sets of permissions that envelop departmental or project-based roles. Enumerating these managed policy attachments for in-scope IAM groups begins with identifying the relevant groups, using iam:ListGroupsForUser or the pair iam:ListGroups and iam:GetGroup, as discussed in previous section. The next action in the chain is iam:ListAttachedGroupPolicies, which reveals the ARNs of the managed policies attached to the identified groups.\\

After enumerating these ARNs, the same concluding steps seen in user principal’s attached managed policy analysis apply: either gather the current active policy version by using iam:ListPolicyVersions or retrieve through the policy details via iam:GetPolicy, then leverage iam:GetPolicyVersion to retrieve the corresponding policy document of the current active version This final step completes the chain by disclosing the complete IAM vision context across the permissions and resources that inherited indirectly from the groups that the user principal belongs to.\\

As an alternative solution, as discussed in previous sections, to resolve the specific case that both the permissions iam:ListPolicyVersions and iam:GetPolicy are not permitted, resulting in the lack of understanding about the current active version of customer-managed policies. SkyEye introduces the “Version Fuzzing Algorithms” which will support the fuzzing capabilities throughout the policy versions of targeting customer-managed policy ARNs, if only iam:GetPolicyVersion is permitted.\\

\subsection{Retrieval of Inline Policies for In-Scope IAM Roles}

\begin{figure}[htbp]
    \centering
    \includegraphics[width=1.0\linewidth]{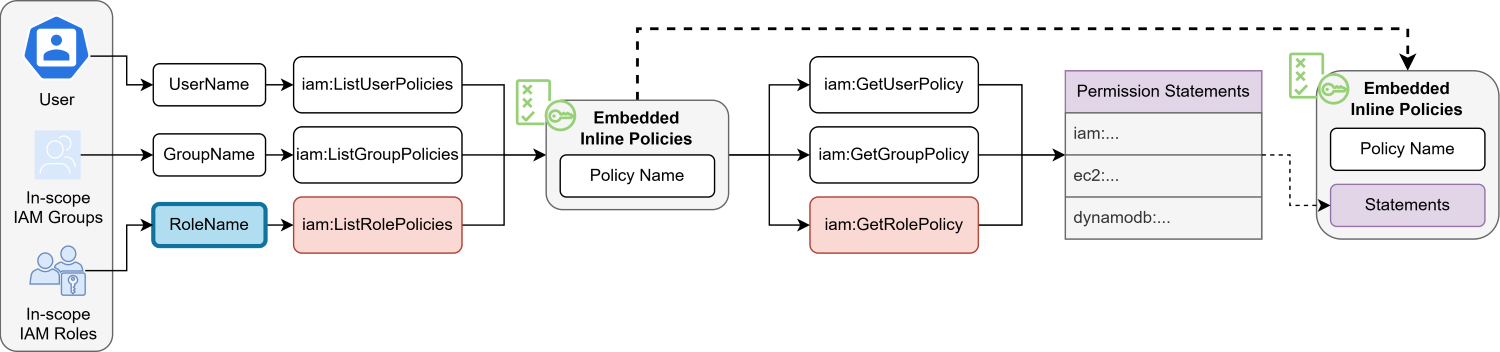}
    \caption{The Retrieval of In-scope IAM Roles' Inline Policies}
    \label{fig:417}
\end{figure}

While users and groups serve as foundational identity constructs within AWS, roles offer a pivotal mechanism by which users, services, or other roles can assume delegated privileges. For an IAM security audit, the scope of roles that a particular user can assume - directly or indirectly - becomes critical, as it potentially augments the user’s effective permissions. Identifying such roles includes determining trusted entity relationships and session token parameters that could extend privileges beyond the user’s nominal baseline. Once these roles are deemed “in-scope”, SkyEye will investigate any inline policies that may confer additional capabilities only found at the role level.\\

The operational chain typically commences with an enumeration of all roles using iam:ListRoles. Although this action returns every role in the account, it is imperative to filter them to identify only those roles that the user principal can assume directly or indirectly, as discussed in the Transitive Cross-Role Enumeration Model (TCREM). Such filtering might rely on analyzing trust policies or gleaning contextual information from the environment (e.g., previously discovered assume-role statements). Once the relevant IAM roles are discovered, the next steps involve retrieving the inline policies of those in-scope IAM roles. The invocation of iam:ListRolePolicies yields the set of inline policy names for each role, followed by iam:GetRolePolicy to obtain the policy documents themselves.\\

Role-based inline policies are paid particular attention by SkyEye framework, as these often grant specialized privileges for tightly scoped runtime scenarios (e.g., a role used by a specific application). If a user principal can assume any such role, that user effectively inherits these permissions. Thus, enumerating these inline policies is crucial for constructing the overarching permission graph. By adopting this systematic chain, no potential extension of privilege remains uncharted.\\

\subsection{Retrieval of Attached Managed Policies for In-Scope IAM Roles}

\begin{figure}[htbp]
    \centering
    \includegraphics[width=1.0\linewidth]{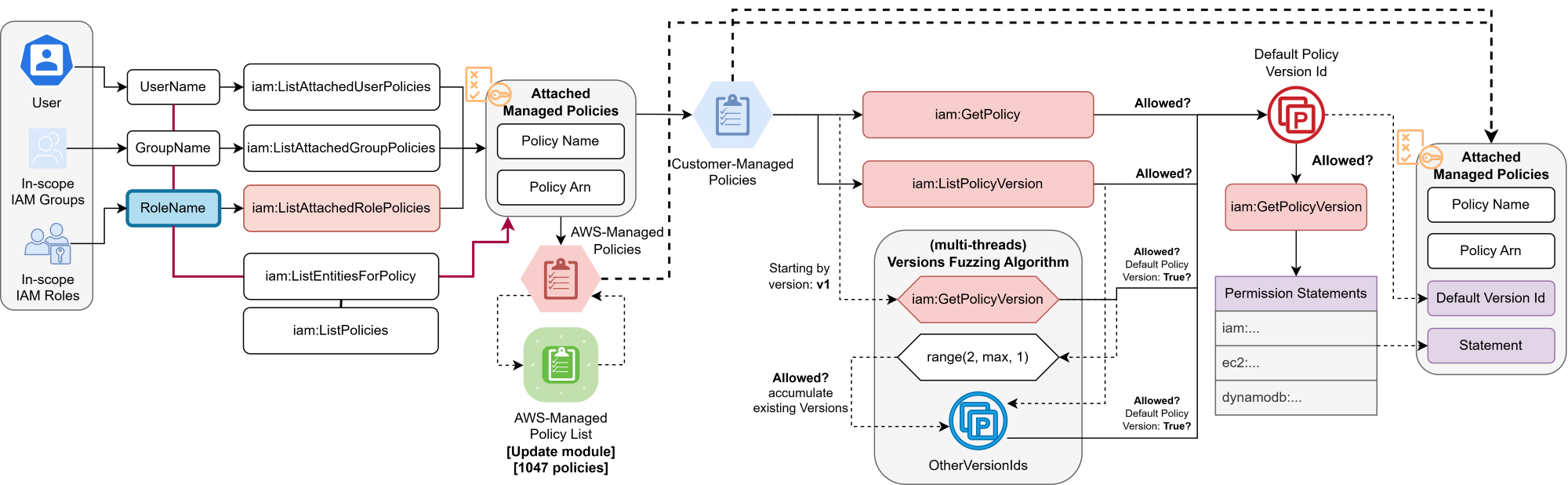}
    \caption{The Retrieval of In-scope IAM Roles' Attached Managed Policies}
    \label{fig:418}
\end{figure}

In addition to uncovering inline policies, attached managed policies at the role level must also be accounted for. Managed policies, whether AWS-managed or customer-managed, are frequently used to simplify the administration of privileges across multiple roles. Hence, a single managed policy can simultaneously grant extensive permissions to various roles, resulting in potential lateral movement opportunities for an adversary within a compromised account.\\

Mirroring the approach used for users and groups, SkyEye framework first establishes the list of relevant in-scope IAM roles via iam:ListRoles, as discussed in the previous section and in Transitive Cross-Role Enumeration Model (TCREM). The next step in the chain focuses on determining which version is currently active. For each policy ARN retrieved from iam:ListAttachedRolePolicies, SkyEye either invokes iam:ListPolicyVersions or iam:GetPolicy to pinpoint the default active version of the policy. Finally, a call to iam:GetPolicyVersion yields the structured JSON policy document for the identified active version of the managed policies. By completing these chains, SkyEye reveals a comprehensive view of the permissions the role confers, ensuring that no hidden privileges or policy misconfigurations are overlooked.\\

As an alternative solution, as discussed in previous sections, to resolve the specific case that both the permissions iam:ListPolicyVersions and iam:GetPolicy are not permitted, resulting in the lack of understanding about the current active version of customer-managed policies. SkyEye introduces the “Version Fuzzing Algorithms” which will support the fuzzing capabilities throughout the policy versions of targeting customer-managed policy ARNs, if only iam:GetPolicyVersion is permitted.\\

\subsection{Alternative Retrieval by iam:GetAccountAuthorizationDetails}

Though prior sections detail explicit chains for enumerating the inline and managed policies of user principals, groups, and roles, SkyEye can also adopt a more holistic strategy using iam:GetAccountAuthorizationDetails. This API call returns a wide range of authorization details encompassing users, groups, roles, and their corresponding inline and attached managed policies in a single output. It thus enables SkyEye to capture a near-comprehensive overview of the account’s IAM configuration without chaining multiple discrete calls.\\

Resource filtering is an essential step once the raw data is received from \texttt{iam:GetAccountAuthorizationDetails}. A large AWS environment might contain hundreds of IAM entities, making it impractical to sift through all permissions manually. Consequently, best practices dictate programmatically narrowing the output to only the relevant targeting user principals, in-scope IAM groups, and in-scope IAM roles (i.e., those that the user principal can assume directly or indirectly).\\

\begin{figure}[htbp]
    \centering
    \includegraphics[width=0.95\linewidth]{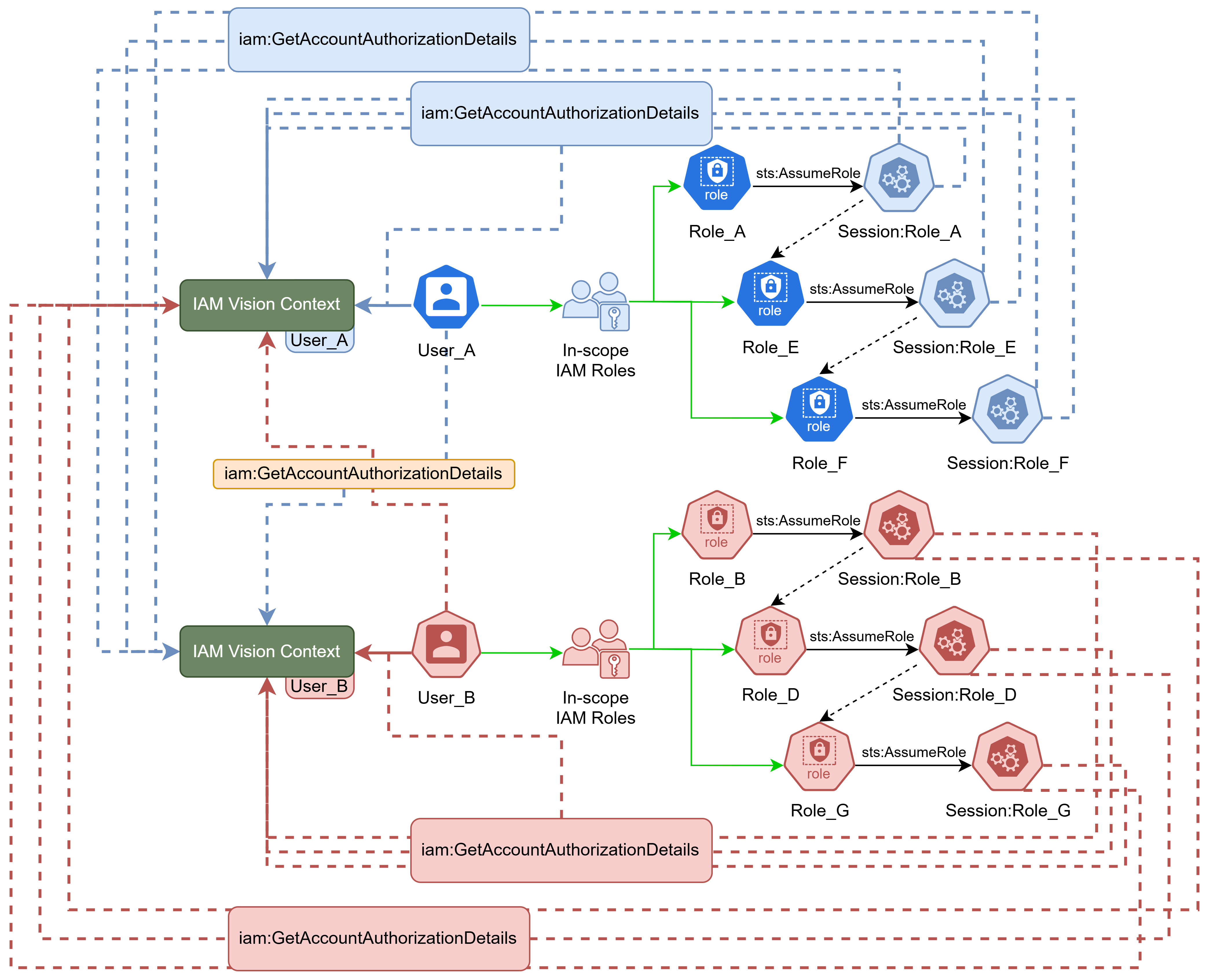}
    \caption{iam:GetAccountAuthorizationDetails integrated into CPIEM and TCREM}
    \label{fig:419}
\end{figure}

As discussed in the Cross-Principal IAM Enumeration Model (CPIEM) and Transitive Cross-Role Enumeration Model (TCREM), during the enumeration process, if the iam:GetAccountAuthorizationDetails permission is detected in run-time at any enumeration stage, to be executable by at least one user principal’s session or role’s session from similar AWS Account Id, the model will immediately terminate all other session come from that AWS Account Id, and utilize the iam:GetAccountAuthorizationDetails permission to retrieve full IAM context of that AWS Account Id, and distribute the correspondent result to the user principal that involved in the IAM enumeration for that AWS Account Id.\\

The benefit of iam:GetAccountAuthorizationDetails is unimaginable. Firstly, it reduces the overhead typically associated with enumerating each principal or policy independently. Secondly, this approach will reduce significantly 95\% of the entire scanning process, and result in a most sufficient IAM outcome for the model while not producing redundant API invocation, potentially leading to detectable traces in logging. Thirdly, its consolidated perspective enables a more robust comparison across multiple policy layers, thereby identifying hidden conflicts and permission redundancies that might otherwise be missed when analyzing each entity in isolation due to insufficient authorization.

\subsection{Inverse Enumeration Model for Attached Managed Policy}

\begin{figure}[htbp]
    \centering
    \includegraphics[width=1.0\linewidth]{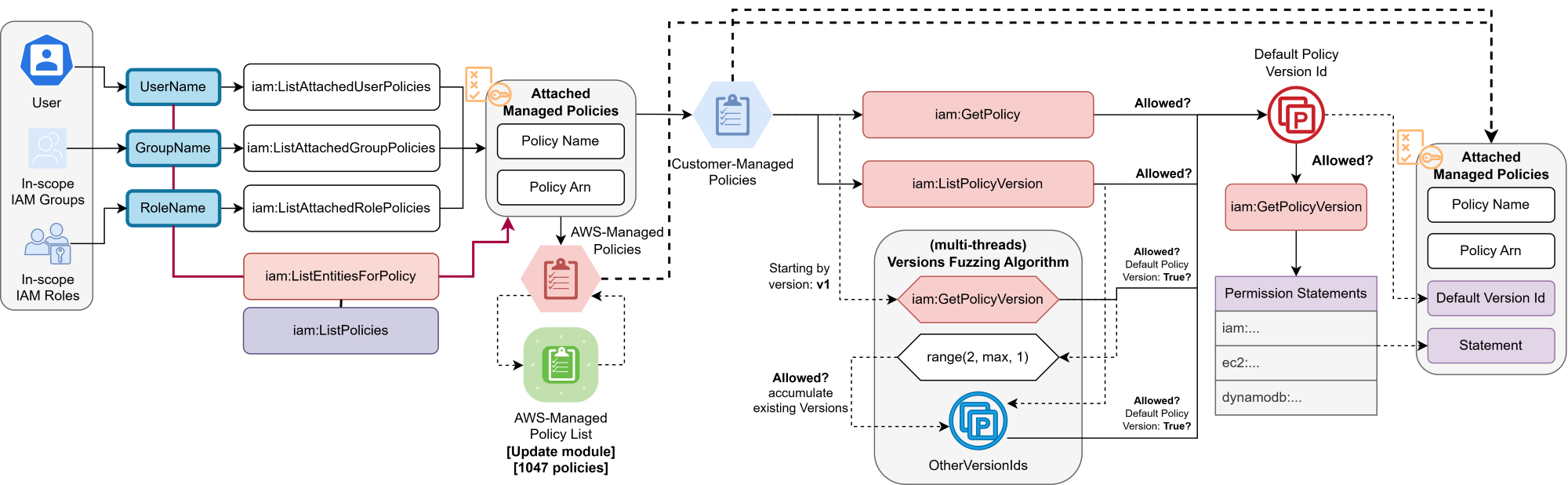}
    \caption{iam:ListEntitiesForPolicy in Attached Managed Policy Enumeration}
    \label{fig:420}
\end{figure}

\begin{figure}[htbp]
    \centering
    \includegraphics[width=1.0\linewidth]{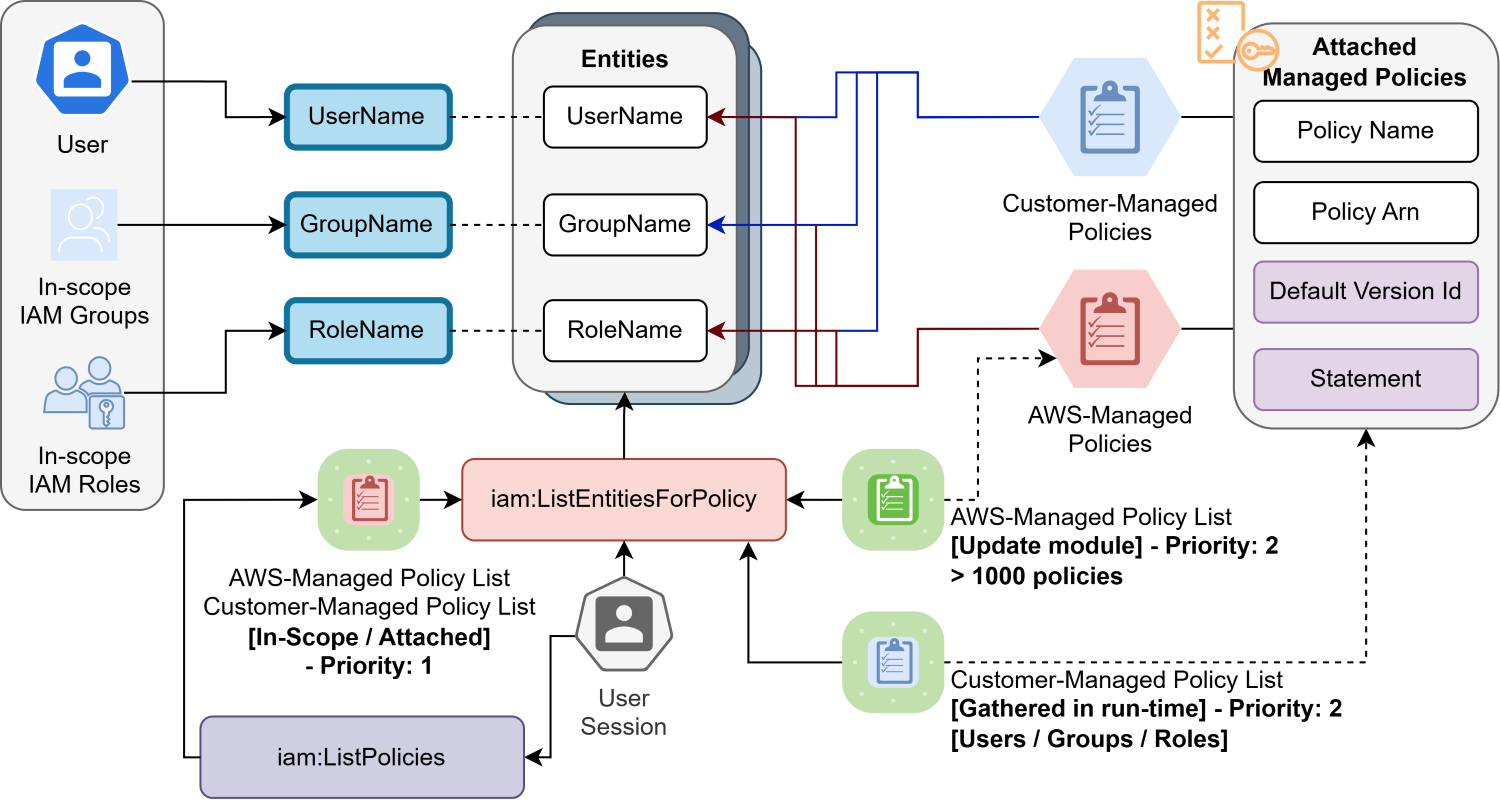}
    \caption{Inverse Enumeration Model by iam:ListEntitiesForPolicy}
    \label{fig:421}
\end{figure}

While forward enumeration (beginning with the principal and progressing to its policies) reveals the most direct route to discovering an entity’s privileges, it inherently risks overlooking configurations if any references to the in-scope principals and entities were not retrievable due to insufficient authorization in earlier steps. Moreover, an extensive set of customer-managed and AWS-managed policies across multiple accounts can introduce complex permission inheritance pathways. In such cases, an inverse enumeration strategy - starting from the policy and mapping back to the principals - can demonstrate superior strengths.\\

The principal mechanism for this inverse enumeration approach is iam:ListEntitiesFor-
Policy, which enumerates all IAM users, groups, and roles attached to a specified policy. When combined with the chain logic from earlier enumerations, this approach can help capture any missing elements in the earlier IAM enumeration result. Specifically, if an assessment by SkyEye uncovers references to attached managed policies that have not yet been sufficient from the overall analysis, SkyEye will try invoking iam:ListPolicies with the parameters: Scope='All', OnlyAttached=True, PolicyUsageFilter ='PermissionsPolicy' (provided if user principal hold sufficient privilege to invoke) and transfer those returned attached managed policies to iam:ListEntitiesForPolicy, to actively compare the iam:ListEntitiesForPolicy result of their corresponding principal attachments by the identified insufficient components to complement to them. If SkyEye was failed to invoke iam:ListPolicies due to insufficient authorization, SkyEye will revisit all publicly AWS-managed policy ARNs or previously-discovered customer-managed policy ARNs and transfer those into iam:ListEntitiesForPolicy call, to complement to the identified insufficient components in a limited scope.\\

\sloppy Technically, iam:ListEntitiesForPolicy acts as the inverse of commands like iam:ListAttachedUserPolicies, iam:ListAttachedGroupPolicies, and iam:ListAttachedRolePolicies. Instead of iterating over each user, group, or role to find its respective policy attachments, the inverse approach enumerates entities from the standpoint of each policy. In large-scale AWS accounts, the synergy between forward and inverse enumeration techniques ensures that SkyEye captures all relationships, even those formed through less conventional resource configurations. This is essential for producing a complete IAM graph, minimizing the possibility of overlooking powerful role or user relationships, or lacking forward enumeration privileges. Combined with advanced analytics on these enumerations, the inverse approach fortifies the capacity to detect, analyze, and mitigate privilege-based vulnerabilities before they evolve into exploitable security weaknesses.\\

\subsection{Deep Comparison Model for Policy Documents of Active Version and Historical Versions}

In AWS, customer-managed policies constitute a crucial mechanism for organizations to tailor and maintain precise access control configurations, thereby ensuring that only the necessary privileges are granted to particular user principals. Notably, AWS supports up to five concurrent policy versions for each customer-managed policy, with one version designated as the default active version. This multi-version approach offers administrators the flexibility to develop and test alternative privilege definitions without disrupting existing workloads. However, as IAM environments grow increasingly complex and policies evolve across multiple revisions, identifying permission changes among these versions becomes essential for maintaining robust security postures.\\

\begin{figure}[htbp]
    \centering
    \includegraphics[width=1.0\linewidth]{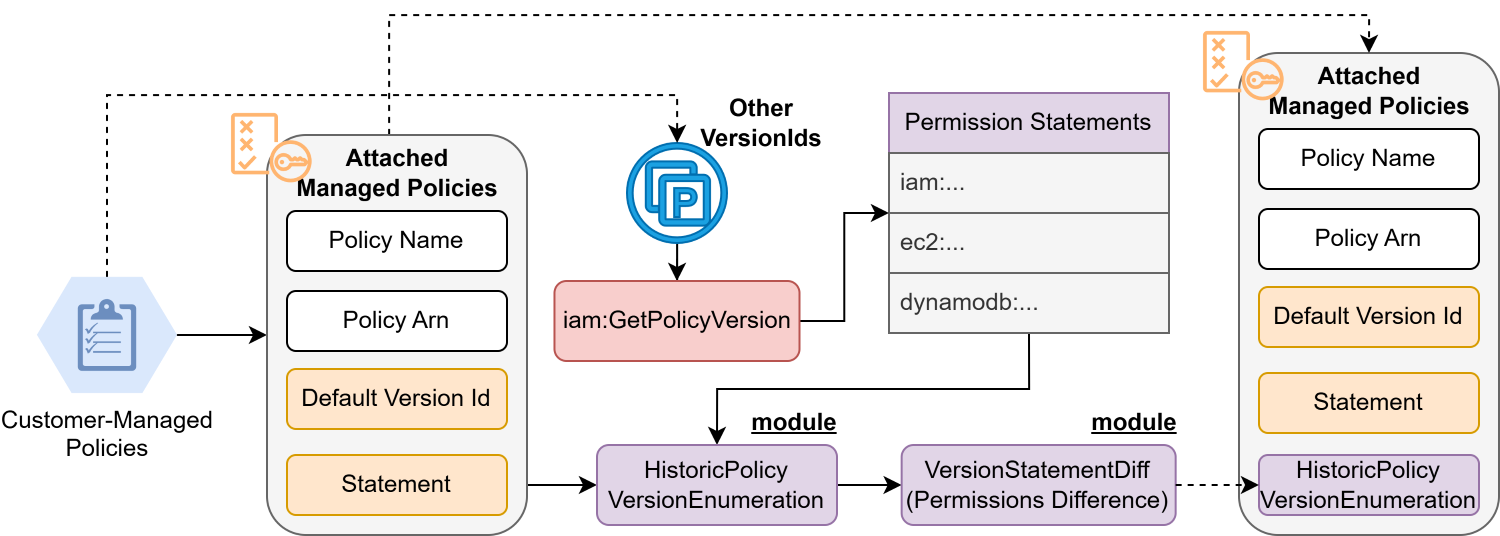}
    \caption{Gathering Policy Documents of Each Customer-Managed Policy Version}
    \label{fig:422}
\end{figure}

\begin{figure}[htbp]
    \centering
    \includegraphics[width=1.0\linewidth]{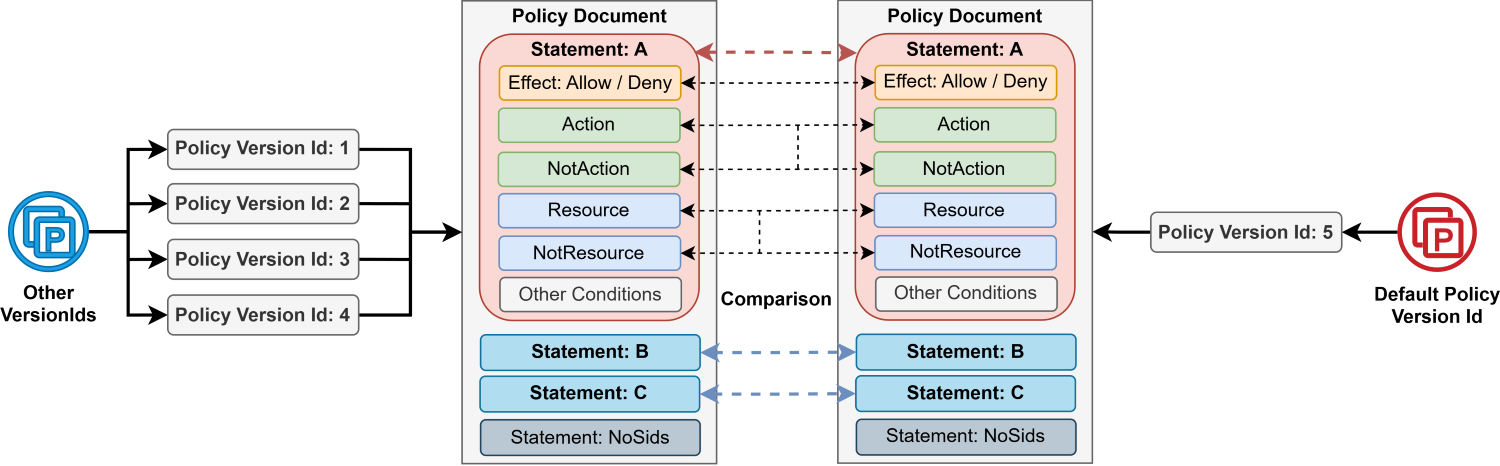}
    \caption{The Core of Deep Comparison Model}
    \label{fig:423}
\end{figure}

To address this challenge, the proposed model systematically compares policy statements between the DefaultPolicyVersionId and each OtherVersionId. By focusing on elements such as Effect (Allow or Deny), Action, NotAction, Resource, and NotResource, the model creates a comprehensive mapping of how privileges shift between versions. This mapping classifies changes into distinct categories: New, NotChange (Kept), and Removed (Old). Thereby highlighting which privileges would be gained, retained, or lost if a future iam:SetDefaultPolicyVersion operation were to activate an older version. Consequently, security professionals can precisely forecast the ramifications of reverting to any previous policy version, enabling informed decisions on whether updates would inadvertently grant excessive permissions or compromise necessary access controls.\\

The significance of this model lies in its capability to provide granular insights that surpass manual policy analysis methods. Traditional diff-based techniques can overlook subtle AWS IAM policy language nuances such as multiple Resource definitions or intricacies in combined NotAction statements. The proposed model not only captures these complexities but also contextualizes them, streamlining the process of identifying privilege escalations and ensuring continuous adherence to the principle of least privilege. This consistent, automated approach assists organizations in establishing a clear audit trail of changes, reducing the risk of unintended permission expansions and compliance violations.\\

Furthermore, this method promotes proactive risk assessment by illustrating the potential effects associated with activating any non-default version. As security teams or penetration testing teams frequently grapple with privilege misconfigurations, the ability to predict precisely which permissions would be introduced or eliminated affords a powerful framework for safeguarding mission-critical infrastructure, or conducting an effective privilege escalation attack vector by penetration testing team. In addition, such systematic enumeration and comparison fosters an integrated security posture, wherein cloud governance aligns with business objectives while preserving compliance standards. Ultimately, this algorithm underscores the dynamic nature of IAM policies, equipping practitioners with an advanced methodology for analyzing multiple policy versions and reinforcing a secure and well-defined access control framework.\\

\section{The Integration of MITRE ATT\&CK Cloud}

The extensible dataset underpinning the SkyEye framework is foundational to its practical utility, as it systematically maps nearly 20,000 AWS actions to corresponding severity-level classifications and contextual adversarial behaviors. Within SkyEye, the capability to detect, classify, and categorize all AWS actions into risk levels ranging from Low, Medium, High, and Critical, to those specifically denoted as PrivEsc-Vector, represents a pivotal advancement in threat exposure. By mapping each AWS action with relevant MITRE ATT\&CK tactics, techniques, and sub-techniques, the framework facilitates granular, multi-dimensional mapping that illuminates how adversaries might exploit specific permissions to achieve objectives such as data exfiltration, persistence within systems, or the sabotage of production workloads. This alignment with the MITRE ATT\&CK cloud matrix not only enhances methodological rigor but also reinforces both automated detection mechanisms and strategic countermeasures by highlighting concrete adversarial behaviors and attack pathways.\\

\begin{figure}[htbp]
    \centering
    \includegraphics[width=1.0\linewidth]{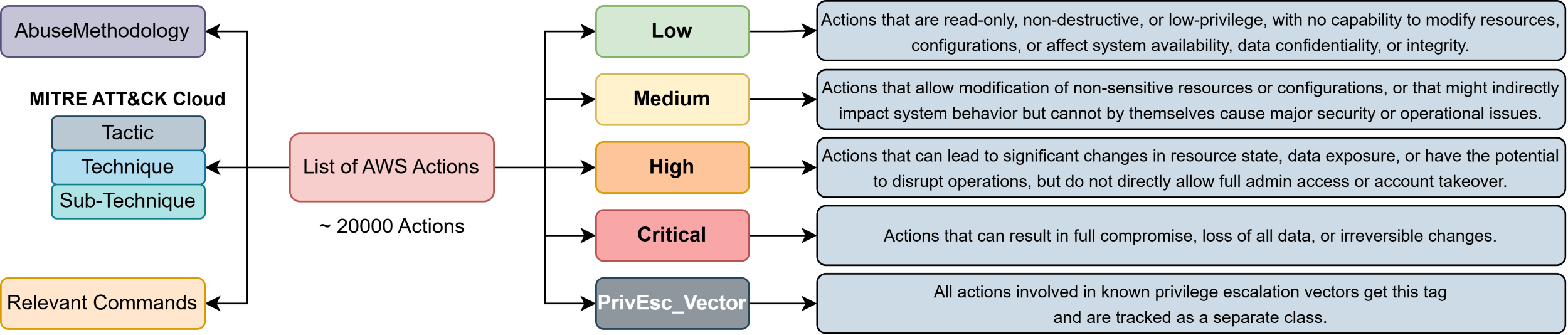}
    \caption{The Integration of Severity-level, Abuse Methodology and MITRE ATT\&CK}
    \label{fig:424}
\end{figure}

A noteworthy aspect of this classification is the thoroughness with which abuse methodologies are delineated for each permission. SkyEye framework maps every AWS action to a structured Abuse Methodology description, articulating how a threat actor might employ that permission to achieve lateral movement, privilege escalation, or data destruction. For instance, a High or Critical classification indicates that an AWS action may allow the modification of critical resources or the near-complete takeover of a specific service, while a PrivEsc-Vector label flags permissions that could directly elevate user privileges beyond their original scope. These detailed references, accompanied by example commands to illustrate the abuse, offer a practical vantage point for the teams to anticipate potential attack vectors and construct effective attack simulation. Such clarity not only highlights which permissions are of particular concern but also enables penetration testing teams to gain a complete situational awareness regarding the environment, or security teams to devise proactive incident response actions.\\

The detailed mapping to MITRE ATT\&CK tactics, techniques, and sub-techniques ensures that the final IAM enumeration result provided by SkyEye framework, is immediately actionable. By labeling each permission with a Tactic code (e.g., Privilege Escalation), Technique code (e.g., T1078 for Valid Accounts), and sub-technique code (as applicable), cloud security engineers or penetration tester can focus on the most salient threats in the targeting cloud environment. This layered approach proves beneficial during compromise assessments, facilitating the correlation of known adversary techniques with existing permissions. Consequently, the classification system bridges the gap between theoretical knowledge of adversary behaviors and the practical realities of maintaining secure cloud deployments.\\

From a defensive perspective, this severity-based categorization guides the development of fine-grained access control policies. Security teams can prioritize the remediation of permissions that have been flagged as Critical or PrivEsc-Vector by restricting or removing them. Additionally, this enables more data-driven policy recommendations, where developers and operations staff can gain better awareness of the privileges they request, thereby aligning their environment with the principle of least privilege. Such alignment reduces the overall attack surface by methodically limiting the exposed hooks that malicious actors might try to exploit. When integrated into continuous deployment pipelines, these disciplined guardrails systematically enforce best practices, promoting a robust security posture.\\

On the other hand, an offensive or red-team perspective leverages the same classification schema for scenario-based testing and vulnerability exploration. By systematically probing permissions labeled as High or Critical or chaining with the identified permissions labeled as Low or Medium, offensive security team can simulate advanced adversary behaviors, thus validating alert mechanisms and identifying real-world paths to privilege escalation. Having explicit example commands to abuse the identified vulnerabilities shortens the feedback loop between reconnaissance and exploitation phases, thereby improving the sophistication and realism of penetration testing exercises. This cyclical process of assessment and remediation ensures that misconfigurations and dangerous permissions are swiftly discovered, cataloged, and neutralized.\\

In general, this capability systematically categorizes and illustrates each AWS action’s inherent risk, associated MITRE ATT\&CK mapping, Abuse Methodology description, and sample abuse commands marks a cornerstone in modern cloud security. The ability to visualize and quantify risk in such depth fosters a decisive advantage for organizations striving to maintain compliance, harden their assets, and prevent potential adversaries. As cloud environments evolve in complexity, this synergy of detailed enumeration, severity classification, and actionable intelligence empowers both defenders and ethical adversaries to make informed and strategic decisions, ultimately fortifying the resilience and integrity of AWS-based infrastructures.\\
\newpage
\chapter{Evaluation}
\section{The Comprehensive Scenario-based Benchmarking}
To rigorously evaluate the effectiveness of advanced IAM enumeration frameworks, we designed a comprehensive scenario-based benchmarking methodology. Our empirical study encompasses a set of twenty-two meticulously crafted scenarios within AWS Identity and Access Management, spanning the core entities of IAM Users, Groups, Roles, and Policies. The primary objective is to systematically compare our proposed framework ("SkyEye") integrated by our proposed core models: CPIEM, TCREM, and several IAM deep enumeration capabilities as demonstrated in the previous chapter, against six established and reputable IAM enumeration frameworks currently available in the field.\\

For each scenario, a controlled AWS environment was provisioned, containing a blend of inline and attached managed policies, nested group memberships, diverse trust policy configurations as presented in the previous section in this chapter. By approaching it with a black-box perspective, standardized AWS credentials of involving user principals with permissions tailored to each scenario, were supplied to every framework under the benchmarking process. The enumeration results produced by each framework were collected and subjected to detailed analysis, focusing on multiple critical dimensions of IAM visibility that are relevant and tied to the targeting AWS credentials.\\

Specifically, the benchmarking process measured the capability of each framework to enumerate deeply: (a) inline polcies and attached managed policies of user principals, in-scope IAM groups which are defined as those to which the enumerated users belong, and in-scope IAM roles which are encompassing roles that a user can directly or indirectly assume via trust relationships. The outputs were further analyzed for their completeness in revealing complete IAM vision context of targeting AWS credentials or user principals, ensuring the comprehensive situational awareness regarding to their permissions and resources that can be interacted with, opening for the understanding of potential pathways in privilege escalation, data exfiltration, resource abuse, and threats to the integrity and confidentiality of the AWS environment.\\

Effectiveness was quantified as the percentage of discovered entities and relationships relative to the known ground truth of each scenario. Our methodology also stressed the frameworks' capacity to surface actionable intelligence, such as identification of latent privilege escalation paths and detection of configurations susceptible to abuse.\\

This scenario-driven, empirical approach ensures that the comparative analysis is both robust and practically relevant. By leveraging real-world AWS configurations and a diverse set of abuse scenarios, we provide a nuanced assessment of strengths and limitations for each framework. The results not only highlight the advancements introduced by our proposed SkyEye framework and its integrated models in IAM enumeration and threat modeling, but also establish a foundational benchmark for future research and tool development in cloud security situational awareness.
\section{Proposed Scenarios}
\subsection*{Scenario [\hyperref[sec:scenario1]{S1}]:}
Scenario 1 represents the most ideal and comprehensive environment for analyzing AWS IAM policy discovery chains. In this setup, a single IAM user (S1\_UserA) is assigned a variety of inline and managed policies, both directly and indirectly through group membership and assumable roles. Each policy provides granular permissions related to IAM actions, including listing and retrieving user, group, and role policies, as well as permissions for other AWS services such as S3, Lambda, and EC2. This scenario allows for an exhaustive demonstration of IAM entitlement enumeration and cross-policy visibility.
\subsection*{Scenario [\hyperref[sec:scenario2]{S2}]:}
Scenario 2 explores a variation of IAM policy enumeration where the action iam:ListGroupsForUser is replaced by the combination of iam:ListGroups and iam:GetGroup. This adjustment reflects environments where group membership must be deduced indirectly. The scenario involves a user (S2\_UserA) with a mix of inline and attached managed policies, group membership, and an assumable role. Permissions span both IAM-related actions and supplemental AWS services (AIOps, IoT, S3, EC2, Lambda), illustrating how diverse policy chains enable visibility and access across AWS resources.
\subsection*{Scenario [\hyperref[sec:scenario3]{S3}]:}
Scenario 3 focuses on the use of iam:GetPolicy instead of iam:ListPolicyVersions for policy enumeration within AWS IAM. This reflects an environment where policy details are accessed directly rather than through version listings. The scenario features a user (S3\_UserA) with various inline and managed policies, group membership, and an assumable role, including permissions across IAM, AIOps, IoT, S3, Lambda, EC2, and Route53 services. This setup demonstrates how entitlement and policy discovery chains adapt when the available IAM actions change.
\subsection*{Scenario [\hyperref[sec:scenario4]{S4}]:}
Scenario 4 demonstrates an IAM environment where both iam:ListPolicyVersions and iam:GetPolicy are unavailable for policy enumeration. This limitation means that direct access to policy version details is not possible, requiring iam:GetPolicyVersions with "Versions Fuzzing Algorithm" which is introduced in this paper to discover and enumerate entitlements of policy document. The scenario still includes a user (S4\_UserA) with inline and attached managed policies, group membership, and an assumable role, illustrating how entitlement visibility chains must adapt when critical enumeration actions are absent.
\subsection*{Scenario [\hyperref[sec:scenario5]{S5}]:}
Scenario 5 explores the use of iam:ListEntitiesForPolicy as an alternative to iam:ListAttachedRolePolicies and iam:ListAttachedGroupPolicies for identifying which users, groups, or roles are attached to a given policy. This adjustment demonstrates how policy-to-entity relationships can be enumerated even when direct attachment-listing actions are restricted. The scenario features a user (S5\_UserA) with comprehensive inline and managed policies, group membership, and an assumable role, all enriched with permissions spanning IAM, AIOps, IoT, Bedrock, S3, Lambda, EC2, Route53, Kinesis, and AmazonMQ services. It highlights flexible enumeration strategies in complex IAM environments.
\subsection*{Scenario [\hyperref[sec:scenario6]{S6}]:}
Scenario 6 focuses on environments where all iam:ListAttached*Policies actions (for users, groups, and roles) are unavailable, and enumeration must leverage iam:ListEntitiesForPolicy and iam:ListPolicies. This approach illustrates the ability to enumerate policy attachments and available managed policies by listing all policies and then determining their associations. The user (S6\_UserA) is equipped with rich inline and managed policies, has group membership, and can assume a role, spanning permissions across diverse AWS services. This scenario demonstrates flexible policy discovery strategies when direct attachment-listing actions are missing.
\subsection*{Scenario [\hyperref[sec:scenario7]{S7}]:}
Scenario 7 illustrates an IAM environment where the ideal set of IAM enumeration and discovery permissions is concentrated within a Role, rather than being distributed among Users or Groups. This configuration demonstrates the power of role assumption for policy visibility and entitlement discovery, as the role (S7\_RoleA) possesses comprehensive IAM listing and retrieval privileges. The user (S7\_UserA) and their group have limited direct IAM permissions, focusing more on non-IAM AWS services. This scenario highlights how centralizing enumeration capabilities in a role can facilitate entitlement mapping and cross-account or privilege escalation investigations.
\subsection*{Scenario [\hyperref[sec:scenario8]{S8}]:}
Scenario 8 demonstrates a sophisticated transitive cross-role enumeration model, where the ideal IAM policy discovery and enumeration privileges are distributed across multiple roles in a chained, assumable sequence. Starting with a user (S8\_UserA) who can assume S8\_RoleA, each subsequent role (S8\_RoleB, S8\_RoleC, S8\_RoleD) is assumable by the previous one, forming a privilege escalation path (S8\_UserA $\rightarrow$ S8\_RoleA $\rightarrow$ S8\_RoleC $\rightarrow$ S8\_RoleD). Policy discovery actions are spread throughout these roles, requiring the user to traverse multiple assumptions to achieve full entitlement visibility. This scenario models real-world advanced enumeration and privilege escalation tactics in complex AWS environments.
\subsection*{Scenario [\hyperref[sec:scenario9]{S9}]:}
Scenario 9 highlights the impact of granting the iam:GetAccountAuthorizationDetails permission directly to a user or group. This action provides comprehensive visibility into all IAM users, groups, roles, and their associated policies within an AWS account, enabling a single API call to enumerate most entitlements and relationships. The scenario features a user (S9\_UserA) with this powerful permission and a mix of other service-level permissions distributed across user, group, and assumable role constructs. This setup emphasizes how a single IAM action can streamline and centralize entitlement discovery for security reviews or audits.
\subsection*{Scenario [\hyperref[sec:scenario10]{S10}]:}
Scenario 10 demonstrates the impact of granting the powerful iam:GetAccountAuthorizationDetails permission within a role (rather than directly to a user or group). When a user (S10\_UserA) assumes S10\_RoleA, they gain the ability to enumerate nearly all IAM entities and their policies for the account through a single API call. This centralized entitlement discovery mechanism is enhanced by additional permissions spread across the user, group, and role, showcasing how key privileges embedded in assumable roles can facilitate comprehensive security reviews or privilege escalation.
\subsection*{Scenario [\hyperref[sec:scenario11]{S11}]:}
Scenario 11 demonstrates a transitive cross-role enumeration model where the highly privileged \texttt{iam:GetAccountAuthorizationDetails} permission is only available at the end of a chained sequence of assumable roles. The initial user and groups have no direct entitlement discovery privileges, but by assuming a series of roles (S11\_UserA $\rightarrow$ S11\_RoleA $\rightarrow$ S11\_RoleB $\rightarrow$ S11\_RoleC $\rightarrow$ S11\_RoleD), the user ultimately obtains the ability to enumerate all IAM entities and their relationships within the AWS account via a single API call. This scenario highlights advanced techniques for privilege escalation and account-wide IAM visibility in complex AWS environments.
\subsection*{Scenario [\hyperref[sec:scenario12]{S12}]:}
Scenario 12 demonstrates an enumeration model in which all key IAM policy and entity discovery permissions are concentrated in a role rather than directly assigned to users or groups. In this scenario, all \texttt{iam:ListAttached*Policies}, \texttt{iam:ListPolicyVersions}, and \texttt{iam:GetPolicy} actions are absent. Instead, the role (S12\_RoleA) leverages \texttt{iam:ListEntitiesForPolicy} and \texttt{iam:ListPolicies} for inverse enumeration of policy's attachment. The \texttt{iam:ListPolicies} action is scoped to customer-managed policies, and the mapping between policies and their attached entities is achieved inversely via \texttt{iam:ListEntitiesForPolicy}. Moreover, it highlights the importance of "Versions Fuzzing Algorithm" as demonstrated previously in the paper, to expose the policy's version without requiring the sufficient IAM permissions. This model mirrors real-world least-privilege or audit scenarios, emphasizing indirect but effective entitlement enumeration without being permitted tp discover attachment or policy versions.
\subsection*{Scenario [\hyperref[sec:scenario13]{S13}]:}
Scenario 13 models an ideal entitlement enumeration environment using a transitive cross-role enumeration model, where all key IAM enumeration permissions are distributed among a chain of assumable roles, rather than granted directly to users or groups. Critically, the scenario removes both \texttt{iam:ListPolicyVersions} and \texttt{iam:GetPolicy}, requiring the use of the "Versions Fuzzing Algorithm" as previously introduced in this paper for discovering and retrieving policy versions. This forces enumeration tooling to rely on indirect methods (such as sequentially attempting to retrieve versions via \texttt{iam:GetPolicyVersion}) to reconstruct the set of policy document versions in the absence of permissions allowing the retrieval of policy's versions. The scenario demonstrates how even without explicit permissions, full entitlement visibility can be achieved through creative enumeration and role chaining.
\subsection*{Scenario [\hyperref[sec:scenario14]{S14}]:}
Scenario 14 presents again an advanced transitive cross-role enumeration model, where all key IAM enumeration and policy discovery permissions are distributed across a chain of assumable roles rather than assigned to users or groups. This scenario removes all direct attachment listing permissions (\texttt{iam:ListAttached*Policies}), as well as policy version listing permissions: (\texttt{iam:ListPolicyVersions}, \texttt{iam:GetPolicy}). Instead, enumeration relies on \texttt{iam:ListEntitiesForPolicy} and \texttt{iam:ListPolicies}. The mapping of policies to entities is performed using an inverse enumeration model by firstly listing policies, then deterministically mapping to principals using \texttt{iam:ListEntitiesForPolicy}. For policy version enumeration and retrieval, the "Versions Fuzzing Algorithm" described in this paper is required, allowing for creative discovery of policy versions even without direct listing permissions. This scenario highlights both the flexibility and the complexity of modern transitive cross-role enumeration model and the flexibility in enumeration capabilities to expose the complete IAM vision context, even in least-privilege situation.
\subsection*{Scenario [\hyperref[sec:scenario15]{S15}]:}
Scenario 15 demonstrates an effective utilization of cross-principal IAM enumeration model, in which multiple users possess distinct but complementary IAM permissions. Instead of a single principal holding all enumeration rights, each user (S15\_UserA, S15\_UserB, S15\_UserC, S15\_UserD) is granted specific, partial IAM discovery actions. Only by combining the permissions of all users can a complete picture of entitlements, including inline and attached managed policies for IAM users, in-scope IAM groups, and in-scope IAM roles, to be fully revealed. This scenario models real-world environments where attackers or auditors could identify multiple AWS credentials in the reconnaissance, and correlate the permissions from several user principals to fully enumerate the complete IAM vision context of each user.
\subsection*{Scenario [\hyperref[sec:scenario16]{S16}]:}
Scenario 16 demonstrates the most comprehensive enumeration approach by combining the cross-principal IAM enumeration model with the transitive cross-role enumeration model. In this scenario, multiple users (S16\_UserA, S16\_UserB, S16\_UserC, S16\_UserD) are each granted distinct but complementary IAM permissions. These users are also associated with intricate chains of assumable roles, where each role exposes only a subset of the required IAM discovery actions. Only by aggregating the permissions from all users and traversing their respective role chains can a complete entitlement map, including inline and attached managed policies for user principals, in-scope IAM groups, and in-scope IAM roles, to be fully discovered. This scenario reflects real-world situations where security teams or adversaries must chain together disparate IAM permissions from multiple credentials and role assumption paths to reveal the true and complete IAM vision context across gathered AWS credentials.
\subsection*{Scenario [\hyperref[sec:scenario17]{S17}]:}
Scenario 17 demonstrates an effective utilization of the cross-principal IAM enumeration model. In this scenario, one among several users (S17\_UserC, via group policy) has the powerful \texttt{iam:GetAccountAuthorizationDetails} permission, which enables comprehensive enumeration of all IAM entities and their policies. The remaining users (S17\_UserA, S17\_UserB, S17\_UserD) do not have any IAM operational permissions in term of discovery, therefore cannot achieve full IAM visibility by itself. Only by correlating permissions from S17\_UserC - \texttt{iam:GetAccountAuthorizationDetails} to retrieve a complete view of their IAM vision context, including inline and attached managed policies for IAM users, in-scope IAM groups, and in-scope IAM rolesbe. This reflects real-world reconnaissance or audit cases where multiple credentials are discovered and at least one credential has the powerful permission \texttt{iam:GetAccountAuthorizationDetails} to support other user principals revealing their own IAM vision context.
\subsection*{Scenario [\hyperref[sec:scenario18]{S18}]:}
Scenario 18 demonstrates a hybrid enumeration case study that leverages both cross-principal IAM enumeration model and transitive cross-role enumeration model. Multiple users (S18\_UserA, S18\_UserB, S18\_UserC, S18\_UserD) are present, each with different and complementary permissions. Each user also maintains one or more associated roles, forming role assumption chains. In this scenario, the highly privileged \texttt{iam:GetAccountAuthorizationDetails} action is granted at the end of a transitive role chain for one user (S18\_UserD), but not directly to any user. As a result, obtaining a complete vision of AWS IAM entitlements (including inline and attached managed policies for users, groups, and roles) requires aggregating actions across all users and traversing the role chain to reach the powerful entitlement of discovery action. This scenario reflects real-world audit or red team situations where the combination of multiple credentials and deep role traversal is necessary to achieve exhaustive IAM visibility.
\subsection*{Scenario [\hyperref[sec:scenario19]{S19}]:}
Scenario 19 demonstrates the cross-principal IAM enumeration model in which multiple users (S19\_UserA, S19\_UserB, S19\_UserC, S19\_UserD) have distinct but complementary IAM discovery and enumeration permissions. In this scenario, critical actions \texttt{iam:ListPolicyVersions} and \texttt{iam:GetPolicy} are intentionally removed, requiring enumeration by utilizing the "Versions Fuzzing Algorithm" which is previously introduced in this paper, iteratively calling \texttt{iam:GetPolicyVersion}, to discover and retrieve policy versions and documents. Only by combining the permissions of all gathered AWS credentials, can reveal the full set of entitlements (inline and attached managed policies for IAM users, in-scope IAM groups, and in-scope IAM roles). This scenario reflects real-world security or audit cases where direct access to policy metadata is lacking and advanced enumeration logic is necessary to reconstruct the complete IAM context.\\
\subsection*{Scenario [\hyperref[sec:scenario20]{S20}]:}
Scenario 20 combines the transitive cross-role enumeration model and the cross-principal IAM enumeration model. Multiple users (S20\_UserA, S20\_UserB, S20\_UserC, S20\_UserD) each possess distinct but complementary IAM permissions. These users are associated with various assumable roles, forming deep role chains. In this scenario, direct access to \texttt{iam:ListPolicyVersions} and \texttt{iam:GetPolicy} is removed from all principals, enforcing the use of the "Versions Fuzzing Algorithm" (as previously introduced in this paper) with \texttt{iam:GetPolicyVersion} to enumerate managed policy versions. Only by combining permissions from all users and traversing their role chains can reveal a complete picture of AWS IAM entitlements (including inline and attached managed policies for IAM users, in-scope IAM groups, and in-scope IAM roles). This scenario reflects advanced audit or red team use cases where privilege is distributed and enumeration must be both cooperative and technically creative.\\
\subsection*{Scenario [\hyperref[sec:scenario21]{S21}]:}
Scenario 21 demonstrates the use of the \texttt{iam:SimulatePrincipalPolicy} permission, which allows simulation of the effective permissions for any AWS action granted to IAM users, groups, and roles. In this scenario, \texttt{iam:SimulatePrincipalPolicy} is granted to a user via an attached managed policy, enabling them to simulate policies across their own user identity, their group membership, and along an assumable transitive role chain. This scenario reflects the power of simulation APIs to reveal the effective permissions model without requiring direct enumeration of every inline and attached policy, and is especially relevant for audit and security review workflows that need to understand real-world access for complex principal chains.
\subsection*{Scenario [\hyperref[sec:scenario22]{S22}]:}
Scenario 22 highlights the powerful technique of fuzzing AWS read-only actions for reconnaissance and resource enumeration, without relying on any IAM-specific enumeration action. In this environment, the user and associated principals lack IAM discovery permissions, but possess a variety of descriptive and listing actions across AWS services (e.g., \texttt{s3:ListBuckets}, \texttt{ec2:DescribeInstances}, \texttt{lambda:ListFunctions}). By systematically invoking these read-only actions, an operator can enumerate resources, understand the account's cloud footprint, and indirectly reveal significant information about the environment. This scenario demonstrates the importance of not overlooking non-IAM permissions when assessing privilege escalation and lateral movement risks.

\subsection{Weighting Methodology for Proposed Scenarios}

\begin{table}[htbp]
\centering
\begin{tabular}{|l|r|}
\hline
\textbf{User Principal} & \textbf{1.00} \\
\hline
-- \textbf{[User]} Inline Policies (PolicyName) & 0.06 \\
\quad -- Detail Policy Documents (Policy Statement) & 0.06 \\
-- \textbf{[User]} Attached Managed Policies (PolicyName + PolicyARN) & 0.06 \\
\quad -- Default Policy Version Id & 0.06 \\
\quad \quad -- Detail Policy Documents (Policy Statement) & 0.06 \\
\hline
-- \textbf{In-scope IAM Groups} (Groups that the user belongs to) & 0.05 \\
\hline
\quad -- \textbf{[Group]} Inline Policies (PolicyName) & 0.06 \\
\quad \quad -- Detail Policy Documents (Policy Statement) & 0.06 \\
\quad -- \textbf{[Group]} Attached Managed Policies (PolicyName + PolicyARN) & 0.06 \\
\quad \quad -- Default Policy Version Id & 0.06 \\
\quad \quad \quad -- Detail Policy Documents (Policy Statement) & 0.06 \\
\hline
-- \textbf{In-scope IAM Roles} (Roles that user can assume directly or in-directly) & 0.05 \\
\hline
\quad -- \textbf{[Role]} Inline Policies (PolicyName) & 0.06 \\
\quad \quad -- Detail Policy Documents (Policy Statement) & 0.06 \\
\quad -- \textbf{[Role]} Attached Managed Policies (PolicyName + PolicyARN) & 0.06 \\
\quad \quad -- Default Policy Version Id & 0.06 \\
\quad \quad \quad -- Detail Policy Documents (Policy Statement) & 0.06 \\
\hline
\end{tabular}
\caption{Benchmarking Weights}
\end{table}

To ensure a granular, equitable, and reproducible comparative assessment of IAM enumeration capabilities, each of the 22 scenarios is structured with a precise weighting schema that reflects the hierarchical and relational complexity of AWS IAM entities. For scenarios involving a single user (Scenarios 1–14), the evaluation framework assigns discrete percentage weights to each discovery dimension, encompassing inline policies, attached managed policies, group and role associations, and detailed extraction of policy documents. Specifically, the enumeration of inline policy names and their detailed policy documents accounts for 0.06 each, while attached managed policies and their corresponding default policy version and detail policy documents similarly contribute 0.06 each. The identification of in-scope IAM groups and roles, as well as the ability to enumerate their associated inline and managed policies, is weighted at 0.05 and 0.06 per objects, respectively. These weights cumulatively ensure that each scenario sums to 1, providing a holistic metric of enumeration coverage.In scenarios where there are multiple objects within a benchmarking category such as in-scope IAM groups, in-scope IAM roles, inline policies, or attached managed policies, the total assigned weight for that category is distributed equally among all objects in the category. Specifically, the weight allotted to each category is divided by the number of objects present, ensuring that the sum of the scores for all objects in that category does not exceed the maximum designated weight for the category. This approach maintains consistency and fairness in benchmarking, regardless of the number of objects enumerated within each evaluative dimension.\\

For scenarios featuring multiple users (Scenarios 15–20), the methodology applies the same weighting criteria; however, all percentages are divided equally among the involved users. For example, when four users participate in a scenario, each user's enumeration facets are apportioned such that the aggregate coverage across all users remains at 1. This proportional allocation maintains fairness in comparative analysis and accounts for the increased complexity introduced by multi-user environments. For the framework that does not support the multiple-user scanning, we will run separately multiple credentials for that framework to ensure the fairness in benchmarking.\\

Scenario 21 leverages the "iam:SimulatePrincipalPolicy" API to test a framework's ability to enumerate effective permissions via policy simulation. Here, the weighting is adjusted to emphasize the intersection of user and group policy simulation, with in-scope IAM group association, separated inline policies, consolidation of managed policies, and their details in default policy version and policy documents weighted at 0.05 (group association), 0.06 (separated inline policies) and 0.12 (consolidated managed policies) respectively. For the role trust relationship and their policy details in inline policies and managed policies, they are weighted similar to the methodology for scenario 1-14 as discussed previously. This scenario highlights each framework's proficiency in aggregating and simulating permissions across composite IAM relationships, reflecting real-world access evaluation workflows.\\

Scenario 22 introduces a fuzzing-based approach, wherein the outputs of permission fuzzing are directly compared to a pre-established set of expected permissions. This scenario departs from fixed percentage weights, instead focusing on the completeness and accuracy of discovered permissions relative to the scenario baseline.\\

This meticulous weighting methodology, applied consistently across all scenarios, underpins the validity and reproducibility of our benchmarking results. By quantifying discovery coverage with fine-grained percentages, the framework enables nuanced differentiation between competing enumeration tools and models, illuminating strengths and weaknesses at every layer of IAM complexity.

\newpage
\subsection{Calculation Methodology}

Delving deeper into calculation methodology, it is necessary to ensure objectivity and reproducibility in benchmarking IAM enumeration frameworks, we formalize the calculation of enumeration coverage using precise mathematical notation. Let \( S \) denote the set of all scenarios, indexed by \( s \), with \( N_s \) representing the total number of in-scope entities or relationships to be enumerated in scenario \( s \). For each framework \( F \), let \( E_{F,s,i} \) be a binary indicator where \( E_{F,s,i} = 1 \) if entity or relationship \( i \) in scenario \( s \) is correctly enumerated by framework \( F \), and \( E_{F,s,i} = 0 \) otherwise.\\

Each entity or relationship \( i \) in scenario \( s \) is assigned a scenario-specific weight \( w_{s,i} \) (e.g., 0.06, 0.12, 0.05, etc.), such that the sum over all \( i \) in \( s \) satisfies:
\[
\sum_{i=1}^{N_s} w_{s,i} = 1
\]

The coverage score \( C_{F,s} \) for framework \( F \) in scenario \( s \) is then computed as:
\[
C_{F,s} = \sum_{i=1}^{N_s} w_{s,i} \cdot E_{F,s,i}
\]
where \( 0 \leq C_{F,s} \leq 1 \), representing the proportion of correctly enumerated entities, weighted by their scenario-specific importance.\\

For scenarios involving multiple users (e.g., \( U_s \) users in scenario \( s \)), the weights for user-specific entities are divided by \( U_s \) to ensure the total weights for all users sum to 1:
\[
w_{s,i}^{(u)} = \frac{w_{s,i}}{U_s}
\]
where \( w_{s,i}^{(u)} \) denotes the weight assigned to entity \( i \) for user \( u \).\\

For the scenario 22, let \( P_s \) denote the set of all permissions expected in the scenario, and \( P_{F,s} \) denote the set of permissions discovered by framework \( F \). The coverage score is calculated as:
\[
C_{F,s}^{\text{fuzz}} = \frac{|P_{F,s} \cap P_s|}{|P_s|}
\]

The overall coverage score \( C_F \) for framework \( F \) across all scenarios is the average:
\[
C_F = \frac{1}{|S|} \sum_{s \in S} C_{F,s}
\]

This formalization enables a transparent and rigorous benchmarking of enumeration effectiveness, accommodating both scenario-specific weighting and diverse entity distributions, including nested and multi-user scenarios.

\newpage
\subsection{Benchmarking Table}
In this section, we will perform an assessment for the capability of enumerating IAM vision context between our framework ("SkyEye") and other frameworks, applying proposed scenarios with the previously-discussed methodology in weighting and calculation.\\

\vspace{3mm}
\centerline{\textsuperscript{a}SkyEye \hspace{5mm}\textsuperscript{b}CloudPEASS \cite{CloudPEASS}\hspace{5mm}\textsuperscript{c}ScoutSuite \cite{ScoutSuite}\hspace{5mm}\textsuperscript{d}CloudFox \cite{CloudFox}\hspace{5mm}\textsuperscript{e}PACU \cite{Pacu2021}\hspace{5mm}\textsuperscript{f}enumerate-iam \cite{enumerate-iam}}

\begin{table}[htbp]
\hspace{-1.2cm}
\footnotesize
\begin{tabular}{|l|c|c|c|c|c|c|}
\hline
\textbf{Scenario} & \textbf{Framework\textsuperscript{a}} & \textbf{Framework\textsuperscript{b}} & \textbf{Framework\textsuperscript{c}} & \textbf{Framework\textsuperscript{d}} & \textbf{Framework\textsuperscript{e}} & \textbf{Framework\textsuperscript{f}} \\
\hline
Scenario [\hyperref[sec:scenario1]{S1}] & 1.00 & 0.65 & 0.05 & 0.00 & 0.16 & 0.05 \\
\hline
Scenario [\hyperref[sec:scenario2]{S2}] & 1.00 & 0.30 & 0.10 & 0.025 & 0.10 & 0.10 \\
\hline
Scenario [\hyperref[sec:scenario3]{S3}] & 1.00 & 0.41 & 0.06 & 0.031 & 0.45 & 0.16 \\
\hline
Scenario [\hyperref[sec:scenario4]{S4}] & 1.00 & 0.41 & 0.52 & 0.29 & 0.21 & 0.16 \\
\hline
Scenario [\hyperref[sec:scenario5]{S5}] & 0.895 & 0.47 & 0.00 & 0.05 & 0.16 & 0.05 \\
\hline
Scenario [\hyperref[sec:scenario6]{S6}] & 1.00 & 0.29 & 0.00 & 0.00 & 0.16 & 0.00 \\
\hline
Scenario [\hyperref[sec:scenario7]{S7}] & 1.00 & 0.00 & 0.00 & 0.05 & 0.05 & 0.00 \\
\hline
Scenario [\hyperref[sec:scenario8]{S8}] & 1.00 & 0.00 & 0.05 & 0.05 & 0.05 & 0.05 \\
\hline
Scenario [\hyperref[sec:scenario9]{S9}] & 1.00 & 0.00 & 0.00 & 0.00 & 1.00 & 1.00 \\
\hline
Scenario [\hyperref[sec:scenario10]{S10}] & 1.00 & 0.00 & 0.00 & 0.05 & 0.05 & 0.05\\
\hline
Scenario [\hyperref[sec:scenario11]{S11}] & 1.00 & 0.00 & 0.05 & 0.05 & 0.05 & 0.05 \\
\hline
Scenario [\hyperref[sec:scenario12]{S12}] & 1.00 & 0.00 & 0.00 & 0.05 & 0.05 & 0.05\\
\hline
Scenario [\hyperref[sec:scenario13]{S13}] & 1.00 & 0.00 & 0.00 & 0.05 & 0.05 & 0.05 \\
\hline
Scenario [\hyperref[sec:scenario14]{S14}] & 1.00 & 0.00 & 0.00 & 0.05 & 0.05 & 0.05 \\
\hline
Scenario [\hyperref[sec:scenario15]{S15}] & 1.00 & 0.0425 & 0.04 & 0.04 & 0.04 & 0.00 \\
\hline
Scenario [\hyperref[sec:scenario16]{S16}] & 1.00 & 0.0275 & 0.00 & 0.00 & 0.0625 & 0.00\\
\hline
Scenario [\hyperref[sec:scenario17]{S17}] & 1.00 & 0.00 & 0.00 & 0.00 & 0.25 & 0.25 \\
\hline
Scenario [\hyperref[sec:scenario18]{S18}] & 1.00 & 0.00 & 0.00 & 0.00 & 0.00 & 0.00 \\
\hline
Scenario [\hyperref[sec:scenario19]{S19}] & 1.00 & 0.0425 & 0.18 & 0.00 & 0.3125 & 0.00 \\
\hline
Scenario [\hyperref[sec:scenario20]{S20}] & 1.00 & 0.0275 & 0.00 & 0.00 & 0.0125 & 0.013 \\
\hline
Scenario [\hyperref[sec:scenario21]{S21}] & 1.00 & 0.65 & 0.00 & 0.00 & 0.05 & 0.00 \\
\hline
Scenario [\hyperref[sec:scenario22]{S22}] & 0.90 & 0.00 & 0.00 & 0.00 & 0.00 & 0.90 \\
\hline
\end{tabular}
\caption{\centering The benchmarking across SkyEye and 6 published frameworks by 22 proposed scenarios}
\end{table}

Table~\ref{tab:skyeye-benchmarking-table} presents the results of our scenario-driven benchmarking study, applying the previously described weighting and calculation methodology across 22 diverse AWS IAM scenarios. The table quantifies each framework’s ability to enumerate the complete IAM vision context, as defined by the ground truth of each scenario, and highlights the empirical strengths and limitations of SkyEye against six established frameworks: \textsuperscript{a}SkyEye, \textsuperscript{b}CloudPEASS~\cite{CloudPEASS}, \textsuperscript{c}ScoutSuite~\cite{ScoutSuite}, \textsuperscript{d}CloudFox~\cite{CloudFox}, \textsuperscript{e}PACU~\cite{Pacu2021}, and \textsuperscript{f}enumerate-iam~\cite{enumerate-iam}.

\paragraph{SkyEye\textsuperscript{a}:}
SkyEye consistently achieves perfect or near-perfect coverage ($1.00$) across all scenarios, with marginal reduction ($0.895$) observed in Scenario S5 (Intended design with the minimum utilizable resources) and ($0.90$) observed in Scenario S22 (fuzzing-based permission enumeration). This empirical dominance reflects the integration of deep enumeration models (CPIEM, TCREM) and advanced algorithms (e.g., “Versions Fuzzing Algorithm”, "Inverse Enumeration", "Comprehensive Retrieval", "IAM Simulation", "IAM Fuzzing"), enabling exhaustive discovery of inline and managed policies, group and role associations, and historical policy documents, even in highly restrictive or indirect enumeration environments. SkyEye’s robustness is further demonstrated in complex transitive and multi-principall scenarios (S8--S20), where distributed privileges and chained role assumptions are required for full context reconstruction. It is important to note that the observed sub-maximal performance of SkyEye in Scenario S5 ($0.895$) and Scenario S22 ($0.90$) does not reflect an inherent limitation or deficiency of the framework itself. Rather, these results are a direct consequence of the deliberate scenario design, in which the available IAM permissions and accessible resources were intentionally constrained. These scenarios were architected to restrict the enumeration surface such that, by construction, complete exposure of the full IAM context was unattainable, even by an ideal or theoretically omniscient framework. The coverage scores achieved by SkyEye in S5 and S22 thus represent the maximal, ground-truth-aligned enumeration possible given the imposed environmental constraints, and should be interpreted as scenario-imposed upper bounds rather than indicators of framework capability.

\paragraph{CloudPEASS\textsuperscript{b}:}
CloudPEASS exhibits moderate performance in scenarios where direct enumeration actions are available (e.g., S1--S6, S21), with coverage declining sharply in advanced, transitive, or distributed permission contexts (S7--S20). The tool’s architecture favors straightforward enumeration paths and lacks the advanced chaining, inverse mapping, and creative inference mechanisms required for full discovery under least-privilege or multi-principal conditions. Consequently, its utility is limited in environments that deviate from ideal enumeration configurations.

\paragraph{ScoutSuite\textsuperscript{c}, CloudFox\textsuperscript{d}, PACU\textsuperscript{e}, enumerate-iam\textsuperscript{f}:}
These frameworks demonstrate substantially reduced coverage across most scenarios, with only isolated instances of higher scores when a small set of privileges (e.g., \texttt{iam:GetAccountAuthorizationDetails} in S9) enables bulk enumeration. Their enumeration logic is generally optimized for direct policy and entity listing actions, resulting in limited performance when such actions are absent, distributed, or require creative inference. The lack of advanced transitive role traversal, multi-user correlation, and versions fuzzing capabilities constrains their effectiveness in complex, real-world AWS IAM environments.

\begin{itemize}
    \item \textbf{S1--S7:} SkyEye achieves perfect coverage, leveraging all available enumeration actions and creative fallback algorithms; other frameworks’ scores are highly sensitive to the presence or absence of direct IAM listing actions.
    \item \textbf{S8--S14:} SkyEye’s unique capacity for chained, transitive, and inverse enumeration is unrivaled; other frameworks are unable to reconstruct the full IAM context in these advanced scenarios.
    \item \textbf{S15--S20:} Only SkyEye can aggregate across multiple users and role chains to achieve holistic enumeration; all others fail to scale to distributed privilege models.
    \item \textbf{S21--S22:} SkyEye’s support for simulation and fuzzing enables high coverage even where enumeration must be inferred from effective permissions or read-only action outputs.
\end{itemize}

The results firmly establish SkyEye as the state-of-the-art for IAM enumeration and situational awareness in AWS environments. Its architectural innovations with novel enumeration models, creative inference, transitive and distributed privilege aggregation, not only outperform existing frameworks but also set a reproducible benchmark for future research and development. These findings underscore the limitations of current open-source enumeration frameworks in capturing the full spectrum of real-world privilege relationships, and motivate the continued advancement of autonomous, context-aware enumeration techniques for cloud security.
\newpage
\chapter{Future Works}
While this paper has established the significance and efficacy of the SkyEye framework and its proposed models: Cross-Principal IAM Enumeration Model (CPIEM), Transitive Cross-Role Enumeration Model (TCREM), advanced IAM deep enumeration capabilities, and the systematic integration of MITRE ATT\&CK mapping and severity-level classification; there remain several promising avenues for future research and development to further advance the state of IAM enumeration and cloud security analytics.\\

The current instantiation of the SkyEye framework is tailored primarily to the AWS ecosystem. However, as cloud adoption diversifies, organizations increasingly rely on heterogeneous cloud environments, often leveraging Microsoft Azure, Google Cloud Platform (GCP), and other providers in parallel. Each platform introduces unique identity constructs, policy languages, and privilege management paradigms. Therefore, a critical direction for future work is the adaptation and generalization of the SkyEye models to support comprehensive, multi-cloud IAM enumeration. This would require the development of abstraction layers capable of normalizing disparate IAM representations and integrating platform-specific enumeration primitives, while preserving the core models and capabilities of the framework.\\

The integration of MITRE ATT\&CK matrix and severity-level classification in SkyEye offers a structured approach to contextualizing IAM exposures. Future work should deepen this integration by incorporating automated adversarial simulation modules capable of emulating advanced persistent threat (APT) behaviors, lateral movement techniques, and privilege escalation scenarios specific to each cloud provider. Furthermore, machine learning techniques could be explored to refine risk scoring models, leveraging empirical attack patterns and real-world incident datasets to prioritize enumeration targets and remediation actions more effectively.\\

Moreover, while the SkyEye framework is mainly developed for offensive security, it is recommended to utilize the proposed models of SkyEye framework in developing a framework that primarily focuses on defensive security aspect which can recommend and potentially automate remediation steps for detected IAM misconfigurations. This could involve the use of policy synthesis algorithms, guided least-privilege recommendations, and automated policy deployment routines, all governed by robust change management and audit controls. Such capabilities would close the loop from detection to remediation, further reducing the risk of privilege-based cloud compromise.
\newpage
\chapter{Conclusion}
This paper has introduced SkyEye, a comprehensive framework for advanced IAM enumeration in AWS cloud environments, along with its proposed models: the Cross-Principal IAM Enumeration Model (CPIEM), the Transitive Cross-Role Enumeration Model (TCREM), and a suite of deep IAM enumeration capabilities. SkyEye provides a robust mechanism for cooperative multi-principal IAM enumeration simultaneously between multiple principals that uncovers the privilege relationships, transitive trust paths, and hidden permissions that prior-art frameworks and models were unable to achieve due to the limitations of single-principal enumeration. Furthermore, SkyEye systematically integrates risk scoring and mapping to the MITRE ATT\&CK Cloud matrix, enabling the identification of potential vectors for privilege escalation within cloud infrastructures.\\

By decomposing and formalizing the enumeration chains necessary for the exhaustive discovery of IAM users, groups, roles, and policies; and through the development of the CPIEM and TCREM models; SkyEye addresses the limitations of conventional IAM enumeration frameworks, significantly reducing false negatives and enhancing the overall accuracy of cloud security assessments. In addition, the framework’s ability to incorporate both forward and inverse enumeration techniques, coupled with situational awareness enabled by deep comparison of policy states and versioning, ensures a holistic understanding of complex dynamic IAM configurations in real-world cloud environments.\\

The practical utility of SkyEye is underscored by its extensible dataset, which systematically maps nearly 20,000 AWS actions to the corresponding risk-level classifications and clearly describes the techniques leveraged by threat actors. This extensive mapping, in conjunction with SkyEye’s alignment with industry-standard adversarial models, enables the framework to provide actionable insights for both offensive and defensive security operations. Furthermore, as organizations increasingly adopt multi-cloud strategies, the architectural flexibility of the SkyEye framework lays a robust foundation for broader applicability, facilitating comprehensive cross-platform enumeration and adaptive threat modeling. Taken together, these attributes position SkyEye as a significant framework to advance the state of cloud security research and practice.\\

Nevertheless, the evolution of cloud identity and access management presents ongoing challenges. The paper has outlined promising directions for future research, including multi-cloud support, dynamic adversarial simulation, automated remediation, and the integration of machine learning for risk prioritization. By advancing both the technical rigor and operational depth of IAM enumeration, SkyEye represents a substantive contribution to the domain of cloud security, equipping practitioners and researchers with the frameworks needed to proactively identify, assess, and mitigate privilege-based risks in rapidly evolving cloud ecosystems.\\

In summary, SkyEye not only re-invents the standard for IAM enumeration and risk classification but also establishes a blueprint for future innovation by eliminating the limitations of conventional IAM enumeration strategy, and bridging the gap between detection and defense in the pursuit of resilient, trustworthy cloud environments.
\bibliographystyle{IEEEtran}
\bibliography{reference}

\begin{thebibliography}{10}
\providecommand{\url}[1]{#1}
\csname url@samestyle\endcsname
\providecommand{\newblock}{\relax}
\providecommand{\bibinfo}[2]{#2}
\providecommand{\BIBentrySTDinterwordspacing}{\spaceskip=0pt\relax}
\providecommand{\BIBentryALTinterwordstretchfactor}{4}
\providecommand{\BIBentryALTinterwordspacing}{\spaceskip=\fontdimen2\font plus
\BIBentryALTinterwordstretchfactor\fontdimen3\font minus \fontdimen4\font\relax}
\providecommand{\BIBforeignlanguage}[2]{{%
\expandafter\ifx\csname l@#1\endcsname\relax
\typeout{** WARNING: IEEEtran.bst: No hyphenation pattern has been}%
\typeout{** loaded for the language `#1'. Using the pattern for}%
\typeout{** the default language instead.}%
\else
\language=\csname l@#1\endcsname
\fi
#2}}
\providecommand{\BIBdecl}{\relax}
\BIBdecl

\bibitem{Pacu2021}
\BIBentryALTinterwordspacing
R.~S. Labs, ``Pacu: The aws exploitation framework,'' \url{https://github.com/RhinoSecurityLabs/pacu}, 2021, accessed: 2025-06-13. [Online]. Available: \url{https://github.com/RhinoSecurityLabs/pacu}
\BIBentrySTDinterwordspacing

\bibitem{CloudPEASS}
carlospolop, ``Cloudpeass - privilege escalation awesome scripts suite for cloud (aws, azure, gcp),'' \url{https://github.com/carlospolop/CloudPEASS}, 2025, accessed: 2025-06-13.

\bibitem{enumerate-iam}
andresriancho, ``Enumerate-iam: Security tool to enumerate and analyze iam permissions in aws environments,'' \url{https://github.com/andresriancho/enumerate-iam}, 2019, accessed: 2025-06-13.

\bibitem{CloudFox}
B.~Fox, ``Cloudfox: Automating situational awareness for cloud penetration tests on aws, azure, and gcp,'' \url{https://github.com/BishopFox/cloudfox}, 2022, accessed: 2025-06-13.

\bibitem{ScoutSuite}
N.~Group, ``Scoutsuite: Multi-cloud security auditing tool,'' \url{https://github.com/nccgroup/ScoutSuite}, 2018, accessed: 2025-06-13.

\bibitem{StratusRedTeam}
DataDog, ``Stratus red team: Adversary emulation for the cloud, in the cloud,'' \url{https://github.com/DataDog/stratus-red-team}, 2022, accessed: 2025-06-13.

\bibitem{Cloudsplaining}
Salesforce, ``Cloudsplaining: Aws iam security assessment tool,'' \url{https://github.com/salesforce/cloudsplaining}, 2024, accessed: 2025-06-13.

\bibitem{spark_2023}
Y.~Hu, W.~Wang, S.~Khurshid, K.~L. McMillan, and M.~Tiwari, ``Fixing privilege escalations in cloud access control with maxsat and graph neural networks,'' \url{https://spark.ece.utexas.edu/pubs/ASE-23-yang.pdf}, accessed: 2025-06-12.

\bibitem{capital_one_2022}
CapitalOne, ``Information on the capital one cyber incident,'' \url{https://www.capitalone.com/digital/facts2019}, accessed: 2025-06-12.

\bibitem{wired_2019}
L.~H. Newman, ``Everything we know about the capital one hacking case so far,'' \url{https://www.wired.com/story/capital-one-paige-thompson-case-hacking-spree}, accessed: 2025-06-12.

\bibitem{lapsus_2022}
Krebsonsecurity, ``A closer look at the lapsus\$ data extortion group,'' \url{https://krebsonsecurity.com/2022/03/a-closer-look-at-the-lapsus-data-extortion-group}, accessed: 2025-06-12.

\bibitem{bbc_2022}
J.~Tidy, ``Lapsus\$: Oxford teen accused of being multi-millionaire cyber-criminal,'' \url{https://www.bbc.com/news/technology-60864283}, accessed: 2025-06-12.

\bibitem{aws_security_blog_2023}
R.~Park, B.~Gorman, C.~Kundapur, and Z.~Miller, ``How to use aws certificate manager to enforce certificate issuance controls,'' \url{https://aws.amazon.com/blogs/security/how-to-use-aws-certificate-manager-to-enforce-certificate-issuance-controls}, accessed: 2025-06-12.

\bibitem{cornel_model_checking_2023}
\BIBentryALTinterwordspacing
A.~Gouglidis, A.~Kagia, and V.~C. Hu, ``Model checking access control policies: A case study using google cloud iam,'' \emph{arXiv preprint arXiv:2303.16688}, 2023, accessed: 2025-06-12. [Online]. Available: \url{https://arxiv.org/abs/2303.16688}
\BIBentrySTDinterwordspacing

\bibitem{detecting_anomalous_misconfig_2022}
T.~van Ede, N.~Khasuntsev, B.~Steen, and A.~Continella, ``Detecting anomalous misconfigurations in aws identity and access management policies,'' \url{https://dl.acm.org/doi/10.1145/3560810.3564264}, accessed: 2025-06-12.

\bibitem{usenix_2023}
I.~Shevrin and O.~Margalit, ``Detecting multi-step iam attacks in aws environments via model checking,'' \url{https://www.usenix.org/system/files/usenixsecurity23-shevrin.pdf}, accessed: 2025-06-12.

\bibitem{cornel_efficient_iam_2023}
Y.~Hu, W.~Wang, S.~Khurshid, and M.~Tiwari, ``Efficient iam greybox penetration testing,'' \url{https://arxiv.org/abs/2304.14540}, accessed: 2025-06-12.

\bibitem{ibm_cost_data_breach_2023}
I.~Security, ``Cost of a data breach report 2023,'' \url{https://d110erj175o600.cloudfront.net/wp-content/uploads/2023/07/25111651/Cost-of-a-Data-Breach-Report-2023.pdf}, accessed: 2025-06-13.

\bibitem{enisa_threat_landscape_2024}
E.~U.~A. for Cybersecurity~(ENISA), ``Enisa threat landscape 2024,'' \url{https://www.enisa.europa.eu/sites/default/files/2024-11/ENISA%20Threat%20Landscape%202024_0.pdf}, accessed: 2025-06-13.

\bibitem{aws_iam_user_guide}
``Aws identity and access management user guide,'' \url{https://docs.aws.amazon.com/IAM/latest/UserGuide/introduction.html}, accessed: 2025-06-11.

\bibitem{aws_iam_federation}
``Aws iam identity providers and federation,'' \url{https://docs.aws.amazon.com/IAM/latest/UserGuide/id_roles_providers.html}, accessed: 2025-06-11.

\bibitem{aws_organizations_iam}
``Aws organizations and iam,'' \url{https://docs.aws.amazon.com/organizations/latest/userguide/orgs_manage_accounts_access.html}, accessed: 2025-06-11.

\bibitem{aws_iam_policies}
``Aws iam policies,'' \url{https://docs.aws.amazon.com/IAM/latest/UserGuide/access_policies.html}, accessed: 2025-06-11.

\bibitem{aws_policy_evaluation_logic}
``Policy evaluation logic,'' \url{https://docs.aws.amazon.com/IAM/latest/UserGuide/reference_policies_evaluation-logic.html}, accessed: 2025-06-11.

\bibitem{aws_iam_users}
``Aws iam users,'' \url{https://docs.aws.amazon.com/IAM/latest/UserGuide/id_users.html}, accessed: 2025-06-11.

\bibitem{aws_iam_groups}
``Aws iam groups,'' \url{https://docs.aws.amazon.com/IAM/latest/UserGuide/id_groups.html}, accessed: 2025-06-11.

\bibitem{aws_iam_roles}
``Aws iam roles,'' \url{https://docs.aws.amazon.com/IAM/latest/UserGuide/id_roles.html}, accessed: 2025-06-11.

\bibitem{iam_json_policy_reference}
``Iam json policy reference,'' \url{https://docs.aws.amazon.com/IAM/latest/UserGuide/reference_policies.html}, accessed: 2025-06-11.

\bibitem{aws_managed_policies_and_inline_policies}
``Aws managed policies and inline policies,'' \url{https://docs.aws.amazon.com/IAM/latest/UserGuide/access_policies_managed-vs-inline.html}, accessed: 2025-06-11.

\bibitem{versioning_iam_policy}
``Versioning iam policies,'' \url{https://docs.aws.amazon.com/IAM/latest/UserGuide/access_policies_managed-versioning.html}, accessed: 2025-06-11.

\bibitem{permissions_boundaries_iam_entities}
``Permissions boundaries for iam entities,'' \url{https://docs.aws.amazon.com/IAM/latest/UserGuide/access_policies_boundaries.html}, accessed: 2025-06-11.

\bibitem{aws_service_control_policies}
``Service control policies (scps),'' \url{https://docs.aws.amazon.com/organizations/latest/userguide/orgs_manage_policies_scps.html}, accessed: 2025-06-11.

\bibitem{aws_accounts}
``Aws accounts,'' \url{https://docs.aws.amazon.com/IAM/latest/UserGuide/getting-started-account-iam.html}, accessed: 2025-06-11.

\bibitem{aws_organizations_ous}
``Managing organizational units (ous) with aws organizations,'' \url{https://docs.aws.amazon.com/organizations/latest/userguide/orgs_manage_ous.html}, accessed: 2025-06-11.

\bibitem{aws_organizations_intro}
``What is aws organizations?'' \url{https://docs.aws.amazon.com/organizations/latest/userguide/orgs_introduction.html}, accessed: 2025-06-11.

\bibitem{aws_cross_account_roles}
``Delegate access across aws accounts using iam roles,'' \url{https://docs.aws.amazon.com/IAM/latest/UserGuide/tutorial_cross-account-with-roles.html}, accessed: 2025-06-11.

\bibitem{aws_delegated_admin}
``Delegated administrator for aws organizations,'' \url{https://docs.aws.amazon.com/organizations/latest/userguide/orgs_delegate_policies.html}, accessed: 2025-06-11.

\bibitem{aws_iam_policy_elements}
``Iam json policy element reference,'' \url{https://docs.aws.amazon.com/IAM/latest/UserGuide/reference_policies_elements.html}, accessed: 2025-06-11.

\bibitem{aws_iam_condition_logic}
``Iam json policy elements: Condition,'' \url{https://docs.aws.amazon.com/IAM/latest/UserGuide/reference_policies_elements_condition.html}, accessed: 2025-06-11.

\bibitem{aws_abac}
``Define permissions based on attributes with abac authorization,'' \url{https://docs.aws.amazon.com/IAM/latest/UserGuide/introduction_attribute-based-access-control.html}, accessed: 2025-06-11.

\bibitem{aws_cli}
``Aws command line interface,'' \url{https://docs.aws.amazon.com/cli/latest/userguide/cli-chap-welcome.html}, accessed: 2025-06-11.

\bibitem{aws_apis}
``Aws apis,'' \url{https://docs.aws.amazon.com/general/latest/gr/Welcome.html#aws-apis}, accessed: 2025-06-11.

\bibitem{aws_temp_security_credentials}
``Temporary security credentials in iam,'' \url{https://docs.aws.amazon.com/IAM/latest/UserGuide/id_credentials_temp.html}, accessed: 2025-06-11.

\bibitem{aws_assume_role_methods}
``Methods to assume a role,'' \url{https://docs.aws.amazon.com/IAM/latest/UserGuide/id_roles_manage-assume.html}, accessed: 2025-06-11.

\bibitem{aws_identity_providers_federation}
``Identity providers and federation,'' \url{https://docs.aws.amazon.com/IAM/latest/UserGuide/id_roles_providers.html}, accessed: 2025-06-11.

\bibitem{aws_mfa_iam}
``Aws multi-factor authentication in iam,'' \url{https://docs.aws.amazon.com/IAM/latest/UserGuide/id_credentials_mfa.html}, accessed: 2025-06-11.

\bibitem{aws_mfa_types_identity_center}
``Available mfa types for iam identity center,'' \url{https://docs.aws.amazon.com/en_us/singlesignon/latest/userguide/mfa-types.html}, accessed: 2025-06-11.

\end{thebibliography}
\addcontentsline{toc}{chapter}{Appendix}
\newpage
\chapter*{Appendix}
\section*{Chapter 5 resources}
This appendices of chapter 5 contains all the resources related to the scenario-based benchmarking between our SkyEye framework and other published frameworks or tools.
{
\scriptsize 
\begin{multicols}{2}

\subsection*{Scenario 1:}
\phantomsection
\label{sec:scenario1}
\textbf{User: S1\_UserA}
\begin{itemize}[itemsep=1pt, topsep=1pt, left=0pt]
    \item \textbf{Inline Policies:}
    \begin{itemize}[itemsep=1pt, topsep=1pt, left=0pt]
        \item \textbf{S1\_IP\_UserA:}
        \begin{itemize}[itemsep=1pt, topsep=1pt, left=0pt]
            \item \texttt{iam:ListGroupsForUser} (G)
            \item \texttt{iam:ListAttachedUserPolicies} (UP)
            \item \texttt{iam:GetUserPolicy} (UI)
        \end{itemize}
    \end{itemize}
    \item \textbf{Attached Managed Policies:}
    \begin{itemize}[itemsep=1pt, topsep=1pt, left=0pt]
        \item \textbf{S1\_AMP\_PolicyA:}
        \begin{itemize}[itemsep=1pt, topsep=1pt, left=0pt]
            \item \texttt{iam:ListUserPolicies} (UI)
            \item \texttt{iam:ListAttachedGroupPolicies} (GP)
            \item \texttt{iam:GetRolePolicy} (RI)
        \end{itemize}
        \item \textbf{S1\_AMP\_PolicyB:}
        \begin{itemize}[itemsep=1pt, topsep=1pt, left=0pt]
            \item \texttt{iam:GetGroupPolicy} (GI)
            \item \texttt{iam:ListGroupPolicies} (GI)
        \end{itemize}
    \end{itemize}
    \item \textbf{Group: S1\_GroupA (Includes: S1\_UserA)}
    \begin{itemize}[itemsep=1pt, topsep=1pt, left=0pt]
        \item \textbf{Inline Policies:}
        \begin{itemize}[itemsep=1pt, topsep=1pt, left=0pt]
            \item \textbf{S1\_IP\_GroupA:}
            \begin{itemize}[itemsep=1pt, topsep=1pt, left=0pt]
                \item \texttt{iam:ListPolicyVersions} (P)
                \item \texttt{iam:ListRolePolicies} (RI)
            \end{itemize}
        \end{itemize}
        \item \textbf{Attached Managed Policies:}
        \begin{itemize}[itemsep=1pt, topsep=1pt, left=0pt]
            \item \textbf{S1\_AMP\_PolicyC:}
            \begin{itemize}[itemsep=1pt, topsep=1pt, left=0pt]
                \item \texttt{iam:GetPolicyVersion} (P)
                \item \texttt{iam:ListAttachedRolePolicies} (RP)
                \item \texttt{iam:ListRoles} (R)
            \end{itemize}
        \end{itemize}
    \end{itemize}
    \item \textbf{Role: S1\_RoleA (Assumable by: S1\_UserA)}
    \begin{itemize}[itemsep=1pt, topsep=1pt, left=0pt]
        \item \textbf{Inline Policies:}
        \begin{itemize}[itemsep=1pt, topsep=1pt, left=0pt]
            \item \textbf{S1\_IP\_RoleA:}
            \begin{itemize}[itemsep=1pt, topsep=1pt, left=0pt]
                \item \texttt{s3:CreateBucket}
                \item \texttt{lambda:CreateFunction}
                \item \texttt{ec2:RunInstances}
            \end{itemize}
        \end{itemize}
        \item \textbf{Attached Managed Policies:}
        \begin{itemize}[itemsep=1pt, topsep=1pt, left=0pt]
            \item \textbf{AmazonEKSServicePolicy (AWS)}
        \end{itemize}
    \end{itemize}
\end{itemize}

\subsection*{Scenario 2:}
\phantomsection
\label{sec:scenario2}
\textbf{User: S2\_UserA}
\begin{itemize}[itemsep=1pt, topsep=1pt, left=0pt]
    \item \textbf{Inline Policies:}
    \begin{itemize}[itemsep=1pt, topsep=1pt, left=0pt]
        \item \textbf{S2\_IP\_UserA:}
        \begin{itemize}[itemsep=1pt, topsep=1pt, left=0pt]
            \item \texttt{iam:ListGroups} (G)
            \item \texttt{iam:ListAttachedUserPolicies} (UP)
            \item \texttt{iam:GetUserPolicy} (UI)
        \end{itemize}
    \end{itemize}
    \item \textbf{Attached Managed Policies:}
    \begin{itemize}[itemsep=1pt, topsep=1pt, left=0pt]
        \item \textbf{S2\_AMP\_PolicyA:}
        \begin{itemize}[itemsep=1pt, topsep=1pt, left=0pt]
            \item \texttt{iam:ListUserPolicies} (UI)
            \item \texttt{iam:ListAttachedGroupPolicies} (GP)
            \item \texttt{iam:GetRolePolicy} (RI)
        \end{itemize}
        \item \textbf{S2\_AMP\_PolicyB:}
        \begin{itemize}[itemsep=1pt, topsep=1pt, left=0pt]
            \item \texttt{iam:GetGroupPolicy} (GI)
            \item \texttt{iam:ListGroupPolicies} (GI)
            \item \texttt{iam:GetGroup} (G)
        \end{itemize}
        \item \textbf{S2\_AMP\_PolicyC:}
        \begin{itemize}[itemsep=1pt, topsep=1pt, left=0pt]
            \item \texttt{aiops:CreateInvestigation}
            \item \texttt{iot:CreateThing}
        \end{itemize}
    \end{itemize}
    \item \textbf{Group: S2\_GroupA (Includes: S2\_UserA)}
    \begin{itemize}[itemsep=1pt, topsep=1pt, left=0pt]
        \item \textbf{Inline Policies:}
        \begin{itemize}[itemsep=1pt, topsep=1pt, left=0pt]
            \item \textbf{S2\_IP\_GroupA:}
            \begin{itemize}[itemsep=1pt, topsep=1pt, left=0pt]
                \item \texttt{iam:ListPolicyVersions} (P)
                \item \texttt{iam:ListRolePolicies} (RI)
            \end{itemize}
        \end{itemize}
        \item \textbf{Attached Managed Policies:}
        \begin{itemize}[itemsep=1pt, topsep=1pt, left=0pt]
            \item \textbf{S2\_AMP\_PolicyD:}
            \begin{itemize}[itemsep=1pt, topsep=1pt, left=0pt]
                \item \texttt{iam:GetPolicyVersion} (P)
                \item \texttt{iam:ListAttachedRolePolicies} (RP)
                \item \texttt{iam:ListRoles} (R)
            \end{itemize}
        \end{itemize}
    \end{itemize}
    \item \textbf{Role: S2\_RoleA (Assumable by: S2\_UserA)}
    \begin{itemize}[itemsep=1pt, topsep=1pt, left=0pt]
        \item \textbf{Inline Policies:}
        \begin{itemize}[itemsep=1pt, topsep=1pt, left=0pt]
            \item \textbf{S2\_IP\_RoleA:}
            \begin{itemize}[itemsep=1pt, topsep=1pt, left=0pt]
                \item \texttt{s3:CreateBucket}
                \item \texttt{lambda:CreateFunction}
                \item \texttt{ec2:RunInstances}
                \item \texttt{s3:ListBucket}
                \item \texttt{ec2:DescribeInstances}
            \end{itemize}
        \end{itemize}
        \item \textbf{Attached Managed Policies:}
        \begin{itemize}[itemsep=1pt, topsep=1pt, left=0pt]
            \item \textbf{AmazonS3TablesFullAccess (AWS)}
        \end{itemize}
    \end{itemize}
\end{itemize}

\subsection*{Scenario 3:}
\phantomsection
\label{sec:scenario3}
\textbf{User: S3\_UserA}
\begin{itemize}[itemsep=1pt, topsep=1pt, left=0pt]
    \item \textbf{Inline Policies:}
    \begin{itemize}[itemsep=1pt, topsep=1pt, left=0pt]
        \item \textbf{S3\_IP\_UserA:}
        \begin{itemize}[itemsep=1pt, topsep=1pt, left=0pt]
            \item \texttt{iam:ListGroups} (G)
            \item \texttt{iam:ListAttachedUserPolicies} (UP)
            \item \texttt{iam:GetUserPolicy} (UI)
        \end{itemize}
    \end{itemize}
    \item \textbf{Attached Managed Policies:}
    \begin{itemize}[itemsep=1pt, topsep=1pt, left=0pt]
        \item \textbf{S3\_AMP\_PolicyA:}
        \begin{itemize}[itemsep=1pt, topsep=1pt, left=0pt]
            \item \texttt{iam:ListUserPolicies} (UI)
            \item \texttt{iam:ListAttachedGroupPolicies} (GP)
            \item \texttt{iam:GetRolePolicy} (RI)
        \end{itemize}
        \item \textbf{S3\_AMP\_PolicyB:}
        \begin{itemize}[itemsep=1pt, topsep=1pt, left=0pt]
            \item \texttt{iam:GetGroupPolicy} (GI)
            \item \texttt{iam:ListGroupPolicies} (GI)
            \item \texttt{iam:GetGroup} (G)
        \end{itemize}
        \item \textbf{S3\_AMP\_PolicyD:}
        \begin{itemize}[itemsep=1pt, topsep=1pt, left=0pt]
            \item \texttt{aiops:CreateInvestigation}
            \item \texttt{iot:CreateThing}
        \end{itemize}
    \end{itemize}
    \item \textbf{Group: S3\_GroupA (Includes: S3\_UserA)}
    \begin{itemize}[itemsep=1pt, topsep=1pt, left=0pt]
        \item \textbf{Inline Policies:}
        \begin{itemize}[itemsep=1pt, topsep=1pt, left=0pt]
            \item \textbf{S3\_IP\_GroupA:}
            \begin{itemize}[itemsep=1pt, topsep=1pt, left=0pt]
                \item \texttt{iam:ListRolePolicies} (RI)
                \item \texttt{iam:GetPolicy} (P)
            \end{itemize}
        \end{itemize}
        \item \textbf{Attached Managed Policies:}
        \begin{itemize}[itemsep=1pt, topsep=1pt, left=0pt]
            \item \textbf{S3\_AMP\_PolicyC:}
            \begin{itemize}[itemsep=1pt, topsep=1pt, left=0pt]
                \item \texttt{iam:GetPolicyVersion} (P)
                \item \texttt{iam:ListAttachedRolePolicies} (RP)
                \item \texttt{iam:ListRoles} (R)
            \end{itemize}
        \end{itemize}
    \end{itemize}
    \item \textbf{Role: S3\_RoleA (Assumable by: S3\_UserA)}
    \begin{itemize}[itemsep=1pt, topsep=1pt, left=0pt]
        \item \textbf{Inline Policies:}
        \begin{itemize}[itemsep=1pt, topsep=1pt, left=0pt]
            \item \textbf{S3\_IP\_RoleA:}
            \begin{itemize}[itemsep=1pt, topsep=1pt, left=0pt]
                \item \texttt{s3:CreateBucket}
                \item \texttt{lambda:CreateFunction}
                \item \texttt{ec2:RunInstances}
                \item \texttt{s3:ListBucket}
                \item \texttt{ec2:DescribeInstances}
            \end{itemize}
        \end{itemize}
        \item \textbf{Attached Managed Policies:}
        \begin{itemize}[itemsep=1pt, topsep=1pt, left=0pt]
            \item \textbf{AmazonRoute53ReadOnlyAccess (AWS)}
        \end{itemize}
    \end{itemize}
\end{itemize}

\subsection*{Scenario 4:}
\phantomsection
\label{sec:scenario4}
\textbf{User: S4\_UserA}
\begin{itemize}[itemsep=1pt, topsep=1pt, left=0pt]
    \item \textbf{Inline Policies:}
    \begin{itemize}[itemsep=1pt, topsep=1pt, left=0pt]
        \item \textbf{S4\_IP\_UserA:}
        \begin{itemize}[itemsep=1pt, topsep=1pt, left=0pt]
            \item \texttt{iam:ListGroups} (G)
            \item \texttt{iam:ListAttachedUserPolicies} (UP)
            \item \texttt{iam:GetUserPolicy} (UI)
        \end{itemize}
    \end{itemize}
    \item \textbf{Attached Managed Policies:}
    \begin{itemize}[itemsep=1pt, topsep=1pt, left=0pt]
        \item \textbf{S4\_AMP\_PolicyA:}
        \begin{itemize}[itemsep=1pt, topsep=1pt, left=0pt]
            \item \texttt{iam:ListUserPolicies} (UI)
            \item \texttt{iam:ListAttachedGroupPolicies} (GP)
            \item \texttt{iam:GetRolePolicy} (RI)
        \end{itemize}
        \item \textbf{S4\_AMP\_PolicyB:}
        \begin{itemize}[itemsep=1pt, topsep=1pt, left=0pt]
            \item \texttt{iam:GetGroupPolicy} (GI)
            \item \texttt{iam:ListGroupPolicies} (GI)
            \item \texttt{iam:GetGroup} (G)
        \end{itemize}
        \item \textbf{S4\_AMP\_PolicyC:}
        \begin{itemize}[itemsep=1pt, topsep=1pt, left=0pt]
            \item \texttt{aiops:CreateInvestigation}
            \item \texttt{iot:CreateThing}
        \end{itemize}
    \end{itemize}
    \item \textbf{Group: S4\_GroupA (Includes: S4\_UserA)}
    \begin{itemize}[itemsep=1pt, topsep=1pt, left=0pt]
        \item \textbf{Inline Policies:}
        \begin{itemize}[itemsep=1pt, topsep=1pt, left=0pt]
            \item \textbf{S4\_IP\_GroupA:}
            \begin{itemize}[itemsep=1pt, topsep=1pt, left=0pt]
                \item \texttt{iam:ListRolePolicies} (RI)
            \end{itemize}
        \end{itemize}
        \item \textbf{Attached Managed Policies:}
        \begin{itemize}[itemsep=1pt, topsep=1pt, left=0pt]
            \item \textbf{S4\_AMP\_PolicyD:}
            \begin{itemize}[itemsep=1pt, topsep=1pt, left=0pt]
                \item \texttt{iam:GetPolicyVersion} (P)
                \item \texttt{iam:ListAttachedRolePolicies} (RP)
                \item \texttt{iam:ListRoles} (R)
            \end{itemize}
        \end{itemize}
    \end{itemize}
    \item \textbf{Role: S4\_RoleA (Assumable by: S4\_UserA)}
    \begin{itemize}[itemsep=1pt, topsep=1pt, left=0pt]
        \item \textbf{Inline Policies:}
        \begin{itemize}[itemsep=1pt, topsep=1pt, left=0pt]
            \item \textbf{S4\_IP\_RoleA:}
            \begin{itemize}[itemsep=1pt, topsep=1pt, left=0pt]
                \item \texttt{s3:CreateBucket}
                \item \texttt{lambda:CreateFunction}
                \item \texttt{ec2:RunInstances}
                \item \texttt{s3:ListBucket}
                \item \texttt{ec2:DescribeInstances}
            \end{itemize}
        \end{itemize}
        \item \textbf{Attached Managed Policies:}
        \begin{itemize}[itemsep=1pt, topsep=1pt, left=0pt]
            \item \textbf{AmazonRoute53ReadOnlyAccess (AWS)}
        \end{itemize}
    \end{itemize}
\end{itemize}

\subsection*{Scenario 5:}
\phantomsection
\label{sec:scenario5}
\textbf{User: S5\_UserA}
\begin{itemize}[itemsep=1pt, topsep=1pt, left=0pt]
    \item \textbf{Inline Policies:}
    \begin{itemize}[itemsep=1pt, topsep=1pt, left=0pt]
        \item \textbf{S5\_IP\_UserA:}
        \begin{itemize}[itemsep=1pt, topsep=1pt, left=0pt]
            \item \texttt{iam:ListGroupsForUser} (G)
            \item \texttt{iam:ListAttachedUserPolicies} (UP)
            \item \texttt{iam:GetUserPolicy} (UI)
            \item \texttt{iam:ListUserPolicies} (UI)
            \item \texttt{iam:GetRolePolicy} (RI)
            \item \texttt{iam:GetGroupPolicy} (GI)
            \item \texttt{iam:ListGroupPolicies} (GI)
            \item \texttt{iam:ListEntitiesForPolicy} (P)
        \end{itemize}
    \end{itemize}
    \item \textbf{Attached Managed Policies:}
    \begin{itemize}[itemsep=1pt, topsep=1pt, left=0pt]
        \item \textbf{S5\_AMP\_PolicyA:}
        \begin{itemize}[itemsep=1pt, topsep=1pt, left=0pt]
            \item \texttt{aiops:UpdateInvestigation}
            \item \texttt{iot:AttachThingPrincipal}
        \end{itemize}
        \item \textbf{S5\_AMP\_PolicyB:}
        \begin{itemize}[itemsep=1pt, topsep=1pt, left=0pt]
            \item \texttt{iot:DeleteThing}
            \item \texttt{bedrock:DeleteGuardrail}
        \end{itemize}
    \end{itemize}
    \item \textbf{Group: S5\_GroupA (Includes S5\_UserA)}
    \begin{itemize}[itemsep=1pt, topsep=1pt, left=0pt]
        \item \textbf{Inline Policies:}
        \begin{itemize}[itemsep=1pt, topsep=1pt, left=0pt]
            \item \textbf{S5\_IP\_GroupA:}
            \begin{itemize}[itemsep=1pt, topsep=1pt, left=0pt]
                \item \texttt{iam:ListRolePolicies} (RI)
                \item \texttt{iam:ListPolicyVersions} (P)
                \item \texttt{iam:GetPolicyVersion} (P)
                \item \texttt{iam:ListRoles} (R)
            \end{itemize}
        \end{itemize}
        \item \textbf{Attached Managed Policies:}
        \begin{itemize}[itemsep=1pt, topsep=1pt, left=0pt]
            \item \textbf{S5\_AMP\_PolicyC:}
            \begin{itemize}[itemsep=1pt, topsep=1pt, left=0pt]
                \item \texttt{aiops:CreateInvestigation}
                \item \texttt{iot:CreateThing}
            \end{itemize}
            \item \textbf{S5\_AMP\_PolicyA:}
            \begin{itemize}[itemsep=1pt, topsep=1pt, left=0pt]
                \item \texttt{aiops:CreateInvestigationResource}
                \item \texttt{qapps:CreateLibraryItemReview}
            \end{itemize}
            \item \textbf{AmazonMQFullAccess (AWS)}
        \end{itemize}
    \end{itemize}
    \item \textbf{Role: S5\_RoleA (Assumable by: S5\_UserA)}
    \begin{itemize}[itemsep=1pt, topsep=1pt, left=0pt]
        \item \textbf{Inline Policies:}
        \begin{itemize}[itemsep=1pt, topsep=1pt, left=0pt]
            \item \textbf{S5\_IP\_RoleA:}
            \begin{itemize}[itemsep=1pt, topsep=1pt, left=0pt]
                \item \texttt{s3:CreateBucket}
                \item \texttt{lambda:CreateFunction}
                \item \texttt{ec2:RunInstances}
                \item \texttt{s3:ListBucket}
                \item \texttt{ec2:DescribeInstances}
            \end{itemize}
        \end{itemize}
        \item \textbf{Attached Managed Policies:}
        \begin{itemize}[itemsep=1pt, topsep=1pt, left=0pt]
            \item \textbf{AmazonRoute53ReadOnlyAccess (AWS)}
            \item \textbf{AmazonKinesisFullAccess (AWS)}
            \item \textbf{S5\_AMP\_PolicyD:}
            \begin{itemize}[itemsep=1pt, topsep=1pt, left=0pt]
                \item \texttt{private-networks:ActivateDeviceIdentifier}
                \item \texttt{auditmanager:UpdateAssessment}
            \end{itemize}
            \item \textbf{S5\_AMP\_PolicyB:}
            \begin{itemize}[itemsep=1pt, topsep=1pt, left=0pt]
                \item \texttt{iot:CancelJob}
                \item \texttt{fis:CreateExperimentTemplate}
            \end{itemize}
        \end{itemize}
    \end{itemize}
\end{itemize}

\subsection*{Scenario 6:}
\phantomsection
\label{sec:scenario6}
\textbf{User: S6\_UserA}
\begin{itemize}[itemsep=1pt, topsep=1pt, left=0pt]
    \item \textbf{Inline Policies:}
    \begin{itemize}[itemsep=1pt, topsep=1pt, left=0pt]
        \item \textbf{S6\_IP\_UserA:}
        \begin{itemize}[itemsep=1pt, topsep=1pt, left=0pt]
            \item \texttt{iam:ListGroupsForUser} (G)
            \item \texttt{iam:GetUserPolicy} (UI)
            \item \texttt{iam:ListUserPolicies} (UI)
            \item \texttt{iam:GetRolePolicy} (RI)
            \item \texttt{iam:GetGroupPolicy} (GI)
            \item \texttt{iam:ListGroupPolicies} (GI)
            \item \texttt{iam:ListPolicies} (P)
            \item \texttt{iam:ListEntitiesForPolicy} (P)
        \end{itemize}
    \end{itemize}
    \item \textbf{Attached Managed Policies:}
    \begin{itemize}[itemsep=1pt, topsep=1pt, left=0pt]
        \item \textbf{S6\_AMP\_PolicyA:}
        \begin{itemize}[itemsep=1pt, topsep=1pt, left=0pt]
            \item \texttt{aiops:UpdateInvestigation}
            \item \texttt{iot:AttachThingPrincipal}
        \end{itemize}
        \item \textbf{S6\_AMP\_PolicyB:}
        \begin{itemize}[itemsep=1pt, topsep=1pt, left=0pt]
            \item \texttt{iot:DeleteThing}
            \item \texttt{bedrock:DeleteGuardrail}
        \end{itemize}
    \end{itemize}
    \item \textbf{Group: S6\_GroupA (Includes S6\_UserA)}
    \begin{itemize}[itemsep=1pt, topsep=1pt, left=0pt]
        \item \textbf{Inline Policies:}
        \begin{itemize}[itemsep=1pt, topsep=1pt, left=0pt]
            \item \textbf{S6\_IP\_GroupA:}
            \begin{itemize}[itemsep=1pt, topsep=1pt, left=0pt]
                \item \texttt{iam:ListRolePolicies} (RI)
                \item \texttt{iam:ListPolicyVersions} (P)
                \item \texttt{iam:GetPolicyVersion} (P)
                \item \texttt{iam:ListRoles} (R)
            \end{itemize}
        \end{itemize}
        \item \textbf{Attached Managed Policies:}
        \begin{itemize}[itemsep=1pt, topsep=1pt, left=0pt]
            \item \textbf{S6\_AMP\_PolicyC:}
            \begin{itemize}[itemsep=1pt, topsep=1pt, left=0pt]
                \item \texttt{aiops:CreateInvestigation}
                \item \texttt{iot:CreateThing}
            \end{itemize}
            \item \textbf{S6\_AMP\_PolicyD:}
            \begin{itemize}[itemsep=1pt, topsep=1pt, left=0pt]
                \item \texttt{aiops:CreateInvestigationResource}
                \item \texttt{qapps:CreateLibraryItemReview}
            \end{itemize}
            \item \textbf{AmazonMQFullAccess (AWS)}
        \end{itemize}
    \end{itemize}
    \item \textbf{Role: S6\_RoleA (Assumable by: S6\_UserA)}
    \begin{itemize}[itemsep=1pt, topsep=1pt, left=0pt]
        \item \textbf{Inline Policies:}
        \begin{itemize}[itemsep=1pt, topsep=1pt, left=0pt]
            \item \textbf{S6\_IP\_RoleA:}
            \begin{itemize}[itemsep=1pt, topsep=1pt, left=0pt]
                \item \texttt{s3:CreateBucket}
                \item \texttt{lambda:CreateFunction}
                \item \texttt{ec2:RunInstances}
                \item \texttt{s3:ListBucket}
                \item \texttt{ec2:DescribeInstances}
            \end{itemize}
        \end{itemize}
        \item \textbf{Attached Managed Policies:}
        \begin{itemize}[itemsep=1pt, topsep=1pt, left=0pt]
            \item \textbf{AmazonRoute53ReadOnlyAccess (AWS)}
            \item \textbf{AmazonKinesisFullAccess (AWS)}
            \item \textbf{S6\_AMP\_PolicyE:}
            \begin{itemize}[itemsep=1pt, topsep=1pt, left=0pt]
                \item \texttt{private-networks:ActivateDeviceIdentifier}
                \item \texttt{auditmanager:UpdateAssessment}
            \end{itemize}
            \item \textbf{S6\_AMP\_PolicyF:}
            \begin{itemize}[itemsep=1pt, topsep=1pt, left=0pt]
                \item \texttt{iot:CancelJob}
                \item \texttt{fis:CreateExperimentTemplate}
            \end{itemize}
        \end{itemize}
    \end{itemize}
\end{itemize}

\subsection*{Scenario 7:}
\phantomsection
\label{sec:scenario7}
\textbf{User: S7\_UserA}
\begin{itemize}[itemsep=1pt, topsep=1pt, left=0pt]
    \item \textbf{Inline Policies:}
    \begin{itemize}[itemsep=1pt, topsep=1pt, left=0pt]
        \item \textbf{S7\_IP\_UserA:}
        \begin{itemize}[itemsep=1pt, topsep=1pt, left=0pt]
            \item \texttt{aiops:CreateInvestigation}
            \item \texttt{iot:CreateThing}
        \end{itemize}
    \end{itemize}
    \item \textbf{Attached Managed Policies:}
    \begin{itemize}[itemsep=1pt, topsep=1pt, left=0pt]
        \item \textbf{S7\_AMP\_PolicyA:}
        \begin{itemize}[itemsep=1pt, topsep=1pt, left=0pt]
            \item \texttt{iot:DeleteThing}
            \item \texttt{bedrock:DeleteGuardrail}
        \end{itemize}
        \item \textbf{S7\_AMP\_PolicyB:}
        \begin{itemize}[itemsep=1pt, topsep=1pt, left=0pt]
            \item \texttt{bedrock:InvokeAgent}
            \item \texttt{bedrock:UpdateFlow}
        \end{itemize}
    \end{itemize}
    \item \textbf{Group: S7\_GroupA (Includes S7\_UserA)}
    \begin{itemize}[itemsep=1pt, topsep=1pt, left=0pt]
        \item \textbf{Inline Policies:}
        \begin{itemize}[itemsep=1pt, topsep=1pt, left=0pt]
            \item \textbf{S7\_IP\_GroupA:}
            \begin{itemize}[itemsep=1pt, topsep=1pt, left=0pt]
                \item \texttt{iam:ListRoles} (R)
                \item \texttt{s3:CreateBucket}
                \item \texttt{lambda:CreateFunction}
                \item \texttt{ec2:RunInstances}
            \end{itemize}
        \end{itemize}
        \item \textbf{Attached Managed Policies:}
        \begin{itemize}[itemsep=1pt, topsep=1pt, left=0pt]
            \item \textbf{AmazonEKSServicePolicy (AWS)}
            \item \textbf{S7\_AMP\_PolicyC:}
            \begin{itemize}[itemsep=1pt, topsep=1pt, left=0pt]
                \item \texttt{s3:ListBucket}
                \item \texttt{ec2:DescribeInstances}
            \end{itemize}
        \end{itemize}
    \end{itemize}
    \item \textbf{Role: S7\_RoleA (Assumable by: S7\_UserA)}
    \begin{itemize}[itemsep=1pt, topsep=1pt, left=0pt]
        \item \textbf{Inline Policies:}
        \begin{itemize}[itemsep=1pt, topsep=1pt, left=0pt]
            \item \textbf{S7\_IP\_RoleA:}
            \begin{itemize}[itemsep=1pt, topsep=1pt, left=0pt]
                \item \texttt{iam:ListGroupsForUser} (G)
                \item \texttt{iam:ListAttachedUserPolicies} (UP)
                \item \texttt{iam:GetUserPolicy} (UI)
                \item \texttt{iam:ListUserPolicies} (UI)
                \item \texttt{iam:ListAttachedGroupPolicies} (GP)
                \item \texttt{iam:GetRolePolicy} (RI)
            \end{itemize}
        \end{itemize}
        \item \textbf{Attached Managed Policies:}
        \begin{itemize}[itemsep=1pt, topsep=1pt, left=0pt]
            \item \textbf{S7\_AMP\_PolicyD:}
            \begin{itemize}[itemsep=1pt, topsep=1pt, left=0pt]
                \item \texttt{iam:GetGroupPolicy} (GI)
                \item \texttt{Iam:ListGroupPolicies} (GI)
                \item \texttt{iam:ListPolicyVersions} (P)
                \item \texttt{iam:ListRolePolicies} (RI)
                \item \texttt{iam:GetPolicyVersion} (P)
                \item \texttt{iam:ListAttachedRolePolicies} (RP)
            \end{itemize}
        \end{itemize}
    \end{itemize}
\end{itemize}

\subsection*{Scenario 8:}
\phantomsection
\label{sec:scenario8}
\textbf{User: S8\_UserA}
\begin{itemize}[itemsep=1pt, topsep=1pt, left=0pt]
    \item \textbf{Inline Policies:}
    \begin{itemize}[itemsep=1pt, topsep=1pt, left=0pt]
        \item \textbf{S8\_IP\_UserA:}
        \begin{itemize}[itemsep=1pt, topsep=1pt, left=0pt]
            \item \texttt{aiops:CreateInvestigation}
            \item \texttt{iot:CreateThing}
        \end{itemize}
    \end{itemize}
    \item \textbf{Attached Managed Policies:}
    \begin{itemize}[itemsep=1pt, topsep=1pt, left=0pt]
        \item \textbf{S8\_AMP\_PolicyA:}
        \begin{itemize}[itemsep=1pt, topsep=1pt, left=0pt]
            \item \texttt{iot:DeleteThing}
            \item \texttt{bedrock:DeleteGuardrail}
        \end{itemize}
        \item \textbf{S8\_AMP\_PolicyB:}
        \begin{itemize}[itemsep=1pt, topsep=1pt, left=0pt]
            \item \texttt{bedrock:InvokeAgent}
            \item \texttt{bedrock:UpdateFlow}
        \end{itemize}
    \end{itemize}
    \item \textbf{Group: S8\_GroupA (Includes S8\_UserA)}
    \begin{itemize}[itemsep=1pt, topsep=1pt, left=0pt]
        \item \textbf{Inline Policies:}
        \begin{itemize}[itemsep=1pt, topsep=1pt, left=0pt]
            \item \textbf{S8\_IP\_GroupA:}
            \begin{itemize}[itemsep=1pt, topsep=1pt, left=0pt]
                \item \texttt{iam:ListRoles} (R)
                \item \texttt{s3:CreateBucket}
                \item \texttt{lambda:CreateFunction}
                \item \texttt{ec2:RunInstances}
            \end{itemize}
        \end{itemize}
        \item \textbf{Attached Managed Policies:}
        \begin{itemize}[itemsep=1pt, topsep=1pt, left=0pt]
            \item \textbf{S8\_AMP\_PolicyC:}
            \begin{itemize}[itemsep=1pt, topsep=1pt, left=0pt]
                \item \texttt{s3:ListBucket}
                \item \texttt{ec2:DescribeInstances}
            \end{itemize}
        \end{itemize}
    \end{itemize}
    \item \textbf{Role: S8\_RoleA (Assumable by: S8\_UserA)}
    \begin{itemize}[itemsep=1pt, topsep=1pt, left=0pt]
        \item \textbf{Inline Policies:}
        \begin{itemize}[itemsep=1pt, topsep=1pt, left=0pt]
            \item \textbf{S8\_IP\_RoleA:}
            \begin{itemize}[itemsep=1pt, topsep=1pt, left=0pt]
                \item \texttt{iam:ListGroupsForUser} (G)
                \item \texttt{iam:ListAttachedUserPolicies} (UP)
                \item \texttt{iam:GetUserPolicy} (UI)
            \end{itemize}
        \end{itemize}
        \item \textbf{Attached Managed Policies:}
        \begin{itemize}[itemsep=1pt, topsep=1pt, left=0pt]
            \item \textbf{S8\_AMP\_PolicyD:}
            \begin{itemize}[itemsep=1pt, topsep=1pt, left=0pt]
                \item \texttt{ssm:CancelCommand}
                \item \texttt{codeguru:GetCodeGuruFreeTrialSummary}
            \end{itemize}
        \end{itemize}
    \end{itemize}
    \item \textbf{Role: S8\_RoleB (Assumable by: S8\_RoleA)}
    \begin{itemize}[itemsep=1pt, topsep=1pt, left=0pt]
        \item \textbf{Inline Policies:}
        \begin{itemize}[itemsep=1pt, topsep=1pt, left=0pt]
            \item \textbf{S8\_IP\_RoleB:}
            \begin{itemize}[itemsep=1pt, topsep=1pt, left=0pt]
                \item \texttt{iam:ListUserPolicies} (UI)
                \item \texttt{iam:ListAttachedGroupPolicies} (GP)
                \item \texttt{iam:GetRolePolicy} (RI)
            \end{itemize}
        \end{itemize}
        \item \textbf{Attached Managed Policies:}
        \begin{itemize}[itemsep=1pt, topsep=1pt, left=0pt]
            \item \textbf{AmazonEKSServicePolicy (AWS)}
            \item \textbf{S8\_AMP\_PolicyE:}
            \begin{itemize}[itemsep=1pt, topsep=1pt, left=0pt]
                \item \texttt{ec2:AllocateAddress}
                \item \texttt{ec2:BundleInstance}
            \end{itemize}
        \end{itemize}
    \end{itemize}
    \item \textbf{Role: S8\_RoleC (Assumable by: S8\_RoleB)}
    \begin{itemize}[itemsep=1pt, topsep=1pt, left=0pt]
        \item \textbf{Inline Policies:}
        \begin{itemize}[itemsep=1pt, topsep=1pt, left=0pt]
            \item \textbf{S8\_IP\_RoleC:}
            \begin{itemize}[itemsep=1pt, topsep=1pt, left=0pt]
                \item \texttt{controltower:CreateManagedAccount}
                \item \texttt{nimble:CreateStudio}
            \end{itemize}
        \end{itemize}
        \item \textbf{Attached Managed Policies:}
        \begin{itemize}[itemsep=1pt, topsep=1pt, left=0pt]
            \item \textbf{S8\_AMP\_PolicyF:}
            \begin{itemize}[itemsep=1pt, topsep=1pt, left=0pt]
                \item \texttt{iam:ListAttachedRolePolicies} (RP)
                \item \texttt{iam:GetGroupPolicy} (GI)
                \item \texttt{Iam:ListGroupPolicies} (GI)
            \end{itemize}
        \end{itemize}
    \end{itemize}
    \item \textbf{Role: S8\_RoleD (Assumable by: S8\_RoleC)}
    \begin{itemize}[itemsep=1pt, topsep=1pt, left=0pt]
        \item \textbf{Inline Policies:}
        \begin{itemize}[itemsep=1pt, topsep=1pt, left=0pt]
            \item \textbf{S8\_IP\_RoleD:}
            \begin{itemize}[itemsep=1pt, topsep=1pt, left=0pt]
                \item \texttt{tax:GetExemptions}
                \item \texttt{s3-object-lambda:GetObjectAcl}
                \item \texttt{qapps:CreateLibraryItemReview}
            \end{itemize}
        \end{itemize}
        \item \textbf{Attached Managed Policies:}
        \begin{itemize}[itemsep=1pt, topsep=1pt, left=0pt]
            \item \textbf{S8\_AMP\_PolicyG:}
            \begin{itemize}[itemsep=1pt, topsep=1pt, left=0pt]
                \item \texttt{iam:ListPolicyVersions} (P)
                \item \texttt{iam:ListRolePolicies} (RI)
                \item \texttt{iam:GetPolicyVersion} (P)
            \end{itemize}
        \end{itemize}
    \end{itemize}
\end{itemize}

\subsection*{Scenario 9}
\phantomsection
\label{sec:scenario9}
\textbf{User: S9\_UserA}
\begin{itemize}[itemsep=1pt, topsep=1pt, left=0pt]
    \item \textbf{Inline Policies:}
    \begin{itemize}[itemsep=1pt, topsep=1pt, left=0pt]
        \item \textbf{S9\_IP\_UserA:}
        \begin{itemize}[itemsep=1pt, topsep=1pt, left=0pt]
            \item \texttt{iam:GetAccountAuthorizationDetails} (All)
        \end{itemize}
    \end{itemize}
    \item \textbf{Attached Managed Policies:}
    \begin{itemize}[itemsep=1pt, topsep=1pt, left=0pt]
        \item \textbf{S9\_AMP\_PolicyA:}
        \begin{itemize}[itemsep=1pt, topsep=1pt, left=0pt]
            \item \texttt{aiops:CreateInvestigation}
            \item \texttt{iot:CreateThing}
        \end{itemize}
    \end{itemize}
    \item \textbf{Group: S9\_GroupA (Includes S9\_UserA)}
    \begin{itemize}[itemsep=1pt, topsep=1pt, left=0pt]
        \item \textbf{Inline Policies:}
        \begin{itemize}[itemsep=1pt, topsep=1pt, left=0pt]
            \item \textbf{S9\_IP\_GroupA:}
            \begin{itemize}[itemsep=1pt, topsep=1pt, left=0pt]
                \item \texttt{iot:DeleteThing}
                \item \texttt{bedrock:DeleteGuardrail}
            \end{itemize}
        \end{itemize}
        \item \textbf{Attached Managed Policies:}
        \begin{itemize}[itemsep=1pt, topsep=1pt, left=0pt]
            \item \textbf{AmazonKinesisFullAccess (AWS)}
            \item \textbf{S9\_AMP\_PolicyB:}
            \begin{itemize}[itemsep=1pt, topsep=1pt, left=0pt]
                \item \texttt{bedrock:InvokeAgent}
                \item \texttt{bedrock:UpdateFlow}
            \end{itemize}
        \end{itemize}
    \end{itemize}
    \item \textbf{Role: S9\_RoleA (Assumable by: S9\_UserA)}
    \begin{itemize}[itemsep=1pt, topsep=1pt, left=0pt]
        \item \textbf{Inline Policies:}
        \begin{itemize}[itemsep=1pt, topsep=1pt, left=0pt]
            \item \textbf{S9\_IP\_RoleA:}
            \begin{itemize}[itemsep=1pt, topsep=1pt, left=0pt]
                \item \texttt{s3:CreateBucket}
                \item \texttt{lambda:CreateFunction}
                \item \texttt{ec2:RunInstances}
            \end{itemize}
        \end{itemize}
        \item \textbf{Attached Managed Policies:}
        \begin{itemize}[itemsep=1pt, topsep=1pt, left=0pt]
            \item \textbf{AmazonEKSServicePolicy (AWS)}
        \end{itemize}
    \end{itemize}
\end{itemize}

\subsection*{Scenario 10:}
\phantomsection
\label{sec:scenario10}
\textbf{User: S10\_UserA}
\begin{itemize}[itemsep=1pt, topsep=1pt, left=0pt]
    \item \textbf{Inline Policies:}
    \begin{itemize}[itemsep=1pt, topsep=1pt, left=0pt]
        \item \textbf{S10\_IP\_UserA:}
        \begin{itemize}[itemsep=1pt, topsep=1pt, left=0pt]
            \item \texttt{tax:GetExemptions}
            \item \texttt{s3-object-lambda:GetObjectAcl}
        \end{itemize}
    \end{itemize}
    \item \textbf{Attached Managed Policies:}
    \begin{itemize}[itemsep=1pt, topsep=1pt, left=0pt]
        \item \textbf{S10\_AMP\_PolicyA:}
        \begin{itemize}[itemsep=1pt, topsep=1pt, left=0pt]
            \item \texttt{aiops:CreateInvestigation}
            \item \texttt{iot:CreateThing}
        \end{itemize}
    \end{itemize}
    \item \textbf{Group: S10\_GroupA (Includes S10\_UserA)}
    \begin{itemize}[itemsep=1pt, topsep=1pt, left=0pt]
        \item \textbf{Inline Policies:}
        \begin{itemize}[itemsep=1pt, topsep=1pt, left=0pt]
            \item \textbf{S10\_IP\_GroupA:}
            \begin{itemize}[itemsep=1pt, topsep=1pt, left=0pt]
                \item \texttt{iot:DeleteThing}
                \item \texttt{bedrock:DeleteGuardrail}
            \end{itemize}
        \end{itemize}
        \item \textbf{Attached Managed Policies:}
        \begin{itemize}[itemsep=1pt, topsep=1pt, left=0pt]
            \item \textbf{AmazonKinesisFullAccess (AWS)}
            \item \textbf{S10\_AMP\_PolicyB:}
            \begin{itemize}[itemsep=1pt, topsep=1pt, left=0pt]
                \item \texttt{bedrock:InvokeAgent}
                \item \texttt{bedrock:UpdateFlow}
                \item \texttt{iam:ListRoles} (R)
            \end{itemize}
        \end{itemize}
    \end{itemize}
    \item \textbf{Role: S10\_RoleA (Assumable by: S10\_UserA)}
    \begin{itemize}[itemsep=1pt, topsep=1pt, left=0pt]
        \item \textbf{Inline Policies:}
        \begin{itemize}[itemsep=1pt, topsep=1pt, left=0pt]
            \item \textbf{S10\_IP\_RoleA:}
            \begin{itemize}[itemsep=1pt, topsep=1pt, left=0pt]
                \item \texttt{iam:GetAccountAuthorizationDetails} (All)
            \end{itemize}
        \end{itemize}
        \item \textbf{Attached Managed Policies:}
        \begin{itemize}[itemsep=1pt, topsep=1pt, left=0pt]
            \item \textbf{AmazonEKSServicePolicy (AWS)}
            \item \textbf{S10\_AMP\_PolicyC:}
            \begin{itemize}[itemsep=1pt, topsep=1pt, left=0pt]
                \item \texttt{s3:CreateBucket}
                \item \texttt{lambda:CreateFunction}
                \item \texttt{ec2:RunInstances}
            \end{itemize}
        \end{itemize}
    \end{itemize}
\end{itemize}

\subsection*{Scenario 11:}
\phantomsection
\label{sec:scenario11}
\textbf{User: S11\_UserA}
\begin{itemize}[itemsep=1pt, topsep=1pt, left=0pt]
    \item \textbf{Inline Policies:}
    \begin{itemize}[itemsep=1pt, topsep=1pt, left=0pt]
        \item \textbf{S11\_IP\_UserA:}
        \begin{itemize}[itemsep=1pt, topsep=1pt, left=0pt]
            \item \texttt{tax:GetExemptions}
            \item \texttt{s3-object-lambda:GetObjectAcl}
        \end{itemize}
    \end{itemize}
    \item \textbf{Attached Managed Policies:}
    \begin{itemize}[itemsep=1pt, topsep=1pt, left=0pt]
        \item \textbf{S11\_AMP\_PolicyA:}
        \begin{itemize}[itemsep=1pt, topsep=1pt, left=0pt]
            \item \texttt{aiops:CreateInvestigation}
            \item \texttt{iot:CreateThing}
            \item \texttt{iam:ListRoles} (R)
        \end{itemize}
    \end{itemize}
    \item \textbf{Group: S11\_GroupA (Includes S11\_UserA)}
    \begin{itemize}[itemsep=1pt, topsep=1pt, left=0pt]
        \item \textbf{Inline Policies:}
        \begin{itemize}[itemsep=1pt, topsep=1pt, left=0pt]
            \item \textbf{S11\_IP\_GroupA:}
            \begin{itemize}[itemsep=1pt, topsep=1pt, left=0pt]
                \item \texttt{iot:DeleteThing}
                \item \texttt{bedrock:DeleteGuardrail}
            \end{itemize}
        \end{itemize}
        \item \textbf{Attached Managed Policies:}
        \begin{itemize}[itemsep=1pt, topsep=1pt, left=0pt]
            \item \textbf{AmazonKinesisFullAccess (AWS)}
            \item \textbf{S11\_AMP\_PolicyB:}
            \begin{itemize}[itemsep=1pt, topsep=1pt, left=0pt]
                \item \texttt{bedrock:InvokeAgent}
                \item \texttt{bedrock:UpdateFlow}
            \end{itemize}
        \end{itemize}
    \end{itemize}
    \item \textbf{Role: S11\_RoleA (Assumable by: S11\_UserA)}
    \begin{itemize}[itemsep=1pt, topsep=1pt, left=0pt]
        \item \textbf{Inline Policies:}
        \begin{itemize}[itemsep=1pt, topsep=1pt, left=0pt]
            \item \textbf{S11\_IP\_RoleA:}
            \begin{itemize}[itemsep=1pt, topsep=1pt, left=0pt]
                \item \texttt{ssm:CancelCommand}
                \item \texttt{codeguru:GetCodeGuruFreeTrialSummary}
            \end{itemize}
        \end{itemize}
        \item \textbf{Attached Managed Policies:}
        \begin{itemize}[itemsep=1pt, topsep=1pt, left=0pt]
            \item \textbf{AmazonEKSServicePolicy (AWS)}
            \item \textbf{S11\_AMP\_PolicyC:}
            \begin{itemize}[itemsep=1pt, topsep=1pt, left=0pt]
                \item \texttt{s3:CreateBucket}
                \item \texttt{lambda:CreateFunction}
                \item \texttt{ec2:RunInstances}
            \end{itemize}
        \end{itemize}
    \end{itemize}
    \item \textbf{Role: S11\_RoleB (Assumable by: S11\_RoleA)}
    \begin{itemize}[itemsep=1pt, topsep=1pt, left=0pt]
        \item \textbf{Inline Policies:}
        \begin{itemize}[itemsep=1pt, topsep=1pt, left=0pt]
            \item \textbf{S11\_IP\_RoleB:}
            \begin{itemize}[itemsep=1pt, topsep=1pt, left=0pt]
                \item \texttt{detective:AcceptInvitation}
                \item \texttt{transfer:CreateAccess}
            \end{itemize}
        \end{itemize}
        \item \textbf{Attached Managed Policies:}
        \begin{itemize}[itemsep=1pt, topsep=1pt, left=0pt]
            \item \textbf{AmazonEKSServicePolicy (AWS)}
            \item \textbf{S11\_AMP\_PolicyE:}
            \begin{itemize}[itemsep=1pt, topsep=1pt, left=0pt]
                \item \texttt{ec2:AllocateAddress}
                \item \texttt{ec2:BundleInstance}
            \end{itemize}
        \end{itemize}
    \end{itemize}
    \item \textbf{Role: S11\_RoleC (Assumable by: S11\_RoleB)}
    \begin{itemize}[itemsep=1pt, topsep=1pt, left=0pt]
        \item \textbf{Inline Policies:}
        \begin{itemize}[itemsep=1pt, topsep=1pt, left=0pt]
            \item \textbf{S11\_IP\_RoleC:}
            \begin{itemize}[itemsep=1pt, topsep=1pt, left=0pt]
                \item \texttt{controltower:CreateManagedAccount}
                \item \texttt{nimble:CreateStudio}
            \end{itemize}
        \end{itemize}
        \item \textbf{Attached Managed Policies:}
        \begin{itemize}[itemsep=1pt, topsep=1pt, left=0pt]
            \item \textbf{S11\_AMP\_PolicyF:}
            \begin{itemize}[itemsep=1pt, topsep=1pt, left=0pt]
                \item \texttt{cloud9:CreateEnvironmentMembership}
                \item \texttt{cloud9:CreateEnvironmentSSH}
            \end{itemize}
        \end{itemize}
    \end{itemize}
    \item \textbf{Role: S11\_RoleD (Assumable by: S11\_RoleC)}
    \begin{itemize}[itemsep=1pt, topsep=1pt, left=0pt]
        \item \textbf{Inline Policies:}
        \begin{itemize}[itemsep=1pt, topsep=1pt, left=0pt]
            \item \textbf{S11\_IP\_RoleD:}
            \begin{itemize}[itemsep=1pt, topsep=1pt, left=0pt]
                \item \texttt{tax:GetExemptions}
                \item \texttt{s3-object-lambda:GetObjectAcl}
            \end{itemize}
        \end{itemize}
        \item \textbf{Attached Managed Policies:}
        \begin{itemize}[itemsep=1pt, topsep=1pt, left=0pt]
            \item \textbf{S11\_AMP\_PolicyG:}
            \begin{itemize}[itemsep=1pt, topsep=1pt, left=0pt]
                \item \texttt{iam:GetAccountAuthorizationDetails} (All)
            \end{itemize}
        \end{itemize}
    \end{itemize}
\end{itemize}

\subsection*{Scenario 12:}
\phantomsection
\label{sec:scenario12}
\textbf{User: S12\_UserA}
\begin{itemize}[itemsep=1pt, topsep=1pt, left=0pt]
    \item \textbf{Inline Policies:}
    \begin{itemize}[itemsep=1pt, topsep=1pt, left=0pt]
        \item \textbf{S12\_IP\_UserA:}
        \begin{itemize}[itemsep=1pt, topsep=1pt, left=0pt]
            \item \texttt{aiops:CreateInvestigation}
            \item \texttt{iot:CreateThing}
        \end{itemize}
    \end{itemize}
    \item \textbf{Attached Managed Policies:}
    \begin{itemize}[itemsep=1pt, topsep=1pt, left=0pt]
        \item \textbf{S12\_AMP\_PolicyA:}
        \begin{itemize}[itemsep=1pt, topsep=1pt, left=0pt]
            \item \texttt{iot:DeleteThing}
            \item \texttt{bedrock:DeleteGuardrail}
        \end{itemize}
        \item \textbf{S12\_AMP\_PolicyB:}
        \begin{itemize}[itemsep=1pt, topsep=1pt, left=0pt]
            \item \texttt{bedrock:InvokeAgent}
            \item \texttt{bedrock:UpdateFlow}
        \end{itemize}
    \end{itemize}
    \item \textbf{Group: S12\_GroupA (Includes S12\_UserA)}
    \begin{itemize}[itemsep=1pt, topsep=1pt, left=0pt]
        \item \textbf{Inline Policies:}
        \begin{itemize}[itemsep=1pt, topsep=1pt, left=0pt]
            \item \textbf{S12\_IP\_GroupA:}
            \begin{itemize}[itemsep=1pt, topsep=1pt, left=0pt]
                \item \texttt{iam:ListRoles} (R)
                \item \texttt{s3:CreateBucket}
                \item \texttt{lambda:CreateFunction}
                \item \texttt{ec2:RunInstances}
            \end{itemize}
        \end{itemize}
        \item \textbf{Attached Managed Policies:}
        \begin{itemize}[itemsep=1pt, topsep=1pt, left=0pt]
            \item \textbf{AmazonEKSServicePolicy (AWS)}
            \item \textbf{S12\_AMP\_PolicyC:}
            \begin{itemize}[itemsep=1pt, topsep=1pt, left=0pt]
                \item \texttt{s3:ListBucket}
                \item \texttt{ec2:DescribeInstances}
            \end{itemize}
        \end{itemize}
    \end{itemize}
    \item \textbf{Role: S12\_RoleA (Assumable by: S12\_UserA)}
    \begin{itemize}[itemsep=1pt, topsep=1pt, left=0pt]
        \item \textbf{Inline Policies:}
        \begin{itemize}[itemsep=1pt, topsep=1pt, left=0pt]
            \item \textbf{S12\_IP\_RoleA:}
            \begin{itemize}[itemsep=1pt, topsep=1pt, left=0pt]
                \item \texttt{iam:ListGroupsForUser} (G)
                \item \texttt{iam:GetUserPolicy} (UI)
                \item \texttt{iam:ListUserPolicies} (UI)
                \item \texttt{iam:GetRolePolicy} (RI)
                \item \texttt{iam:ListEntitiesForPolicy} (P)
                \item \texttt{iam:ListPolicies} (P)
            \end{itemize}
        \end{itemize}
        \item \textbf{Attached Managed Policies:}
        \begin{itemize}[itemsep=1pt, topsep=1pt, left=0pt]
            \item \textbf{AmazonRoute53ReadOnlyAccess (AWS)}
            \item \textbf{S12\_AMP\_PolicyD:}
            \begin{itemize}[itemsep=1pt, topsep=1pt, left=0pt]
                \item \texttt{iam:GetGroupPolicy} (GI)
                \item \texttt{Iam:ListGroupPolicies} (GI)
                \item \texttt{iam:ListRolePolicies} (RI)
                \item \texttt{iam:GetPolicyVersion} (P)
            \end{itemize}
        \end{itemize}
    \end{itemize}
\end{itemize}

\subsection*{Scenario 13:}
\phantomsection
\label{sec:scenario13}
\textbf{User: S13\_UserA}
\begin{itemize}[itemsep=1pt, topsep=1pt, left=0pt]
    \item \textbf{Inline Policies:}
    \begin{itemize}[itemsep=1pt, topsep=1pt, left=0pt]
        \item \textbf{S13\_IP\_UserA:}
        \begin{itemize}[itemsep=1pt, topsep=1pt, left=0pt]
            \item \texttt{aiops:CreateInvestigation}
            \item \texttt{iot:CreateThing}
        \end{itemize}
    \end{itemize}
    \item \textbf{Attached Managed Policies:}
    \begin{itemize}[itemsep=1pt, topsep=1pt, left=0pt]
        \item \textbf{S13\_AMP\_PolicyA:}
        \begin{itemize}[itemsep=1pt, topsep=1pt, left=0pt]
            \item \texttt{iot:DeleteThing}
            \item \texttt{bedrock:DeleteGuardrail}
        \end{itemize}
        \item \textbf{S13\_AMP\_PolicyB:}
        \begin{itemize}[itemsep=1pt, topsep=1pt, left=0pt]
            \item \texttt{bedrock:InvokeAgent}
            \item \texttt{bedrock:UpdateFlow}
        \end{itemize}
    \end{itemize}
    \item \textbf{Group: S13\_GroupA (Includes S13\_UserA)}
    \begin{itemize}[itemsep=1pt, topsep=1pt, left=0pt]
        \item \textbf{Inline Policies:}
        \begin{itemize}[itemsep=1pt, topsep=1pt, left=0pt]
            \item \textbf{S13\_IP\_GroupA:}
            \begin{itemize}[itemsep=1pt, topsep=1pt, left=0pt]
                \item \texttt{iam:ListRoles} (R)
                \item \texttt{s3:CreateBucket}
                \item \texttt{lambda:CreateFunction}
                \item \texttt{ec2:RunInstances}
            \end{itemize}
        \end{itemize}
        \item \textbf{Attached Managed Policies:}
        \begin{itemize}[itemsep=1pt, topsep=1pt, left=0pt]
            \item \textbf{S13\_AMP\_PolicyC:}
            \begin{itemize}[itemsep=1pt, topsep=1pt, left=0pt]
                \item \texttt{s3:ListBucket}
                \item \texttt{ec2:DescribeInstances}
            \end{itemize}
        \end{itemize}
    \end{itemize}
    \item \textbf{Role: S13\_RoleA (Assumable by: S13\_UserA)}
    \begin{itemize}[itemsep=1pt, topsep=1pt, left=0pt]
        \item \textbf{Inline Policies:}
        \begin{itemize}[itemsep=1pt, topsep=1pt, left=0pt]
            \item \textbf{S13\_IP\_RoleA:}
            \begin{itemize}[itemsep=1pt, topsep=1pt, left=0pt]
                \item \texttt{iam:ListGroupsForUser} (G)
                \item \texttt{iam:ListAttachedUserPolicies} (UP)
                \item \texttt{iam:GetUserPolicy} (UI)
            \end{itemize}
        \end{itemize}
        \item \textbf{Attached Managed Policies:}
        \begin{itemize}[itemsep=1pt, topsep=1pt, left=0pt]
            \item \textbf{S13\_AMP\_PolicyD:}
            \begin{itemize}[itemsep=1pt, topsep=1pt, left=0pt]
                \item \texttt{ssm:CancelCommand}
                \item \texttt{codeguru:GetCodeGuruFreeTrialSummary}
            \end{itemize}
        \end{itemize}
    \end{itemize}
    \item \textbf{Role: S13\_RoleB (Assumable by: S13\_RoleA)}
    \begin{itemize}[itemsep=1pt, topsep=1pt, left=0pt]
        \item \textbf{Inline Policies:}
        \begin{itemize}[itemsep=1pt, topsep=1pt, left=0pt]
            \item \textbf{S13\_IP\_RoleB:}
            \begin{itemize}[itemsep=1pt, topsep=1pt, left=0pt]
                \item \texttt{iam:ListUserPolicies} (UI)
                \item \texttt{iam:ListAttachedGroupPolicies} (GP)
                \item \texttt{iam:GetRolePolicy} (RI)
            \end{itemize}
        \end{itemize}
        \item \textbf{Attached Managed Policies:}
        \begin{itemize}[itemsep=1pt, topsep=1pt, left=0pt]
            \item \textbf{AmazonEKSServicePolicy (AWS)}
            \item \textbf{S13\_AMP\_PolicyE:}
            \begin{itemize}[itemsep=1pt, topsep=1pt, left=0pt]
                \item \texttt{ec2:AllocateAddress}
                \item \texttt{ec2:BundleInstance}
            \end{itemize}
        \end{itemize}
    \end{itemize}
    \item \textbf{Role: S13\_RoleC (Assumable by: S13\_RoleB)}
    \begin{itemize}[itemsep=1pt, topsep=1pt, left=0pt]
        \item \textbf{Inline Policies:}
        \begin{itemize}[itemsep=1pt, topsep=1pt, left=0pt]
            \item \textbf{S13\_IP\_RoleC:}
            \begin{itemize}[itemsep=1pt, topsep=1pt, left=0pt]
                \item \texttt{controltower:CreateManagedAccount}
                \item \texttt{nimble:CreateStudio}
            \end{itemize}
        \end{itemize}
        \item \textbf{Attached Managed Policies:}
        \begin{itemize}[itemsep=1pt, topsep=1pt, left=0pt]
            \item \textbf{S13\_AMP\_PolicyF:}
            \begin{itemize}[itemsep=1pt, topsep=1pt, left=0pt]
                \item \texttt{iam:ListAttachedRolePolicies} (RP)
                \item \texttt{iam:GetGroupPolicy} (GI)
                \item \texttt{Iam:ListGroupPolicies} (GI)
            \end{itemize}
        \end{itemize}
    \end{itemize}
    \item \textbf{Role: S13\_RoleD (Assumable by: S13\_RoleC)}
    \begin{itemize}[itemsep=1pt, topsep=1pt, left=0pt]
        \item \textbf{Inline Policies:}
        \begin{itemize}[itemsep=1pt, topsep=1pt, left=0pt]
            \item \textbf{S13\_IP\_RoleD:}
            \begin{itemize}[itemsep=1pt, topsep=1pt, left=0pt]
                \item \texttt{tax:GetExemptions}
                \item \texttt{s3-object-lambda:GetObjectAcl}
            \end{itemize}
        \end{itemize}
        \item \textbf{Attached Managed Policies:}
        \begin{itemize}[itemsep=1pt, topsep=1pt, left=0pt]
            \item \textbf{S13\_AMP\_PolicyG:}
            \begin{itemize}[itemsep=1pt, topsep=1pt, left=0pt]
                \item \texttt{iam:ListRolePolicies} (RI)
                \item \texttt{iam:GetPolicyVersion} (P)
            \end{itemize}
        \end{itemize}
    \end{itemize}
\end{itemize}

\subsection*{Scenario 14:}
\phantomsection
\label{sec:scenario14}
\textbf{User: S14\_UserA}
\begin{itemize}[itemsep=1pt, topsep=1pt, left=0pt]
    \item \textbf{Inline Policies:}
    \begin{itemize}[itemsep=1pt, topsep=1pt, left=0pt]
        \item \textbf{S14\_IP\_UserA:}
        \begin{itemize}[itemsep=1pt, topsep=1pt, left=0pt]
            \item \texttt{aiops:CreateInvestigation}
            \item \texttt{iot:CreateThing}
        \end{itemize}
    \end{itemize}
    \item \textbf{Attached Managed Policies:}
    \begin{itemize}[itemsep=1pt, topsep=1pt, left=0pt]
        \item \textbf{S14\_AMP\_PolicyA:}
        \begin{itemize}[itemsep=1pt, topsep=1pt, left=0pt]
            \item \texttt{iot:DeleteThing}
            \item \texttt{bedrock:DeleteGuardrail}
        \end{itemize}
        \item \textbf{S14\_AMP\_PolicyB:}
        \begin{itemize}[itemsep=1pt, topsep=1pt, left=0pt]
            \item \texttt{bedrock:InvokeAgent}
            \item \texttt{bedrock:UpdateFlow}
        \end{itemize}
    \end{itemize}
    \item \textbf{Group: S14\_GroupA (Includes S14\_UserA)}
    \begin{itemize}[itemsep=1pt, topsep=1pt, left=0pt]
        \item \textbf{Inline Policies:}
        \begin{itemize}[itemsep=1pt, topsep=1pt, left=0pt]
            \item \textbf{S14\_IP\_GroupA:}
            \begin{itemize}[itemsep=1pt, topsep=1pt, left=0pt]
                \item \texttt{iam:ListRoles} (R)
                \item \texttt{s3:CreateBucket}
                \item \texttt{lambda:CreateFunction}
                \item \texttt{ec2:RunInstances}
            \end{itemize}
        \end{itemize}
        \item \textbf{Attached Managed Policies:}
        \begin{itemize}[itemsep=1pt, topsep=1pt, left=0pt]
            \item \textbf{S14\_AMP\_PolicyC:}
            \begin{itemize}[itemsep=1pt, topsep=1pt, left=0pt]
                \item \texttt{s3:ListBucket}
                \item \texttt{ec2:DescribeInstances}
            \end{itemize}
        \end{itemize}
    \end{itemize}
    \item \textbf{Role: S14\_RoleA (Assumable by: S14\_UserA)}
    \begin{itemize}[itemsep=1pt, topsep=1pt, left=0pt]
        \item \textbf{Inline Policies:}
        \begin{itemize}[itemsep=1pt, topsep=1pt, left=0pt]
            \item \textbf{S14\_IP\_RoleA:}
            \begin{itemize}[itemsep=1pt, topsep=1pt, left=0pt]
                \item \texttt{iam:ListGroupsForUser} (G)
                \item \texttt{iam:GetUserPolicy} (UI)
            \end{itemize}
        \end{itemize}
        \item \textbf{Attached Managed Policies:}
        \begin{itemize}[itemsep=1pt, topsep=1pt, left=0pt]
            \item \textbf{AmazonKinesisFullAccess (AWS)}
            \item \textbf{S14\_AMP\_PolicyD:}
            \begin{itemize}[itemsep=1pt, topsep=1pt, left=0pt]
                \item \texttt{ssm:CancelCommand}
                \item \texttt{codeguru:GetCodeGuruFreeTrialSummary}
            \end{itemize}
        \end{itemize}
    \end{itemize}
    \item \textbf{Role: S14\_RoleB (Assumable by: S14\_RoleA)}
    \begin{itemize}[itemsep=1pt, topsep=1pt, left=0pt]
        \item \textbf{Inline Policies:}
        \begin{itemize}[itemsep=1pt, topsep=1pt, left=0pt]
            \item \textbf{S14\_IP\_RoleB:}
            \begin{itemize}[itemsep=1pt, topsep=1pt, left=0pt]
                \item \texttt{iam:ListUserPolicies} (UI)
                \item \texttt{iam:GetRolePolicy} (RI)
            \end{itemize}
        \end{itemize}
        \item \textbf{Attached Managed Policies:}
        \begin{itemize}[itemsep=1pt, topsep=1pt, left=0pt]
            \item \textbf{AmazonEKSServicePolicy (AWS)}
            \item \textbf{S14\_AMP\_PolicyE:}
            \begin{itemize}[itemsep=1pt, topsep=1pt, left=0pt]
                \item \texttt{ec2:AllocateAddress}
                \item \texttt{ec2:BundleInstance}
            \end{itemize}
        \end{itemize}
    \end{itemize}
    \item \textbf{Role: S14\_RoleC (Assumable by: S14\_RoleB)}
    \begin{itemize}[itemsep=1pt, topsep=1pt, left=0pt]
        \item \textbf{Inline Policies:}
        \begin{itemize}[itemsep=1pt, topsep=1pt, left=0pt]
            \item \textbf{S14\_IP\_RoleC:}
            \begin{itemize}[itemsep=1pt, topsep=1pt, left=0pt]
                \item \texttt{controltower:CreateManagedAccount}
                \item \texttt{nimble:CreateStudio}
            \end{itemize}
        \end{itemize}
        \item \textbf{Attached Managed Policies:}
        \begin{itemize}[itemsep=1pt, topsep=1pt, left=0pt]
            \item \textbf{AmazonRoute53ReadOnlyAccess (AWS)}
            \item \textbf{S14\_AMP\_PolicyF:}
            \begin{itemize}[itemsep=1pt, topsep=1pt, left=0pt]
                \item \texttt{iam:GetGroupPolicy} (GI)
                \item \texttt{Iam:ListGroupPolicies} (GI)
            \end{itemize}
        \end{itemize}
    \end{itemize}
    \item \textbf{Role: S14\_RoleD (Assumable by: S14\_RoleC)}
    \begin{itemize}[itemsep=1pt, topsep=1pt, left=0pt]
        \item \textbf{Inline Policies:}
        \begin{itemize}[itemsep=1pt, topsep=1pt, left=0pt]
            \item \textbf{S14\_IP\_RoleD:}
            \begin{itemize}[itemsep=1pt, topsep=1pt, left=0pt]
                \item \texttt{tax:GetExemptions}
                \item \texttt{s3-object-lambda:GetObjectAcl}
            \end{itemize}
        \end{itemize}
        \item \textbf{Attached Managed Policies:}
        \begin{itemize}[itemsep=1pt, topsep=1pt, left=0pt]
            \item \textbf{S14\_AMP\_PolicyG:}
            \begin{itemize}[itemsep=1pt, topsep=1pt, left=0pt]
                \item \texttt{iam:ListRolePolicies} (RI)
                \item \texttt{iam:GetPolicyVersion} (P)
                \item \texttt{iam:ListEntitiesForPolicy} (P)
                \item \texttt{iam:ListPolicies} (P)
            \end{itemize}
        \end{itemize}
    \end{itemize}
\end{itemize}

\subsection*{Scenario 15:}
\phantomsection
\label{sec:scenario15}
\textbf{User: S15\_UserA}
\begin{itemize}[itemsep=1pt, topsep=1pt, left=0pt]
    \item \textbf{Inline Policies:}
    \begin{itemize}[itemsep=1pt, topsep=1pt, left=0pt]
        \item \textbf{S15\_IP\_UserA:}
        \begin{itemize}[itemsep=1pt, topsep=1pt, left=0pt]
            \item \texttt{iam:ListGroupsForUser} (G)
            \item \texttt{iam:ListAttachedUserPolicies} (UP)
            \item \texttt{iam:GetUserPolicy} (UI)
        \end{itemize}
    \end{itemize}
    \item \textbf{Attached Managed Policies:}
    \begin{itemize}[itemsep=1pt, topsep=1pt, left=0pt]
        \item \textbf{S15\_AMP\_PolicyA:}
        \begin{itemize}[itemsep=1pt, topsep=1pt, left=0pt]
            \item \texttt{iot:DeleteThing}
            \item \texttt{bedrock:DeleteGuardrail}
        \end{itemize}
        \item \textbf{S15\_AMP\_PolicyB:}
        \begin{itemize}[itemsep=1pt, topsep=1pt, left=0pt]
            \item \texttt{bedrock:InvokeAgent}
            \item \texttt{bedrock:UpdateFlow}
        \end{itemize}
    \end{itemize}
    \item \textbf{Group: S15\_UserA\_GroupA (Includes S15\_UserA)}
    \begin{itemize}[itemsep=1pt, topsep=1pt, left=0pt]
        \item \textbf{Inline Policies:}
        \begin{itemize}[itemsep=1pt, topsep=1pt, left=0pt]
            \item \textbf{S15\_IP\_UserA\_GroupA:}
            \begin{itemize}[itemsep=1pt, topsep=1pt, left=0pt]
                \item \texttt{iam:ListRoles} (R)
                \item \texttt{s3:CreateBucket}
                \item \texttt{lambda:CreateFunction}
                \item \texttt{ec2:RunInstances}
            \end{itemize}
        \end{itemize}
        \item \textbf{Attached Managed Policies:}
        \begin{itemize}[itemsep=1pt, topsep=1pt, left=0pt]
            \item \textbf{S15\_AMP\_PolicyC:}
            \begin{itemize}[itemsep=1pt, topsep=1pt, left=0pt]
                \item \texttt{private-networks:ActivateDeviceIdentifier}
                \item \texttt{auditmanager:UpdateAssessment}
            \end{itemize}
        \end{itemize}
    \end{itemize}
    \item \textbf{Role: S15\_UserA\_RoleA (Assumable by: S15\_UserA)}
    \begin{itemize}[itemsep=1pt, topsep=1pt, left=0pt]
        \item \textbf{Inline Policies:}
        \begin{itemize}[itemsep=1pt, topsep=1pt, left=0pt]
            \item \textbf{S15\_IP\_UserA\_RoleA:}
            \begin{itemize}[itemsep=1pt, topsep=1pt, left=0pt]
                \item \texttt{ec2:AllocateAddress}
                \item \texttt{controltower:CreateManagedAccount}
            \end{itemize}
        \end{itemize}
        \item \textbf{Attached Managed Policies:}
        \begin{itemize}[itemsep=1pt, topsep=1pt, left=0pt]
            \item \textbf{S15\_AMP\_PolicyT:}
            \begin{itemize}[itemsep=1pt, topsep=1pt, left=0pt]
                \item \texttt{ssm:CancelCommand}
                \item \texttt{globalaccelerator:CreateAccelerator}
            \end{itemize}
        \end{itemize}
    \end{itemize}
\end{itemize}
\textbf{User: S15\_UserB}
\begin{itemize}[itemsep=1pt, topsep=1pt, left=0pt]
    \item \textbf{Inline Policies:}
    \begin{itemize}[itemsep=1pt, topsep=1pt, left=0pt]
        \item \textbf{S15\_IP\_UserB:}
        \begin{itemize}[itemsep=1pt, topsep=1pt, left=0pt]
            \item \texttt{iam:ListUserPolicies} (UI)
            \item \texttt{iam:ListAttachedGroupPolicies} (GP)
            \item \texttt{iam:GetRolePolicy} (RI)
        \end{itemize}
    \end{itemize}
    \item \textbf{Attached Managed Policies:}
    \begin{itemize}[itemsep=1pt, topsep=1pt, left=0pt]
        \item \textbf{S15\_AMP\_PolicyD:}
        \begin{itemize}[itemsep=1pt, topsep=1pt, left=0pt]
            \item \texttt{codecatalyst:DeleteConnection}
            \item \texttt{codeguru:GetCodeGuruFreeTrialSummary}
        \end{itemize}
    \end{itemize}
    \item \textbf{Group: S15\_UserB\_GroupA (Includes S15\_UserB)}
    \begin{itemize}[itemsep=1pt, topsep=1pt, left=0pt]
        \item \textbf{Inline Policies:}
        \begin{itemize}[itemsep=1pt, topsep=1pt, left=0pt]
            \item \textbf{S15\_IP\_UserB:}
            \begin{itemize}[itemsep=1pt, topsep=1pt, left=0pt]
                \item \texttt{iam:ListPolicyVersions} (P)
            \end{itemize}
        \end{itemize}
        \item \textbf{Attached Managed Policies:}
        \begin{itemize}[itemsep=1pt, topsep=1pt, left=0pt]
            \item \textbf{S15\_AMP\_PolicyX:}
            \begin{itemize}[itemsep=1pt, topsep=1pt, left=0pt]
                \item \texttt{globalaccelerator:DeleteAccelerator}
                \item \texttt{codedeploy:BatchGetApplications}
            \end{itemize}
        \end{itemize}
    \end{itemize}
    \item \textbf{Role: S15\_UserB\_RoleA (Assumable by: S15\_UserB)}
    \begin{itemize}[itemsep=1pt, topsep=1pt, left=0pt]
        \item \textbf{Inline Policies:}
        \begin{itemize}[itemsep=1pt, topsep=1pt, left=0pt]
            \item \textbf{S15\_IP\_UserB\_RoleA:}
            \begin{itemize}[itemsep=1pt, topsep=1pt, left=0pt]
                \item \texttt{ram:CreatePermission}
                \item \texttt{ec2:BundleInstance}
            \end{itemize}
        \end{itemize}
        \item \textbf{Attached Managed Policies:}
        \begin{itemize}[itemsep=1pt, topsep=1pt, left=0pt]
            \item \textbf{AmazonEKSServicePolicy (AWS)}
            \item \textbf{S15\_AMP\_PolicyE:}
            \begin{itemize}[itemsep=1pt, topsep=1pt, left=0pt]
                \item \texttt{s3:ListBucket}
                \item \texttt{ec2:DescribeInstances}
            \end{itemize}
        \end{itemize}
    \end{itemize}
\end{itemize}
\textbf{User: S15\_UserC}
\begin{itemize}[itemsep=1pt, topsep=1pt, left=0pt]
    \item \textbf{Inline Policies:}
    \begin{itemize}[itemsep=1pt, topsep=1pt, left=0pt]
        \item \textbf{S15\_IP\_UserC:}
        \begin{itemize}[itemsep=1pt, topsep=1pt, left=0pt]
            \item \texttt{sdb:BatchPutAttributes}
            \item \texttt{nimble:CreateStudio}
        \end{itemize}
    \end{itemize}
    \item \textbf{Attached Managed Policies:}
    \begin{itemize}[itemsep=1pt, topsep=1pt, left=0pt]
        \item \textbf{S15\_AMP\_PolicyF:}
        \begin{itemize}[itemsep=1pt, topsep=1pt, left=0pt]
            \item \texttt{iam:ListAttachedRolePolicies} (RP)
            \item \texttt{iam:GetGroupPolicy} (GI)
            \item \texttt{Iam:ListGroupPolicies} (GI)
        \end{itemize}
    \end{itemize}
    \item \textbf{Group: S15\_UserC\_GroupA (Includes S15\_UserC)}
    \begin{itemize}[itemsep=1pt, topsep=1pt, left=0pt]
        \item \textbf{Inline Policies:}
        \begin{itemize}[itemsep=1pt, topsep=1pt, left=0pt]
            \item \textbf{S15\_IP\_UserC\_GroupA:}
            \begin{itemize}[itemsep=1pt, topsep=1pt, left=0pt]
                \item \texttt{tax:GetExemptions}
                \item \texttt{s3-object-lambda:GetObjectAcl}
            \end{itemize}
        \end{itemize}
        \item \textbf{Attached Managed Policies:}
        \begin{itemize}[itemsep=1pt, topsep=1pt, left=0pt]
            \item \textbf{S15\_AMP\_PolicyG:}
            \begin{itemize}[itemsep=1pt, topsep=1pt, left=0pt]
                \item \texttt{iam:ListRolePolicies} (RI)
                \item \texttt{iam:GetPolicyVersion} (P)
            \end{itemize}
        \end{itemize}
    \end{itemize}
    \item \textbf{Role: S15\_UserC\_RoleA (Assumable by: S15\_UserC)}
    \begin{itemize}[itemsep=1pt, topsep=1pt, left=0pt]
        \item \textbf{Inline Policies:}
        \begin{itemize}[itemsep=1pt, topsep=1pt, left=0pt]
            \item \textbf{S15\_IP\_UserC\_RoleA:}
            \begin{itemize}[itemsep=1pt, topsep=1pt, left=0pt]
                \item \texttt{ram:CreateResourceShare}
                \item \texttt{scn:DescribeInstance}
            \end{itemize}
        \end{itemize}
    \end{itemize}
\end{itemize}
\textbf{User: S15\_UserD}
\begin{itemize}[itemsep=1pt, topsep=1pt, left=0pt]
    \item \textbf{Inline Policies:}
    \begin{itemize}[itemsep=1pt, topsep=1pt, left=0pt]
        \item \textbf{S15\_IP\_UserD:}
        \begin{itemize}[itemsep=1pt, topsep=1pt, left=0pt]
            \item \texttt{codecatalyst:CreateIdentityCenterApplication}
            \item \texttt{codedeploy:UpdateApplication}
        \end{itemize}
    \end{itemize}
    \item \textbf{Attached Managed Policies:}
    \begin{itemize}[itemsep=1pt, topsep=1pt, left=0pt]
        \item \textbf{S15\_AMP\_PolicyY:}
        \begin{itemize}[itemsep=1pt, topsep=1pt, left=0pt]
            \item \texttt{ram:GetPermission}
            \item \texttt{scn:CreateDataLakeDataset}
        \end{itemize}
    \end{itemize}
    \item \textbf{Group: S15\_UserD\_GroupA (Includes S15\_UserD)}
    \begin{itemize}[itemsep=1pt, topsep=1pt, left=0pt]
        \item \textbf{Inline Policies:}
        \begin{itemize}[itemsep=1pt, topsep=1pt, left=0pt]
            \item \textbf{S15\_IP\_UserD\_GroupA:}
            \begin{itemize}[itemsep=1pt, topsep=1pt, left=0pt]
                \item \texttt{iotanalytics:CreatePipeline}
                \item \texttt{aiops:CreateInvestigationResource}
            \end{itemize}
        \end{itemize}
        \item \textbf{Attached Managed Policies:}
        \begin{itemize}[itemsep=1pt, topsep=1pt, left=0pt]
            \item \textbf{S15\_AMP\_PolicyU:}
            \begin{itemize}[itemsep=1pt, topsep=1pt, left=0pt]
                \item \texttt{iotanalytics:DescribeChannel}
                \item \texttt{codedeploy:CreateDeployment}
            \end{itemize}
        \end{itemize}
    \end{itemize}
    \item \textbf{Role: S15\_UserD\_RoleA (Assumable by: S15\_UserD)}
    \begin{itemize}[itemsep=1pt, topsep=1pt, left=0pt]
        \item \textbf{Inline Policies:}
        \begin{itemize}[itemsep=1pt, topsep=1pt, left=0pt]
            \item \textbf{S15\_IP\_UserD\_RoleA:}
            \begin{itemize}[itemsep=1pt, topsep=1pt, left=0pt]
                \item \texttt{ram:PromotePermissionCreatedFromPolicy}
                \item \texttt{scn:CreateBillOfMaterialsImportJob}
            \end{itemize}
        \end{itemize}
        \item \textbf{Attached Managed Policies:}
        \begin{itemize}[itemsep=1pt, topsep=1pt, left=0pt]
            \item \textbf{AmazonRoute53ReadOnlyAccess (AWS)}
            \item \textbf{S15\_AMP\_PolicyZ:}
            \begin{itemize}[itemsep=1pt, topsep=1pt, left=0pt]
                \item \texttt{iotanalytics:DeleteDataset}
                \item \texttt{qapps:CreateLibraryItemReview}
            \end{itemize}
        \end{itemize}
    \end{itemize}
\end{itemize}

\subsection*{Scenario 16:}
\phantomsection
\label{sec:scenario16}
\textbf{User: S16\_UserA}
\begin{itemize}[itemsep=1pt, topsep=1pt, left=0pt]
    \item \textbf{Inline Policies:}
    \begin{itemize}[itemsep=1pt, topsep=1pt, left=0pt]
        \item \textbf{S16\_IP\_UserA:}
        \begin{itemize}[itemsep=1pt, topsep=1pt, left=0pt]
            \item \texttt{iam:ListGroupsForUser} (G)
        \end{itemize}
    \end{itemize}
    \item \textbf{Attached Managed Policies:}
    \begin{itemize}[itemsep=1pt, topsep=1pt, left=0pt]
        \item \textbf{S16\_AMP\_PolicyA:}
        \begin{itemize}[itemsep=1pt, topsep=1pt, left=0pt]
            \item \texttt{iot:DeleteThing}
            \item \texttt{bedrock:DeleteGuardrail}
        \end{itemize}
        \item \textbf{S16\_AMP\_PolicyB:}
        \begin{itemize}[itemsep=1pt, topsep=1pt, left=0pt]
            \item \texttt{bedrock:InvokeAgent}
            \item \texttt{bedrock:UpdateFlow}
        \end{itemize}
    \end{itemize}
    \item \textbf{Group: S16\_UserA\_GroupA (Includes S16\_UserA)}
    \begin{itemize}[itemsep=1pt, topsep=1pt, left=0pt]
        \item \textbf{Inline Policies:}
        \begin{itemize}[itemsep=1pt, topsep=1pt, left=0pt]
            \item \textbf{S16\_IP\_UserA\_GroupA:}
            \begin{itemize}[itemsep=1pt, topsep=1pt, left=0pt]
                \item \texttt{iam:ListRoles} (R)
                \item \texttt{s3:CreateBucket}
                \item \texttt{lambda:CreateFunction}
                \item \texttt{ec2:RunInstances}
            \end{itemize}
        \end{itemize}
        \item \textbf{Attached Managed Policies:}
        \begin{itemize}[itemsep=1pt, topsep=1pt, left=0pt]
            \item \textbf{S16\_AMP\_PolicyC:}
            \begin{itemize}[itemsep=1pt, topsep=1pt, left=0pt]
                \item \texttt{private-networks:ActivateDeviceIdentifier}
                \item \texttt{auditmanager:UpdateAssessment}
            \end{itemize}
        \end{itemize}
    \end{itemize}
    \item \textbf{Role: S16\_UserA\_RoleA (Assumable by: S16\_UserA)}
    \begin{itemize}[itemsep=1pt, topsep=1pt, left=0pt]
        \item \textbf{Inline Policies:}
        \begin{itemize}[itemsep=1pt, topsep=1pt, left=0pt]
            \item \textbf{S16\_IP\_UserA\_RoleA:}
            \begin{itemize}[itemsep=1pt, topsep=1pt, left=0pt]
                \item \texttt{iam:ListAttachedUserPolicies} (UP)
                \item \texttt{ec2:AllocateAddress}
                \item \texttt{controltower:CreateManagedAccount}
            \end{itemize}
        \end{itemize}
        \item \textbf{Attached Managed Policies:}
        \begin{itemize}[itemsep=1pt, topsep=1pt, left=0pt]
            \item \textbf{S16\_AMP\_PolicyT:}
            \begin{itemize}[itemsep=1pt, topsep=1pt, left=0pt]
                \item \texttt{ssm:CancelCommand}
                \item \texttt{globalaccelerator:CreateAccelerator}
            \end{itemize}
        \end{itemize}
    \end{itemize}
    \item \textbf{Role: S16\_UserA\_RoleB (Assumable by: S16\_UserA\_RoleA)}
    \begin{itemize}[itemsep=1pt, topsep=1pt, left=0pt]
        \item \textbf{Inline Policies:}
        \begin{itemize}[itemsep=1pt, topsep=1pt, left=0pt]
            \item \textbf{S16\_IP\_UserA\_RoleB:}
            \begin{itemize}[itemsep=1pt, topsep=1pt, left=0pt]
                \item \texttt{iam:GetUserPolicy} (UI)
                \item \texttt{iotanalytics:UpdateDatastore}
                \item \texttt{iotanalytics:UpdatePipeline}
            \end{itemize}
        \end{itemize}
        \item \textbf{Attached Managed Policies:}
        \begin{itemize}[itemsep=1pt, topsep=1pt, left=0pt]
            \item \textbf{AmazonEKSServicePolicy (AWS)}
        \end{itemize}
    \end{itemize}
\end{itemize}
\textbf{User: S16\_UserB}
\begin{itemize}[itemsep=1pt, topsep=1pt, left=0pt]
    \item \textbf{Inline Policies:}
    \begin{itemize}[itemsep=1pt, topsep=1pt, left=0pt]
        \item \textbf{S16\_IP\_UserB:}
        \begin{itemize}[itemsep=1pt, topsep=1pt, left=0pt]
            \item \texttt{iam:ListUserPolicies} (UI)
        \end{itemize}
    \end{itemize}
    \item \textbf{Attached Managed Policies:}
    \begin{itemize}[itemsep=1pt, topsep=1pt, left=0pt]
        \item \textbf{S16\_AMP\_PolicyD:}
        \begin{itemize}[itemsep=1pt, topsep=1pt, left=0pt]
            \item \texttt{codecatalyst:DeleteConnection}
            \item \texttt{codeguru:GetCodeGuruFreeTrialSummary}
        \end{itemize}
    \end{itemize}
    \item \textbf{Group: S16\_UserB\_GroupA (Includes S16\_UserB)}
    \begin{itemize}[itemsep=1pt, topsep=1pt, left=0pt]
        \item \textbf{Inline Policies:}
        \begin{itemize}[itemsep=1pt, topsep=1pt, left=0pt]
            \item \textbf{S16\_IP\_UserB:}
            \begin{itemize}[itemsep=1pt, topsep=1pt, left=0pt]
                \item \texttt{iam:ListPolicyVersions} (P)
            \end{itemize}
        \end{itemize}
        \item \textbf{Attached Managed Policies:}
        \begin{itemize}[itemsep=1pt, topsep=1pt, left=0pt]
            \item \textbf{S16\_AMP\_PolicyX:}
            \begin{itemize}[itemsep=1pt, topsep=1pt, left=0pt]
                \item \texttt{globalaccelerator:DeleteAccelerator}
                \item \texttt{codedeploy:BatchGetApplications}
            \end{itemize}
        \end{itemize}
    \end{itemize}
    \item \textbf{Role: S16\_UserB\_RoleA (Assumable by: S16\_UserB)}
    \begin{itemize}[itemsep=1pt, topsep=1pt, left=0pt]
        \item \textbf{Inline Policies:}
        \begin{itemize}[itemsep=1pt, topsep=1pt, left=0pt]
            \item \textbf{S16\_IP\_UserB\_RoleA:}
            \begin{itemize}[itemsep=1pt, topsep=1pt, left=0pt]
                \item \texttt{ram:CreatePermission}
                \item \texttt{cloudfront:AssociateAlias}
                \item \texttt{iam:ListAttachedGroupPolicies} (GP)
            \end{itemize}
        \end{itemize}
        \item \textbf{Attached Managed Policies:}
        \begin{itemize}[itemsep=1pt, topsep=1pt, left=0pt]
            \item \textbf{S16\_AMP\_PolicyE:}
            \begin{itemize}[itemsep=1pt, topsep=1pt, left=0pt]
                \item \texttt{s3:ListBucket}
                \item \texttt{ec2:DescribeInstances}
            \end{itemize}
        \end{itemize}
    \end{itemize}
    \item \textbf{Role: S16\_UserB\_RoleB (Assumable by: S16\_UserB\_RoleA)}
    \begin{itemize}[itemsep=1pt, topsep=1pt, left=0pt]
        \item \textbf{Inline Policies:}
        \begin{itemize}[itemsep=1pt, topsep=1pt, left=0pt]
            \item \textbf{S16\_IP\_UserB\_RoleB:}
            \begin{itemize}[itemsep=1pt, topsep=1pt, left=0pt]
                \item \texttt{connect:ActivateEvaluationForm}
                \item \texttt{connect:AssociateBot}
            \end{itemize}
        \end{itemize}
        \item \textbf{Attached Managed Policies:}
        \begin{itemize}[itemsep=1pt, topsep=1pt, left=0pt]
            \item \textbf{AmazonEKSServicePolicy (AWS)}
        \end{itemize}
    \end{itemize}
    \item \textbf{Role: S16\_UserB\_RoleC (Assumable by: S16\_UserB\_RoleB)}
    \begin{itemize}[itemsep=1pt, topsep=1pt, left=0pt]
        \item \textbf{Inline Policies:}
        \begin{itemize}[itemsep=1pt, topsep=1pt, left=0pt]
            \item \textbf{S16\_IP\_UserB\_RoleC:}
            \begin{itemize}[itemsep=1pt, topsep=1pt, left=0pt]
                \item \texttt{iam:GetRolePolicy} (RI)
            \end{itemize}
        \end{itemize}
        \item \textbf{Attached Managed Policies:}
        \begin{itemize}[itemsep=1pt, topsep=1pt, left=0pt]
            \item \textbf{AmazonRoute53ReadOnlyAccess (AWS)}
        \end{itemize}
    \end{itemize}
\end{itemize}
\textbf{User: S16\_UserC}
\begin{itemize}[itemsep=1pt, topsep=1pt, left=0pt]
    \item \textbf{Inline Policies:}
    \begin{itemize}[itemsep=1pt, topsep=1pt, left=0pt]
        \item \textbf{S16\_IP\_UserC:}
        \begin{itemize}[itemsep=1pt, topsep=1pt, left=0pt]
            \item \texttt{sdb:BatchPutAttributes}
            \item \texttt{nimble:CreateStudio}
        \end{itemize}
    \end{itemize}
    \item \textbf{Attached Managed Policies:}
    \begin{itemize}[itemsep=1pt, topsep=1pt, left=0pt]
        \item \textbf{S16\_AMP\_PolicyF:}
        \begin{itemize}[itemsep=1pt, topsep=1pt, left=0pt]
            \item \texttt{iam:ListAttachedRolePolicies} (RP)
            \item \texttt{iam:GetGroupPolicy} (GI)
            \item \texttt{Iam:ListGroupPolicies} (GI)
        \end{itemize}
    \end{itemize}
    \item \textbf{Group: S16\_UserC\_GroupA (Includes S16\_UserC)}
    \begin{itemize}[itemsep=1pt, topsep=1pt, left=0pt]
        \item \textbf{Inline Policies:}
        \begin{itemize}[itemsep=1pt, topsep=1pt, left=0pt]
            \item \textbf{S16\_IP\_UserC\_GroupA:}
            \begin{itemize}[itemsep=1pt, topsep=1pt, left=0pt]
                \item \texttt{cloudfront:CreateKeyGroup}
                \item \texttt{cloudfront:CreateKeyValueStore}
            \end{itemize}
        \end{itemize}
        \item \textbf{Attached Managed Policies:}
        \begin{itemize}[itemsep=1pt, topsep=1pt, left=0pt]
            \item \textbf{S16\_AMP\_PolicyG:}
            \begin{itemize}[itemsep=1pt, topsep=1pt, left=0pt]
                \item \texttt{cloudfront:CreatePublicKey}
                \item \texttt{cloudfront:CopyDistribution}
            \end{itemize}
        \end{itemize}
    \end{itemize}
    \item \textbf{Role: S16\_UserC\_RoleA (Assumable by: S16\_UserC)}
    \begin{itemize}[itemsep=1pt, topsep=1pt, left=0pt]
        \item \textbf{Inline Policies:}
        \begin{itemize}[itemsep=1pt, topsep=1pt, left=0pt]
            \item \textbf{S16\_IP\_UserC\_RoleA:}
            \begin{itemize}[itemsep=1pt, topsep=1pt, left=0pt]
                \item \texttt{iam:ListRolePolicies} (RI)
            \end{itemize}
        \end{itemize}
        \item \textbf{Attached Managed Policies:}
        \begin{itemize}[itemsep=1pt, topsep=1pt, left=0pt]
            \item \textbf{S16\_AMP\_PolicyO:}
            \begin{itemize}[itemsep=1pt, topsep=1pt, left=0pt]
                \item \texttt{ram:CreateResourceShare}
            \end{itemize}
        \end{itemize}
    \end{itemize}
    \item \textbf{Role: S16\_UserC\_RoleB (Assumable by: S16\_UserC\_RoleA)}
    \begin{itemize}[itemsep=1pt, topsep=1pt, left=0pt]
        \item \textbf{Inline Policies:}
        \begin{itemize}[itemsep=1pt, topsep=1pt, left=0pt]
            \item \textbf{S16\_IP\_UserC\_RoleB:}
            \begin{itemize}[itemsep=1pt, topsep=1pt, left=0pt]
                \item \texttt{cloudfront:DisassociateDistributionWebACL}
            \end{itemize}
        \end{itemize}
        \item \textbf{Attached Managed Policies:}
        \begin{itemize}[itemsep=1pt, topsep=1pt, left=0pt]
            \item \textbf{AmazonRoute53ReadOnlyAccess (AWS)}
        \end{itemize}
    \end{itemize}
    \item \textbf{Role: S16\_UserC\_RoleC (Assumable by: S16\_UserC\_RoleB)}
    \begin{itemize}[itemsep=1pt, topsep=1pt, left=0pt]
        \item \textbf{Inline Policies:}
        \begin{itemize}[itemsep=1pt, topsep=1pt, left=0pt]
            \item \textbf{S16\_IP\_UserC\_RoleC:}
            \begin{itemize}[itemsep=1pt, topsep=1pt, left=0pt]
                \item \texttt{s3-object-lambda:GetObjectAcl}
                \item \texttt{iam:GetPolicyVersion} (P)
            \end{itemize}
        \end{itemize}
        \item \textbf{Attached Managed Policies:}
        \begin{itemize}[itemsep=1pt, topsep=1pt, left=0pt]
            \item \textbf{S16\_AMP\_PolicyJ:}
            \begin{itemize}[itemsep=1pt, topsep=1pt, left=0pt]
                \item \texttt{tax:GetExemptions}
                \item \texttt{ec2:BundleInstance}
            \end{itemize}
        \end{itemize}
    \end{itemize}
\end{itemize}
\textbf{User: S16\_UserD}
\begin{itemize}[itemsep=1pt, topsep=1pt, left=0pt]
    \item \textbf{Inline Policies:}
    \begin{itemize}[itemsep=1pt, topsep=1pt, left=0pt]
        \item \textbf{S16\_IP\_UserD:}
        \begin{itemize}[itemsep=1pt, topsep=1pt, left=0pt]
            \item \texttt{sns:CreatePlatformEndpoint}
            \item \texttt{sns:CreatePlatformApplication}
        \end{itemize}
    \end{itemize}
    \item \textbf{Attached Managed Policies:}
    \begin{itemize}[itemsep=1pt, topsep=1pt, left=0pt]
        \item \textbf{S16\_AMP\_PolicyY:}
        \begin{itemize}[itemsep=1pt, topsep=1pt, left=0pt]
            \item \texttt{sns:SetTopicAttributes}
            \item \texttt{sns:CreateTopic}
        \end{itemize}
    \end{itemize}
    \item \textbf{Group: S16\_UserD\_GroupA (Includes S16\_UserD)}
    \begin{itemize}[itemsep=1pt, topsep=1pt, left=0pt]
        \item \textbf{Inline Policies:}
        \begin{itemize}[itemsep=1pt, topsep=1pt, left=0pt]
            \item \textbf{S16\_IP\_UserD\_GroupA:}
            \begin{itemize}[itemsep=1pt, topsep=1pt, left=0pt]
                \item \texttt{elasticbeanstalk:AssociateEnvironmentOperationsRole}
                \item \texttt{elasticbeanstalk:DescribeApplications}
            \end{itemize}
        \end{itemize}
        \item \textbf{Attached Managed Policies:}
        \begin{itemize}[itemsep=1pt, topsep=1pt, left=0pt]
            \item \textbf{S16\_AMP\_PolicyU:}
            \begin{itemize}[itemsep=1pt, topsep=1pt, left=0pt]
                \item \texttt{elasticbeanstalk:RemoveTags}
                \item \texttt{elasticbeanstalk:TerminateEnvironment}
            \end{itemize}
        \end{itemize}
    \end{itemize}
    \item \textbf{Role: S16\_UserD\_RoleA (Assumable by: S16\_UserD)}
    \begin{itemize}[itemsep=1pt, topsep=1pt, left=0pt]
        \item \textbf{Inline Policies:}
        \begin{itemize}[itemsep=1pt, topsep=1pt, left=0pt]
            \item \textbf{S16\_IP\_UserD\_RoleA:}
            \begin{itemize}[itemsep=1pt, topsep=1pt, left=0pt]
                \item \texttt{sns:Publish}
                \item \texttt{sns:DeleteTopic}
            \end{itemize}
        \end{itemize}
        \item \textbf{Attached Managed Policies:}
        \begin{itemize}[itemsep=1pt, topsep=1pt, left=0pt]
            \item \textbf{AmazonRoute53ReadOnlyAccess (AWS)}
            \item \textbf{S16\_AMP\_PolicyZ:}
            \begin{itemize}[itemsep=1pt, topsep=1pt, left=0pt]
                \item \texttt{elasticbeanstalk:DeletePlatformVersion}
                \item \texttt{elasticbeanstalk:DescribeEvents}
            \end{itemize}
        \end{itemize}
    \end{itemize}
\end{itemize}

\subsection*{Scenario 17:}
\phantomsection
\label{sec:scenario17}
\textbf{User: S17\_UserA}
\begin{itemize}[itemsep=1pt, topsep=1pt, left=0pt]
    \item \textbf{Inline Policies:}
    \begin{itemize}[itemsep=1pt, topsep=1pt, left=0pt]
        \item \textbf{S17\_IP\_UserA:}
        \begin{itemize}[itemsep=1pt, topsep=1pt, left=0pt]
            \item \texttt{cloudfront:CreatePublicKey}
            \item \texttt{cloudfront:CopyDistribution}
        \end{itemize}
    \end{itemize}
    \item \textbf{Attached Managed Policies:}
    \begin{itemize}[itemsep=1pt, topsep=1pt, left=0pt]
        \item \textbf{S17\_AMP\_PolicyA:}
        \begin{itemize}[itemsep=1pt, topsep=1pt, left=0pt]
            \item \texttt{iot:DeleteThing}
            \item \texttt{bedrock:DeleteGuardrail}
        \end{itemize}
        \item \textbf{S17\_AMP\_PolicyB:}
        \begin{itemize}[itemsep=1pt, topsep=1pt, left=0pt]
            \item \texttt{bedrock:InvokeAgent}
            \item \texttt{bedrock:UpdateFlow}
        \end{itemize}
    \end{itemize}
    \item \textbf{Group: S17\_UserA\_GroupA (Includes S17\_UserA)}
    \begin{itemize}[itemsep=1pt, topsep=1pt, left=0pt]
        \item \textbf{Inline Policies:}
        \begin{itemize}[itemsep=1pt, topsep=1pt, left=0pt]
            \item \textbf{S17\_IP\_UserA\_GroupA:}
            \begin{itemize}[itemsep=1pt, topsep=1pt, left=0pt]
                \item \texttt{s3:CreateBucket}
                \item \texttt{lambda:CreateFunction}
                \item \texttt{ec2:RunInstances}
            \end{itemize}
        \end{itemize}
        \item \textbf{Attached Managed Policies:}
        \begin{itemize}[itemsep=1pt, topsep=1pt, left=0pt]
            \item \textbf{S17\_AMP\_PolicyC:}
            \begin{itemize}[itemsep=1pt, topsep=1pt, left=0pt]
                \item \texttt{private-networks:ActivateDeviceIdentifier}
                \item \texttt{auditmanager:UpdateAssessment}
            \end{itemize}
        \end{itemize}
    \end{itemize}
    \item \textbf{Role: S17\_UserA\_RoleA (Assumable by: S17\_UserA)}
    \begin{itemize}[itemsep=1pt, topsep=1pt, left=0pt]
        \item \textbf{Inline Policies:}
        \begin{itemize}[itemsep=1pt, topsep=1pt, left=0pt]
            \item \textbf{S17\_IP\_UserA\_RoleA:}
            \begin{itemize}[itemsep=1pt, topsep=1pt, left=0pt]
                \item \texttt{ec2:AllocateAddress}
                \item \texttt{controltower:CreateManagedAccount}
            \end{itemize}
        \end{itemize}
        \item \textbf{Attached Managed Policies:}
        \begin{itemize}[itemsep=1pt, topsep=1pt, left=0pt]
            \item \textbf{S17\_AMP\_PolicyT:}
            \begin{itemize}[itemsep=1pt, topsep=1pt, left=0pt]
                \item \texttt{ssm:CancelCommand}
                \item \texttt{globalaccelerator:CreateAccelerator}
            \end{itemize}
        \end{itemize}
    \end{itemize}
\end{itemize}
\textbf{User: S17\_UserB}
\begin{itemize}[itemsep=1pt, topsep=1pt, left=0pt]
    \item \textbf{Inline Policies:}
    \begin{itemize}[itemsep=1pt, topsep=1pt, left=0pt]
        \item \textbf{S17\_IP\_UserB:}
        \begin{itemize}[itemsep=1pt, topsep=1pt, left=0pt]
            \item \texttt{elasticbeanstalk:DeletePlatformVersion}
            \item \texttt{elasticbeanstalk:DescribeEvents}
        \end{itemize}
    \end{itemize}
    \item \textbf{Attached Managed Policies:}
    \begin{itemize}[itemsep=1pt, topsep=1pt, left=0pt]
        \item \textbf{S17\_AMP\_PolicyD:}
        \begin{itemize}[itemsep=1pt, topsep=1pt, left=0pt]
            \item \texttt{codecatalyst:DeleteConnection}
            \item \texttt{codeguru:GetCodeGuruFreeTrialSummary}
        \end{itemize}
    \end{itemize}
    \item \textbf{Group: S17\_UserB\_GroupA (Includes S17\_UserB)}
    \begin{itemize}[itemsep=1pt, topsep=1pt, left=0pt]
        \item \textbf{Inline Policies:}
        \begin{itemize}[itemsep=1pt, topsep=1pt, left=0pt]
            \item \textbf{S17\_IP\_UserB:}
            \begin{itemize}[itemsep=1pt, topsep=1pt, left=0pt]
                \item \texttt{sns:Publish}
                \item \texttt{sns:DeleteTopic}
            \end{itemize}
        \end{itemize}
        \item \textbf{Attached Managed Policies:}
        \begin{itemize}[itemsep=1pt, topsep=1pt, left=0pt]
            \item \textbf{S17\_AMP\_PolicyX:}
            \begin{itemize}[itemsep=1pt, topsep=1pt, left=0pt]
                \item \texttt{globalaccelerator:DeleteAccelerator}
                \item \texttt{codedeploy:BatchGetApplications}
            \end{itemize}
        \end{itemize}
    \end{itemize}
    \item \textbf{Role: S17\_UserB\_RoleA (Assumable by: S17\_UserB)}
    \begin{itemize}[itemsep=1pt, topsep=1pt, left=0pt]
        \item \textbf{Inline Policies:}
        \begin{itemize}[itemsep=1pt, topsep=1pt, left=0pt]
            \item \textbf{S17\_IP\_UserB\_RoleA:}
            \begin{itemize}[itemsep=1pt, topsep=1pt, left=0pt]
                \item \texttt{ram:CreatePermission}
                \item \texttt{ec2:BundleInstance}
            \end{itemize}
        \end{itemize}
        \item \textbf{Attached Managed Policies:}
        \begin{itemize}[itemsep=1pt, topsep=1pt, left=0pt]
            \item \textbf{AmazonEKSServicePolicy (AWS)}
            \item \textbf{S17\_AMP\_PolicyE:}
            \begin{itemize}[itemsep=1pt, topsep=1pt, left=0pt]
                \item \texttt{s3:ListBucket}
                \item \texttt{ec2:DescribeInstances}
            \end{itemize}
        \end{itemize}
    \end{itemize}
\end{itemize}
\textbf{User: S17\_UserC}
\begin{itemize}[itemsep=1pt, topsep=1pt, left=0pt]
    \item \textbf{Inline Policies:}
    \begin{itemize}[itemsep=1pt, topsep=1pt, left=0pt]
        \item \textbf{S17\_IP\_UserC:}
        \begin{itemize}[itemsep=1pt, topsep=1pt, left=0pt]
            \item \texttt{sdb:BatchPutAttributes}
            \item \texttt{nimble:CreateStudio}
        \end{itemize}
    \end{itemize}
    \item \textbf{Attached Managed Policies:}
    \begin{itemize}[itemsep=1pt, topsep=1pt, left=0pt]
        \item \textbf{S17\_AMP\_PolicyF:}
        \begin{itemize}[itemsep=1pt, topsep=1pt, left=0pt]
            \item \texttt{ce:CreateAnomalyMonitor}
        \end{itemize}
    \end{itemize}
    \item \textbf{Group: S17\_UserC\_GroupA (Includes S17\_UserC)}
    \begin{itemize}[itemsep=1pt, topsep=1pt, left=0pt]
        \item \textbf{Inline Policies:}
        \begin{itemize}[itemsep=1pt, topsep=1pt, left=0pt]
            \item \textbf{S17\_IP\_UserC\_GroupA:}
            \begin{itemize}[itemsep=1pt, topsep=1pt, left=0pt]
                \item \texttt{tax:GetExemptions}
                \item \texttt{s3-object-lambda:GetObjectAcl}
            \end{itemize}
        \end{itemize}
        \item \textbf{Attached Managed Policies:}
        \begin{itemize}[itemsep=1pt, topsep=1pt, left=0pt]
            \item \textbf{S17\_AMP\_PolicyG:}
            \begin{itemize}[itemsep=1pt, topsep=1pt, left=0pt]
                \item \texttt{iam:GetAccountAuthorizationDetails (All)}
            \end{itemize}
        \end{itemize}
    \end{itemize}
    \item \textbf{Role: S17\_UserC\_RoleA (Assumable by: S17\_UserC)}
    \begin{itemize}[itemsep=1pt, topsep=1pt, left=0pt]
        \item \textbf{Inline Policies:}
        \begin{itemize}[itemsep=1pt, topsep=1pt, left=0pt]
            \item \textbf{S17\_IP\_UserC\_RoleA:}
            \begin{itemize}[itemsep=1pt, topsep=1pt, left=0pt]
                \item \texttt{ram:CreateResourceShare}
                \item \texttt{scn:DescribeInstance}
            \end{itemize}
        \end{itemize}
    \end{itemize}
\end{itemize}
\textbf{User: S17\_UserD}
\begin{itemize}[itemsep=1pt, topsep=1pt, left=0pt]
    \item \textbf{Inline Policies:}
    \begin{itemize}[itemsep=1pt, topsep=1pt, left=0pt]
        \item \textbf{S17\_IP\_UserD:}
        \begin{itemize}[itemsep=1pt, topsep=1pt, left=0pt]
            \item \texttt{codecatalyst:CreateIdentityCenterApplication}
            \item \texttt{codedeploy:UpdateApplication}
        \end{itemize}
    \end{itemize}
    \item \textbf{Attached Managed Policies:}
    \begin{itemize}[itemsep=1pt, topsep=1pt, left=0pt]
        \item \textbf{S17\_AMP\_PolicyY:}
        \begin{itemize}[itemsep=1pt, topsep=1pt, left=0pt]
            \item \texttt{ram:GetPermission}
            \item \texttt{scn:CreateDataLakeDataset}
        \end{itemize}
    \end{itemize}
    \item \textbf{Group: S17\_UserD\_GroupA (Includes S17\_UserD)}
    \begin{itemize}[itemsep=1pt, topsep=1pt, left=0pt]
        \item \textbf{Inline Policies:}
        \begin{itemize}[itemsep=1pt, topsep=1pt, left=0pt]
            \item \textbf{S17\_IP\_UserD\_GroupA:}
            \begin{itemize}[itemsep=1pt, topsep=1pt, left=0pt]
                \item \texttt{iotanalytics:CreatePipeline}
                \item \texttt{aiops:CreateInvestigationResource}
            \end{itemize}
        \end{itemize}
        \item \textbf{Attached Managed Policies:}
        \begin{itemize}[itemsep=1pt, topsep=1pt, left=0pt]
            \item \textbf{S17\_AMP\_PolicyU:}
            \begin{itemize}[itemsep=1pt, topsep=1pt, left=0pt]
                \item \texttt{iotanalytics:DescribeChannel}
                \item \texttt{codedeploy:CreateDeployment}
            \end{itemize}
        \end{itemize}
    \end{itemize}
    \item \textbf{Role: S17\_UserD\_RoleA (Assumable by: S17\_UserD)}
    \begin{itemize}[itemsep=1pt, topsep=1pt, left=0pt]
        \item \textbf{Inline Policies:}
        \begin{itemize}[itemsep=1pt, topsep=1pt, left=0pt]
            \item \textbf{S17\_IP\_UserD\_RoleA:}
            \begin{itemize}[itemsep=1pt, topsep=1pt, left=0pt]
                \item \texttt{ram:PromotePermissionCreatedFromPolicy}
                \item \texttt{scn:CreateBillOfMaterialsImportJob}
            \end{itemize}
        \end{itemize}
        \item \textbf{Attached Managed Policies:}
        \begin{itemize}[itemsep=1pt, topsep=1pt, left=0pt]
            \item \textbf{AmazonRoute53ReadOnlyAccess (AWS)}
            \item \textbf{S17\_AMP\_PolicyZ:}
            \begin{itemize}[itemsep=1pt, topsep=1pt, left=0pt]
                \item \texttt{iotanalytics:DeleteDataset}
                \item \texttt{qapps:CreateLibraryItemReview}
            \end{itemize}
        \end{itemize}
    \end{itemize}
\end{itemize}

\subsection*{Scenario 18:}
\phantomsection
\label{sec:scenario18}
\textbf{User: S18\_UserA}
\begin{itemize}[itemsep=1pt, topsep=1pt, left=0pt]
    \item \textbf{Inline Policies:}
    \begin{itemize}[itemsep=1pt, topsep=1pt, left=0pt]
        \item \textbf{S18\_IP\_UserA:}
        \begin{itemize}[itemsep=1pt, topsep=1pt, left=0pt]
            \item \texttt{applicationinsights:CreateApplication}
            \item \texttt{applicationinsights:CreateComponent}
        \end{itemize}
    \end{itemize}
    \item \textbf{Attached Managed Policies:}
    \begin{itemize}[itemsep=1pt, topsep=1pt, left=0pt]
        \item \textbf{S18\_AMP\_PolicyA:}
        \begin{itemize}[itemsep=1pt, topsep=1pt, left=0pt]
            \item \texttt{iot:DeleteThing}
            \item \texttt{bedrock:DeleteGuardrail}
        \end{itemize}
        \item \textbf{S18\_AMP\_PolicyB:}
        \begin{itemize}[itemsep=1pt, topsep=1pt, left=0pt]
            \item \texttt{bedrock:InvokeAgent}
            \item \texttt{bedrock:UpdateFlow}
        \end{itemize}
    \end{itemize}
    \item \textbf{Group: S18\_UserA\_GroupA (Includes S18\_UserA)}
    \begin{itemize}[itemsep=1pt, topsep=1pt, left=0pt]
        \item \textbf{Inline Policies:}
        \begin{itemize}[itemsep=1pt, topsep=1pt, left=0pt]
            \item \textbf{S18\_IP\_UserA\_GroupA:}
            \begin{itemize}[itemsep=1pt, topsep=1pt, left=0pt]
                \item \texttt{s3:CreateBucket}
                \item \texttt{lambda:CreateFunction}
                \item \texttt{ec2:RunInstances}
            \end{itemize}
        \end{itemize}
        \item \textbf{Attached Managed Policies:}
        \begin{itemize}[itemsep=1pt, topsep=1pt, left=0pt]
            \item \textbf{S18\_AMP\_PolicyC:}
            \begin{itemize}[itemsep=1pt, topsep=1pt, left=0pt]
                \item \texttt{private-networks:ActivateDeviceIdentifier}
                \item \texttt{auditmanager:UpdateAssessment}
            \end{itemize}
        \end{itemize}
    \end{itemize}
    \item \textbf{Role: S18\_UserA\_RoleA (Assumable by: S18\_UserA)}
    \begin{itemize}[itemsep=1pt, topsep=1pt, left=0pt]
        \item \textbf{Inline Policies:}
        \begin{itemize}[itemsep=1pt, topsep=1pt, left=0pt]
            \item \textbf{S18\_IP\_UserA\_RoleA:}
            \begin{itemize}[itemsep=1pt, topsep=1pt, left=0pt]
                \item \texttt{ec2:AllocateAddress}
                \item \texttt{controltower:CreateManagedAccount}
            \end{itemize}
        \end{itemize}
        \item \textbf{Attached Managed Policies:}
        \begin{itemize}[itemsep=1pt, topsep=1pt, left=0pt]
            \item \textbf{S18\_AMP\_PolicyT:}
            \begin{itemize}[itemsep=1pt, topsep=1pt, left=0pt]
                \item \texttt{ssm:CancelCommand}
                \item \texttt{globalaccelerator:CreateAccelerator}
            \end{itemize}
        \end{itemize}
    \end{itemize}
    \item \textbf{Role: S18\_UserA\_RoleB (Assumable by: S18\_UserA\_RoleA)}
    \begin{itemize}[itemsep=1pt, topsep=1pt, left=0pt]
        \item \textbf{Inline Policies:}
        \begin{itemize}[itemsep=1pt, topsep=1pt, left=0pt]
            \item \textbf{S18\_IP\_UserA\_RoleB:}
            \begin{itemize}[itemsep=1pt, topsep=1pt, left=0pt]
                \item \texttt{iotanalytics:UpdateDatastore}
                \item \texttt{iotanalytics:UpdatePipeline}
            \end{itemize}
        \end{itemize}
        \item \textbf{Attached Managed Policies:}
        \begin{itemize}[itemsep=1pt, topsep=1pt, left=0pt]
            \item \textbf{AmazonEKSServicePolicy (AWS)}
        \end{itemize}
    \end{itemize}
\end{itemize}
\textbf{User: S18\_UserB}
\begin{itemize}[itemsep=1pt, topsep=1pt, left=0pt]
    \item \textbf{Inline Policies:}
    \begin{itemize}[itemsep=1pt, topsep=1pt, left=0pt]
        \item \textbf{S18\_IP\_UserB:}
        \begin{itemize}[itemsep=1pt, topsep=1pt, left=0pt]
            \item \texttt{iam:ListRoles} (R)
        \end{itemize}
    \end{itemize}
    \item \textbf{Attached Managed Policies:}
    \begin{itemize}[itemsep=1pt, topsep=1pt, left=0pt]
        \item \textbf{S18\_AMP\_PolicyD:}
        \begin{itemize}[itemsep=1pt, topsep=1pt, left=0pt]
            \item \texttt{codecatalyst:DeleteConnection}
            \item \texttt{codeguru:GetCodeGuruFreeTrialSummary}
        \end{itemize}
    \end{itemize}
    \item \textbf{Group: S18\_UserB\_GroupA (Includes S18\_UserB)}
    \begin{itemize}[itemsep=1pt, topsep=1pt, left=0pt]
        \item \textbf{Inline Policies:}
        \begin{itemize}[itemsep=1pt, topsep=1pt, left=0pt]
            \item \textbf{S18\_IP\_UserB:}
            \begin{itemize}[itemsep=1pt, topsep=1pt, left=0pt]
                \item \texttt{applicationinsights:UpdateApplication}
            \end{itemize}
        \end{itemize}
        \item \textbf{Attached Managed Policies:}
        \begin{itemize}[itemsep=1pt, topsep=1pt, left=0pt]
            \item \textbf{S18\_AMP\_PolicyX:}
            \begin{itemize}[itemsep=1pt, topsep=1pt, left=0pt]
                \item \texttt{globalaccelerator:DeleteAccelerator}
                \item \texttt{codedeploy:BatchGetApplications}
            \end{itemize}
        \end{itemize}
    \end{itemize}
    \item \textbf{Role: S18\_UserB\_RoleA (Assumable by: S18\_UserB)}
    \begin{itemize}[itemsep=1pt, topsep=1pt, left=0pt]
        \item \textbf{Inline Policies:}
        \begin{itemize}[itemsep=1pt, topsep=1pt, left=0pt]
            \item \textbf{S18\_IP\_UserB\_RoleA:}
            \begin{itemize}[itemsep=1pt, topsep=1pt, left=0pt]
                \item \texttt{ram:CreatePermission}
                \item \texttt{cloudfront:AssociateAlias}
            \end{itemize}
        \end{itemize}
        \item \textbf{Attached Managed Policies:}
        \begin{itemize}[itemsep=1pt, topsep=1pt, left=0pt]
            \item \textbf{S18\_AMP\_PolicyE:}
            \begin{itemize}[itemsep=1pt, topsep=1pt, left=0pt]
                \item \texttt{s3:ListBucket}
                \item \texttt{ec2:DescribeInstances}
            \end{itemize}
        \end{itemize}
    \end{itemize}
    \item \textbf{Role: S18\_UserB\_RoleB (Assumable by: S18\_UserB\_RoleA)}
    \begin{itemize}[itemsep=1pt, topsep=1pt, left=0pt]
        \item \textbf{Inline Policies:}
        \begin{itemize}[itemsep=1pt, topsep=1pt, left=0pt]
            \item \textbf{S18\_IP\_UserB\_RoleB:}
            \begin{itemize}[itemsep=1pt, topsep=1pt, left=0pt]
                \item \texttt{connect:ActivateEvaluationForm}
                \item \texttt{connect:AssociateBot}
            \end{itemize}
        \end{itemize}
        \item \textbf{Attached Managed Policies:}
        \begin{itemize}[itemsep=1pt, topsep=1pt, left=0pt]
            \item \textbf{AmazonEKSServicePolicy (AWS)}
        \end{itemize}
    \end{itemize}
    \item \textbf{Role: S18\_UserB\_RoleC (Assumable by: S18\_UserB\_RoleB)}
    \begin{itemize}[itemsep=1pt, topsep=1pt, left=0pt]
        \item \textbf{Inline Policies:}
        \begin{itemize}[itemsep=1pt, topsep=1pt, left=0pt]
            \item \textbf{S18\_IP\_UserB\_RoleC:}
            \begin{itemize}[itemsep=1pt, topsep=1pt, left=0pt]
                \item \texttt{elasticloadbalancing:CreateLoadBalancer}
            \end{itemize}
        \end{itemize}
        \item \textbf{Attached Managed Policies:}
        \begin{itemize}[itemsep=1pt, topsep=1pt, left=0pt]
            \item \textbf{AmazonRoute53ReadOnlyAccess (AWS)}
        \end{itemize}
    \end{itemize}
\end{itemize}
\textbf{User: S18\_UserC}
\begin{itemize}[itemsep=1pt, topsep=1pt, left=0pt]
    \item \textbf{Inline Policies:}
    \begin{itemize}[itemsep=1pt, topsep=1pt, left=0pt]
        \item \textbf{S18\_IP\_UserC:}
        \begin{itemize}[itemsep=1pt, topsep=1pt, left=0pt]
            \item \texttt{sdb:BatchPutAttributes}
            \item \texttt{nimble:CreateStudio}
        \end{itemize}
    \end{itemize}
    \item \textbf{Attached Managed Policies:}
    \begin{itemize}[itemsep=1pt, topsep=1pt, left=0pt]
        \item \textbf{S18\_AMP\_PolicyF:}
        \begin{itemize}[itemsep=1pt, topsep=1pt, left=0pt]
            \item \texttt{lookoutequipment:DeleteModel}
        \end{itemize}
    \end{itemize}
    \item \textbf{Group: S18\_UserC\_GroupA (Includes S18\_UserC)}
    \begin{itemize}[itemsep=1pt, topsep=1pt, left=0pt]
        \item \textbf{Inline Policies:}
        \begin{itemize}[itemsep=1pt, topsep=1pt, left=0pt]
            \item \textbf{S18\_IP\_UserC\_GroupA:}
            \begin{itemize}[itemsep=1pt, topsep=1pt, left=0pt]
                \item \texttt{cloudfront:CreateKeyGroup}
                \item \texttt{cloudfront:CreateKeyValueStore}
            \end{itemize}
        \end{itemize}
        \item \textbf{Attached Managed Policies:}
        \begin{itemize}[itemsep=1pt, topsep=1pt, left=0pt]
            \item \textbf{S18\_AMP\_PolicyG:}
            \begin{itemize}[itemsep=1pt, topsep=1pt, left=0pt]
                \item \texttt{cloudfront:CreatePublicKey}
                \item \texttt{cloudfront:CopyDistribution}
            \end{itemize}
        \end{itemize}
    \end{itemize}
    \item \textbf{Role: S18\_UserC\_RoleA (Assumable by: S18\_UserC)}
    \begin{itemize}[itemsep=1pt, topsep=1pt, left=0pt]
        \item \textbf{Inline Policies:}
        \begin{itemize}[itemsep=1pt, topsep=1pt, left=0pt]
            \item \textbf{S18\_IP\_UserC\_RoleA:}
            \begin{itemize}[itemsep=1pt, topsep=1pt, left=0pt]
                \item \texttt{lookoutequipment:CreateModel}
            \end{itemize}
        \end{itemize}
        \item \textbf{Attached Managed Policies:}
        \begin{itemize}[itemsep=1pt, topsep=1pt, left=0pt]
            \item \textbf{S18\_AMP\_PolicyO:}
            \begin{itemize}[itemsep=1pt, topsep=1pt, left=0pt]
                \item \texttt{ram:CreateResourceShare}
            \end{itemize}
        \end{itemize}
    \end{itemize}
    \item \textbf{Role: S18\_UserC\_RoleB (Assumable by: S18\_UserC\_RoleA)}
    \begin{itemize}[itemsep=1pt, topsep=1pt, left=0pt]
        \item \textbf{Inline Policies:}
        \begin{itemize}[itemsep=1pt, topsep=1pt, left=0pt]
            \item \textbf{S18\_IP\_UserC\_RoleB:}
            \begin{itemize}[itemsep=1pt, topsep=1pt, left=0pt]
                \item \texttt{cloudfront:DisassociateDistributionWebACL}
            \end{itemize}
        \end{itemize}
        \item \textbf{Attached Managed Policies:}
        \begin{itemize}[itemsep=1pt, topsep=1pt, left=0pt]
            \item \textbf{AmazonRoute53ReadOnlyAccess (AWS)}
        \end{itemize}
    \end{itemize}
    \item \textbf{Role: S18\_UserC\_RoleC (Assumable by: S18\_UserC\_RoleB)}
    \begin{itemize}[itemsep=1pt, topsep=1pt, left=0pt]
        \item \textbf{Inline Policies:}
        \begin{itemize}[itemsep=1pt, topsep=1pt, left=0pt]
            \item \textbf{S18\_IP\_UserC\_RoleC:}
            \begin{itemize}[itemsep=1pt, topsep=1pt, left=0pt]
                \item \texttt{s3-object-lambda:GetObjectAcl}
            \end{itemize}
        \end{itemize}
        \item \textbf{Attached Managed Policies:}
        \begin{itemize}[itemsep=1pt, topsep=1pt, left=0pt]
            \item \textbf{S18\_AMP\_PolicyJ:}
            \begin{itemize}[itemsep=1pt, topsep=1pt, left=0pt]
                \item \texttt{tax:GetExemptions}
                \item \texttt{ec2:BundleInstance}
            \end{itemize}
        \end{itemize}
    \end{itemize}
\end{itemize}
\textbf{User: S18\_UserD}
\begin{itemize}[itemsep=1pt, topsep=1pt, left=0pt]
    \item \textbf{Inline Policies:}
    \begin{itemize}[itemsep=1pt, topsep=1pt, left=0pt]
        \item \textbf{S18\_IP\_UserD:}
        \begin{itemize}[itemsep=1pt, topsep=1pt, left=0pt]
            \item \texttt{sns:CreatePlatformEndpoint}
            \item \texttt{sns:CreatePlatformApplication}
        \end{itemize}
    \end{itemize}
    \item \textbf{Attached Managed Policies:}
    \begin{itemize}[itemsep=1pt, topsep=1pt, left=0pt]
        \item \textbf{S18\_AMP\_PolicyY:}
        \begin{itemize}[itemsep=1pt, topsep=1pt, left=0pt]
            \item \texttt{sns:SetTopicAttributes}
            \item \texttt{sns:CreateTopic}
        \end{itemize}
    \end{itemize}
    \item \textbf{Group: S18\_UserD\_GroupA (Includes S18\_UserD)}
    \begin{itemize}[itemsep=1pt, topsep=1pt, left=0pt]
        \item \textbf{Inline Policies:}
        \begin{itemize}[itemsep=1pt, topsep=1pt, left=0pt]
            \item \textbf{S18\_IP\_UserD\_GroupA:}
            \begin{itemize}[itemsep=1pt, topsep=1pt, left=0pt]
                \item \texttt{elasticbeanstalk:AssociateEnvironmentOperationsRole}
                \item \texttt{elasticbeanstalk:DescribeApplications}
            \end{itemize}
        \end{itemize}
        \item \textbf{Attached Managed Policies:}
        \begin{itemize}[itemsep=1pt, topsep=1pt, left=0pt]
            \item \textbf{S18\_AMP\_PolicyU:}
            \begin{itemize}[itemsep=1pt, topsep=1pt, left=0pt]
                \item \texttt{elasticbeanstalk:RemoveTags}
                \item \texttt{elasticbeanstalk:TerminateEnvironment}
            \end{itemize}
        \end{itemize}
    \end{itemize}
    \item \textbf{Role: S18\_UserD\_RoleA (Assumable by: S18\_UserD)}
    \begin{itemize}[itemsep=1pt, topsep=1pt, left=0pt]
        \item \textbf{Inline Policies:}
        \begin{itemize}[itemsep=1pt, topsep=1pt, left=0pt]
            \item \textbf{S18\_IP\_UserD\_RoleA:}
            \begin{itemize}[itemsep=1pt, topsep=1pt, left=0pt]
                \item \texttt{sns:Publish}
                \item \texttt{sns:DeleteTopic}
            \end{itemize}
        \end{itemize}
        \item \textbf{Attached Managed Policies:}
        \begin{itemize}[itemsep=1pt, topsep=1pt, left=0pt]
            \item \textbf{AmazonRoute53ReadOnlyAccess (AWS)}
            \item \textbf{S18\_AMP\_PolicyZ:}
            \begin{itemize}[itemsep=1pt, topsep=1pt, left=0pt]
                \item \texttt{elasticbeanstalk:DeletePlatformVersion}
                \item \texttt{elasticbeanstalk:DescribeEvents}
            \end{itemize}
        \end{itemize}
    \end{itemize}
    \item \textbf{Role: S18\_UserD\_RoleB (Assumable by: S18\_UserD\_RoleA)}
    \begin{itemize}[itemsep=1pt, topsep=1pt, left=0pt]
        \item \textbf{Inline Policies:}
        \begin{itemize}[itemsep=1pt, topsep=1pt, left=0pt]
            \item \textbf{S18\_IP\_UserD\_RoleB:}
            \begin{itemize}[itemsep=1pt, topsep=1pt, left=0pt]
                \item \texttt{lookoutequipment:DescribeDataset}
                \item \texttt{logs:CreateDelivery}
            \end{itemize}
        \end{itemize}
        \item \textbf{Attached Managed Policies:}
        \begin{itemize}[itemsep=1pt, topsep=1pt, left=0pt]
            \item \textbf{AmazonRoute53ReadOnlyAccess (AWS)}
        \end{itemize}
    \end{itemize}
    \item \textbf{Role: S18\_UserD\_RoleC (Assumable by: S18\_UserD\_RoleB)}
    \begin{itemize}[itemsep=1pt, topsep=1pt, left=0pt]
        \item \textbf{Inline Policies:}
        \begin{itemize}[itemsep=1pt, topsep=1pt, left=0pt]
            \item \textbf{S18\_IP\_UserD\_RoleC:}
            \begin{itemize}[itemsep=1pt, topsep=1pt, left=0pt]
                \item \texttt{lookoutequipment:ListInferenceEvents}
            \end{itemize}
        \end{itemize}
        \item \textbf{Attached Managed Policies:}
        \begin{itemize}[itemsep=1pt, topsep=1pt, left=0pt]
            \item \textbf{AmazonEKSServicePolicy (AWS)}
        \end{itemize}
    \end{itemize}
    \item \textbf{Role: S18\_UserD\_RoleD (Assumable by: S18\_UserD\_RoleC)}
    \begin{itemize}[itemsep=1pt, topsep=1pt, left=0pt]
        \item \textbf{Inline Policies:}
        \begin{itemize}[itemsep=1pt, topsep=1pt, left=0pt]
            \item \textbf{S18\_IP\_UserD\_RoleD:}
            \begin{itemize}[itemsep=1pt, topsep=1pt, left=0pt]
                \item \texttt{drs:CreateSourceNetwork}
                \item \texttt{iam:GetAccountAuthorizationDetails (All)}
            \end{itemize}
        \end{itemize}
        \item \textbf{Attached Managed Policies:}
        \begin{itemize}[itemsep=1pt, topsep=1pt, left=0pt]
            \item \textbf{AmazonRoute53ReadOnlyAccess (AWS)}
        \end{itemize}
    \end{itemize}
\end{itemize}

\subsection*{Scenario 19:}
\phantomsection
\label{sec:scenario19}
\textbf{User: S19\_UserA}
\begin{itemize}[itemsep=1pt, topsep=1pt, left=0pt]
    \item \textbf{Inline Policies:}
    \begin{itemize}[itemsep=1pt, topsep=1pt, left=0pt]
        \item \textbf{S19\_IP\_UserA:}
        \begin{itemize}[itemsep=1pt, topsep=1pt, left=0pt]
            \item \texttt{iam:ListGroupsForUser} (G)
            \item \texttt{iam:ListAttachedUserPolicies} (UP)
            \item \texttt{iam:GetUserPolicy} (UI)
        \end{itemize}
    \end{itemize}
    \item \textbf{Attached Managed Policies:}
    \begin{itemize}[itemsep=1pt, topsep=1pt, left=0pt]
        \item \textbf{S19\_AMP\_PolicyA:}
        \begin{itemize}[itemsep=1pt, topsep=1pt, left=0pt]
            \item \texttt{iot:DeleteThing}
            \item \texttt{bedrock:DeleteGuardrail}
        \end{itemize}
        \item \textbf{S19\_AMP\_PolicyB:}
        \begin{itemize}[itemsep=1pt, topsep=1pt, left=0pt]
            \item \texttt{bedrock:InvokeAgent}
            \item \texttt{bedrock:UpdateFlow}
        \end{itemize}
    \end{itemize}
    \item \textbf{Group: S19\_UserA\_GroupA (Includes S19\_UserA)}
    \begin{itemize}[itemsep=1pt, topsep=1pt, left=0pt]
        \item \textbf{Inline Policies:}
        \begin{itemize}[itemsep=1pt, topsep=1pt, left=0pt]
            \item \textbf{S19\_IP\_UserA\_GroupA:}
            \begin{itemize}[itemsep=1pt, topsep=1pt, left=0pt]
                \item \texttt{iam:ListRoles} (R)
                \item \texttt{s3:CreateBucket}
                \item \texttt{lambda:CreateFunction}
                \item \texttt{ec2:RunInstances}
            \end{itemize}
        \end{itemize}
        \item \textbf{Attached Managed Policies:}
        \begin{itemize}[itemsep=1pt, topsep=1pt, left=0pt]
            \item \textbf{S19\_AMP\_PolicyC:}
            \begin{itemize}[itemsep=1pt, topsep=1pt, left=0pt]
                \item \texttt{private-networks:ActivateDeviceIdentifier}
                \item \texttt{auditmanager:UpdateAssessment}
            \end{itemize}
        \end{itemize}
    \end{itemize}
    \item \textbf{Role: S19\_UserA\_RoleA (Assumable by: S19\_UserA)}
    \begin{itemize}[itemsep=1pt, topsep=1pt, left=0pt]
        \item \textbf{Inline Policies:}
        \begin{itemize}[itemsep=1pt, topsep=1pt, left=0pt]
            \item \textbf{S19\_IP\_UserA\_RoleA:}
            \begin{itemize}[itemsep=1pt, topsep=1pt, left=0pt]
                \item \texttt{ec2:AllocateAddress}
                \item \texttt{controltower:CreateManagedAccount}
            \end{itemize}
        \end{itemize}
        \item \textbf{Attached Managed Policies:}
        \begin{itemize}[itemsep=1pt, topsep=1pt, left=0pt]
            \item \textbf{S19\_AMP\_PolicyT:}
            \begin{itemize}[itemsep=1pt, topsep=1pt, left=0pt]
                \item \texttt{ssm:CancelCommand}
                \item \texttt{globalaccelerator:CreateAccelerator}
            \end{itemize}
        \end{itemize}
    \end{itemize}
\end{itemize}
\textbf{User: S19\_UserB}
\begin{itemize}[itemsep=1pt, topsep=1pt, left=0pt]
    \item \textbf{Inline Policies:}
    \begin{itemize}[itemsep=1pt, topsep=1pt, left=0pt]
        \item \textbf{S19\_IP\_UserB:}
        \begin{itemize}[itemsep=1pt, topsep=1pt, left=0pt]
            \item \texttt{iam:ListUserPolicies} (UI)
            \item \texttt{iam:ListAttachedGroupPolicies} (GP)
            \item \texttt{iam:GetRolePolicy} (RI)
        \end{itemize}
    \end{itemize}
    \item \textbf{Attached Managed Policies:}
    \begin{itemize}[itemsep=1pt, topsep=1pt, left=0pt]
        \item \textbf{S19\_AMP\_PolicyD:}
        \begin{itemize}[itemsep=1pt, topsep=1pt, left=0pt]
            \item \texttt{codecatalyst:DeleteConnection}
            \item \texttt{codeguru:GetCodeGuruFreeTrialSummary}
        \end{itemize}
    \end{itemize}
    \item \textbf{Group: S19\_UserB\_GroupA (Includes S19\_UserB)}
    \begin{itemize}[itemsep=1pt, topsep=1pt, left=0pt]
        \item \textbf{Inline Policies:}
        \begin{itemize}[itemsep=1pt, topsep=1pt, left=0pt]
            \item \textbf{S19\_IP\_UserB:}
            \begin{itemize}[itemsep=1pt, topsep=1pt, left=0pt]
                \item \texttt{drs:CreateConvertedSnapshotForDrs}
            \end{itemize}
        \end{itemize}
        \item \textbf{Attached Managed Policies:}
        \begin{itemize}[itemsep=1pt, topsep=1pt, left=0pt]
            \item \textbf{S19\_AMP\_PolicyX:}
            \begin{itemize}[itemsep=1pt, topsep=1pt, left=0pt]
                \item \texttt{globalaccelerator:DeleteAccelerator}
                \item \texttt{codedeploy:BatchGetApplications}
            \end{itemize}
        \end{itemize}
    \end{itemize}
    \item \textbf{Role: S19\_UserB\_RoleA (Assumable by: S19\_UserB)}
    \begin{itemize}[itemsep=1pt, topsep=1pt, left=0pt]
        \item \textbf{Inline Policies:}
        \begin{itemize}[itemsep=1pt, topsep=1pt, left=0pt]
            \item \textbf{S19\_IP\_UserB\_RoleA:}
            \begin{itemize}[itemsep=1pt, topsep=1pt, left=0pt]
                \item \texttt{ram:CreatePermission}
                \item \texttt{ec2:BundleInstance}
            \end{itemize}
        \end{itemize}
        \item \textbf{Attached Managed Policies:}
        \begin{itemize}[itemsep=1pt, topsep=1pt, left=0pt]
            \item \textbf{AmazonEKSServicePolicy (AWS)}
            \item \textbf{S19\_AMP\_PolicyE:}
            \begin{itemize}[itemsep=1pt, topsep=1pt, left=0pt]
                \item \texttt{s3:ListBucket}
                \item \texttt{ec2:DescribeInstances}
            \end{itemize}
        \end{itemize}
    \end{itemize}
\end{itemize}
\textbf{User: S19\_UserC}
\begin{itemize}[itemsep=1pt, topsep=1pt, left=0pt]
    \item \textbf{Inline Policies:}
    \begin{itemize}[itemsep=1pt, topsep=1pt, left=0pt]
        \item \textbf{S19\_IP\_UserC:}
        \begin{itemize}[itemsep=1pt, topsep=1pt, left=0pt]
            \item \texttt{sdb:BatchPutAttributes}
            \item \texttt{nimble:CreateStudio}
        \end{itemize}
    \end{itemize}
    \item \textbf{Attached Managed Policies:}
    \begin{itemize}[itemsep=1pt, topsep=1pt, left=0pt]
        \item \textbf{S19\_AMP\_PolicyF:}
        \begin{itemize}[itemsep=1pt, topsep=1pt, left=0pt]
            \item \texttt{iam:ListAttachedRolePolicies} (RP)
            \item \texttt{iam:GetGroupPolicy} (GI)
            \item \texttt{Iam:ListGroupPolicies} (GI)
        \end{itemize}
    \end{itemize}
    \item \textbf{Group: S19\_UserC\_GroupA (Includes S19\_UserC)}
    \begin{itemize}[itemsep=1pt, topsep=1pt, left=0pt]
        \item \textbf{Inline Policies:}
        \begin{itemize}[itemsep=1pt, topsep=1pt, left=0pt]
            \item \textbf{S19\_IP\_UserC\_GroupA:}
            \begin{itemize}[itemsep=1pt, topsep=1pt, left=0pt]
                \item \texttt{tax:GetExemptions}
                \item \texttt{s3-object-lambda:GetObjectAcl}
            \end{itemize}
        \end{itemize}
        \item \textbf{Attached Managed Policies:}
        \begin{itemize}[itemsep=1pt, topsep=1pt, left=0pt]
            \item \textbf{S19\_AMP\_PolicyG:}
            \begin{itemize}[itemsep=1pt, topsep=1pt, left=0pt]
                \item \texttt{iam:ListRolePolicies} (RI)
                \item \texttt{iam:GetPolicyVersion} (P)
            \end{itemize}
        \end{itemize}
    \end{itemize}
    \item \textbf{Role: S19\_UserC\_RoleA (Assumable by: S19\_UserC)}
    \begin{itemize}[itemsep=1pt, topsep=1pt, left=0pt]
        \item \textbf{Inline Policies:}
        \begin{itemize}[itemsep=1pt, topsep=1pt, left=0pt]
            \item \textbf{S19\_IP\_UserC\_RoleA:}
            \begin{itemize}[itemsep=1pt, topsep=1pt, left=0pt]
                \item \texttt{ram:CreateResourceShare}
                \item \texttt{scn:DescribeInstance}
            \end{itemize}
        \end{itemize}
    \end{itemize}
\end{itemize}
\textbf{User: S19\_UserD}
\begin{itemize}[itemsep=1pt, topsep=1pt, left=0pt]
    \item \textbf{Inline Policies:}
    \begin{itemize}[itemsep=1pt, topsep=1pt, left=0pt]
        \item \textbf{S19\_IP\_UserD:}
        \begin{itemize}[itemsep=1pt, topsep=1pt, left=0pt]
            \item \texttt{codecatalyst:CreateIdentityCenterApplication}
            \item \texttt{codedeploy:UpdateApplication}
        \end{itemize}
    \end{itemize}
    \item \textbf{Attached Managed Policies:}
    \begin{itemize}[itemsep=1pt, topsep=1pt, left=0pt]
        \item \textbf{S19\_AMP\_PolicyY:}
        \begin{itemize}[itemsep=1pt, topsep=1pt, left=0pt]
            \item \texttt{ram:GetPermission}
            \item \texttt{scn:CreateDataLakeDataset}
        \end{itemize}
    \end{itemize}
    \item \textbf{Group: S19\_UserD\_GroupA (Includes S19\_UserD)}
    \begin{itemize}[itemsep=1pt, topsep=1pt, left=0pt]
        \item \textbf{Inline Policies:}
        \begin{itemize}[itemsep=1pt, topsep=1pt, left=0pt]
            \item \textbf{S19\_IP\_UserD\_GroupA:}
            \begin{itemize}[itemsep=1pt, topsep=1pt, left=0pt]
                \item \texttt{iotanalytics:CreatePipeline}
                \item \texttt{aiops:CreateInvestigationResource}
            \end{itemize}
        \end{itemize}
        \item \textbf{Attached Managed Policies:}
        \begin{itemize}[itemsep=1pt, topsep=1pt, left=0pt]
            \item \textbf{S19\_AMP\_PolicyU:}
            \begin{itemize}[itemsep=1pt, topsep=1pt, left=0pt]
                \item \texttt{iotanalytics:DescribeChannel}
                \item \texttt{codedeploy:CreateDeployment}
            \end{itemize}
        \end{itemize}
    \end{itemize}
    \item \textbf{Role: S19\_UserD\_RoleA (Assumable by: S19\_UserD)}
    \begin{itemize}[itemsep=1pt, topsep=1pt, left=0pt]
        \item \textbf{Inline Policies:}
        \begin{itemize}[itemsep=1pt, topsep=1pt, left=0pt]
            \item \textbf{S19\_IP\_UserD\_RoleA:}
            \begin{itemize}[itemsep=1pt, topsep=1pt, left=0pt]
                \item \texttt{ram:PromotePermissionCreatedFromPolicy}
                \item \texttt{scn:CreateBillOfMaterialsImportJob}
            \end{itemize}
        \end{itemize}
        \item \textbf{Attached Managed Policies:}
        \begin{itemize}[itemsep=1pt, topsep=1pt, left=0pt]
            \item \textbf{AmazonRoute53ReadOnlyAccess (AWS)}
            \item \textbf{S19\_AMP\_PolicyZ:}
            \begin{itemize}[itemsep=1pt, topsep=1pt, left=0pt]
                \item \texttt{iotanalytics:DeleteDataset}
                \item \texttt{qapps:CreateLibraryItemReview}
            \end{itemize}
        \end{itemize}
    \end{itemize}
\end{itemize}

\subsection*{Scenario 20:}
\phantomsection
\label{sec:scenario20}
\textbf{User: S20\_UserA}
\begin{itemize}[itemsep=1pt, topsep=1pt, left=0pt]
    \item \textbf{Inline Policies:}
    \begin{itemize}[itemsep=1pt, topsep=1pt, left=0pt]
        \item \textbf{S20\_IP\_UserA:}
        \begin{itemize}[itemsep=1pt, topsep=1pt, left=0pt]
            \item \texttt{iam:ListGroupsForUser} (G)
        \end{itemize}
    \end{itemize}
    \item \textbf{Attached Managed Policies:}
    \begin{itemize}[itemsep=1pt, topsep=1pt, left=0pt]
        \item \textbf{S20\_AMP\_PolicyA:}
        \begin{itemize}[itemsep=1pt, topsep=1pt, left=0pt]
            \item \texttt{iot:DeleteThing}
            \item \texttt{bedrock:DeleteGuardrail}
        \end{itemize}
        \item \textbf{S20\_AMP\_PolicyB:}
        \begin{itemize}[itemsep=1pt, topsep=1pt, left=0pt]
            \item \texttt{bedrock:InvokeAgent}
            \item \texttt{bedrock:UpdateFlow}
        \end{itemize}
    \end{itemize}
    \item \textbf{Group: S20\_UserA\_GroupA (Includes S20\_UserA)}
    \begin{itemize}[itemsep=1pt, topsep=1pt, left=0pt]
        \item \textbf{Inline Policies:}
        \begin{itemize}[itemsep=1pt, topsep=1pt, left=0pt]
            \item \textbf{S20\_IP\_UserA\_GroupA:}
            \begin{itemize}[itemsep=1pt, topsep=1pt, left=0pt]
                \item \texttt{iam:ListRoles} (R)
                \item \texttt{s3:CreateBucket}
                \item \texttt{lambda:CreateFunction}
                \item \texttt{ec2:RunInstances}
            \end{itemize}
        \end{itemize}
        \item \textbf{Attached Managed Policies:}
        \begin{itemize}[itemsep=1pt, topsep=1pt, left=0pt]
            \item \textbf{S20\_AMP\_PolicyC:}
            \begin{itemize}[itemsep=1pt, topsep=1pt, left=0pt]
                \item \texttt{private-networks:ActivateDeviceIdentifier}
                \item \texttt{auditmanager:UpdateAssessment}
            \end{itemize}
        \end{itemize}
    \end{itemize}
    \item \textbf{Role: S20\_UserA\_RoleA (Assumable by: S20\_UserA)}
    \begin{itemize}[itemsep=1pt, topsep=1pt, left=0pt]
        \item \textbf{Inline Policies:}
        \begin{itemize}[itemsep=1pt, topsep=1pt, left=0pt]
            \item \textbf{S20\_IP\_UserA\_RoleA:}
            \begin{itemize}[itemsep=1pt, topsep=1pt, left=0pt]
                \item \texttt{iam:ListAttachedUserPolicies} (UP)
                \item \texttt{ec2:AllocateAddress}
                \item \texttt{controltower:CreateManagedAccount}
            \end{itemize}
        \end{itemize}
        \item \textbf{Attached Managed Policies:}
        \begin{itemize}[itemsep=1pt, topsep=1pt, left=0pt]
            \item \textbf{S20\_AMP\_PolicyT:}
            \begin{itemize}[itemsep=1pt, topsep=1pt, left=0pt]
                \item \texttt{ssm:CancelCommand}
                \item \texttt{globalaccelerator:CreateAccelerator}
            \end{itemize}
        \end{itemize}
    \end{itemize}
    \item \textbf{Role: S20\_UserA\_RoleB (Assumable by: S20\_UserA\_RoleA)}
    \begin{itemize}[itemsep=1pt, topsep=1pt, left=0pt]
        \item \textbf{Inline Policies:}
        \begin{itemize}[itemsep=1pt, topsep=1pt, left=0pt]
            \item \textbf{S20\_IP\_UserA\_RoleB:}
            \begin{itemize}[itemsep=1pt, topsep=1pt, left=0pt]
                \item \texttt{iam:GetUserPolicy} (UI)
                \item \texttt{iotanalytics:UpdateDatastore}
                \item \texttt{iotanalytics:UpdatePipeline}
            \end{itemize}
        \end{itemize}
        \item \textbf{Attached Managed Policies:}
        \begin{itemize}[itemsep=1pt, topsep=1pt, left=0pt]
            \item \textbf{AmazonEKSServicePolicy (AWS)}
        \end{itemize}
    \end{itemize}
\end{itemize}
\textbf{User: S20\_UserB}
\begin{itemize}[itemsep=1pt, topsep=1pt, left=0pt]
    \item \textbf{Inline Policies:}
    \begin{itemize}[itemsep=1pt, topsep=1pt, left=0pt]
        \item \textbf{S20\_IP\_UserB:}
        \begin{itemize}[itemsep=1pt, topsep=1pt, left=0pt]
            \item \texttt{iam:ListUserPolicies} (UI)
        \end{itemize}
    \end{itemize}
    \item \textbf{Attached Managed Policies:}
    \begin{itemize}[itemsep=1pt, topsep=1pt, left=0pt]
        \item \textbf{S20\_AMP\_PolicyD:}
        \begin{itemize}[itemsep=1pt, topsep=1pt, left=0pt]
            \item \texttt{codecatalyst:DeleteConnection}
            \item \texttt{codeguru:GetCodeGuruFreeTrialSummary}
        \end{itemize}
    \end{itemize}
    \item \textbf{Group: S20\_UserB\_GroupA (Includes S20\_UserB)}
    \begin{itemize}[itemsep=1pt, topsep=1pt, left=0pt]
        \item \textbf{Inline Policies:}
        \begin{itemize}[itemsep=1pt, topsep=1pt, left=0pt]
            \item \textbf{S20\_IP\_UserB:}
            \begin{itemize}[itemsep=1pt, topsep=1pt, left=0pt]
                \item \texttt{drs:DeleteJob}
            \end{itemize}
        \end{itemize}
        \item \textbf{Attached Managed Policies:}
        \begin{itemize}[itemsep=1pt, topsep=1pt, left=0pt]
            \item \textbf{S20\_AMP\_PolicyX:}
            \begin{itemize}[itemsep=1pt, topsep=1pt, left=0pt]
                \item \texttt{globalaccelerator:DeleteAccelerator}
                \item \texttt{codedeploy:BatchGetApplications}
            \end{itemize}
        \end{itemize}
    \end{itemize}
    \item \textbf{Role: S20\_UserB\_RoleA (Assumable by: S20\_UserB)}
    \begin{itemize}[itemsep=1pt, topsep=1pt, left=0pt]
        \item \textbf{Inline Policies:}
        \begin{itemize}[itemsep=1pt, topsep=1pt, left=0pt]
            \item \textbf{S20\_IP\_UserB\_RoleA:}
            \begin{itemize}[itemsep=1pt, topsep=1pt, left=0pt]
                \item \texttt{ram:CreatePermission}
                \item \texttt{cloudfront:AssociateAlias}
                \item \texttt{iam:ListAttachedGroupPolicies} (GP)
            \end{itemize}
        \end{itemize}
        \item \textbf{Attached Managed Policies:}
        \begin{itemize}[itemsep=1pt, topsep=1pt, left=0pt]
            \item \textbf{S20\_AMP\_PolicyE:}
            \begin{itemize}[itemsep=1pt, topsep=1pt, left=0pt]
                \item \texttt{s3:ListBucket}
                \item \texttt{ec2:DescribeInstances}
            \end{itemize}
        \end{itemize}
    \end{itemize}
    \item \textbf{Role: S20\_UserB\_RoleB (Assumable by: S20\_UserB\_RoleA)}
    \begin{itemize}[itemsep=1pt, topsep=1pt, left=0pt]
        \item \textbf{Inline Policies:}
        \begin{itemize}[itemsep=1pt, topsep=1pt, left=0pt]
            \item \textbf{S20\_IP\_UserB\_RoleB:}
            \begin{itemize}[itemsep=1pt, topsep=1pt, left=0pt]
                \item \texttt{connect:ActivateEvaluationForm}
                \item \texttt{connect:AssociateBot}
            \end{itemize}
        \end{itemize}
        \item \textbf{Attached Managed Policies:}
        \begin{itemize}[itemsep=1pt, topsep=1pt, left=0pt]
            \item \textbf{AmazonEKSServicePolicy (AWS)}
        \end{itemize}
    \end{itemize}
    \item \textbf{Role: S20\_UserB\_RoleC (Assumable by: S20\_UserB\_RoleB)}
    \begin{itemize}[itemsep=1pt, topsep=1pt, left=0pt]
        \item \textbf{Inline Policies:}
        \begin{itemize}[itemsep=1pt, topsep=1pt, left=0pt]
            \item \textbf{S20\_IP\_UserB\_RoleC:}
            \begin{itemize}[itemsep=1pt, topsep=1pt, left=0pt]
                \item \texttt{iam:GetRolePolicy} (RI)
            \end{itemize}
        \end{itemize}
        \item \textbf{Attached Managed Policies:}
        \begin{itemize}[itemsep=1pt, topsep=1pt, left=0pt]
            \item \textbf{AmazonRoute53ReadOnlyAccess (AWS)}
        \end{itemize}
    \end{itemize}
\end{itemize}
\textbf{User: S20\_UserC}
\begin{itemize}[itemsep=1pt, topsep=1pt, left=0pt]
    \item \textbf{Inline Policies:}
    \begin{itemize}[itemsep=1pt, topsep=1pt, left=0pt]
        \item \textbf{S20\_IP\_UserC:}
        \begin{itemize}[itemsep=1pt, topsep=1pt, left=0pt]
            \item \texttt{sdb:BatchPutAttributes}
            \item \texttt{nimble:CreateStudio}
        \end{itemize}
    \end{itemize}
    \item \textbf{Attached Managed Policies:}
    \begin{itemize}[itemsep=1pt, topsep=1pt, left=0pt]
        \item \textbf{S20\_AMP\_PolicyF:}
        \begin{itemize}[itemsep=1pt, topsep=1pt, left=0pt]
            \item \texttt{iam:ListAttachedRolePolicies} (RP)
            \item \texttt{iam:GetGroupPolicy} (GI)
            \item \texttt{Iam:ListGroupPolicies} (GI)
        \end{itemize}
    \end{itemize}
    \item \textbf{Group: S20\_UserC\_GroupA (Includes S20\_UserC)}
    \begin{itemize}[itemsep=1pt, topsep=1pt, left=0pt]
        \item \textbf{Inline Policies:}
        \begin{itemize}[itemsep=1pt, topsep=1pt, left=0pt]
            \item \textbf{S20\_IP\_UserC\_GroupA:}
            \begin{itemize}[itemsep=1pt, topsep=1pt, left=0pt]
                \item \texttt{cloudfront:CreateKeyGroup}
                \item \texttt{cloudfront:CreateKeyValueStore}
            \end{itemize}
        \end{itemize}
        \item \textbf{Attached Managed Policies:}
        \begin{itemize}[itemsep=1pt, topsep=1pt, left=0pt]
            \item \textbf{S20\_AMP\_PolicyG:}
            \begin{itemize}[itemsep=1pt, topsep=1pt, left=0pt]
                \item \texttt{cloudfront:CreatePublicKey}
                \item \texttt{cloudfront:CopyDistribution}
            \end{itemize}
        \end{itemize}
    \end{itemize}
    \item \textbf{Role: S20\_UserC\_RoleA (Assumable by: S20\_UserC)}
    \begin{itemize}[itemsep=1pt, topsep=1pt, left=0pt]
        \item \textbf{Inline Policies:}
        \begin{itemize}[itemsep=1pt, topsep=1pt, left=0pt]
            \item \textbf{S20\_IP\_UserC\_RoleA:}
            \begin{itemize}[itemsep=1pt, topsep=1pt, left=0pt]
                \item \texttt{iam:ListRolePolicies} (RI)
            \end{itemize}
        \end{itemize}
        \item \textbf{Attached Managed Policies:}
        \begin{itemize}[itemsep=1pt, topsep=1pt, left=0pt]
            \item \textbf{S20\_AMP\_PolicyO:}
            \begin{itemize}[itemsep=1pt, topsep=1pt, left=0pt]
                \item \texttt{ram:CreateResourceShare}
            \end{itemize}
        \end{itemize}
    \end{itemize}
    \item \textbf{Role: S20\_UserC\_RoleB (Assumable by: S20\_UserC\_RoleA)}
    \begin{itemize}[itemsep=1pt, topsep=1pt, left=0pt]
        \item \textbf{Inline Policies:}
        \begin{itemize}[itemsep=1pt, topsep=1pt, left=0pt]
            \item \textbf{S20\_IP\_UserC\_RoleB:}
            \begin{itemize}[itemsep=1pt, topsep=1pt, left=0pt]
                \item \texttt{cloudfront:DisassociateDistributionWebACL}
            \end{itemize}
        \end{itemize}
        \item \textbf{Attached Managed Policies:}
        \begin{itemize}[itemsep=1pt, topsep=1pt, left=0pt]
            \item \textbf{AmazonRoute53ReadOnlyAccess (AWS)}
        \end{itemize}
    \end{itemize}
    \item \textbf{Role: S20\_UserC\_RoleC (Assumable by: S20\_UserC\_RoleB)}
    \begin{itemize}[itemsep=1pt, topsep=1pt, left=0pt]
        \item \textbf{Inline Policies:}
        \begin{itemize}[itemsep=1pt, topsep=1pt, left=0pt]
            \item \textbf{S20\_IP\_UserC\_RoleC:}
            \begin{itemize}[itemsep=1pt, topsep=1pt, left=0pt]
                \item \texttt{s3-object-lambda:GetObjectAcl}
                \item \texttt{iam:GetPolicyVersion} (P)
            \end{itemize}
        \end{itemize}
        \item \textbf{Attached Managed Policies:}
        \begin{itemize}[itemsep=1pt, topsep=1pt, left=0pt]
            \item \textbf{S20\_AMP\_PolicyJ:}
            \begin{itemize}[itemsep=1pt, topsep=1pt, left=0pt]
                \item \texttt{tax:GetExemptions}
                \item \texttt{ec2:BundleInstance}
            \end{itemize}
        \end{itemize}
    \end{itemize}
\end{itemize}
\textbf{User: S20\_UserD}
\begin{itemize}[itemsep=1pt, topsep=1pt, left=0pt]
    \item \textbf{Inline Policies:}
    \begin{itemize}[itemsep=1pt, topsep=1pt, left=0pt]
        \item \textbf{S20\_IP\_UserD:}
        \begin{itemize}[itemsep=1pt, topsep=1pt, left=0pt]
            \item \texttt{sns:CreatePlatformEndpoint}
            \item \texttt{sns:CreatePlatformApplication}
        \end{itemize}
    \end{itemize}
    \item \textbf{Attached Managed Policies:}
    \begin{itemize}[itemsep=1pt, topsep=1pt, left=0pt]
        \item \textbf{S20\_AMP\_PolicyY:}
        \begin{itemize}[itemsep=1pt, topsep=1pt, left=0pt]
            \item \texttt{sns:SetTopicAttributes}
            \item \texttt{sns:CreateTopic}
        \end{itemize}
    \end{itemize}
    \item \textbf{Group: S20\_UserD\_GroupA (Includes S20\_UserD)}
    \begin{itemize}[itemsep=1pt, topsep=1pt, left=0pt]
        \item \textbf{Inline Policies:}
        \begin{itemize}[itemsep=1pt, topsep=1pt, left=0pt]
            \item \textbf{S20\_IP\_UserD\_GroupA:}
            \begin{itemize}[itemsep=1pt, topsep=1pt, left=0pt]
                \item \texttt{elasticbeanstalk:AssociateEnvironmentOperationsRole}
                \item \texttt{elasticbeanstalk:DescribeApplications}
            \end{itemize}
        \end{itemize}
        \item \textbf{Attached Managed Policies:}
        \begin{itemize}[itemsep=1pt, topsep=1pt, left=0pt]
            \item \textbf{S20\_AMP\_PolicyU:}
            \begin{itemize}[itemsep=1pt, topsep=1pt, left=0pt]
                \item \texttt{elasticbeanstalk:RemoveTags}
                \item \texttt{elasticbeanstalk:TerminateEnvironment}
            \end{itemize}
        \end{itemize}
    \end{itemize}
    \item \textbf{Role: S20\_UserD\_RoleA (Assumable by: S20\_UserD)}
    \begin{itemize}[itemsep=1pt, topsep=1pt, left=0pt]
        \item \textbf{Inline Policies:}
        \begin{itemize}[itemsep=1pt, topsep=1pt, left=0pt]
            \item \textbf{S20\_IP\_UserD\_RoleA:}
            \begin{itemize}[itemsep=1pt, topsep=1pt, left=0pt]
                \item \texttt{sns:Publish}
                \item \texttt{sns:DeleteTopic}
            \end{itemize}
        \end{itemize}
        \item \textbf{Attached Managed Policies:}
        \begin{itemize}[itemsep=1pt, topsep=1pt, left=0pt]
            \item \textbf{AmazonRoute53ReadOnlyAccess (AWS)}
            \item \textbf{S20\_AMP\_PolicyZ:}
            \begin{itemize}[itemsep=1pt, topsep=1pt, left=0pt]
                \item \texttt{elasticbeanstalk:DeletePlatformVersion}
                \item \texttt{elasticbeanstalk:DescribeEvents}
            \end{itemize}
        \end{itemize}
    \end{itemize}
\end{itemize}

\subsection*{Scenario 21:}
\phantomsection
\label{sec:scenario21}
\textbf{User: S21\_UserA}
\begin{itemize}[itemsep=1pt, topsep=1pt, left=0pt]
    \item \textbf{Inline Policies:}
    \begin{itemize}[itemsep=1pt, topsep=1pt, left=0pt]
        \item \textbf{S21\_IP\_UserA:}
        \begin{itemize}[itemsep=1pt, topsep=1pt, left=0pt]
            \item \texttt{aiops:CreateInvestigation}
            \item \texttt{iot:CreateThing}
        \end{itemize}
    \end{itemize}
    \item \textbf{Attached Managed Policies:}
    \begin{itemize}[itemsep=1pt, topsep=1pt, left=0pt]
        \item \textbf{S21\_AMP\_PolicyA:}
        \begin{itemize}[itemsep=1pt, topsep=1pt, left=0pt]
            \item \texttt{iot:DeleteThing}
            \item \texttt{bedrock:DeleteGuardrail}
        \end{itemize}
        \item \textbf{S21\_AMP\_PolicyB:}
        \begin{itemize}[itemsep=1pt, topsep=1pt, left=0pt]
            \item \texttt{bedrock:InvokeAgent}
            \item \texttt{bedrock:UpdateFlow}
            \item \texttt{iam:SimulatePrincipalPolicy} (SPP)
        \end{itemize}
    \end{itemize}
    \item \textbf{Group: S21\_GroupA (Includes S21\_UserA)}
    \begin{itemize}[itemsep=1pt, topsep=1pt, left=0pt]
        \item \textbf{Inline Policies:}
        \begin{itemize}[itemsep=1pt, topsep=1pt, left=0pt]
            \item \textbf{S21\_IP\_GroupA:}
            \begin{itemize}[itemsep=1pt, topsep=1pt, left=0pt]
                \item \texttt{iam:ListRoles} (R)
                \item \texttt{s3:CreateBucket}
                \item \texttt{lambda:CreateFunction}
                \item \texttt{ec2:RunInstances}
            \end{itemize}
        \end{itemize}
        \item \textbf{Attached Managed Policies:}
        \begin{itemize}[itemsep=1pt, topsep=1pt, left=0pt]
            \item \textbf{AmazonRoute53ReadOnlyAccess (AWS)}
            \item \textbf{S21\_AMP\_PolicyC:}
            \begin{itemize}[itemsep=1pt, topsep=1pt, left=0pt]
                \item \texttt{s3:CreateBucket}
                \item \texttt{lambda:CreateFunction}
                \item \texttt{ec2:RunInstances}
                \item \texttt{s3:ListBucket}
                \item \texttt{ec2:DescribeInstances}
            \end{itemize}
        \end{itemize}
    \end{itemize}
    \item \textbf{Role: S21\_RoleA (Assumable by: S21\_UserA)}
    \begin{itemize}[itemsep=1pt, topsep=1pt, left=0pt]
        \item \textbf{Inline Policies:}
        \begin{itemize}[itemsep=1pt, topsep=1pt, left=0pt]
            \item \textbf{S21\_IP\_RoleA:}
            \begin{itemize}[itemsep=1pt, topsep=1pt, left=0pt]
                \item \texttt{s3:ListBucket}
                \item \texttt{ec2:DescribeInstances}
            \end{itemize}
        \end{itemize}
        \item \textbf{Attached Managed Policies:}
        \begin{itemize}[itemsep=1pt, topsep=1pt, left=0pt]
            \item \textbf{S21\_AMP\_PolicyD:}
            \begin{itemize}[itemsep=1pt, topsep=1pt, left=0pt]
                \item \texttt{ssm:CancelCommand}
                \item \texttt{codeguru:GetCodeGuruFreeTrialSummary}
            \end{itemize}
        \end{itemize}
    \end{itemize}
    \item \textbf{Role: S21\_RoleB (Assumable by: S21\_RoleA)}
    \begin{itemize}[itemsep=1pt, topsep=1pt, left=0pt]
        \item \textbf{Inline Policies:}
        \begin{itemize}[itemsep=1pt, topsep=1pt, left=0pt]
            \item \textbf{S21\_IP\_RoleB:}
            \begin{itemize}[itemsep=1pt, topsep=1pt, left=0pt]
                \item \texttt{s3:CreateBucket}
                \item \texttt{lambda:CreateFunction}
            \end{itemize}
        \end{itemize}
        \item \textbf{Attached Managed Policies:}
        \begin{itemize}[itemsep=1pt, topsep=1pt, left=0pt]
            \item \textbf{AmazonEKSServicePolicy (AWS)}
            \item \textbf{S21\_AMP\_PolicyE:}
            \begin{itemize}[itemsep=1pt, topsep=1pt, left=0pt]
                \item \texttt{ec2:AllocateAddress}
                \item \texttt{ec2:BundleInstance}
            \end{itemize}
        \end{itemize}
    \end{itemize}
\end{itemize}

\subsection*{Scenario 22:}
\phantomsection
\label{sec:scenario22}
\textbf{User: S22\_UserA}
\begin{itemize}[itemsep=1pt, topsep=1pt, left=0pt]
    \item \textbf{Inline Policies:}
    \begin{itemize}[itemsep=1pt, topsep=1pt, left=0pt]
        \item \textbf{S22\_IP\_UserA:}
        \begin{itemize}[itemsep=1pt, topsep=1pt, left=0pt]
            \item \texttt{rds:DescribeDBSnapshots}
            \item \texttt{rds:DescribeDBSecurityGroups}
        \end{itemize}
    \end{itemize}
    \item \textbf{Attached Managed Policies:}
    \begin{itemize}[itemsep=1pt, topsep=1pt, left=0pt]
        \item \textbf{S22\_AMP\_PolicyA:}
        \begin{itemize}[itemsep=1pt, topsep=1pt, left=0pt]
            \item \texttt{rds:DescribeDBInstances}
            \item \texttt{rds:DescribeDBClusters}
        \end{itemize}
        \item \textbf{S22\_AMP\_PolicyB:}
        \begin{itemize}[itemsep=1pt, topsep=1pt, left=0pt]
            \item \texttt{ec2:DescribeAddresses}
            \item \texttt{ec2:DescribeBundleTasks}
            \item \texttt{ec2:DescribeInstances}
        \end{itemize}
    \end{itemize}
    \item \textbf{Group: S22\_GroupA (Includes S22\_UserA)}
    \begin{itemize}[itemsep=1pt, topsep=1pt, left=0pt]
        \item \textbf{Inline Policies:}
        \begin{itemize}[itemsep=1pt, topsep=1pt, left=0pt]
            \item \textbf{S22\_IP\_GroupA:}
            \begin{itemize}[itemsep=1pt, topsep=1pt, left=0pt]
                \item \texttt{lambda:ListLayers}
                \item \texttt{lambda:ListFunctions}
            \end{itemize}
        \end{itemize}
        \item \textbf{Attached Managed Policies:}
        \begin{itemize}[itemsep=1pt, topsep=1pt, left=0pt]
            \item \textbf{S22\_AMP\_PolicyC:}
            \begin{itemize}[itemsep=1pt, topsep=1pt, left=0pt]
                \item \texttt{s3:ListBucket}
                \item \texttt{cloudformation:ListStacks}
            \end{itemize}
        \end{itemize}
    \end{itemize}
\end{itemize}

\end{multicols}
} 
\end{document}